%% file: mainthesis-bibtex-2025.tex
 \newcommand{\mydriver}{pdflatex} %Making a PDF directly using pdflatex.

\documentclass[12pt,\mydriver]{thesis}  %12pt is larger than 11pt
\usepackage[margin=1cm]{caption}
\usepackage{supertabular}
\usepackage{tabls}
\usepackage{titlesec}
   \titleformat{\chapter}
      {\normalfont\large}{Chapter \thechapter:}{1em}{}
\usepackage{tikz}
\usepackage{times}
\usepackage[USenglish]{babel}
\usepackage{subcaption}
\usepackage{graphicx}
\usepackage{cite}
\usepackage{lscape}
\usepackage{indentfirst}
\usepackage{latexsym}
\usepackage{multirow}
\usepackage{epstopdf}
\usepackage{wrapfig}
\usepackage{slashbox}
\usepackage{subeqn}
\usepackage{float}
\usepackage{caption}
\usepackage[onehalfspacing]{setspace} %used for footnote spacing
\usepackage{bibentry} %Using this to add capability for in-text full citations.
\nobibliography*
\usepackage{array}
\usepackage{needspace}
\usepackage{slashed}
\usepackage{amssymb}

\usepackage[colorlinks=true,urlcolor=black,linkcolor=blue,citecolor=blue]{hyperref}

\usepackage{tocloft,calc}
\renewcommand{\cftlottitlefont}{\hspace*{\fill}\large}
\renewcommand{\cftafterlottitle}{\hspace*{\fill}}
\renewcommand{\cftloftitlefont}{\hspace*{\fill}\large}
\renewcommand{\cftafterloftitle}{\hspace*{\fill}}
\renewcommand{\cftchapaftersnum}{:\ }
\renewcommand{\cftchappresnum}{\chaptername\space}
\makeatletter
\g@addto@macro\appendix{%
  \addtocontents{toc}{%
    \protect%
  }%
}
\renewcommand{\baselinestretch}{2}
\begin{document}
\pagestyle{empty}
\include{Abstract} %(must be first, required, non-numbered)
\include{Titlepage} %(must follow Abstract, required, non-numbered)
\include{Copyright} %(highly recommended, non-numbered)

%Pages from this point start at lower-case Roman number ii)
\pagestyle{plain} \pagenumbering{roman} \setcounter{page}{2}

% \iffalse
\iftrue
% \phantomsection %create the correct anchor for the bookmark
% \addcontentsline{toc}{chapter}{Preface or Foreword}
% \include{Preface}  %(if present, start at lower-case Roman number ii)

%\phantomsection %create the correct anchor for the bookmark
%\addcontentsline{toc}{chapter}{Foreword}
%\include{Foreword} %(if present, lower-case Roman)

\phantomsection %create the correct anchor for the bookmark
\addcontentsline{toc}{chapter}{Dedication}
\include{Dedication} %(if present, lower-case Roman)

\phantomsection %create the correct anchor for the bookmark
\addcontentsline{toc}{chapter}{Acknowledgements}
\include{Acknowledgements} %(if present, lower-case Roman)
    \cleardoublepage
\fi

\phantomsection %create the correct anchor for the bookmark
\vspace*{-2.5cm}
\addcontentsline{toc}{chapter}{Table of Contents}
    \renewcommand{\contentsname}{Table of Contents}
\renewcommand{\baselinestretch}{1}
\small\normalsize
\tableofcontents %(required, lower-case Roman)
\newpage

\phantomsection %create the correct anchor for the bookmark
\vspace*{-2.5cm}
\addcontentsline{toc}{chapter}{List of Tables}
    \renewcommand{\contentsname}{List of Tables}
\listoftables %(if present, lower-case Roman)
\newpage

\phantomsection %create the correct anchor for the bookmark
\vspace*{-2.5cm}
\addcontentsline{toc}{chapter}{List of Figures}
    \renewcommand{\contentsname}{List of Figures}
\listoffigures %(if present, lower-case Roman)
\newpage

% LIST OF ABBREVIATIONS
\phantomsection %create the correct anchor for the bookmark
\vspace*{-1.5in}
\addcontentsline{toc}{chapter}{List of Abbreviations}
\include{Abbreviations-supertabular}

\newpage
\setlength{\parskip}{0em}
\renewcommand{\baselinestretch}{2}
\small\normalsize

%Pages from this point start at Arabic numeral 1
\setcounter{page}{1}
\pagenumbering{arabic}
\include{Chapter1}
\include{Chapter2}
\include{Chapter3}
\include{Chapter4}

\include{Chapter5}

\include{Chapter6}
\titleformat{\chapter}
{\normalfont\large}{Appendix \thechapter:}{1em}{}

\renewcommand{\baselinestretch}{1}
\small\normalsize
\vspace*{-1in}
%\addcontentsline{toc}{chapter}{Bibliography}
\bibliographystyle{unsrt}
\bibliography{Ch1,Ch2,Ch3,Ch4,Ch5,Ch6} %replace "Galactic,Dottie" 
%with the file name(s) of your bib file(s)
%Use \cite when referencing bibtex files}

\end{document}

%% file: Abstract.tex
%Abstract Page

\hbox{\ } \vspace{.7in}
\renewcommand{\baselinestretch}{1}
\small \normalsize

\begin{center}
\large{{ABSTRACT}}

\vspace{3em}

\end{center}
\hspace{-.15in}
\begin{tabular}{ll}
Title of Dissertation:    & {\large  TOWARDS PRECISION-CONTROLLED }\\
&                     {\large  PARTONIC STRUCTURES} \\
&                     {\large  FROM FIRST PRINCIPLES} \\
\ \\
&                          {\large  Jinchen He} \\
&                           {\large Doctor of Philosophy, 2026} \\
\ \\
Dissertation Directed by: & {\large  Professor Xiangdong Ji} \\
&               {\large  Department of Physics } \\
\end{tabular}

\vspace{3em}

\renewcommand{\baselinestretch}{2}
\large \normalsize

The internal structure of hadrons is governed by nonperturbative Quantum Chromodynamics (QCD) and is being probed with increasing precision by modern high-energy experiments. This dissertation presents first-principles calculations of partonic observables using lattice QCD together with effective field theory methods, with controlled systematic uncertainties, advancing from collinear structure to transverse-momentum-dependent distributions (TMDs) that encode the three-dimensional partonic structure of hadrons.

Within the large momentum effective theory (LaMET) framework, this work presents state-of-the-art calculations of pion distribution amplitudes and systematic studies of nucleon parton distributions, with control of renormalization, excited-state contamination, Fourier-transform systematics, and power corrections. A Coulomb-gauge formulation of quasi-distributions is implemented to simplify ultraviolet structure by avoiding Wilson-line related linear divergences. Systematic effects associated with Gribov copies are quantified and found to be negligible at current statistical precision, supporting the effectiveness of this approach.

Building on these developments, this dissertation reports first-principles lattice determinations of key TMD quantities, including nucleon TMD parton distributions, the Collins-Soper kernel, the intrinsic soft function, and pion TMD observables. These results provide nonperturbative inputs directly relevant to global QCD analyses and to the precision hadron-structure program at current and future facilities, including the Electron-Ion Collider.

In parallel, this dissertation explores machine-learning acceleration of lattice gauge simulations through neural field transformations embedded in Hybrid Monte Carlo. In two-dimensional U(1) tests, the method significantly reduces autocorrelation and improves performance toward finer lattice spacing. Such gains support precision-controlled partonic calculations by enabling finer lattices and larger hadron boosts, thereby reducing LaMET power corrections and improving control of discretization effects. The dissertation also outlines emerging directions using AI agents and foundation models to advance lattice data-analysis workflows under physics-based constraints.

%% file: Titlepage.tex
%Titlepage

\thispagestyle{empty} \hbox{\ }
\vspace{.5in}

\renewcommand{\baselinestretch}{1}
\small\normalsize
\begin{center}

\large{{TOWARDS PRECISION-CONTROLLED \\
PARTONIC STRUCTURES FROM FIRST PRINCIPLES}}\\
\ \\
\ \\
\large{by} \\
\ \\
\large{Jinchen He}%Your full name as it appears in University records.
\ \\
\ \\
\ \\
\ \\
\normalsize
Dissertation submitted to the Faculty of the Graduate School of the \\
University of Maryland, College Park in partial fulfillment \\
of the requirements for the degree of \\
Doctor of Philosophy \\
2026
\end{center}

\vspace{7.5em}

\noindent Advisory Committee: \\
\hbox{\ }\hspace{.5in}Professor Xiangdong Ji, Chair/Advisor \\
\hbox{\ }\hspace{.5in}Professor Thomas Cohen \\
\hbox{\ }\hspace{.5in}Professor Anson Hook \\
\hbox{\ }\hspace{.5in}Professor Zackaria Chacko \\
\hbox{\ }\hspace{.5in}Professor Harry Dankowicz

%% file: Copyright.tex
%Copyright

\thispagestyle{empty}
\hbox{\ }

\vfill
\renewcommand{\baselinestretch}{1}
\small\normalsize

\begin{center}
\large{\copyright \hbox{ }Copyright by\\
Jinchen He  %Type your name as it appears in University records
\\
2026}
\end{center}

\vfill

\newpage

%% file: Dedication.tex
%Dedication
\vspace{.65in}

\renewcommand{\baselinestretch}{2}
\small\normalsize
\hbox{\ }

\vspace*{-.85in}

\begin{center}
\large{Dedication}
\end{center}

\begin{center}
For an intelligible universe.
\end{center}

%% file: Acknowledgements.tex
%Acknowledgments

\renewcommand{\baselinestretch}{2}

\small\normalsize
\hbox{\ }

\vspace*{-.85in}

\begin{center}
\large{Acknowledgments}
\end{center}

\vspace{1ex}

Over the past four years of my Ph.D. journey, I have been fortunate to meet many friends, classmates, mentors, and elders whose support, encouragement, and companionship have enabled me to complete this long and demanding process. What appears on these pages is not the result of individual effort alone. It is the outcome of sustained guidance, collaboration, and trust from a community that has shaped both my academic growth and my personal development.

I would first like to express my deepest gratitude to my advisor, Professor Xiangdong Ji. I have always admired your profound knowledge and deep physical insight. Your guidance has been invaluable to my training as a theoretical physicist. Many of your insightful ideas will continue to influence and guide my future research. The two years I spent at the University of Maryland laid the foundation for my research training. Jinghong Yang, as both a classmate and roommate, was the person with whom I exchanged ideas most frequently. Your sharp thinking and careful consideration often left me humbled, and I am grateful to have met such a thoughtful friend. I hope we will have many more opportunities to collaborate in the future. I also thank Jack Holligan, Chen Yang, and Haocheng Zhang; discussions with you have always been inspiring.

I am sincerely grateful to Dr. Yong Zhao for your meticulous and patient academic guidance and encouragement. Your precision and rigor in physics set an example that I continue to learn from. The two years at Argonne National Laboratory were years of rapid growth for me. I am thankful for the research environment and for the Argonne theory group: Ian, Ale, Anna, Harry, and Bob. I also thank Patrick, Chiara, Bryce, Noah, and Justin. In particular, I would like to thank Joey, Daniel, Josh, Jane, Arthur, Eric, and Bill. I am very glad to have met all of you; you made my time at Argonne both productive and memorable.

I would like to give special thanks to several senior colleagues who provided generous academic support. Yushan Su offered me substantial help and encouragement; your dedication to physics has always impressed me. I thank Rui Zhang, Yuxun Guo, and Xiang Gao for your guidance and support through collaboration and discussion. I also appreciate the pleasant collaborations with Fei Yao and Wei-Yang Liu. I am particularly grateful to Xiangyu Jiang for extensive assistance in coding and software-related matters.

I also acknowledge my collaborators at Brookhaven National Laboratory, Dr. Swagato Mukherjee, Dr. Peter Petreczky, and Dr. Dennis Bollweg, as well as James Osborn and Xiao-Yong Jin at Argonne. I would especially like to thank Andre and Jason at Berkeley Lab. You were my introduction to lattice QCD, and what I learned at CalLat continues to benefit my research to this day.

I am deeply grateful to Professor Wei Wang. It was under your guidance that I first became engaged with partonic physics and LaMET. I greatly respect your pragmatic and rigorous approach to research and teaching. I also thank the teachers and colleagues of the Lattice Parton Collaboration. In particular, I appreciate the guidance of Yi-Bo Yang, Jianhui Zhang, and Xiaonu Xiong. I am especially thankful to Qi-An Zhang, Jun Hua, Jun Zeng, Minhuan Chu, Zhifu Deng, Chang Yang, Fei Huang, and Hang Liu. I will always remember the time we spent working and discussing together, and I sincerely wish you all continued success.

I am indebted to my parents and family in China. Your unwavering support has been a necessary condition for me to complete this Ph.D. program. Studying abroad for many years and being far from home has often left me feeling that I have not accompanied you enough. I wish you health and longevity.

I thank my friends from afar: Boyue Li, Bocheng Zhang, Shenghang Xu, Le Zhang, and Weiqi Shi. Your conversations and companionship have been an essential part of my Ph.D. life.

I am especially grateful to my girlfriend, Yuwei Li. Thank you for accompanying me through the entirety of this Ph.D. program. A relationship sustained across long distances is rare and precious, and I value it deeply.

There are many more people who have contributed to my journey than can be fully listed here. Among the forces that have brought me to this point, some are visible, like the support explicitly acknowledged above. Others are less visible: the expectations of those who know me, the trajectory I have followed within academia, and the curiosity and desire for understanding that drew me to theoretical physics in the first place. I have always regarded the opportunity to pursue theoretical physics as a luxury. I am grateful for the chance to find, within an otherwise ordinary and at times monotonous life, something that reaches beyond it. I am grateful for the opportunity to seek, within a finite lifetime, laws that aspire to permanence. Above all, I am grateful for the opportunity to believe in an intelligible universe.

%% file: Abbreviations-supertabular.tex
%List of Abbreviations

\renewcommand{\baselinestretch}{1}
\small\normalsize
\hbox{\ }

\vspace{-.6in}

\begin{center}
\large{List of Abbreviations}
\end{center}

\vspace{1em}

\begin{supertabular}{ll}
AI & Artificial Intelligence \\
AION-1 & Astronomical Omni-modal Network \\
AMA & All-Mode Averaging \\
CG & Coulomb Gauge \\
CNN & Convolutional Neural Network \\
CSS & Collins--Soper--Sterman \\
DA & Distribution Amplitude \\
DCSB & Dynamical Chiral Symmetry Breaking \\
DGLAP & Dokshitzer--Gribov--Lipatov--Altarelli--Parisi \\
DIS & Deep-Inelastic Scattering \\
DSE & Dyson--Schwinger Equations \\
DVCS & Deeply Virtual Compton Scattering \\
DVMP & Deeply Virtual Meson Production \\
DY & Drell--Yan \\
EIC & Electron--Ion Collider \\
ERBL & Efremov--Radyushkin--Brodsky--Lepage \\
FH & Feynman--Hellmann \\
GEVP & Generalized Eigenvalue Problem \\
GNoME & Graph Networks for Materials Exploration \\
GPD & Generalized Parton Distributions \\
GI & Gauge-Invariant \\
HERA & Hadron-Elektron-Ringanlage \\
HISQ & Highly Improved Staggered Quark \\
HMC & Hybrid Monte Carlo \\
HYP & Hypercubic \\
LaMET & Large Momentum Effective Theory \\
LHC & Large Hadron Collider \\
LGT & Lattice Gauge Theory \\
LL & Leading Logarithmic \\
LO & Leading Order \\
LPC & Lattice Parton Collaboration \\
LRR & Leading Renormalon Resummation \\
MCMC & Markov Chain Monte Carlo \\
MD & Molecular Dynamics \\
ML & Machine Learning \\
NLL & Next-to-Leading Logarithmic \\
NLO & Next-to-Leading Order \\
NNLO & Next-to-Next-to-Leading Order \\
NTHMC & Neural Network Transformed Hybrid Monte Carlo \\
OPE & Operator Product Expansion \\
PDF & Parton Distribution Function \\
QCD & Quantum Chromodynamics \\
RG & Renormalization Group \\
RGR & Renormalization-Group Resummation \\
RI/MOM & Regularization-Independent Momentum Subtraction \\
SCET & Soft-Collinear Effective Theory \\
SIDIS & Semi-Inclusive Deep-Inelastic Scattering \\
SLAC & Stanford Linear Accelerator Center \\
SNR & Signal-to-Noise Ratio \\
TMD & Transverse-Momentum-Dependent \\
TMDs & Transverse-Momentum-Dependent Distributions \\
TMDPDF & Transverse-Momentum-Dependent Parton Distribution Function \\
TMDWF & Transverse-Momentum-Dependent Wave Function \\
TR & Threshold Resummation \\
UV & Ultraviolet \\
\end{supertabular}

%% file: Chapter1.tex
%Chapter 1

\renewcommand{\thechapter}{1}

\vspace*{-1.35in}

\chapter{Introduction}
\label{ch:intro}

\vspace*{-.5in}

From the earliest notions of atoms to the modern picture of quarks and gluons, the history of physics has been guided by reductionism: the idea that complex phenomena can be understood by identifying more elementary constituents and the interactions among them. The path from atomic structure to nuclei, from nuclei to nucleons, and from nucleons to quarks has repeatedly reshaped our view of matter, each successive layer unveiling a new effective scale where smaller degrees of freedom organize the behavior of the whole. In the visible universe, the smallest known structural description of ordinary matter is at the parton level, where quarks and gluons form the internal degrees of freedom of hadrons.

\begin{figure}[!th]
    \begin{center}
    \includegraphics[width=0.6\textwidth]{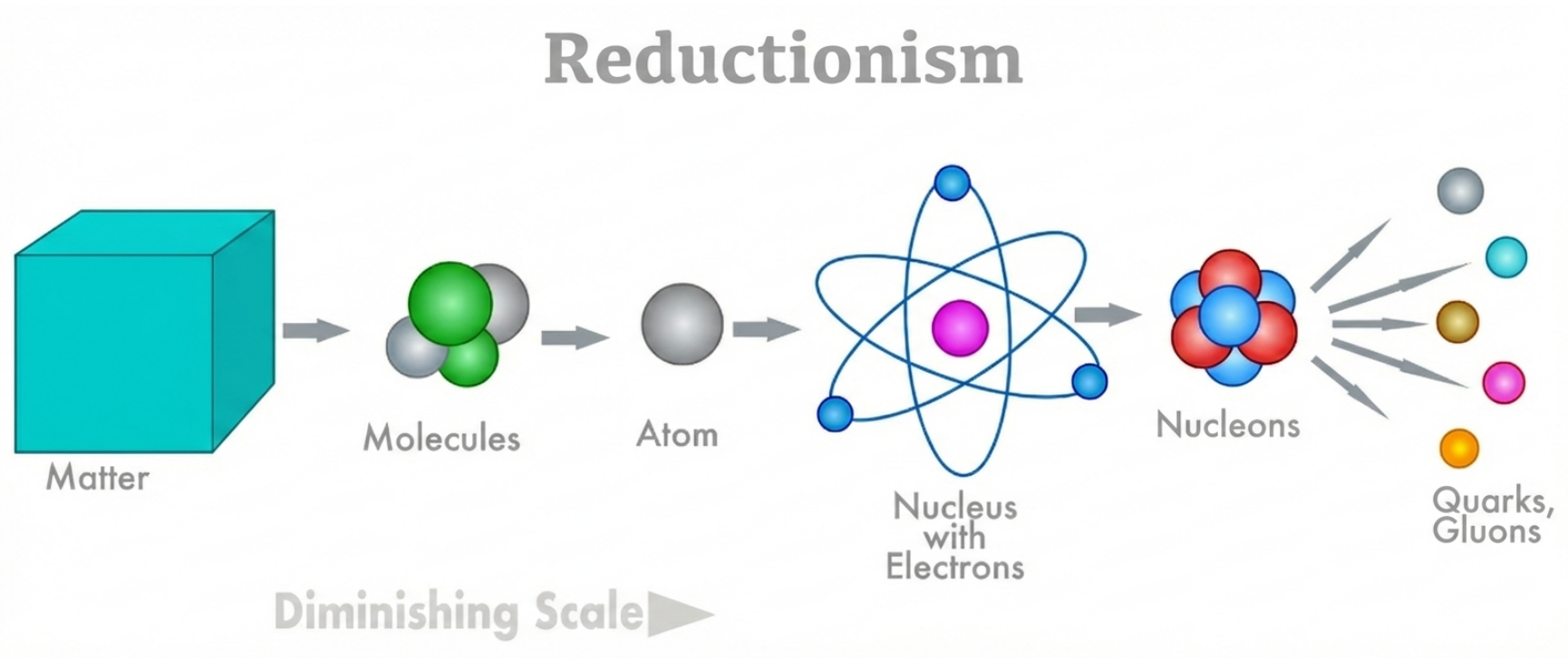}
    \caption[Reductionism]{The concept of reductionism in physics, showing how complex systems can be understood through their simpler constituents.}\label{fig:reductionism}
    \end{center}
\end{figure}

At the same time, reductionism has never been the sole organizing principle. The celebrated essay ``More is Different''~\cite{Anderson:1972pca} argued that new organizing principles can emerge at higher levels of complexity, challenging the idea that a full microscopic description automatically yields macroscopic understanding.
This critique remains influential, especially in strongly coupled systems where emergent behavior is not easily reduced to a few simple constituents.

In the present era, however, the balance between these perspectives is being reshaped by the rapid growth of computing power and algorithms. High-performance computing and simulation now make it possible to solve problems that were previously out of reach, enabling direct numerical access to nonperturbative regimes of quantum field theory. This is particularly true in quantum chromodynamics (QCD), where analytic control is limited and where the strong coupling at low energies renders perturbation theory ineffective. In this sense, computing is not merely a tool but an enabling paradigm: it reopens the possibility that ``more'' can be approached directly, with complex many-body behavior computed from first principles.

These developments sit within the broader framework of the Standard Model, which organizes known fundamental particles and interactions into a unified field-theoretic description. The Standard Model is a renormalizable quantum field theory based on the gauge group 
$SU(3)_c \times SU(2)_L \times U(1)_Y$, describing the strong and electroweak interactions, while gravity is not incorporated within this framework. In a broader physics context one often speaks of four fundamental interactions: electromagnetic, weak, strong, and gravitational. The non-Abelian gauge structure of QCD explains asymptotic freedom, which underlies the partonic picture and the emergence of scaling violations in high-energy scattering~\cite{Gross:1973id,Politzer:1973fx}. The same framework provides the theoretical foundation for factorization in hard processes and for the operator-based definitions of parton distributions~\cite{Collins:2011zzd}.

Experimentally, our knowledge of structure at ever smaller distance scales has been enabled by high-energy facilities and precision detectors. Deep-inelastic scattering at SLAC revealed Bjorken scaling and the parton model~\cite{Bjorken:1968dy,Feynman:1969ej}. Later, colliders such as HERA, the Tevatron, and the Large Hadron Collider (LHC) pushed these probes to higher energies, while dedicated nuclear facilities such as Jefferson Lab mapped the valence region with high precision. The planned Electron--Ion Collider (EIC) will further extend this program by enabling three-dimensional imaging of hadrons and precision studies of parton dynamics in nucleons and nuclei~\cite{Accardi:2012qut,AbdulKhalek:2021gbh}.

Cosmological observations further sharpen the motivation. While dark matter and dark energy dominate the cosmic energy budget, their microscopic nature remains unknown. Within the framework of the Standard Model, the dynamics of the visible sector are predominantly governed by QCD and the electroweak theory. Within this visible sector, the partonic structure of hadrons is the most elementary known organization of matter, and a complete description of partons is essential for connecting fundamental theory with experimental measurements in nuclear and particle physics.

Despite decades of progress, the strong interaction still presents fundamental challenges~\cite{Gross:2022hyw}. Confinement, the emergence of hadrons as relativistic bound states, and the dynamical generation of mass and structure remain only partially understood. Describing a relativistic bound state from first principles is one of the deepest problems in quantum field theory. The internal partonic structure of hadrons encodes these nonperturbative dynamics, yet it is precisely this structure that is most difficult to compute analytically.

This dissertation is motivated by these intertwined themes. It aims to advance first-principles calculations of partonic structure using lattice QCD and modern effective field theory frameworks. The core objective is to connect QCD at the quark--gluon level to measurable parton observables, with controlled systematics and clear theoretical interpretation. 

\section{List of Research Contributions}

The project on lattice calculations of the nucleon axial charge in Chapter~\ref{ch:lattice} is based on the following research contribution:

[1] J.~He, D.~A.~Brantley, C.~C.~Chang, I.~Chernyshev, E.~Berkowitz, D.~Howarth, C.~K{\"o}rber, A.~S.~Meyer, H.~Monge-Camacho and E.~Rinaldi, \textit{et al.}
``Detailed analysis of excited-state systematics in a lattice QCD calculation of gA,''
Phys. Rev. C \textbf{105}, no.6, 065203 (2022)
[arXiv:2104.05226 [hep-lat]].

The project on systematic uncertainties from Gribov copies in lattice calculation of parton distributions in the Coulomb Gauge in Chapter~\ref{ch:systematics} is based on the following research contribution:

[2] X.~Gao, J.~He, R.~Zhang and Y.~Zhao,
``Systematic Uncertainties from Gribov Copies in Lattice Calculation of Parton Distributions in the Coulomb Gauge,''
Chin. Phys. Lett. \textbf{41}, no.12, 121201 (2024)
[arXiv:2408.05910 [hep-lat]].

The projects on lattice calculations of partonic physics from lattice QCD described in Chapter~\ref{ch:systematics} are based on the following research contributions:

[3] J.~Hua \textit{et al.} [Lattice Parton],
``Pion and Kaon Distribution Amplitudes from Lattice QCD,''
Phys. Rev. Lett. \textbf{129}, no.13, 132001 (2022)
[arXiv:2201.09173 [hep-lat]].

[4] J.~C.~He \textit{et al.} [Lattice Parton Collaboration (LPC)],
``Unpolarized transverse momentum dependent parton distributions of the nucleon from lattice QCD,''
Phys. Rev. D \textbf{109}, no.11, 114513 (2024)
[arXiv:2211.02340 [hep-lat]].

[5] M.~H.~Chu \textit{et al.} [Lattice Parton],
``Transverse-momentum-dependent wave functions of the pion from lattice QCD,''
Phys. Rev. D \textbf{109}, no.9, L091503 (2024)
[arXiv:2302.09961 [hep-lat]].

[6] M.~H.~Chu \textit{et al.} [Lattice Parton (LPC)],
``Lattice calculation of the intrinsic soft function and the Collins-Soper kernel,''
JHEP \textbf{08}, 172 (2023)
[arXiv:2306.06488 [hep-lat]].

[7] D.~Bollweg, X.~Gao, J.~He, S.~Mukherjee and Y.~Zhao,
``Transverse-momentum-dependent pion structures from lattice QCD: Collins-Soper kernel, soft factor, TMDWF, and TMDPDF,''
Phys. Rev. D \textbf{112}, no.3, 034501 (2025)
[arXiv:2504.04625 [hep-lat]].

[8] X.~Gao, J.~He, J.~Lin, S.~Mukherjee, P.~Petreczky, R.~Zhang and Y.~Zhao,
``Nucleon Parton Distribution Functions from Boosted Correlations in the Coulomb gauge,''
[arXiv:2602.11283 [hep-lat]].

The projects on applications of machine learning in lattice gauge theory described in Chapter~\ref{ch:ml} are based on the following research contributions:

[9] M.~Zhu, M.~Tian, X.~Yang, T.~Zhou, L.~Yuan, P.~Zhu, E.~Chertkov, S.~Liu, Y.~Du and Z.~Ji, \textit{et al.}
``Probing the Critical Point (CritPt) of AI Reasoning: a Frontier Physics Research Benchmark,''
[arXiv:2509.26574 [cs.AI]].

[10] J.~He, X.~Y.~Jin, J.~C.~Osborn and Y.~Zhao,
``Neural Field Transformations for Hybrid Monte Carlo: Architectural Design and Scaling,''
[arXiv:2511.02018 [hep-lat]].

\section{Outline}

The remainder of this dissertation is organized as follows. Chapter~\ref{ch:partonic} reviews the phenomenology of partonic physics in hadrons, with a focus on collinear distributions and transverse-momentum-dependent distributions (TMDs). Chapter~\ref{ch:lattice} introduces the lattice QCD framework and presents a benchmark calculation of the nucleon axial charge. Chapter~\ref{ch:systematics} describes systematic calculations of partonic observables from lattice QCD, including pion distribution amplitudes (DAs), nucleon parton distribution functions (PDFs), and TMDs for both pion and nucleon. Chapter~\ref{ch:ml} explores applications of machine learning in lattice gauge theory, including applications in configuration generation, data analysis, AI agent and foundation model. Finally, Chapter~\ref{ch:summary} summarizes the thesis and discusses future directions.

%% file: Chapter2.tex
%Chapter 2

\renewcommand{\thechapter}{2}

\vspace*{-1.35in}

\chapter{Partonic Physics in Hadrons}
\label{ch:partonic}

\vspace*{-.5in}

Understanding the internal structure of hadrons in terms of quarks and gluons is a central goal of QCD. Over the past five decades a coherent theoretical framework has been developed to describe how partons behave inside hadrons and how they can be probed experimentally. This chapter provides a brief overview of partonic physics, beginning with the historical foundations of the parton model and QCD factorization. We then discuss collinear parton structures, including PDFs and DAs, followed by TMD structures and their evolution. We then briefly mention the generalized parton distributions (GPDs) and spatial tomography, emphasizing the experimental and phenomenological program that extracts these objects from data and the role of future facilities in completing a three-dimensional picture of hadrons~\cite{Angeles-Martinez:2015sea,Accardi:2012qut,AbdulKhalek:2021gbh}.

\section{From Scaling to QCD: The Parton Framework}

\begin{figure}[!th]
    \begin{center}
    \includegraphics[width=0.6\textwidth]{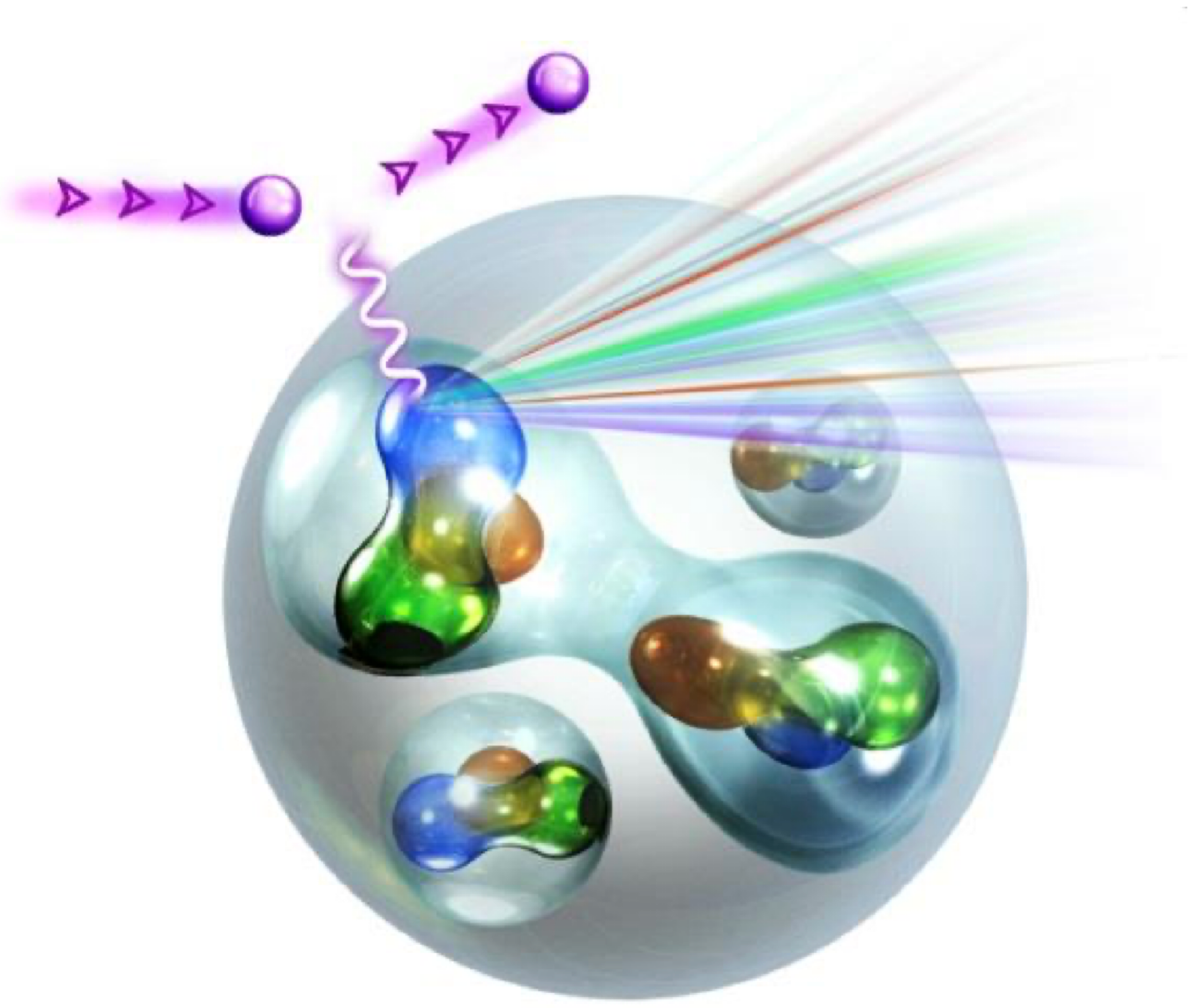}
    \caption[Deep-inelastic scattering]{Diagramatic illustration of the deep-inelastic scattering process. Credit to Ref.~\cite{Gaskell:2015dis}}\label{fig:dis}
    \end{center}
\end{figure}

In the late 1960s, the electron-nucleon deep-inelastic scattering (DIS) experiments at SLAC revealed approximate Bjorken scaling, indicating scattering from point-like constituents inside the proton, as illustrated schematically in Fig.~\ref{fig:dis}. The parton model proposed by Feynman provided a simple interpretation: a fast-moving hadron contains quasi-free partons carrying fractions of its momentum~\cite{Feynman:1969ej}. Bjorken and Paschos showed that such a picture leads to scale-invariant structure functions and sum rules, relating the DIS structure function $F_2(x)$ to momentum fraction distributions of charged partons~\cite{Bjorken:1968dy,Bjorken:1969ja}.

Embedding the parton model into a quantum field theory required asymptotic freedom. The discovery of asymptotic freedom in non-Abelian gauge theories established QCD as the theory of strong interactions and explained the observed scaling violations~\cite{Gross:1973id,Politzer:1973fx}. Asymptotic freedom implies that the strong coupling becomes weak at short distances, with the running coupling (at one-loop accuracy) governed by the renormalization group equation,
\begin{equation}
    \alpha^{(1)}_s(Q^2) = \frac{1}{\beta_0 \ln(Q^2/\Lambda_{\rm QCD}^2)}~,
\end{equation}
so that radiative corrections at high momentum transfer generate logarithmic contributions of the form $\alpha_s \ln(Q^2/\mu^2)$, such terms appear as powers $(\alpha_s \ln Q^2)^n$ and must be resummed. These logarithmic contributions arise from collinear parton splitting processes such as $q \to qg$, $g \to gg$, and $g \to q\bar q$, which redistribute longitudinal momentum among partons and induce a scale dependence of parton distribution functions. 

The QCD evolution of PDFs is described by the Dokshitzer–Gribov–Lipatov–Altarelli–Parisi (DGLAP) equations, derived in the 1970s, which resum these leading logarithmic contributions and account for the logarithmic scaling violations observed in deep-inelastic structure functions~\cite{Gribov:1972ri,Dokshitzer:1977sg,Altarelli:1977zs}.

Another crucial development was Wilson's operator product expansion (OPE) and the concept of factorization~\cite{Wilson:1969zs}. The OPE provides a separation of short-distance effects from long-distance physics in the product of currents, enabling the formal proof that inclusive DIS cross-sections factorize into perturbative coefficients and universal PDFs~\cite{Collins:2011zzd}. This foundation underlies the universality of PDFs across processes such as DIS and Drell--Yan~\cite{Drell:1970yt}, up to higher-twist corrections. In exclusive processes, the analogous collinear object are the DAs, formalized independently by Brodsky--Lepage~\cite{Lepage:1979zb,Lepage:1980fj} and by Efremov--Radyushkin~\cite{Efremov:1978rn,Efremov:1979qk}, leading to the ERBL evolution equations. Together with the DGLAP description of scaling violations in DIS, this framework laid the foundation for modern QCD phenomenology, where global analyses combine a wide range of experimental data to extract universal parton distributions through parameterized fits.

\section{Light-front Quantization}

The experimental discovery of approximate Bjorken scaling and its interpretation within the parton model naturally emphasize the Bjorken limit, defined by $Q^2 \to \infty$ with the Bjorken variable $x_B = \frac{Q^2}{2P\cdot q}$ held fixed. In this regime the virtual photon probes distance scales of order $1/Q$, and the hadronic tensor can be written as the Fourier transform of a current--current correlator~\cite{Collins:2011zzd},
\begin{equation}
    W_{\mu\nu}(P,q)
    = \frac{1}{4\pi}\int d^4\xi\, e^{iq\cdot \xi}
    \langle P | [J_\mu(\xi),J_\nu(0)] | P \rangle ~,
\end{equation}
where $J_\mu(x)=\sum_f e_f\,\bar\psi_f(x)\gamma_\mu\psi_f(x)$ is the electromagnetic current. The state $|P\rangle$ denotes a single-hadron momentum eigenstate with four-momentum $P^\mu$, normalized relativistically as
\begin{equation}
    \langle P'|P\rangle
    =
    2P^0 (2\pi)^3 \delta^{(3)}(\vec{P}-\vec{P}')~.
\end{equation} 
For large $Q$, the rapidly oscillating phase factor $e^{iq\cdot \xi}$ suppresses contributions from large space--time separations, so that the integral is dominated by short distances $|\xi|\sim 1/Q$. More specifically, the current product becomes most singular as the invariant separation approaches the light cone, $\xi^2 \to 0$, implying that the leading contributions arise from space--time intervals near light-like directions.

The dominance of light-like separations admits a natural interpretation within the parton picture. In the Bjorken limit, the hard probe couples primarily to partons carrying a finite fraction of the hadron momentum and moving approximately collinearly with the parent hadron. These partons are nearly on shell compared with the hard scale $Q^2$, and their propagation between the two currents is therefore effectively light-like. Consequently, the dominant contributions to the current--current correlator originate from correlations of quark and gluon fields separated along the light cone. This provides a direct bridge between the phenomenology of DIS and light-cone correlation functions of quark and gluon fields. Indeed, PDFs and DAs are most naturally defined as hadronic matrix elements of nonlocal operators separated along a light-like direction.

From a Hamiltonian perspective, these considerations motivate quantization directly on a light-like hypersurface, following Dirac's front-form dynamics~\cite{Burkardt:1995ct,Brodsky:1997de,Miller:2000kv}. In light-front quantization one introduces light-cone coordinates
\begin{equation}
    x^\pm = \frac{1}{\sqrt{2}}(x^0 \pm x^3)~, \qquad 
    \vec{x}_\perp = (x^1,x^2)~,
\end{equation}
and chooses $x^+$ as the evolution parameter, so that fields are quantized on the light-front hypersurface $x^+=0$. The generator of translations in $x^+$,
\begin{equation}
    P^- = \frac{1}{\sqrt{2}}(P^0 - P^3),
\end{equation}
plays the role of the Hamiltonian.

Dirac classified the generators of the Poincar\'e algebra into \emph{kinematical} and \emph{dynamical} ones according to whether they preserve the quantization hypersurface. Kinematical generators leave the surface $x^+=0$ invariant and therefore do not involve the Hamiltonian in their realization, while dynamical generators transform this hypersurface into another one and consequently depend on the dynamics of the theory. In the front form the kinematical generators include the longitudinal momentum $P^+$, the transverse momenta $P^\perp$, the transverse rotations $J^{12}$, the longitudinal boost $J^{+-}$, and the transverse boosts $J^{+i}$ ($i=1,2$). The remaining generators, $P^-,\, J^{-i}\,(i=1,2)$, are dynamical.

This distribution differs from the conventional equal-time quantization, where fields are quantized on the hypersurface $x^0=0$. In that case the Hamiltonian $P^0$ and the Lorentz boosts $J^{0i}$ are dynamical, while the spatial translations $P^i$ and spatial rotations $J^{ij}$ are kinematical. Consequently, the front form reduces the number of dynamical generators from four to three. This enlargement of the kinematical subgroup greatly simplifies the description of relativistic bound states and boosts along the longitudinal direction.

Another distinctive feature of this framework is that the longitudinal momentum $p^+ = (p^0 + p^3)/\sqrt{2}$ is positive for physical particles, which severely restricts vacuum fluctuations. As a consequence, the light-front vacuum in many theories is considerably simpler than the equal-time vacuum~\cite{Burkardt:1995ct}, and particle number changing processes are encoded directly in the structure of the light-front Hamiltonian rather than in vacuum condensates.

Within light-front quantization, a hadronic state admits a Fock space expansion at fixed light-front time~\cite{Brodsky:1997de},
\begin{equation}
    | P \rangle = \sum_n \int [\mathrm{d}x_i\, \mathrm{d}^2\vec{k}_{\perp i}] \,
    \psi_n(x_i,\vec{k}_{\perp i},\lambda_i) \,
    | n; x_i P^+, \vec{k}_{\perp i}, \lambda_i \rangle ~,
\end{equation}
where $n$ is the number of partons in the Fock component, $\vec{k}_{\perp i}$ are the transverse momenta of the partons, and $\lambda_i$ are the light-front helicities, $x_i = p_i^+/P^+$ are longitudinal momentum fractions satisfying $\sum_i x_i = 1$, and $\psi_n$ are light-front wave functions. These wave functions provide a complete set of amplitudes characterizing hadronic structure in the light-front framework. Because longitudinal boosts act multiplicatively on all $p_i^+$, the momentum fractions $x_i$ remain invariant, and the light-front wave functions furnish a frame-independent generalization of the nonrelativistic wave-function description of bound states~\cite{Burkardt:1995ct}. The intuitive quantum-mechanical picture of a bound state as a superposition of multi-particle configurations thus carries over to relativistic quantum field theory, with the light-front wave functions encoding the probability amplitudes for finding partonic constituents with specified longitudinal momentum fractions and transverse momenta.

This perspective gives a transparent amplitude interpretation of both PDFs and DAs. PDFs can be expressed as diagonal matrix elements of nonlocal light-cone operators and are related to squared light-front wave functions summed over spectator configurations. At leading twist, they correspond to probability densities in longitudinal momentum fraction, modulo the necessary gauge links ensuring color gauge invariance. DAs arise in exclusive processes as projections of light-front wave functions onto states with vanishing transverse separation at leading power. In particular, the leading-twist meson DA is obtained from the valence Fock component by integrating over transverse momenta up to the factorization scale, thereby encoding the longitudinal momentum distribution of the minimal quark--antiquark configuration. The evolution equations governing PDFs and DAs can then be understood as renormalization-group equations for these light-cone operators, reflecting the scale dependence of the underlying light-front amplitudes.

Despite these appealing features, light-front quantization does not eliminate the dynamical complexity of relativistic bound states. The simplicity of the vacuum is accompanied by subtleties associated with constrained degrees of freedom and zero modes, corresponding to field configurations with vanishing longitudinal momentum $p^+ = 0$~\cite{Ji:2021znw}. Such modes can contribute to matrix elements and are closely related to issues of spontaneous symmetry breaking, gauge invariance, and the realization of vacuum structure. In modern language, light-front quantization may be viewed as an effective description in which small-$k^+$ modes are integrated out, with their effects absorbed into renormalization factors~\cite{Ji:2021znw}. In practice, zero modes and rapidity divergences complicate both analytical treatments and numerical approaches based on truncated Fock spaces. Moreover, solving the light-front Hamiltonian eigenvalue problem remains a nonperturbative challenge equivalent in difficulty to solving QCD itself. In this sense, light-front quantization reformulates rather than resolves the fundamental problem of describing relativistic bound states.

Nevertheless, the light-front framework provides a unifying language that connects the parton model intuition emerging from the Bjorken limit with the operator definitions of PDFs and DAs in QCD. It clarifies why light-cone correlations dominate high-energy processes, furnishes a boost-invariant description of hadronic structure, and supplies an amplitude-level interpretation of the universal distributions that enter factorization theorems. In the following sections, we make these connections explicit by introducing the operator definitions and evolution properties of PDFs and DAs in QCD.

\section{Collinear Distributions}

QCD factorization separates short-distance perturbative dynamics from long-distance nonperturbative structure in high-energy processes. In processes characterized by a large momentum transfer $Q \gg \Lambda_{\rm QCD}$, the cross-section can be written schematically as a convolution of perturbatively calculable hard-scattering kernels with universal hadronic matrix elements. In the collinear framework, transverse momenta of partons are integrated out up to the factorization scale $\mu$, and the dominant nonperturbative objects are PDFs for inclusive reactions and DAs for exclusive reactions. Their operator definitions make explicit their light-cone structure and ensure gauge invariance through appropriate Wilson lines.

For a quark of flavor $q$ in a hadron $H$ with momentum $P^\mu$, the leading-twist collinear PDF is defined as the forward matrix element of a gauge-invariant nonlocal light-cone operator,
\begin{equation}
    f_{q/H,\Gamma}(x;\mu)
    =
    \int \frac{d\lambda}{2\pi}\,
    e^{-i\lambda x}
    \,
    \langle P S|
    \bar\psi_q(\xi^-)\,
    W(\xi^-, 0)\,
    \Gamma\,
    \psi_q(0)
    |P S\rangle
    \Big|_{\overline{\rm MS},\,\mu},
\end{equation}
where $|PS\rangle$ denotes a proton plane-wave state with momentum 
$P^\mu = (P^t,0,0,P^z)$ and spin $S$. The Wilson line 
$W(\xi^-,0)$ runs along the light-like direction to ensure color gauge invariance. 
Light-cone coordinates are defined as $\xi^\pm = (t \pm z)/\sqrt{2}$, 
and the variable $\lambda = P^+ \xi^-$ renders the Fourier transform dimensionless. 
The distribution is renormalized in the $\overline{\rm MS}$ scheme at scale $\mu$.

When spin degrees of freedom are considered, the cross-section depends on the polarization of the parton relative to the nucleon spin, leading to different types of spin-dependent PDFs. 
At leading twist, three independent collinear quark PDFs appear. 
For unpolarized, helicity, and transversity channels, the Dirac structures are chosen as $\Gamma = \gamma^+$, $\gamma^+\gamma^5$, and $i\sigma^{+\perp}\gamma^5$, respectively. 
These correspond to the number density, longitudinal spin density, and transverse spin density of quarks in a longitudinally or transversely polarized nucleon. 
Analogous gauge-invariant definitions apply to gluon distributions, constructed from light-cone correlators of the gluon field strength tensor.

In inclusive deep-inelastic scattering (DIS), $e H \to e X$, the cross-section at large momentum transfer $Q^2$ admits a leading-twist factorization into perturbatively calculable hard-scattering coefficients and universal PDFs~\cite{Collins:1989gx}:
\begin{equation}
    \sigma_{eH}(x,Q^2)
    =
    \sum_a
    \int_x^1 d\xi\,
    f_{a/H}(\xi;\mu_F)\,
    \hat{\sigma}_{ea}\!\left(\frac{x}{\xi},Q^2;\mu_F,\mu_R\right) ~,
\end{equation}
where $f_{a/H}(\xi;\mu_F)$ denotes the PDF for a parton of type $a$ carrying longitudinal momentum fraction $\xi$ of the hadron $H$ at factorization scale $\mu_F$. The quantity $\hat{\sigma}_{ea}$ represents the perturbatively calculable partonic cross section for electron--parton scattering, evaluated at renormalization scale $\mu_R$. 

The factorization scale $\mu_F$ separates long-distance, nonperturbative physics encoded in the PDFs from short-distance dynamics contained in the hard-scattering kernel, while $\mu_R$ governs the running of the strong coupling in the perturbative expansion. Although these scales are, in principle, independent, a common practical choice is $\mu_F = \mu_R = \mu$. As mentioned above, due to the running of the strong coupling and the presence of collinear divergences, the PDFs acquire a scale dependence governed by the DGLAP evolution equation,
\begin{equation}
    \frac{\partial f(x; \mu)}{\partial \ln \mu}=\int_x^1 \frac{d \nu}{|\nu|} \mathcal{P}(\nu; \mu) f(x / \nu; \mu) ~,
\end{equation}
where $\mathcal{P}(\nu; \mu)$ is the DGLAP kernel, also known as the splitting function, that encodes the probability of a parton splitting into two partons with momentum fractions $\nu$ and $1-\nu$. The DGLAP kernel has been calculated to 5-loop in perturbation theory~\cite{Herzog:2017ohr}.

A crucial consequence of QCD factorization is the universality of PDFs: the same distributions extracted from DIS appear in other hard processes involving the same hadron. For example, in Drell--Yan lepton-pair production~\cite{ATLAS:2016gic}, electroweak boson production~\cite{ATLAS:2021qnl}, and inclusive jet production~\cite{CMS:2016lna}, the cross sections factorize into convolutions of the same PDFs with process-dependent hard-scattering kernels. This universality enables global analyses in which a wide range of experimental data sets are combined to determine PDFs with controlled uncertainties, which can then be used to make precision predictions for other reactions.

Over the past decade, global analyses of proton PDFs have entered a high-precision era, largely driven by the final combined HERA I+II inclusive DIS measurements and by the wealth of differential data from LHC Runs I and II. Modern global fits are now performed consistently at next-to-next-to-leading order (NNLO) in QCD for all processes that play a significant role in constraining PDFs, often supplemented by electroweak corrections and refined treatments of correlated experimental systematics. These developments have reduced PDF uncertainties in phenomenologically crucial regions of Bjorken-$x$, while simultaneously exposing nontrivial tensions among datasets and methodological differences between fitting groups.

Several recent global PDF determinations, including CT18~\cite{Hou:2019efy}, MSHT20~\cite{Bailey:2020ooq}, and NNPDF4.0~\cite{NNPDF:2021njg}, play a central role in current LHC phenomenology. These analyses, together with related global efforts, provide the main inputs to the PDF4LHC21 combination~\cite{PDF4LHCWorkingGroup:2022cjn}, which establishes a common framework for PDF usage and uncertainty estimation in LHC studies. Although these analyses share a common experimental foundation---including HERA inclusive and heavy-flavor DIS, fixed-target Drell--Yan data, Tevatron measurements, and an extensive set of LHC observables such as inclusive jets, vector-boson production, and top-quark pair production---their results exhibit meaningful differences in specific flavor sectors and kinematic regions.

\begin{figure}[t]
    \centering
    \includegraphics[width=0.8\textwidth]{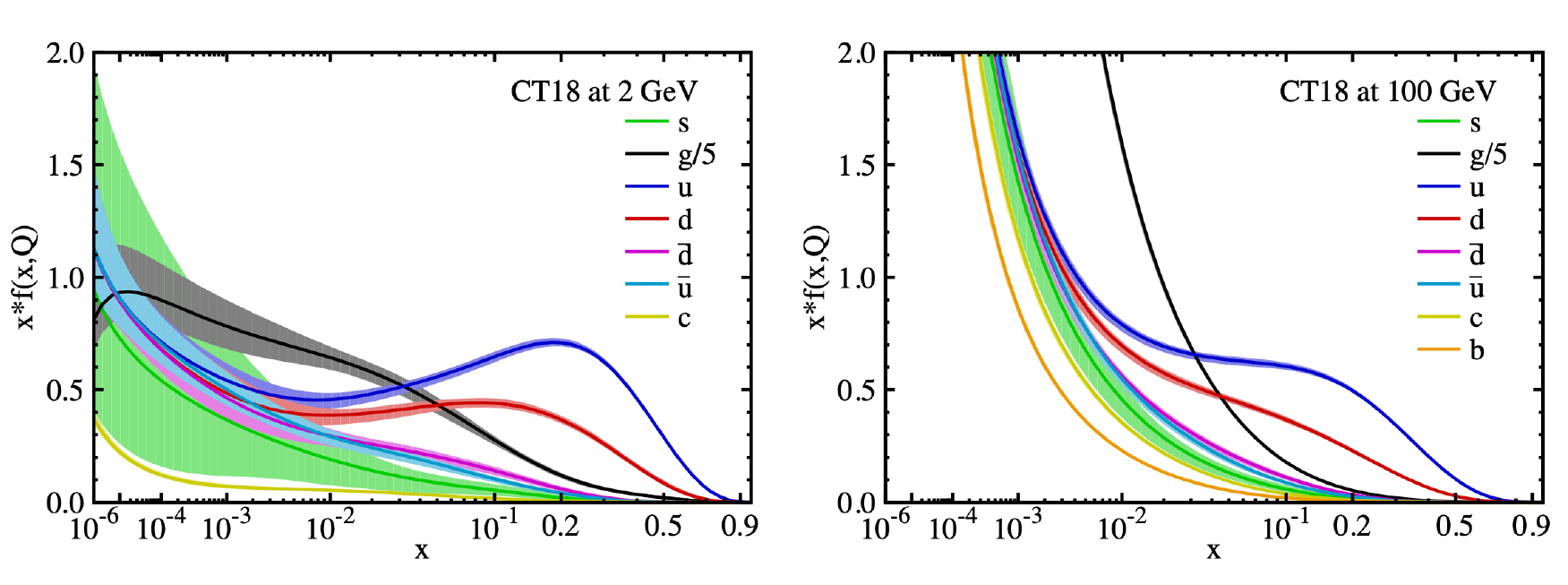}
    \caption[The CT18 NNLO PDFs]{The CT18 PDFs at Q=2 GeV (left) and Q=100 GeV(right)~\cite{Hou:2019efy}.}
    \label{fig:CT18_PDFs}
\end{figure}

\begin{figure}[t]
    \centering
    \includegraphics[width=0.8\textwidth]{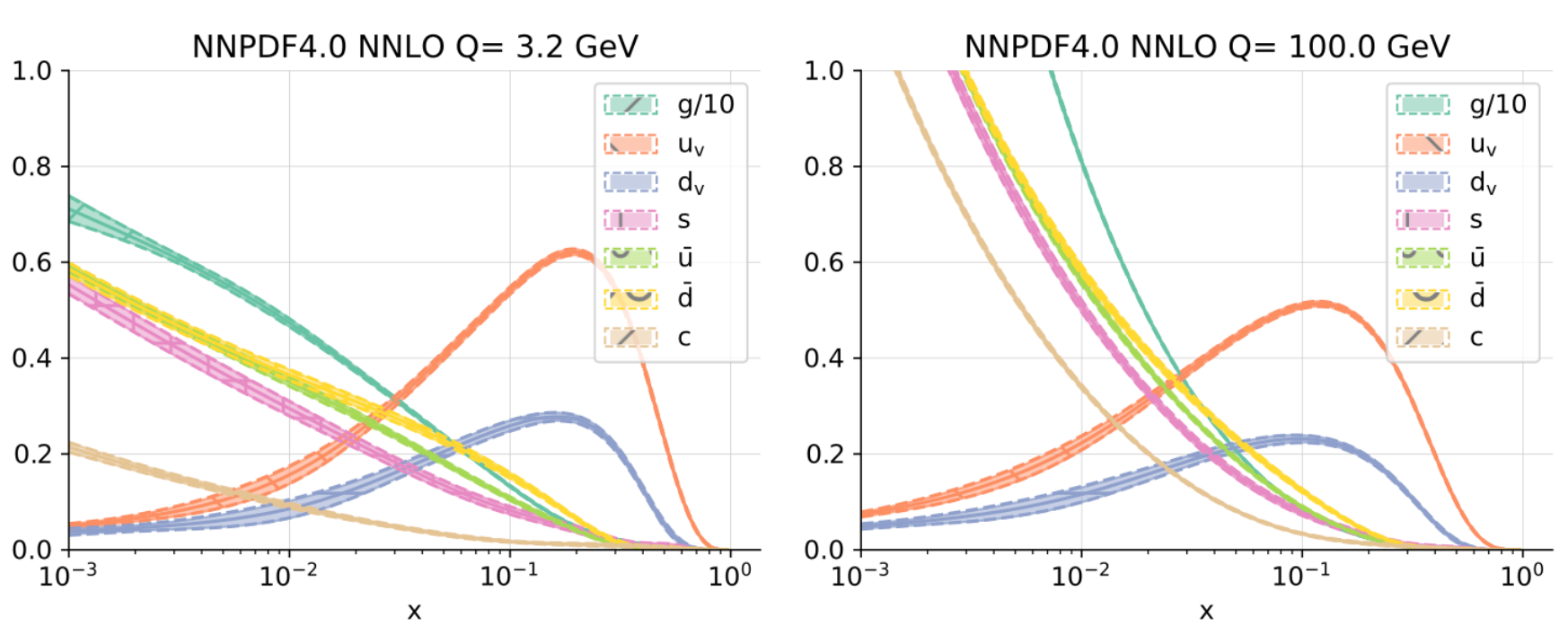}
    \caption[The NNPDF4.0 NNLO PDFs]{The NNPDF4.0 NNLO PDFs at Q= 3.2 GeV (left) and Q= 100 GeV (right)~\cite{NNPDF:2021njg}.}
    \label{fig:NNPDF40_PDFs}
\end{figure}

The CT18 analysis represents a major update of the CT14~\cite{Dulat:2015mca} generation and places strong emphasis on incorporating high-precision LHC differential measurements. The resulting NNLO PDFs show improved constraints on the gluon distribution in the region $10^{-3} \lesssim x \lesssim 0.3$, driven primarily by inclusive jet and top-pair production data, and a revised strange-quark distribution influenced by LHC $W/Z$ measurements. As illustrated in Fig.~\ref{fig:CT18_PDFs}, the CT18 gluon uncertainty at intermediate $x$ is significantly reduced compared to earlier generations, while the strange-quark distribution exhibits a noticeable enhancement relative to CT14 in certain $x$ regions. At small $x$, the combined HERA data continue to dominate the constraints. 

The NNPDF4.0 determination~\cite{NNPDF:2021njg} extends the global dataset further and reports a general reduction of PDF uncertainties relative to earlier NNPDF releases. Its phenomenological picture is largely consistent with CT18, featuring a well-constrained gluon at intermediate $x$ and an enhanced strange sea compared to older fits, as illustrated in Fig.~\ref{fig:NNPDF40_PDFs}. Nevertheless, visible spreads among central values persist, particularly in the detailed flavor decomposition of the light sea and in the large-$x$ gluon region. The PDF4LHC21 combination~\cite{PDF4LHCWorkingGroup:2022cjn}, based on CT18, MSHT20, and NNPDF3.1, reflects this overall level of agreement while retaining the residual spread among global fits as a measure of systematic differences. 

Taken together, these modern global analyses of unpolarized PDFs exhibit substantial consistency across most kinematic regions relevant for LHC phenomenology, while still displaying non-negligible variations in specific flavor sectors, in particular the strange distribution and the large-$x$ gluon. These residual differences highlight the importance of complementary first-principles inputs, such as lattice QCD calculations, which offer an independent avenue to constrain and ultimately validate aspects of proton structure beyond the phenomenological global-fit framework. This complementarity is even more pronounced in the case of polarized PDFs, where the experimental constraints are considerably more limited and global fits rely on a much smaller dataset, thereby increasing the potential impact of first-principles determinations.

In comparison, DAs describe longitudinal momentum sharing among valence partons in hard exclusive processes, such as the $B$ meson decay $B \to \pi \pi$ and $B \to l \nu \pi$~\cite{Beneke:2001ev,Beneke:1999br}. For a pseudoscalar meson $M$ with momentum $p^\mu$, the leading-twist light-cone DA $\phi_M(x;\mu)$ is defined as 
\begin{eqnarray} 
    \int \frac{d\xi^-}{2\pi} e^{ixp^+ \xi^-}\langle 0 | \bar \psi_1(0) n\!\!\!\slash \gamma_5 U(0,\xi^-)\psi_2(\xi^-) |M(P)\rangle = i f_M(p \cdot n) \phi_{M}(x;\mu) ~, 
    \label{eq:da_definition}
\end{eqnarray} 
where $U(0,\xi^-) = P\,{\rm exp}\big[ig_s\int_{\xi_-}^0 ds ~ {n}\cdot A(sn)\big]$ is the path-ordered gauge link defined along the minus light-cone direction $n$ ($n^2=0$), and $f_M$ the decay constant. 

Similar to the factorization-scale dependence of PDFs, DAs depend on the renormalization scale $\mu$ through the ERBL evolution equation,
\begin{equation}
    \frac{\partial \phi(x; \mu)}{\partial \ln \mu}
    = \int_0^1 d y \, V(x, y; \mu)\, \phi(y; \mu) \, ,
\end{equation}
where $V(x,y;\mu)$ is the ERBL kernel that governs the evolution of a quark-antiquark pair carrying momentum fractions $(y,1-y)$ into a pair with fractions $(x,1-x)$.

As discussed above, DAs are probability amplitudes and are conceptually analogous to wave functions in quantum mechanics. Unlike PDFs, however, DAs describe the correlated longitudinal momentum distribution of a quark--antiquark pair inside a meson. Consequently, experimental observables do not probe the DA at a fixed value of $x$; instead, the DA enters physical amplitudes through a convolution with a perturbatively calculable hard-scattering kernel. Most phenomenological constraints on DAs are obtained from exclusive processes such as the $\pi \to \gamma \gamma^*$ transition form factor~\cite{Gao:2021iqq} and electromagnetic form factors~\cite{Melic:1998qr}. For instance, the pion transition form factor $F_{\pi\gamma}(Q^2)$ can be written schematically as a convolution of the hard kernel $T_H(x,Q^2;\mu)$ with the pion DA $\phi_\pi(x;\mu)$. Early measurements of this observable by the CLEO collaboration~\cite{CLEO:1997fho} up to $Q^2 \sim 9~\mathrm{GeV}^2$ were broadly consistent with perturbative QCD expectations based on the asymptotic form $\phi_\pi(x)=6x(1-x)$.

However, the BABAR experiment~\cite{BaBar:2009rrj}, which extended the measurement to $Q^2 \sim 40~\mathrm{GeV}^2$, reported a continued growth of $Q^2 F_{\pi\gamma}(Q^2)$ at large $Q^2$, exceeding the asymptotic prediction and suggesting a significantly broader or endpoint-enhanced pion DA. This unexpected trend triggered extensive theoretical investigations and debate regarding possible modifications of the DA shape, higher-twist effects, or limitations of collinear factorization in the accessible kinematic regime. Subsequent measurements by the Belle collaboration~\cite{Belle:2012wwz} did not confirm such a strong rise and were instead more consistent with perturbative QCD expectations, partially alleviating the so-called ``BABAR puzzle''~\cite{Dorokhov:2010bz}. As a result, experimental data constrain only specific moments or weighted integrals of the DA rather than its full $x$-dependence. To bridge this gap, various nonperturbative approaches have been developed to model the shape of the DA, including QCD sum rules~\cite{Chernyak:1981zz,Chernyak:1983ej,Braun:1988qv} and Dyson--Schwinger equation studies~\cite{Chang:2013pq}. Nevertheless, these approaches typically rely on model assumptions or truncations, which makes first-principles calculations of $x$-dependent DAs particularly valuable.

\section{Momentum Tomography: TMDs}

While collinear parton distributions provide a one-dimensional description of hadron structure in terms of longitudinal momentum fractions, many high-energy processes are sensitive to the intrinsic transverse motion of partons. When the transverse momentum $q_\perp \equiv |\vec{q}_\perp| = \sqrt{q_x^2 + q_y^2}$ of an observed final state is parametrically smaller than the hard scale, $\Lambda_{\rm QCD} \lesssim q_\perp \ll Q$, a more differential description is required. In this regime, QCD factorization must retain transverse momentum degrees of freedom, leading to the framework of TMDs. These objects encode a multidimensional momentum-space structure and therefore provide access to momentum tomography beyond the collinear approximation.

The necessity of this refinement can be understood from the general logic of QCD factorization, whose validity relies on a systematic separation of scales. In inclusive observables characterized by a single hard scale $Q$, collinear factorization exploits the hierarchy $Q \gg \Lambda_{\rm QCD}$ and integrates over transverse momenta up to the hard scale, thereby absorbing long-distance physics into universal collinear PDFs. However, when the observable is differential in a small transverse momentum $q_\perp \ll Q$, an additional scale enters the problem. This small $q_\perp$ acts as a new infrared-sensitive scale, and the simple separation between hard and collinear modes is no longer sufficient. Large logarithms of the form $\ln(Q/q_\perp)$ emerge in perturbation theory, signaling the breakdown of fixed-order expansions within the collinear framework.

To restore predictive power, the factorization structure must be reorganized to consistently separate hard, collinear, and soft contributions at the level of transverse momentum. This leads to TMD factorization, in which the nonperturbative physics is encoded in TMDs, accompanied by an explicit soft factor and appropriate rapidity regularization and evolution. In this way, the appearance of a new hierarchical scale necessitates a corresponding refinement of the factorization theorem and the definition of the underlying nonperturbative functions.

\begin{figure}[t]
    \centering
    \includegraphics[width=0.6\textwidth]{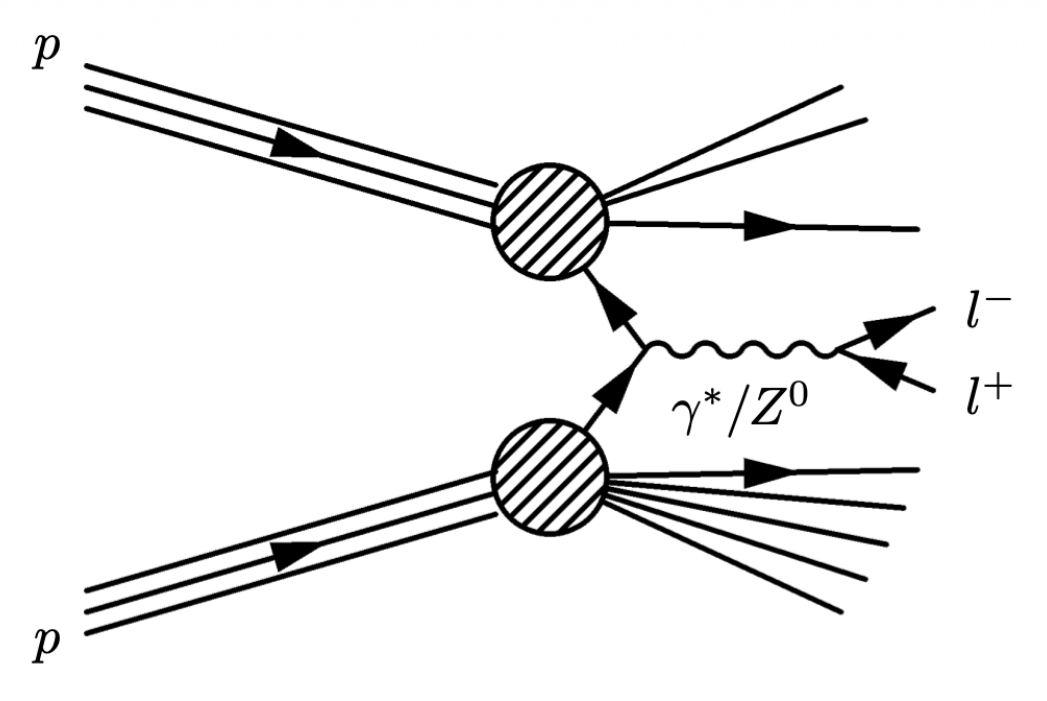}
    \caption[The Drell--Yan process]{The diagrammatic illustration of the Drell--Yan process. Credit to Ref.~\cite{Avkhadiev:2025vsx}.}
    \label{fig:DrellYan}
\end{figure}

More specifically, the physical motivation for TMD factorization can be illustrated using the Drell--Yan process, $pp \to l^+l^- X$. In this reaction, we consider the light-cone components of the lepton pair momentum $q^\mu = (q^+, q^-, \vec{q}_\perp)$, where $q^+ = (q^0 + q^3)/\sqrt{2}$ and $q^- = (q^0 - q^3)/\sqrt{2}$. The invariant mass $Q$ of the lepton pair sets the hard scale, while the measured transverse momentum $q_\perp$ of the pair probes the transverse motion of the annihilating partons. The diagrammatic illustration of this process is shown in Fig.~\ref{fig:DrellYan}. When $q_\perp \sim Q$, the transverse momentum arises perturbatively from hard emissions and the standard collinear factorization formula suffices. In contrast, when $q_\perp \ll Q$, the cross-section becomes sensitive to soft and collinear radiation as well as to intrinsic transverse momentum inside the incoming hadrons. In this region, the differential cross-section can be organized as~\cite{Boussarie:2023izj}
\begin{equation}
    \frac{d\sigma}{dQ^2\, dy\, d^2 \vec{q}_\perp}
    =
    \left(W(Q,\vec{q}_\perp)
    +
    Y(Q,\vec{q}_\perp) \right) 
    \left[ 1 + \mathcal{O}\!\left(\frac{\Lambda_{\rm QCD}^2}{Q^2}\right) \right] ~,
\end{equation}
where $y \equiv \frac{1}{2} \ln \left( \frac{q^+}{q^-} \right)$ defines the rapidity of the lepton pair, the $W$-term resums logarithms dominant in the small-$q_\perp$ region, while the $Y$-term accounts for the regular fixed-order contributions that remain finite as $q_\perp \to 0$. At leading power for $q_\perp \ll Q$, the $W$-term factorizes as
\begin{equation}
    W(Q,\vec{q}_\perp)
    =
    \sum_q
    H_{q\bar q}(Q;\mu)
    \int d^2 \vec{b}_\perp
    e^{i \vec{b}_\perp \cdot \vec{q}_\perp}
    f_{q/H_1}(x_1, \vec{b}_\perp; \mu, \zeta_1)
    f_{\bar q/H_2}(x_2, \vec{b}_\perp; \mu, \zeta_2)~,
\end{equation}
where $H_{q\bar q}(Q;\mu)$ is the perturbatively calculable hard function, and $f_{q/H}(x,\vec{b}_\perp;\mu,\zeta)$ denotes the renormalized unpolarized TMD PDF in impact-parameter $b_\perp \equiv |\vec{b}_\perp|$ space, Fourier conjugate to transverse momentum. The variables $\zeta_{1,2}$ are Collins--Soper scales associated with rapidity regularization, satisfying $\zeta_1 \zeta_2 = Q^4$. This structure, originally derived in the Collins--Soper--Sterman (CSS) formalism and subsequently refined in modern formulations, establishes the separation of short-distance physics at scale $Q$ from long-distance transverse dynamics encoded in universal TMDs~\cite{Boussarie:2023izj}.

The necessity of working in transverse coordinate space arises from the fact that large logarithms of the form $\ln(Q^2/q_\perp^2)$ appear at fixed order when $q_\perp \ll Q$. In $b_\perp$ space these logarithms exponentiate and can be resummed systematically through renormalization group evolution. The small-$b_\perp$ region corresponds to perturbative transverse momenta $q_\perp \sim b_\perp^{-1} \gg \Lambda_{\rm QCD}$, where TMDs can be matched onto collinear PDFs via an OPE~\cite{Collins:1981uw,Collins:1984kg},
\begin{equation}
    f_{q/H}(x,\vec{b}_\perp;\mu,\zeta)
    =
    \sum_i
    \int_x^1 \frac{d\xi}{\xi}
    C_{q \leftarrow i}\!\left(\frac{x}{\xi}, \vec{b}_\perp; \mu, \zeta \right)
    f_{i/H}(\xi;\mu)
    +
    \mathcal{O}\!\left(b^2_\perp \Lambda^2_{\rm QCD}\right)~.
\end{equation}
The coefficient functions $C_{q \leftarrow i}$ are perturbatively calculable. In this way, TMDs interpolate between perturbative transverse dynamics at small $b_\perp$ and genuinely nonperturbative physics at large $b_\perp$, providing a unified description across momentum scales. 

\begin{figure}[t]
    \centering
    \includegraphics[width=0.8\textwidth]{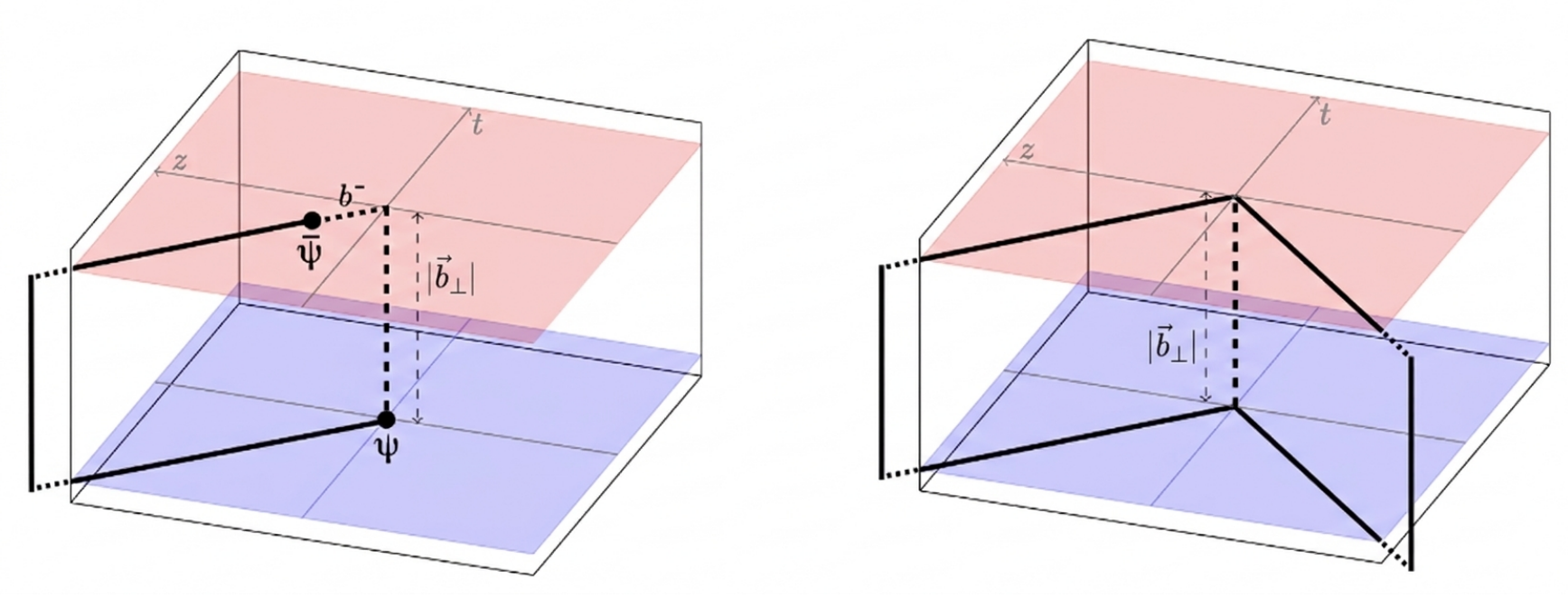}
    \caption[Graphs of the Wilson line structure]{Graphs of the Wilson line structure of bare unsubtracted TMD correlator (left) and bare soft function (right). Credit to Ref.~\cite{Boussarie:2023izj}.}
    \label{fig:staple_wilson_link}
\end{figure}

At the operator level, TMD PDFs are defined as gauge-invariant nonlocal matrix elements of quark or gluon fields separated along both light-cone and transverse directions. For a quark of flavor $q$ in a proton moving predominantly along the $+$ light-cone direction, the bare unsubtracted TMD correlator in coordinate space takes the schematic form
\begin{eqnarray}
    &&f_{q/H}^{0,~\mathrm{unsub}}(x,\vec{b}_\perp,P^+,\epsilon,\tau)
    = \int \frac{db^-}{2\pi}
    e^{-ix P^+ b^-} \nonumber\\
    &&\qquad \qquad \qquad \times \left\langle P S \left|
    \left[ \bar \psi^0_q(b^-,\vec{b}_\perp)\,
    \mathcal{W}^\dagger_{\rm staple}(b^-,\vec{b}_\perp)
    \Gamma
    \mathcal{W}_{\rm staple}(0)
    \psi^0_q(0) \right]_\tau
    \right|P S \right\rangle ~,
\end{eqnarray}
where the staple-shaped Wilson lines $\mathcal{W}_{\rm staple}$ extend to light-cone infinity and include transverse gauge links to ensure full gauge invariance. The diagrammatic illustration of the staple-shaped Wilson lines is shown in the left panel of Fig.~\ref{fig:staple_wilson_link}. In contrast to collinear PDFs, TMD correlators contain rapidity divergences associated with gluons of arbitrarily large rapidity. These divergences are not regulated by dimensional regularization alone and require an additional rapidity regulator, so we have $\epsilon$ and $\tau$ as the dimensional regulator and rapidity regulator, respectively. After combining the unsubtracted bare TMD matrix element with a bare soft factor $S^0$, a soft subtraction factor $S^{0 \mathrm{subt}}$ that subtracts double counting of soft radiation, and performing ultraviolet renormalization in the $\overline{\rm MS}$ scheme, one obtains the properly defined TMD PDF $f_{q/H}(x,\vec{b}_\perp;\mu,\zeta)$~\cite{Boussarie:2023izj}. The bare soft factor $S^0$ is defined as a vacuum matrix element of Wilson lines,
\begin{equation}
    S^0(\vec{b}_\perp, \epsilon, \tau)
    =
    \frac{1}{N_c}
    \left\langle 0 \left|
    \left[ \mathcal{W}_n^\dagger(\vec{b}_\perp)
    \mathcal{W}_{\bar n}(\vec{b}_\perp)
    \mathcal{W}_{\bar n}^\dagger(0)
    \mathcal{W}_n(0) \right]_\tau
    \right| 0 \right\rangle ~,
\end{equation}
where the shape of Wilson lines is presented in the right panel of Fig.~\ref{fig:staple_wilson_link}. Regarding the soft subtraction factor $S^{0 \mathrm{subt}}$, in SCET these subtractions are known as zero-bin subtractions~\cite{Manohar:2006nz} and arise from ensuring fluctuations encoded by collinear fields do not have singular overlap with those of the soft fields, thus also ensuring there is no double counting of infrared regions. More details about the subtraction factor can be found in Ref.~\cite{Boussarie:2023izj}.

A central structural difference relative to collinear PDFs is therefore the presence of two evolution scales. In addition to the usual renormalization scale $\mu$, TMDs depend on a rapidity scale $\zeta$, whose variation is governed by the Collins--Soper equation~\cite{Collins:1981va},
\begin{equation}
    \frac{\partial}{\partial \ln \sqrt{\zeta}}
    f_{q/H}(x,\vec{b}_\perp,\mu,\zeta)
    =
    K(b_\perp,\mu)\,
    f_{q/H}(x,\vec{b}_\perp,\mu,\zeta)~,
\end{equation}
where $K(b_\perp,\mu)$ is the Collins--Soper kernel. Together with the renormalization group equation in $\mu$, this yields a coupled evolution system that resums Sudakov logarithms and controls the scale dependence of TMD cross-sections. In contrast to DGLAP evolution, which mixes different momentum fractions $x$, Collins--Soper evolution is multiplicative in $x$ and acts at fixed transverse separation $b_\perp$.

When spin degrees of freedom are included, the TMD framework reveals a much richer structure than in the collinear case. At leading twist, the quark TMD correlator for a spin-$1/2$ nucleon can be decomposed into eight independent functions, corresponding to different correlations between parton transverse momentum and nucleon spin. In addition to the unpolarized distribution, one encounters helicity and transversity TMDs, as well as time-reversal-odd functions such as the Sivers and Boer--Mulders distributions. The presence of staple-shaped Wilson lines implies that certain TMDs are process dependent, differing by an overall sign between semi-inclusive deep-inelastic scattering and Drell--Yan. This so-called modified universality arises from the opposite directions of the gauge links in the two processes and is consistent with time-reversal invariance once Wilson lines are properly taken into account.

The phenomenological extraction of TMDs combines data from Drell--Yan, semi-inclusive deep-inelastic scattering, and back-to-back hadron production in $e^+e^-$ annihilation. In each case, the measured small transverse momentum provides sensitivity to intrinsic transverse motion and to spin-momentum correlations. The interplay between perturbative matching at small $b_\perp$, nonperturbative modeling at large $b_\perp$, and coupled $(\mu,\zeta)$ evolution equations allows for global analyses of unpolarized and polarized TMDs, analogous in spirit to global fits of collinear PDFs but with substantially richer kinematic structure.

In summary, TMD factorization extends the one-dimensional parton model into three-dimensional momentum space. By retaining transverse degrees of freedom and properly accounting for soft radiation and rapidity evolution, it provides a theoretically consistent and phenomenologically powerful framework for describing processes at small transverse momentum.

\section{Spatial Tomography: GPDs}

While collinear PDFs describe the longitudinal momentum structure of partons and TMDs extend this picture into three-dimensional momentum space, GPDs provide complementary access to the spatial structure of hadrons. In particular, GPDs correlate the longitudinal momentum of partons with the transverse spatial distribution of their color charge, thereby enabling a spatial tomography of the nucleon.

GPDs appear in hard exclusive processes in which the hadron remains intact and the full final state is reconstructed. The paradigmatic examples are deeply virtual Compton scattering (DVCS)~\cite{Ji:1996ek,Ji:1996nm}, and deeply
virtual meson production (DVMP)~\cite{Radyushkin:1996ru,Collins:1996fb}. Unlike collinear PDFs, GPDs depend on additional kinematic variables that encode the non-forward nature of the process. In the forward limit, where the momentum transfer to the hadron vanishes, GPDs reduce to ordinary PDFs. On the other hand, integration over the momentum fraction $x$ relates GPDs to elastic form factors. In this way, GPDs interpolate smoothly between inclusive parton distributions and exclusive hadronic observables, unifying momentum and spatial information within a single framework.

A particularly important aspect of GPDs is that their Fourier transform with respect to the transverse momentum transfer provides impact-parameter dependent distributions. In a frame where the longitudinal momentum transfer is fixed, this transformation yields parton densities in transverse coordinate space, offering a probabilistic interpretation of the transverse spatial distribution of partons at fixed longitudinal momentum fraction. This connection underlies the interpretation of GPDs as tools for spatial imaging of the nucleon.

Furthermore, GPDs contain information on the total angular momentum carried by quarks and gluons through Ji's sum rule~\cite{Ji:1996ek}, which relates the second moment of specific GPD combinations to the total partonic contribution to the nucleon spin. This establishes a direct link between exclusive scattering measurements and the decomposition of nucleon spin in QCD.

Experimentally, GPDs are constrained by measurements of DVCS and exclusive meson production at fixed-target facilities and colliders. While the kinematic coverage remains limited compared to inclusive DIS, ongoing and future programs, in particular at high-luminosity electron-ion facilities, are expected to significantly improve constraints on the multidimensional structure encoded in GPDs.

\section{Conclusion}

In this chapter, we have reviewed how the study of partonic structure has evolved into a coherent and multidimensional description of hadrons within QCD. Collinear PDFs quantify the longitudinal momentum and spin structure and underpin precision collider phenomenology; TMDs resolve intrinsic transverse motion and spin-momentum correlations, revealing nonperturbative dynamics at the few-hundred-MeV scale; and GPDs correlate momentum and transverse spatial distributions, enabling access to impact-parameter tomography and to energy-momentum tensor form factors that encode mass, angular momentum, and mechanical properties. Together, these distributions form a unified framework for hadron tomography, supported by QCD factorization, increasingly precise global analyses, and complementary first-principles input from lattice QCD. With ongoing measurements at facilities such as Jefferson Lab and the Large Hadron Collider, and future advances anticipated at the Electron-Ion Collider, partonic physics stands as a mature and rapidly progressing program aimed at a quantitative understanding of how quarks and gluons generate the observable properties of hadrons.

%% file: Chapter3.tex
%Chapter 1

\renewcommand{\thechapter}{3}

\vspace*{-1.35in}

\chapter{First-Principles Calculation of Strong Interaction: Lattice QCD}
\label{ch:lattice}

\vspace*{-.5in}

Lattice QCD emerged from the need to address quark confinement and to formulate QCD beyond the domain of perturbation theory. In 1974, Kenneth G. Wilson proposed the discretization of non-Abelian gauge theories on a space-time lattice and, in his seminal paper ``Confinement of Quarks”~\cite{Wilson:1974sk}, demonstrated through a strong-coupling expansion that lattice Yang-Mills theory exhibits an area law behavior for large Wilson loops. This area law implies a linearly rising potential between static color sources at large separations, thereby providing a concrete realization of confinement within a gauge-invariant, nonperturbative framework. The recovery of continuum QCD from the lattice formulation was clarified later through the development of renormalization-group arguments and lattice perturbation theory. Asymptotic freedom~\cite{Gross:1973id,Politzer:1973fx} implies that the continuum limit corresponds to the ultraviolet fixed point at vanishing bare coupling, and it was subsequently established that lattice gauge theory is renormalizable and belongs to the same universality class as continuum Yang--Mills theory~\cite{Symanzik:1983dc,Luscher:1985zq,Reisz:1988kk}. Although this formulation established the conceptual foundation of lattice gauge theory, practical calculations of hadronic observables required the later development of efficient numerical algorithms and sufficient computational resources.

By 1979-1980, the development of Monte Carlo simulation techniques transformed Wilson's proposal into a practical computational framework. Michael Creutz and collaborators carried out the first large-scale numerical simulations of non-Abelian gauge theory on the lattice, including SU(2) Yang-Mills theory, and directly measured Wilson loops and the static quark-antiquark potential~\cite{Creutz:1980zw}. The observed linear rise of the potential at large separations provided numerical evidence for confinement, consistent with the area-law behavior anticipated in the strong-coupling expansion. Shortly thereafter, it was demonstrated that hadron spectroscopy could be addressed from first principles: early calculations of meson and baryon masses by Weingarten, and by Hamber and Parisi, showed that lattice QCD could reproduce key features of the hadron spectrum~\cite{Weingarten:1982qe,Hamber:1981um,Hamber:1981zn}. These pioneering works established lattice gauge theory as a viable ab initio approach to nonperturbative QCD.

From the 1980s onward, progress in algorithms and computing power significantly broadened the scope of lattice QCD. The formulation of efficient algorithms to include dynamical fermions—most notably the Hybrid Monte Carlo (HMC) method~\cite{Duane:1987de}—made it possible to incorporate the fermion determinant in a controlled manner. Over time, simulations progressed from the quenched approximation to fully dynamical calculations with 2, 2+1, and eventually 2+1+1 quark flavors. By the 2000s and 2010s, lattice ensembles at or near physical quark masses and large volumes became feasible~\cite{Aoki:2019cca}. As a result, modern lattice QCD calculation provides high-precision determinations of hadronic observables, ranging from spectroscopy and decay constants to form factors and weak matrix elements, after systematic extrapolations to the continuum limit, infinite volume, and physical quark masses.

In recent years, lattice QCD has further expanded beyond traditional spectroscopy and matrix elements to address the internal structure of hadrons in terms of quarks and gluons. The proposal of the large momentum effective theory (LaMET)~\cite{Ji:2013dva,Ji:2014gla,Ji:2020ect,Ji:2024oka} opened a new avenue for computing $x$-dependence of parton distribution functions directly on the lattice through quasi-distributions defined at finite hadron momenta. Building on this framework, current lattice calculations now achieve controlled theoretical uncertainties for PDFs in the moderate-$x$ region\cite{Zhao:2025oto}. Exploratory studies of TMDs and GPDs are also underway, marking a transition from static properties to a multidimensional, partonic description of hadrons within first-principles QCD. More detailed discussions on lattice calculations of partonic structure will be presented in Chapter~\ref{ch:systematics}.

\section{Lattice QCD as a First-Principles Framework}

Lattice QCD provides a nonperturbative definition of QCD by promoting the formal path integral of the continuum theory to a mathematically well-defined statistical system. After Wick rotation to Euclidean spacetime, vacuum expectation values take the form
\begin{eqnarray}
    \langle \mathcal{O} \rangle
    =
    \frac{1}{Z}
    \int \mathcal{D}U\,\mathcal{D}\bar\psi\,\mathcal{D}\psi\;
    \mathcal{O}\,e^{-S_E[U,\bar\psi,\psi]}~,
\end{eqnarray}
which can be interpreted as an ensemble average over gauge-field configurations weighted by $e^{-S_E}$. Discretizing spacetime into a hypercubic lattice of spacing $a$ introduces an ultraviolet regulator while preserving exact gauge invariance: the gauge field $A_\mu(x)$ is replaced by link variables
\begin{equation}
    U_\mu(x)=\exp[iagA_\mu(x)]~,
\end{equation}
living on links between neighboring sites. The simplest gauge action is the Wilson plaquette action~\cite{Wilson:1974sk}, constructed from elementary $1\times1$ Wilson loops (plaquette), which reproduces the Yang--Mills action in the classical continuum limit. Together with a discretized fermion action, this defines a regulated version of QCD in which all ultraviolet divergences are removed by the finite cutoff $1/a$.

Conceptually, lattice QCD realizes Wilson's renormalization-group picture: long-distance physics is insensitive to the microscopic details of the regulator. The lattice spacing plays the role of a short-distance cutoff, while physical observables depend only on the correlation length measured in lattice units. Asymptotic freedom implies that the continuum limit corresponds to tuning the bare coupling toward the ultraviolet fixed point at $g_0 \to 0$, such that the physical correlation length becomes much larger than the lattice spacing. In this regime, different discretizations of the action lie in the same universality class and yield identical continuum predictions after renormalization. The continuum QCD Lagrangian thus emerges as the critical theory describing the long-distance behavior of the lattice system.

Beyond providing a regulator, the lattice formulation makes the theory computationally accessible. Because the Euclidean path integral resembles a partition function of a four-dimensional statistical system, importance-sampling Monte Carlo methods can be employed to generate representative gauge-field configurations. Observables are computed as ensemble averages over these configurations, without reliance on perturbative expansions. Nonperturbative phenomena such as confinement, spontaneous chiral symmetry breaking, and topological fluctuations arise dynamically from the underlying gauge dynamics rather than from model assumptions.

Moreover, the framework is systematically improvable. Discretization effects can be reduced through optimized discretization schemes, such as Symanzik improvement~\cite{Symanzik:1983dc,Symanzik:1983gh}, and continuum extrapolation by performing simulations at multiple lattice spacings. Finite-volume effects are controlled by increasing the physical volume and are quantitatively understood through effective field theory. Quark masses can be tuned to their physical values, and renormalization procedures connect lattice operators to continuum schemes. Modern simulations therefore analyze multiple ensembles varying $a$, $L$, and quark masses in order to perform controlled extrapolations to the continuum, infinite-volume, and physical-mass limits.

In summary, lattice QCD provides a first-principles, nonperturbative formulation of the strong interaction directly from the QCD Lagrangian. By defining the theory with a gauge-invariant ultraviolet regulator and evaluating the Euclidean path integral numerically, it converts the formal definition of quantum field theory into a controllable statistical system. The framework is systematically improvable through well-defined extrapolations. Crucially, no model assumptions about confinement or chiral symmetry breaking are imposed. In this sense, lattice QCD furnishes a quantitatively reliable bridge between the fundamental gauge theory and observable hadronic physics in the nonperturbative regime.

\section{Fermion Formulations on the Lattice}

While gauge fields admit a geometrically natural discretization in terms of link variables, the formulation of fermions on the lattice is considerably more subtle. The difficulty already arises at the level of the free Dirac field and is rooted in the so-called fermion doubling problem~\cite{Karsten:1980wd}. A naive transcription of the continuum Dirac action to the lattice, obtained by replacing derivatives with symmetric finite differences, leads in four dimensions to sixteen fermionic modes in the continuum limit instead of a single physical species.

To see the origin of the problem, consider the Euclidean lattice action for a free fermion obtained by replacing $\partial_\mu$ with the antihermitian symmetric difference operator. In momentum space, the inverse propagator takes the form
\begin{equation}
    D(p) = i\sum_\mu \gamma_\mu \tilde{p}_\mu + M~,
    \qquad
    \tilde{p}_\mu = \frac{1}{a}\sin(p_\mu a)~,
\end{equation}
with $p_\mu$ restricted to the Brillouin zone $(-\pi/a,\pi/a]$. While $\tilde{p}_\mu \to p_\mu$ for small momenta, the sine function also vanishes at the edges of the Brillouin zone. As a consequence, the propagator develops additional poles near $p_\mu = \pi/a$ (and permutations thereof). In $d$ space-time dimensions this yields $2^d$ fermionic excitations in the continuum limit, fifteen of which are lattice artefacts in four dimensions. These extra modes are not removable by simply discarding high-momentum contributions, since they are associated with symmetry transformations of the lattice action that relate the different corners of the Brillouin zone.

The doubling phenomenon is closely tied to fundamental structural requirements. The Nielsen-Ninomiya theorem~\cite{Nielsen:1981hk} states that, under the assumptions of locality, translational invariance, and hermiticity, a lattice fermion formulation with exact chiral symmetry necessarily contains fermion doublers. Therefore, eliminating the unwanted species requires relaxing at least one of these conditions. Most practical formulations break chiral symmetry explicitly at nonzero lattice spacing, with the understanding that it is recovered in the continuum limit.

The Wilson formulation~\cite{Wilson:1975id} achieves this by adding a second-derivative term (the Wilson term) to the naive action. The Wilson-Dirac operator reads
\begin{equation}
    D_W = i\sum_\mu \gamma_\mu \tilde{p}_\mu
    + M
    + \frac{r}{a}\sum_\mu \left(1 - \cos(p_\mu a)\right)~,
\end{equation}
where $r$ is the Wilson parameter. The additional term acts as a momentum-dependent mass that vanishes for small momenta but grows as $1/a$ near the edges of the Brillouin zone. Consequently, the doublers acquire masses of order the cutoff and decouple in the continuum limit. The price paid is the explicit breaking of chiral symmetry at finite lattice spacing, which leads to additive mass renormalization and necessitates fine tuning of the bare mass parameter. 

Discretization errors in Wilson fermions are generically of $\mathcal{O}(a)$, which can be reduced by Symanzik improvement~\cite{Symanzik:1983dc,Symanzik:1983gh}. In particular, the clover-improved Wilson action (Sheikholeslami-Wohlert improvement)~\cite{Sheikholeslami:1985ij} introduces an additional dimension-five operator proportional to $\sigma_{\mu\nu}F_{\mu\nu}$, which cancels $\mathcal{O}(a)$ effects in on-shell quantities at tree level and, with appropriate tuning, beyond tree level. Clover fermions are widely used in large-scale simulations due to their balance between theoretical control and computational cost.

An alternative strategy is provided by staggered (Kogut-Susskind) fermions~\cite{Kogut:1974ag,Susskind:1976jm}. Instead of lifting the doubler masses, the staggered approach reduces the number of fermionic degrees of freedom per lattice site by distributing spin components over neighboring sites. Through a spin-diagonalization transformation, the naive Dirac structure is replaced by position-dependent phases, leaving a single Grassmann variable per site. In four dimensions, this procedure reduces the sixteen naive species to four so-called ``tastes.'' The resulting action preserves a remnant $U(1)\times U(1)$ chiral symmetry at vanishing bare mass, which protects against additive mass renormalization. However, because four tastes remain in the continuum limit, practical simulations with $N_f=2$ or $2+1$ flavors require the rooting procedure, in which the fermion determinant is replaced by an appropriate fractional power. Although supported by substantial numerical evidence, the theoretical justification of rooting has been the subject of extensive discussion.

Formulations that preserve exact lattice chiral symmetry at finite $a$ are obtained by implementing the Ginsparg-Wilson relation~\cite{Ginsparg:1981bj}. Overlap fermions~\cite{Neuberger:1997fp} provide an explicit realization of this relation via a nonlocal Dirac operator constructed from the sign function of a Hermitian Wilson operator. Domain-wall fermions~\cite{Kaplan:1992bt,Shamir:1993zy} introduce an additional fifth dimension, in which chiral modes are localized on opposite boundaries; in the limit of infinite fifth dimension, they approach overlap fermions. These formulations possess an exact lattice-modified chiral symmetry and automatically reproduce the correct axial anomaly. They are therefore particularly well suited for studies of chiral symmetry breaking, topological properties, and flavor-singlet observables. Their principal drawback is the significantly increased computational cost compared to Wilson or staggered fermions.

All of these discretizations represent distinct regularizations of continuum QCD. Provided that the lattice action is local and gauge invariant, universality guarantees convergence to the same continuum theory as $a\to 0$. Nevertheless, the pattern of lattice artefacts, the realization of chiral symmetry, and the computational efficiency differ markedly among formulations. Systematic continuum extrapolations and cross-checks between different fermion actions therefore constitute an essential component of precision lattice QCD. The coexistence of multiple fermion formulations is not merely a historical artifact but a practical and conceptual asset, enabling controlled assessments of discretization effects and strengthening the reliability of lattice results as first-principles determinations of hadronic observables.

\section{Monte Carlo Algorithms for Lattice Gauge Ensembles}

The generation of lattice gauge ensembles amounts to constructing a stochastic process whose stationary distribution coincides with the Euclidean path-integral measure,
\begin{equation}
    \pi[U] \propto e^{-S_g[U]} 
\end{equation}
for pure gauge theories, or
\begin{equation}
    \pi[U] \propto e^{-S_g[U]} \det D[U] 
\end{equation}
for theories including dynamical fermions. Here, $S_g[U]$ is the discretized gauge action, and $\det D[U]$ arises from integrating out the fermion fields. The goal is to sample gauge-field configurations $U$ according to $\pi[U]$, enabling the computation of observables as ensemble averages.
In practice, this is achieved through Markov Chain Monte Carlo (MCMC) algorithms, which define a transition kernel $K(U \to U')$ such that $\pi[U]$ is invariant under the evolution of the chain. Two fundamental requirements ensure correctness of the sampling procedure: ergodicity and stationarity. Ergodicity guarantees that all physically relevant regions of configuration space can be reached from any initial configuration in a finite number of steps, while stationarity ensures that $\pi[U]$ is preserved under the transition dynamics. A stationary distribution $\pi[U]$ satisfies the global balance condition,
\begin{equation}
    \pi[U] = \int \mathcal{D}U' \,\pi[U'] K(U' \to U)~.
\end{equation}
A sufficient, though not strictly necessary, condition for stationarity is detailed balance,
\begin{equation}
    \pi[U] K(U \to U') = \pi[U'] K(U' \to U)~,
\end{equation}
which implies that the Markov chain is reversible with respect to $\pi[U]$. In addition to satisfying these formal requirements, practical algorithms must minimize autocorrelation times and computational cost, particularly as the lattice spacing decreases and critical slowing down becomes pronounced.

From this viewpoint, different Monte Carlo algorithms correspond to different constructions of the transition kernel. In the Metropolis--Hastings framework, the kernel is built from a proposal distribution $q(U \to U')$ together with an accept--reject step. Choosing the acceptance probability as
\begin{equation}
    P_{\mathrm{acc}}(U \to U')
    =
    \min\!\left(
        1,
        \frac{\pi[U'] \, q(U' \to U)}
             {\pi[U] \, q(U \to U')}
    \right)
\end{equation}
ensures that the resulting kernel satisfies detailed balance. For symmetric proposals, $q(U \to U') = q(U' \to U)$, this reduces to the familiar Metropolis form
\begin{equation}
    P_{\mathrm{acc}}
    =
    \min\!\left(1, e^{-[S(U')-S(U)]}\right)~.
\end{equation}
In early lattice gauge simulations, such proposals often modified a single link or a small set of links at each step. While conceptually simple and straightforward to implement for pure gauge simulations, local update schemes suffer from strong autocorrelations as the correlation length increases, leading to critical slowing down near the continuum limit. With dynamical fermions, the nonlocal dependence of $\det D[U]$ renders naive local Metropolis proposals computationally inefficient.

Overrelaxation updates~\cite{Adler:1987ce,Adler:1981sn} provide a complementary strategy by constructing deterministic, approximately microcanonical transformations that preserve (or nearly preserve) the local action. In the formulation of Ref.~\cite{Brown:1987rra}, the overrelaxed heat-bath algorithm can be continuously interpolated between the conventional stochastic heat bath ($\omega=1$) and a microcanonical limit ($\omega=2$), in which the update becomes a deterministic constant-action reflection. Such updates move configurations along (nearly) constant-action surfaces and thereby reduce the diffusive, random-walk behavior characteristic of purely stochastic local updates. However, purely deterministic overrelaxation updates were observed to produce incorrect expectation values due to lack of ergodicity~\cite{Brown:1987rra}. For this reason, overrelaxation steps must be alternated with conventional stochastic updates (such as heat-bath or Metropolis sweeps) to ensure ergodicity of the Markov chain. Although highly effective for pure gauge theories, overrelaxation alone does not incorporate the nonlocal fermion determinant and therefore does not address the fundamental computational challenges posed by dynamical fermions. It thus does not constitute a complete algorithm in that setting.

One of the dominant methods for simulations with dynamical fermions is Hybrid Monte Carlo (HMC)~\cite{Duane:1987de}, which introduces an extended phase space in order to construct global proposals. Auxiliary conjugate momenta $P$ are assigned to the gauge fields, and an artificial Hamiltonian is defined as
\begin{equation}
    H(P,U)=\frac{1}{2}P^2+S(U)~,
\end{equation}
so that the joint distribution in phase space factorizes as $\Pi(P,U)\propto e^{-H(P,U)}=e^{-\frac12 P^2}e^{-S(U)}$. Because the kinetic term depends only on $P$ while the potential term depends only on $U$, integrating out the Gaussian momenta reproduces the desired gauge-field distribution $\pi[U]\propto e^{-S(U)}$. Hamilton's equations are then integrated along fictitious molecular-dynamics trajectories using a reversible and symplectic integrator. In the limit of exact integration the Hamiltonian would be conserved and $\Delta H=0$, implying unit acceptance; in practice, discretization errors lead to small violations of energy conservation, and a final Metropolis accept--reject step with probability $\min(1,e^{-\Delta H})$ restores exact sampling.

The essential insight of HMC is that it resolves the usual tension between sampling efficiency and acceptance rate. In conventional local updates, small changes yield high acceptance but explore configuration space slowly, whereas large updates typically generate large $\Delta S$ and are rejected. In contrast, HMC produces global, collective changes in $U$ through deterministic molecular-dynamics evolution that approximately conserves the Hamiltonian. The acceptance rate is therefore governed by the integrator error in $\Delta H$, rather than directly by the magnitude of $\Delta S$, allowing large moves in configuration space while maintaining high acceptance. Combined with stochastic momentum refreshment to ensure ergodicity, this mechanism greatly suppresses random-walk behavior and makes HMC substantially more efficient than purely local algorithms for simulations with dynamical fermions.

An alternative class of algorithms is based on stochastic differential equations, most notably Langevin dynamics~\cite{Parisi:1980ys}. In this approach, the gauge fields are evolved in a fictitious stochastic time $t$ according to the Langevin equation
\begin{equation}
    dU_t = -\nabla S[U_t]\, dt + \sqrt{2}\, dW_t~,
\end{equation}
where the first term is a deterministic \emph{drift} term proportional to the functional gradient of the action, driving the system toward regions of lower action, and $dW_t$ denotes a Wiener process representing Gaussian white noise. The noise satisfies $\langle dW_t \rangle = 0$ and $\langle dW_t dW_{t'} \rangle = \delta(t-t') dt$, ensuring a balance between deterministic relaxation and stochastic diffusion. The probability density $P(U,t)$ for configurations evolving under this stochastic process obeys the associated Fokker--Planck equation,
\begin{equation}
    \frac{\partial P(U,t)}{\partial t}
    =
    \nabla \!\cdot\!
    \left(
        \nabla S[U]\, P(U,t)
        +
        \nabla P(U,t)
    \right),
\end{equation}
whose stationary solution is precisely $P_{\mathrm{eq}}(U)\propto e^{-S(U)}$. Thus, in continuous stochastic time, Langevin evolution samples the desired path-integral measure without the need for a Metropolis accept--reject step. In practice, however, the stochastic differential equation must be discretized in finite time steps, typically leading to systematic step-size errors unless a corrective accept--reject mechanism is introduced, as in the stochastic quantization framework. While conceptually simple and local in structure, and appealing due to the absence of global momentum variables, Langevin-type algorithms are generally less efficient and less robust than HMC for real-action lattice QCD simulations, particularly at fine lattice spacings.

More recently, generalized MCMC strategies have been explored to mitigate critical slowing down and topological freezing in lattice gauge theory. While detailed balance is sufficient for correctness, it is not strictly necessary, and nonreversible Markov chains that preserve global balance have been shown in statistical sampling theory to reduce diffusive behavior by breaking time-reversal symmetry~\cite{turitsyn2011irreversible}. In the lattice context, algorithmic modifications aimed at improving topological tunneling and long-distance decorrelation include, for example, open boundary conditions~\cite{Luscher:2010we} and related developments~\cite{Schaefer:2010hu}. Although fully developed irreversible formulations for four-dimensional QCD remain under active investigation, such approaches reflect a broader effort to alter the dynamical structure of the Markov process in order to alleviate critical slowing down without changing the target distribution. Another direction introduces learned, invertible transformations of the gauge fields, such as normalizing flows~\cite{Albergo:2019eim,Albergo:2021vyo}, which map configurations to a latent space where the distribution is simpler and proposals are more global. In such approaches, an invertible map $U \mapsto \tilde U$ with computable Jacobian is used to reshape the sampling manifold, and a Metropolis correction step ensures exactness of the resulting Markov chain. By effectively preconditioning configuration space and flattening energy barriers, these methods aim to enhance mixing and reduce autocorrelation times. Although promising and actively investigated in lattice field theory contexts, they introduce additional algorithmic and computational complexity and remain under ongoing development.

In summary, lattice ensemble generation can be understood as the construction of an ergodic Markov process with invariant distribution $\pi[U] \propto e^{-S[U]}$. Local Metropolis-type algorithms emphasize simplicity but suffer from critical slowing down; Langevin dynamics offers a stochastic differential perspective with limited practical efficiency; and Hybrid Monte Carlo, through its global Hamiltonian evolution and exact accept-reject correction, has emerged as the standard framework for dynamical fermion simulations. Ongoing algorithmic research continues to address the remaining challenges of scaling toward finer lattice spacings and larger physical volumes.

\section{Benchmark Calculations: Axial Charge}

One example is worth a thousand arguments. In this section, a benchmark calculation is presented to demonstrate the efficacy of lattice QCD and the control of relevant systematics. We focus on the nucleon axial charge $g_A$, which has historically shown tension with experiment mainly due to excited-state contaminations. This benchmark shows that these systematic effects can be quantified and controlled, yielding a reliable determination of a key single-hadron matrix element.

\subsection{Nucleon Axial Charge}

The nucleon axial charge, $g_A$, is a fundamental quantity in hadron structure and weak interactions, governing neutron $\beta$ decay and providing a precision benchmark for lattice QCD. Experimentally, the ratio of the nucleon's axial to vector charges, $g_A/g_V$, is known at the sub-percent level from polarized ultracold-neutron measurements, including the UCNA result, $1.2772(20)$~\cite{UCNA:2012fhw,UCNA:2017obv}, and the PERKEO II result, $1.2761^{+14}_{-17}$~\cite{Mund:2012fq}. In the Standard Model, $g_V$ is protected by vector-current conservation and is unity up to small isospin-breaking corrections~\cite{Ademollo:1964sr,Donoghue:1990ti}; consequently, $g_A/g_V$ directly constrains the axial structure of the nucleon.

\begin{figure}[th!]
    \centering
    \includegraphics[width=.6\linewidth]{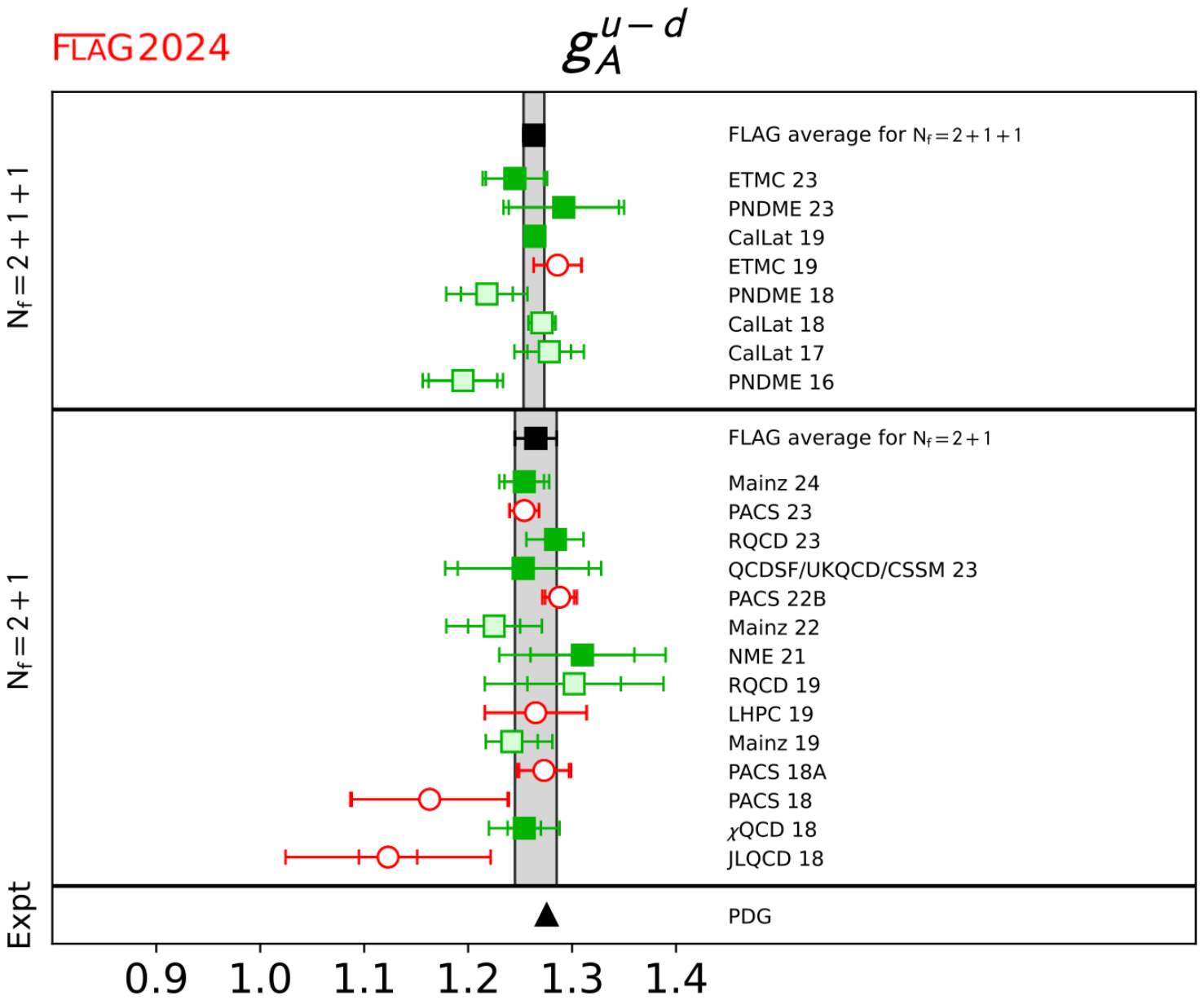}
    \caption[Lattice results and FLAG averages for the isovector axial charge $g_A^{u-d}$]{Lattice results and FLAG averages for the isovector axial charge $g_A^{u-d}$ for 2 + 1 and 2 + 1 + 1 flavour calculations from Ref~\cite{FlavourLatticeAveragingGroupFLAG:2024oxs}. Also shown is the experimental result as quoted in the PDG~\cite{ParticleDataGroup:2022pth}. 
    \label{fig:flag_ga}}
    \end{figure}

Historically, lattice determinations of $g_A$ tended to lie below experiment. Recent calculations, however, have reached a level of precision and systematic control sufficient for inclusion in the latest FLAG review~\cite{FlavourLatticeAveragingGroupFLAG:2024oxs}. The current status of lattice results and FLAG averages for the isovector axial charge $g_A^{u-d}$ in $2+1$ and $2+1+1$ flavor simulations is shown in Fig.~\ref{fig:flag_ga}. As shown in the figure, there remains some tension in the recent lattice calculations. It is now broadly understood that excited-state contamination in nucleon three-point functions is a leading source of the lattice calculations. The analysis of Ref.~\cite{He:2021yvm} was therefore designed to quantify this contamination and to establish a robust extraction strategy for $g_A$.

The calculation was performed on an ensemble with lattice spacing $a \approx 0.09~\mathrm{fm}$ and pion mass $m_\pi \approx 310~\mathrm{MeV}$, using Möbius domain-wall valence fermions on a Highly Improved Staggered Quark (HISQ) sea. The approximate chiral symmetry of the valence action implies that the axial and vector renormalization constants are equal up to small corrections, allowing the renormalized axial charge to be obtained from the ratio of bare matrix elements. This reduces renormalization-related systematics and isolates excited-state effects as the dominant uncertainty.

A central feature of this calculation is the use of a large set of source--sink separations, $t_{\mathrm{sep}}/a \in \{2,3,\dots,14\}$, corresponding to physical separations between approximately $0.18$ and $1.22$ fm. Conventional nucleon matrix-element calculations typically employ only a few values of $t_{\mathrm{sep}}$ concentrated near $1$ fm or larger, where statistical noise becomes severe. In contrast, the present strategy exploits precise short- and intermediate-time data combined with controlled multi-state fits.

The lattice QCD calculation of $g_A$ and $g_V$ relies on the analysis of two-point and three-point correlation functions. The two-point function, $C_2(t)$, encodes the propagation of a nucleon state from source to sink, defined as
\begin{eqnarray}
    &C_2\left(t_{\mathrm {sep }}\right) =\sum_{\vec{x}}\langle\Omega| N\left(t_{\mathrm {sep }}, \vec{x}\right) N^{\dagger}(0, \vec{0})|\Omega\rangle \\
    &=\sum_{n=0}^{\infty}\left|z_n\right|^2 e^{-E_n t_{\mathrm {sep }}}  =\left|z_0\right|^2 e^{-E_0 t_{\mathrm {sep }}}\left[1+\sum_{n=1}^{\infty}\left|r_n\right|^2 e^{-\Delta_{n 0} t_{\mathrm {sep }}}\right] ~.
    \label{eq:ga_2pt}
\end{eqnarray}
In this expression, we assume that the overlap factors, $z_n=\langle\Omega|N|n\rangle$, used to create ($N^\dagger$) and annihilate ($N$) the states with quantum numbers of the nucleon from the vacuum ($|\Omega\rangle$) are conjugate to each other.
\footnote{We are using the non-relativistic normalization $\langle n | m\rangle = \delta_{nm}$ and $\mathbf{1} \equiv | \Omega \rangle \langle \Omega | +\sum_{n=0}^\infty |n\rangle \langle n|$.
We assume that the contributions from the thermal terms that are exponentially suppressed from the full temporal extent are sufficiently suppressed that they can be ignored.}
In the last line, we have defined the energy splitting and ratio of overlap factors:
\begin{eqnarray}
\Delta_{mn} = E_m - E_n~,\quad r_n = \frac{z_n}{z_0}\, .
\end{eqnarray}
The three-point function, $C_{\Gamma}(t_{\mathrm{sep}},\tau)$, includes an insertion of the axial or vector current at an intermediate time $\tau$. In the limit of zero momentum and zero-momentum transfer, it is defined as
\begin{eqnarray}
    C_\Gamma(t_{\rm sep},\tau)
    &=& \sum_{\vec{y},\vec{x}}
    \langle \Omega| N(t_{\rm sep},\vec{y})\, j_\Gamma(\tau,\vec{x})\, N^\dagger(0,\vec{0}) | \Omega\rangle \\
    &=& \sum_n |z_n|^2\, g^\Gamma_{nn}\, e^{-E_n t_{\rm sep}} 
    + 2\sum_{n<m} z_n z^\dagger_m\, g^\Gamma_{nm}\,
    e^{-(E_n +\frac{\Delta_{mn}}{2}) t_{\rm sep}} \\
    &&\quad \times 
    \cosh \left[ \Delta_{mn}\left(\tau -\frac{t_{\rm sep}}{2}\right)\right] ~,
    \label{eq:ga_3pt}
\end{eqnarray}
where the matrix elements of interest are given by
\begin{equation}
    g^\Gamma_{mn} = \langle m| j_\Gamma |n\rangle \, .
\end{equation}
In this limit, the correlation functions are real, so $z^\dagger_n = z_n$ and $g^\Gamma_{nm} = g^\Gamma_{mn}$.

The axial and vector matrix elements can be extracted from the ratio
\begin{equation}
    R_{\Gamma}(t_{\mathrm{sep}},\tau) = \frac{C_{\Gamma}(t_{\mathrm{sep}},\tau)}{C_2(t_{\mathrm{sep}})} ~,
\end{equation}
where $\Gamma = A_3$ or $V_4$ denotes the axial or vector current, $C_{\Gamma}$ is the three-point function with current insertion at time $\tau$, and $C_2$ is the nucleon two-point function. In the limit of large $t_{\mathrm{sep}}$ and $\tau$ away from the source and sink, this ratio approaches the desired matrix element $g_A$ or $g_V$. Physically, deviations from the asymptotic value arise from intermediate excited states, most prominently $N\pi$ scattering states and their interactions. A key observation of Ref.~\cite{He:2021yvm} is that transition matrix elements connecting the ground state to excited states dominate the contamination. These contributions decay parametrically as $e^{-\Delta t_{\mathrm{sep}}/2}$, whereas diagonal excited-state contributions scale as $e^{-\Delta t_{\mathrm{sep}}}$, as shown in the Eq.~(10) of Ref.~\cite{He:2021yvm}. As a consequence, moderately increasing $t_{\mathrm{sep}}$ does not eliminate the leading systematic.

To further suppress excited states, the analysis also employs Feynman--Hellmann (FH) correlators constructed from summed insertion correlation functions. The FH correlator is defined as
\begin{eqnarray}
    &\mathrm{FH}_{\Gamma}\left(t_{\mathrm{sep}}, \tau_c, d t\right) = \sum_{\tau=\tau_c}^{t_{\mathrm{sep}}-\tau_c} \frac{R_{\Gamma}\left(t_{\mathrm{sep}}+d t, \tau\right)-R_{\Gamma}\left(t_{\mathrm{sep}}, \tau\right)}{d t}  =\frac{1}{d t}\left[\frac{\Sigma_{\Gamma}\left(t_{\mathrm{sep}}+d t, \tau_c\right)}{C_2\left(t_{\mathrm{sep}}+d t\right)}-\frac{\Sigma_{\Gamma}\left(t_{\mathrm{sep}}, \tau_c\right)}{C_2\left(t_{\mathrm{sep}}\right)}\right] ~,\\
    &\mathrm{FH}_{\Gamma}\left(t_{\mathrm{sep}}, \tau_c\right) \equiv \mathrm{FH}_{\Gamma}\left(t_{\mathrm{sep}}, \tau_c, d t = 1\right) ~.
\end{eqnarray}
In this formulation, the dominant transition contributions are suppressed by the full energy gap, as shown in the Eq.~(12) of Ref.~\cite{He:2021yvm}, resulting in significantly improved ground-state isolation at shorter separations. 
\begin{figure}[th!]
    \centering
    \includegraphics[width=.4\linewidth]{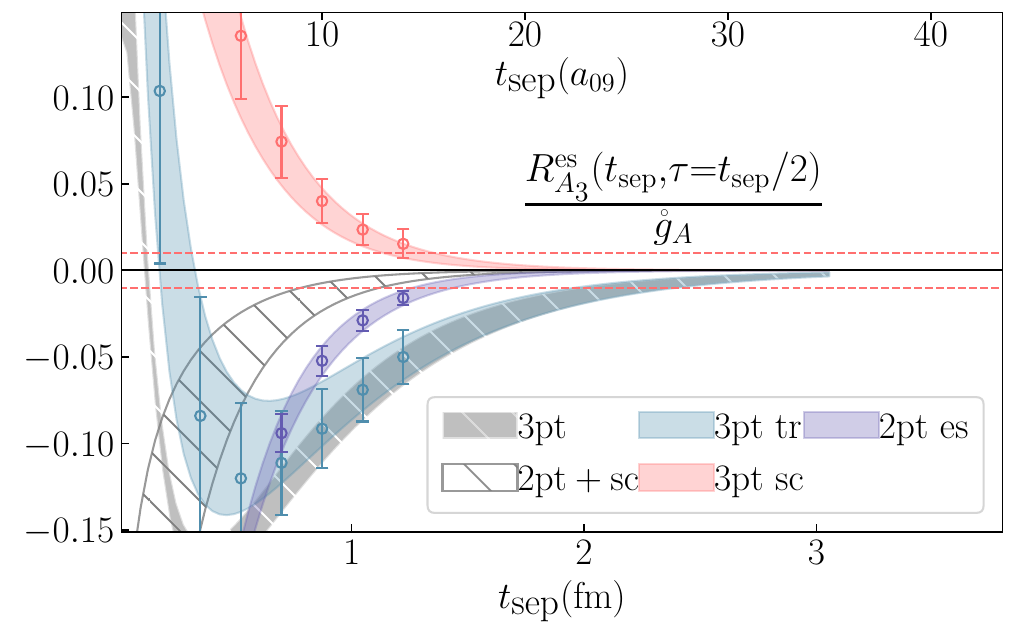}
    \includegraphics[width=.4\linewidth]{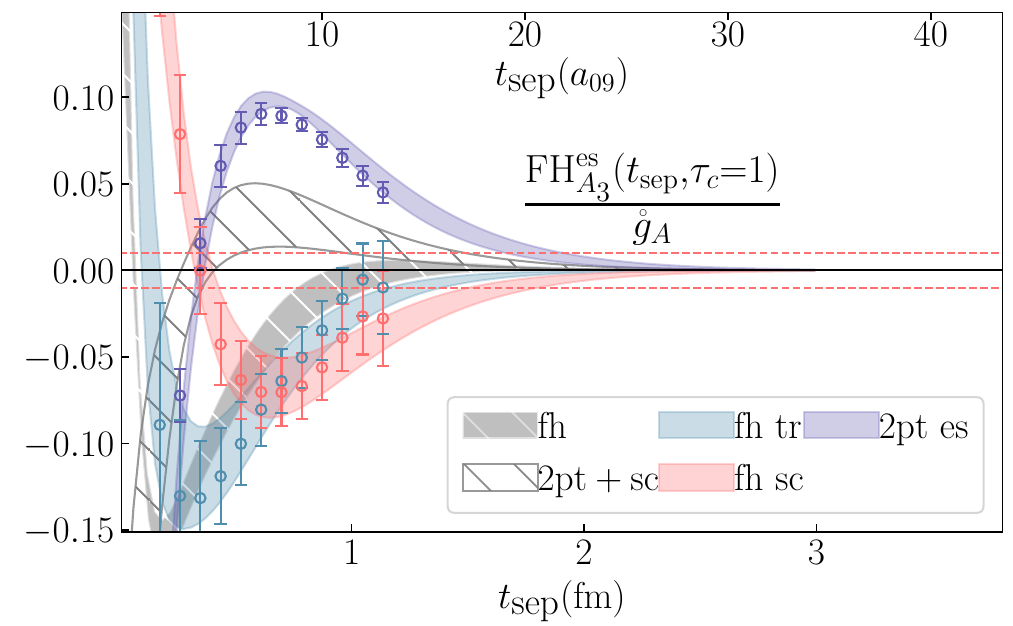}
    \caption[Decomposition of excited state contributions to the axial charge extraction]
    {
    Ratio of excited state contributions to the ground state, $g_A$, for $R_{A_3}(t_{\rm sep},\tau)$ (left, 3pt) and ${\rm FH}_{A_3}(t_{\rm sep},\tau_c=1)$ (right, fh).
    Each panel contains three types of data points: the $n$-to-$n$ scattering (sc) excited states, $C_\Gamma^{\rm sc}/(g_A C_2)$,
    the $n$-to-$m$ transition (tr) excited states, $C_\Gamma^{\rm tr}/ (g_A C_2)$,
    and the excited states from the two-point (2pt es) function, $-C_2^{\rm es}/C_2$ (and the corresponding Feynman-Hellmann versions of these ratios in the right figure).
    The decomposition of the numerical data into these three classes of excited state contamination depends upon our posterior reconstruction of the correlation functions (2pt, 3pt and FH).
    Because we are looking down the middle of $R_{A_3}$ (along $\tau=t_{\rm sep}/2$), we only show data at even $t_{\rm sep}$s in the left panel.
    In addition to plotting the corresponding posterior contribution from each class of excited state, we plot two hatched bands.
    The white hatched band is the sum of the 3pt sc and 2pt es contributions, which are observed to largely cancel.
    The gray hatched band is the sum of all excited state contributions (the full posterior distribution, minus the ground state, normalized by the ground state).
    The 3pt sc contribution is positive for $R_{A_3}(t_{\rm sep},\tau)$ and it becomes negative for ${\rm FH}_{A_3}(t_{\rm sep},\tau_c=1)$ at $t_{\rm sep}>0.5$~fm.
    The opposite is true for the 2pt es contribution.
    The sum of the scattering and two-point excited state contributions is opposite in sign to the transition excited state contributions, unlike for $R_{A_3}(t_{\rm sep},\tau)$ in which they are the same sign, leading to an even stronger suppression of the total excited state contributions beyond the expected suppression, as described in the text. The horizontal red dashed lines represent the threshold of a 1\% contribution of excited state contamination to the ground state value. 
    \label{fig:es_decomposition}}
\end{figure}
Detailed decompositions of excited-state contamination in ratio and FH correlators are presented in Fig.~\ref{fig:es_decomposition}. In the left panel, we see that the scattering excited state contribution (the upper (red) band/data) and the contribution from the excited states coming from the two-point function (the middle (purple) band/data) are roughly equal and opposite in sign.
The hatched curve, which is mostly white with gray hash lines, is the sum of these two contributions, and it lies between them.
We observe that the sum of these two sources of excited state contributions decay to a 1\% correction (the horizontal dashed (red) lines) at $t_{\rm sep}\approx 1$~fm.
In contrast, the transition (3pt tr) excited states depicted by the lowest (blue) band/data do not decay to the 1\% level until $t_{\rm{sep}}\approx2.2$~fm. 
In the right panel, we show the same excited state breakdown for the Feynman-Hellmann correlation function.
The sign of the scattering and two-point excited state contributions both change compared to $R_{A_3}(t_{\rm sep},\tau=t_{\rm sep}/2)$.
It is also interesting to note that the magnitude of the transition excited state contributions at $t_{\rm sep}\approx1$~fm go from $\approx7.5\%$ for $R_{A_3}(t_{\rm sep},\tau=t_{\rm sep}/2)$ to $\approx2.5\%$ for $FH_{A_3}(t_{\rm sep},\tau_c=1)$.
Besides, for $R_{A_3}(t_{\rm sep},\tau=t_{\rm sep}/2)$, the sum of the scattering and two-point excited state contributions is the same sign as the sum of transition excited states, while for FH$_{A_3}(t_{\rm sep},\tau_c=1)$ it is the opposite sign at intermediate and large $t_{\rm sep}$.
Thus, the total excited state contamination decays to the 1\% level for the FH correlation function at $t_{\rm sep}\approx1$~fm while this does not happen until $t_{\rm sep}>2$~fm for $R_{A_3}(t_{\rm sep},\tau=t_{\rm sep}/2)$, which shows the significant improvement of the FH construction in suppressing excited states.

The extraction of $g_A$ is performed through a fully correlated joint analysis of the two-point function, the axial and vector three-point ratios, and the corresponding FH correlators. Rather than determining the excited-state spectrum from the two-point function alone, all correlation functions are fitted simultaneously. This approach is essential because local three-quark interpolating operators do not uniquely isolate individual excited states; fitting correlators independently can therefore introduce biases in the ground-state matrix element. 
\begin{figure}[th!]
    \centering
    \includegraphics[width=.4\linewidth]{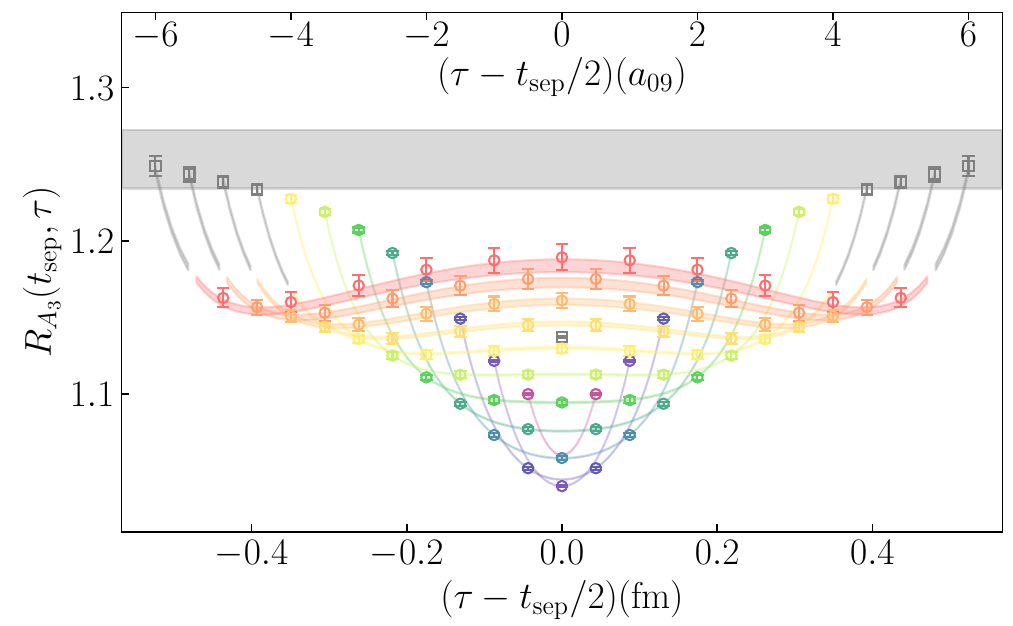}
    \includegraphics[width=.4\linewidth]{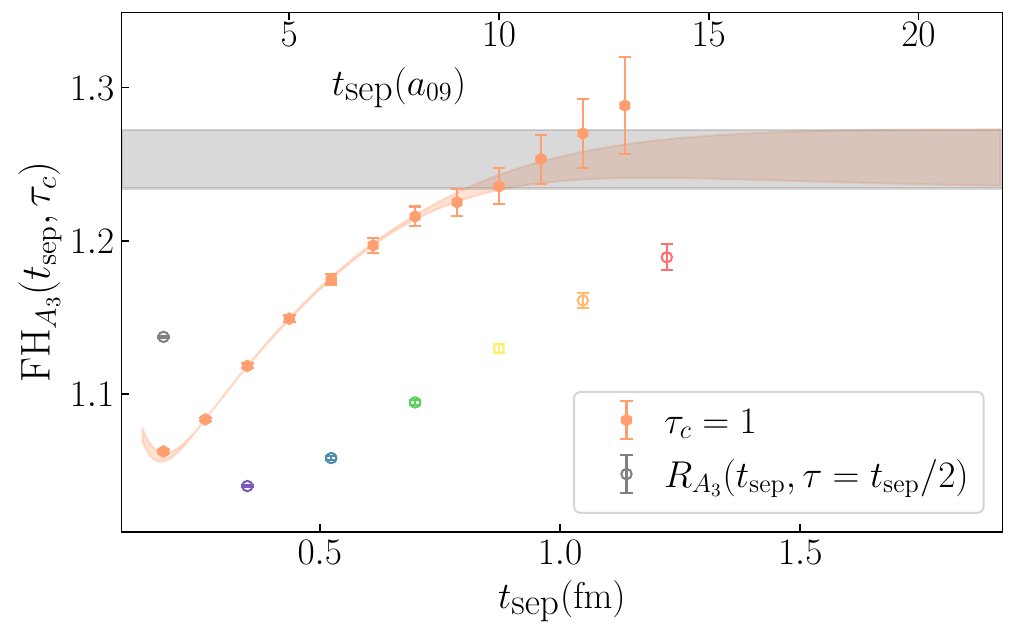}
    \caption[Psychedelic Moose and FH ratio plot for the axial charge extraction]
    {
    Left: (Psychedelic Moose/Water Buffalo plot)
    We plot the numerical results of $R_{A_3}(t_{\rm sep},\tau)$ for $t_{\rm sep}=2$ (the single gray point in the middle) to $t_{\rm sep}=14$, the top (red) dataset.
    In addition, we plot the resulting posterior description of the correlation function from our 5-state fit as the (correspondingly colored) fit bands.
    The (gray) squares in the upper left/right of the ``moose antlers'' are not included in the analysis, as indicated by the break in the fit band from the inner region.
    The horizontal (gray) band is the ground state matrix element, $\mathring{g}_A$.
    Right: The ${\rm FH}_{A_3}(t_{\rm sep},\tau_c=1)$ numerical data (filled orange symbols) is plotted along with the posterior description of the correlation function from the joint analysis.
    As a comparison, for even values of $t_{\rm sep}$, we plot $R_{A_3}(t_{\rm sep},\tau=t_{\rm sep}/2)$ with open (colored) symbols (the numcerical data in the middle of the ``moose'').
    The time-axis is converted from lattice units (top) to fm (bottom) using our scale-setting~\cite{Miller:2020evg}.
    \label{fig:psychedelic_moose}}
\end{figure}
The joint fit results as well as the data points are presented in Fig.~\ref{fig:psychedelic_moose}. In the left panel (the ``Psychedelic Moose''), we plot the numerical results for the ratio of the three-point function generated with the $A_3=\bar{q}\gamma_3\gamma_5 \tau_3 q$ current, divided by the two-point function at the given value of $t_{\rm sep}$.
The fit quality is good and visually one can see that the model accurately describes the numerical results over the full range of $t_{\rm sep}$ and $\tau$ used in the analysis.
The horizontal (gray) band is the value of the ground state matrix element $\mathring{g}_A$.
In the right panel we plot the ${\rm FH}_{A_3}(t_{\rm sep},\tau_c=1)$ data that are used in the joint fit as well as the resulting posterior distribution of this correlation function.

An important aspect of the analysis is the demonstration of stability under systematic variations. The extracted value of $g_A$ remains consistent when varying the number of excited states included in the fit, modifying the prior model for energy gaps, truncating early- or late-time data, or restricting the analysis to subsets of source--sink separations. This stability indicates that the excited-state uncertainty is quantitatively captured within the fit.

Beyond the determination of $g_A$ on this ensemble, this study establishes a methodological framework for nucleon structure calculations. It shows that reliable control of excited states requires a sufficiently large lever arm in $t_{\mathrm{sep}}$ combined with joint multi-correlator fits, and that FH correlators offer a practical avenue for improved ground-state isolation.

%% file: Chapter4.tex
%Chapter 1

\renewcommand{\thechapter}{4}

\vspace*{-1.35in}

\chapter{Systematics Controlled Calculations of Partonic Physics from Lattice QCD}
\label{ch:systematics}

\vspace*{-.5in}

\section{Large Momentum Effective Theory}

If we think of the operator definition of PDFs, it is a light-cone correlation function with light-like separation. However, any real-time evolution is not directly accessible in lattice QCD due to the sign problem. More specifically, if we try to do the Wick rotation on a light-like separation, we will end up with a null separation in Euclidean space, which is not sufficient to define a nontrivial correlation function. Therefore, the early stage applications of lattice QCD to partonic physics were limited to the calculations of low moments of PDFs~\cite{Lin:2017snn}, which can be expressed as local operators. However, the calculation of higher moments is hindered by the power-divergent mixing with lower-dimensional operators, making it impractical to access the full $x$-dependence of PDFs.

A few years ago, a general approach to calculate $x$-dependent parton distributions based on Feynman's original idea about partons: They are the infinite-momentum limit of static properties of the proton at large momentum, and therefore are intrinsically Euclidean quantities accessible through lattice QCD~\cite{Feynman:1969ej}. This approach is known as LaMET~\cite{Ji:2013dva,Ji:2014gla,Ji:2020ect,Ji:2024oka}, since a rigorous connection between the infinite-momentum frame partons and quarks and gluons at a finite momentum requires the matching of the UV modes with large momentum in effective field theory and systematic power counting. The key idea of LaMET is to construct a class of Euclidean correlation functions, called quasi-distributions, which are defined with spatially separated operators and can be computed on the lattice. These quasi-distributions depend on the hadron momentum $P^z$ and approach the light-cone distributions in the infinite-momentum limit. By performing a perturbative matching calculation, one can relate the quasi-distributions to the physical PDFs, allowing for a systematic extraction of partonic physics from lattice QCD calculations at finite momentum. So far, LaMET has been widely applied to calculate various partonic observables, including PDFs, GPDs, TMDs, as well as the light-cone wave functions~\cite{Gao:2026hix,Xiong:2013bka,Lin:2014zya,Ji:2014hxa,Ji:2014lra,Sufian:2014jma,Ji:2015qla,Xiong:2015nua,Alexandrou:2016eyt,Chen:2016utp,Lin:2016qia,Yang:2016nfc,Yang:2016plb,Alexandrou:2017dzj,Alexandrou:2017huk,Alexandrou:2017qpu,Chen:2017mie,Chen:2017mzz,Constantinou:2017sej,Ishikawa:2017iym,Ji:2017oey,Ji:2017rah,Lin:2017ani,Stewart:2017tvs,Xiong:2017jtn,Zhang:2017bzy,Zhang:2017zfe,Alexandrou:2018eet,Alexandrou:2018pbm,Alexandrou:2018yuy,Chen:2018xof,Ebert:2018gzl,Fan:2018dxu,Izubuchi:2018srq,Ji:2018hvs,LatticeParton:2018gjr,Lin:2018pvv,Liu:2018hxv,Liu:2018tox,Zhang:2018diq,Zhang:2018nsy,Zhang:2018rls,Zhao:2018fyu,Alexandrou:2019dax,Alexandrou:2019lfo,Chai:2019rer,Chen:2019lcm,Constantinou:2019vyb,Ebert:2019tvc,Liu:2019urm,Shanahan:2019zcq,Wang:2019msf,Zhang:2019qiq,Alexandrou:2020qtt,Alexandrou:2020uyt,Alexandrou:2020zbe,Bhattacharya:2020cen,Bhattacharya:2020jfj,Chai:2020nxw,Chen:2020arf,Chen:2020iqi,Chen:2020ody,Ebert:2020gxr,Fan:2020nzz,Gao:2020ito,Hua:2020gnw,Ji:2020brr,Ji:2020ect,Ji:2020jeb,LatticeParton:2020uhz,Lin:2020fsj,Lin:2020rxa,Lin:2020ssv,Shanahan:2020zxr,Shugert:2020tgq,Vladimirov:2020ofp,Zhang:2020dkn,Zhang:2020gaj,Zhang:2020rsx,Alexandrou:2021bbo,Alexandrou:2021oih,Bhattacharya:2021moj,Bhattacharya:2021rua,Constantinou:2021nbn,Dodson:2021rdq,Gao:2021hxl,Gao:2021dbh,LatticePartonLPC:2021gpi,Li:2021wvl,Lin:2021brq,Lin:2021ukf,Scapellato:2021uke,Schlemmer:2021aij,Shanahan:2021tst,Bhattacharya:2022aob,Constantinou:2022fqt,Deng:2022gzi,Ebert:2022fmh,Gao:2022iex,Gao:2022uhg,LatticeParton:2022xsd,LatticeParton:2022zqc,LatticePartonCollaborationLPC:2022myp,LatticePartonLPC:2022eev,Scapellato:2022mai,Schindler:2022eva,Zhang:2022xuw,Alexandrou:2023ucc,Avkhadiev:2023poz,Bhattacharya:2023jsc,Bhattacharya:2023nmv,Bhattacharya:2023tik,Cichy:2023dgk,Deng:2023csv,Gao:2023ktu,Gao:2023lny,Holligan:2023jqh,Holligan:2023rex,Ji:2023pba,LatticeParton:2023xdl,LatticePartonLPC:2023pdv,Lin:2023gxz,Zhao:2023ptv,Liu:2023onm,Avkhadiev:2024mgd,Baker:2024zcd,Bollweg:2024zet,Chen:2024rgi,Cloet:2024vbv,Ding:2024saz,Gao:2024fbh,Good:2024iur,Han:2024min,Holligan:2024umc,Holligan:2024wpv,Ji:2024hit,LatticeParton:2024mxp,LatticeParton:2024vck,LatticeParton:2024zko,Miller:2024yfw,Mukherjee:2024xie,Spanoudes:2024kpb,Zhang:2024omt,Bollweg:2025ecn,Bollweg:2025iol,Han:2025odf,Holligan:2025ydm,LPC:2025spt,Zhang:2025hvf,Ji:2025mvk,Ji:2015jwa,Chen:2016fxx,Zhang:2018ggy,Braun:2018brg,Wang:2019tgg,Ji:2021uvr,Su:2022fiu,Zhu:2022bja,Yao:2022vtp,Pang:2024sdl,Han:2024cht}. 

In this chapter, we will briefly review the recent developments within LaMET, with a focus on the novel Coulomb gauge formulation of quasi-distributions within LaMET framework. We will also present some example calculations of pion and nucleon partonic distributions using LaMET, including both collinear and TMD structures.

\subsection{Recent Developments in LaMET}

\begin{figure}[t]
    \centering
    \includegraphics[width=0.6\textwidth]{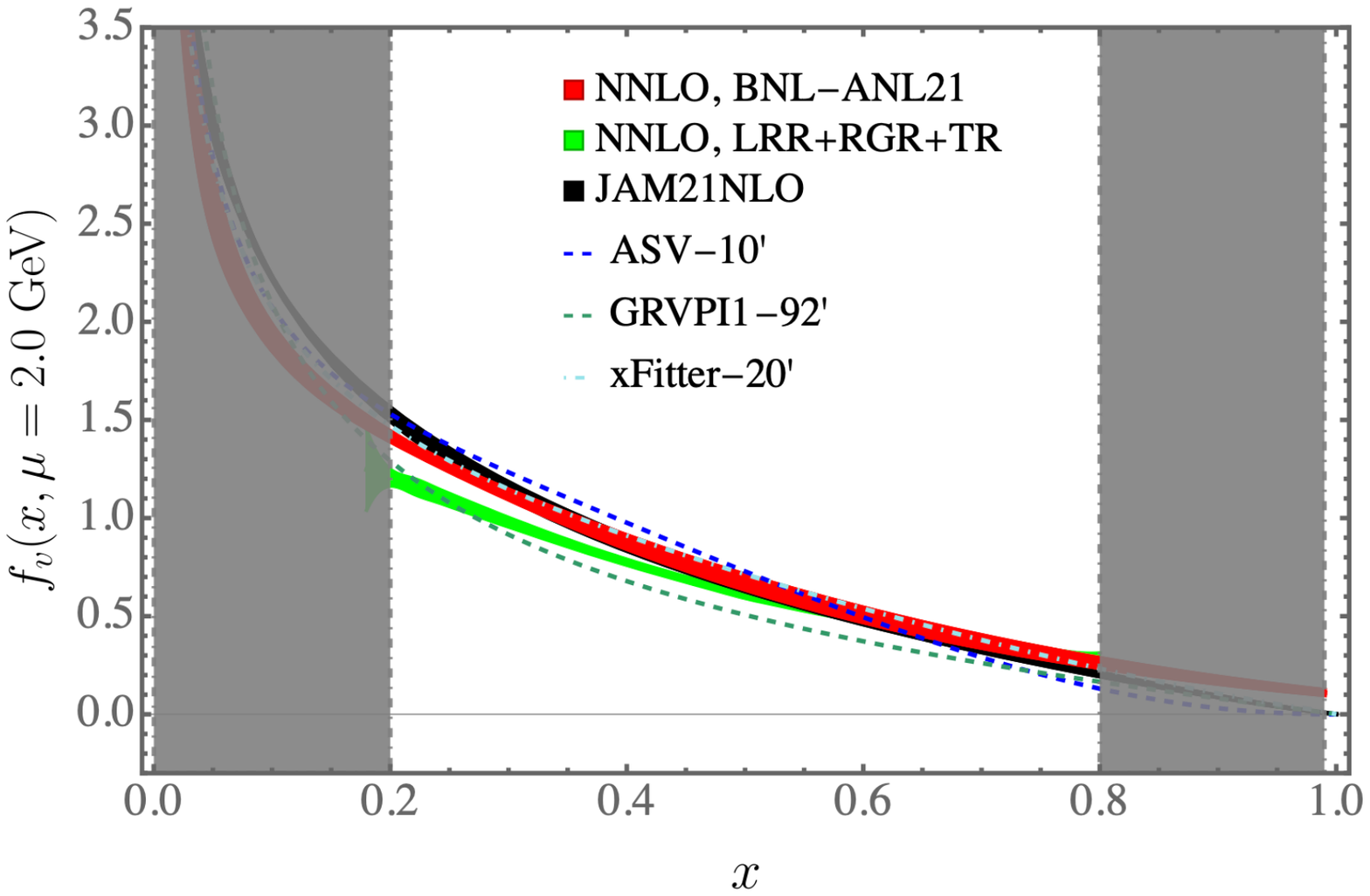}
    \caption[Pion valence quark PDF]{Illustration of a pion valence quark PDF extracted within the LaMET framework via calculations on a lattice with spacing $a=0.04$ fm and pion mass 300 MeV at momentum $P^z = 1.94~\mathrm{GeV}$ using NNLO matching with LRR, RGR, and TR.}
    \label{fig:pion_pdf_precise}
\end{figure}

Since its original proposal, the LaMET has undergone substantial developments. These developments have significantly improved the precision from theoretical side. Firstly, several nonperturbative renormalization strategies were proposed, including regularization-independent momentum-subtraction (RI/MOM) schemes~\cite{Constantinou:2017sej,Stewart:2017tvs,Alexandrou:2017huk,Chen:2017mzz} and ratio-based schemes~\cite{Orginos:2017kos,Braun:2018brg,Li:2020xml,Fan:2020nzz}, as well as the hybrid renormalization framework~\cite{Ji:2020brr}, which leads to the later self-renormalization approach~\cite{LatticePartonLPC:2021gpi}.

In parallel, advances in perturbative matching have played a central role. The introduction of leading renormalon resummation (LRR)~\cite{Holligan:2023rex,Zhang:2023bxs} clarified ambiguities in the matching kernel and eliminated spurious linear power corrections in the LaMET expansion. Higher-order matching calculations~\cite{Li:2020xml,Chen:2020ody,Cheng:2024wyu,Ji:2025mvk} and the implementation of renormalization-group and threshold resummations (RGR and TR)~\cite{Su:2022fiu,Ji:2023pba,Ji:2024hit} further enabled systematic resummation of large logarithms arising in the limits $x \to 0$ and $x \to 1$. As a result, the perturbative matching kernel is now under quantitative control across a wide kinematic range.

With these theoretical ingredients in place, the remaining power corrections suppressed by inverse powers of the hadron momentum can be estimated through controlled infinite-momentum extrapolation. Consequently, modern LaMET calculations achieve well-defined theoretical uncertainties for $x \in [x_{\min}, x_{\max}]$, rendering the framework suitable for precision phenomenology. An illustrative example is shown in Fig.~\ref{fig:pion_pdf_precise}, where the pion valence quark PDF is calculated on a lattice with spacing $a=0.04$ fm and pion mass 300 MeV at momentum $P^z = 1.94~\mathrm{GeV}$ using next-to-next-to-leading order (NNLO) matching supplemented by LRR, RGR, and TR. The lattice prediction exhibits quantitative agreement with global analyses in the intermediate region $0.4 < x < 0.6$, despite relying on entirely independent theoretical and methodological inputs.

\subsection{Quasi Distributions in Coulomb Gauge}

It should be noted that the operator definition of quasi-distributions in LaMET framework is not unique. Instead, all operators projecting out the same light-front physics form a universality class~\cite{Ji:2020ect,Hatta:2013gta}. While can be proven that all operators in the same universality class will yield the same light-front physics in the infinite-momentum limit, they can have different UV structures at finite momentum, and therefore different renormalization and power corrections. In particular, the original formulation of LaMET is based on gauge-invariant nonlocal operators with Wilson lines, which are known to have complicated renormalization properties due to the presence of linear divergences associated with the Wilson lines. A novel formulation of quasi-distributions in Coulomb gauge has been proposed recently~\cite{Gao:2023lny,Zhao:2023ptv}, which avoids the use of Wilson lines and therefore has a much simpler renormalization structure. 

In this formulation, the quasi-distributions are defined as nonlocal quark or gluon correlators in Coulomb gauge without Wilson lines, which can be expressed gauge-invariantly as the correlators of Coulomb-gauge-dressed fields. Thanks to the absence of Wilson lines, the renormalization of these operators is multiplicative and does not involve linear divergences, making it more straightforward to implement on the lattice. The perturbative matching between the Coulomb gauge quasi-distributions and the light-cone distributions has been derived using Soft Collinear Effective Theory (SCET)~\cite{Bauer:2000ew,Bauer:2000yr,Bauer:2001ct,Bauer:2001yt} and verified at one-loop order.

\subsubsection{Gribov Copies}

After the proposal of the Coulomb gauge formulation of quasi-distributions, there are concerns on whether the CG fixing process introduces extra systematic uncertainties that cannot be controlled. The main concern is about the effects from Gribov copies~\cite{Gribov:1977wm,Singer:1978dk}, which are multiple gauge configurations that satisfy the same gauge-fixing condition. 

To understand the origin of these ambiguities more precisely, it is useful to revisit the general framework of gauge fixing in non-Abelian gauge theories. In the path-integral quantization of Yang–Mills theory, the naive functional integration over gauge fields is ill-defined because gauge-equivalent configurations are overcounted. The Faddeev–Popov procedure resolves this redundancy by inserting a gauge-fixing condition together with the associated determinant, thereby isolating one representative from each gauge orbit. Implicit in this construction is the assumption that the gauge condition intersects each gauge orbit uniquely and that the corresponding Faddeev–Popov determinant remains positive. In the context of perturbation theory, one expands around small gauge fields, then the Faddeev–Popov operator is dominated by the Laplacian term (take Landau gauge as an example), which has no zero eigenvalues. Consequently, no nontrivial zero modes arise, and the Gribov copies are irrelevant. However, this assumption is not automatically guaranteed beyond perturbation theory. In particular, the existence of zero modes of the Faddeev–Popov operator is connected to the presence of multiple solutions to the gauge-fixing condition, i.e. Gribov copies~\cite{Vandersickel:2012tz}. 

On the lattice, gauge fixing is implemented numerically, typically by minimizing a gauge-fixing functional. In the case of Coulomb gauge, the condition is imposed independently on each time slice. The minimization procedure generally finds a local minimum of the gauge-fixing functional with a finite tolerance, rather than the absolute minimum. Different local minima correspond to different Gribov copies within the same gauge orbit. The numerical effects introduced by Gribov copies can be categorized into two primary types: lattice Gribov noise and measurement distortion~\cite{Giusti:2001xf}. Lattice Gribov noise arises from residual gauge freedoms and is related to the distribution of the target quantity across all Gribov copies. This noise is indistinguishable from the statistical uncertainties that arise from variations between lattice configurations~\cite{Martinelli:1993dq, Paciello:1994gs}. Consequently, the predominant concern regarding Gribov copies is the resultant measurement distortion. The distortion is related to the choice of the representative within various Gribov copies in gauge-fixing, i.e., the gauge-fixing strategy. A biased strategy can result in deviations from the true value, even though the Gribov noise remains indistinguishable from the statistical uncertainties. For example, selecting the copy with the maximal value of the target quantity as the representative for each configuration will clearly produce a distorted measurement. Therefore, it is necessary to quantify measurement distortion, for which a standard approach within the literature is to take various gauge-fixing strategies for comparative analysis. The resultant discrepancies among these strategies give an estimation of the measurement distortion.

\begin{figure}[!th]
    \centering
    \includegraphics[width=.4\textwidth]{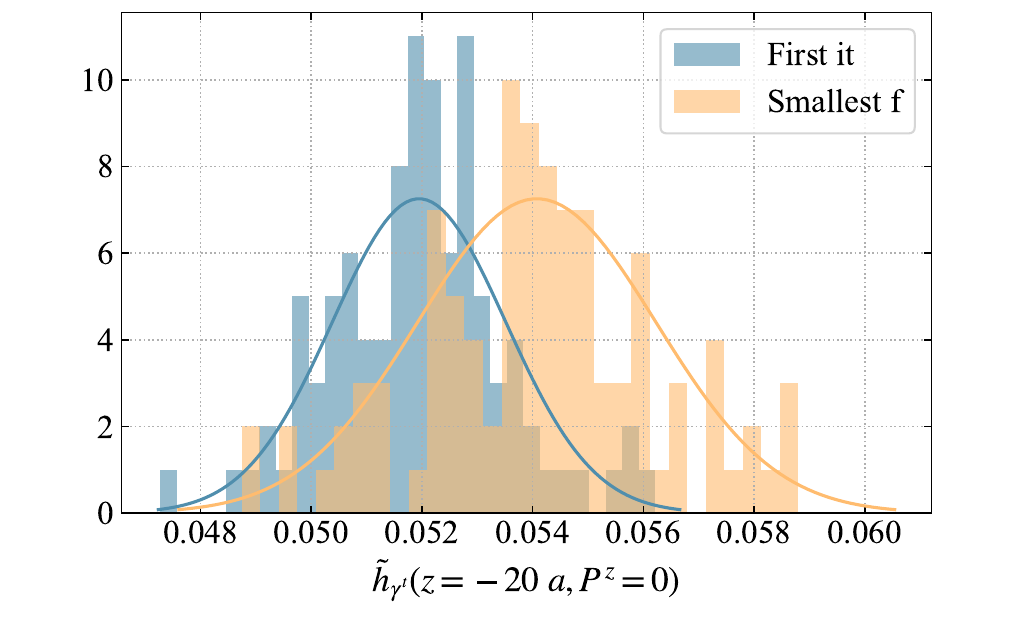}
    \includegraphics[width=.4\textwidth]{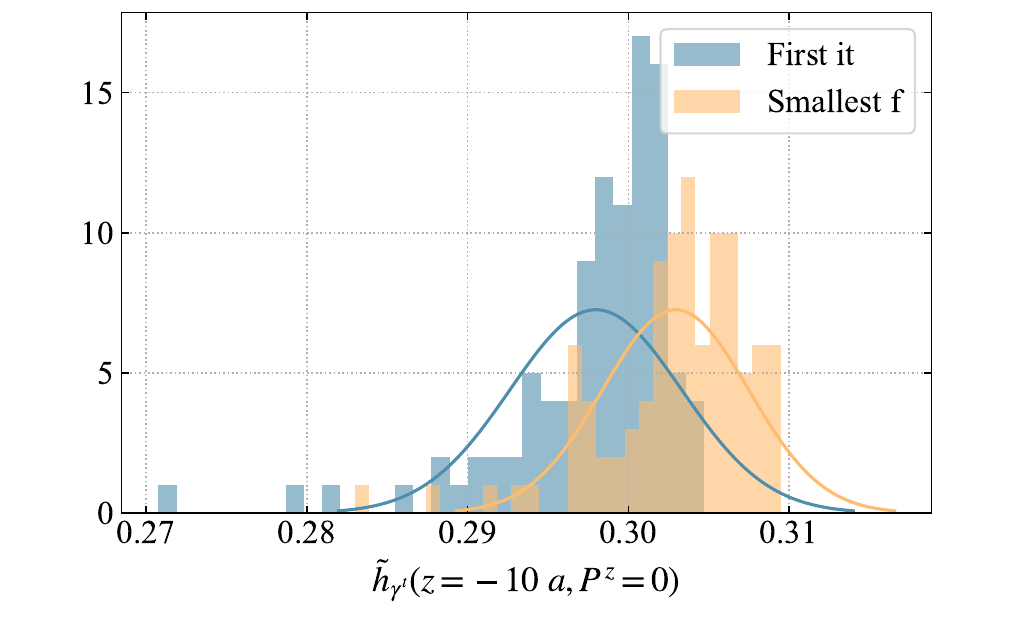}
    \includegraphics[width=.4\textwidth]{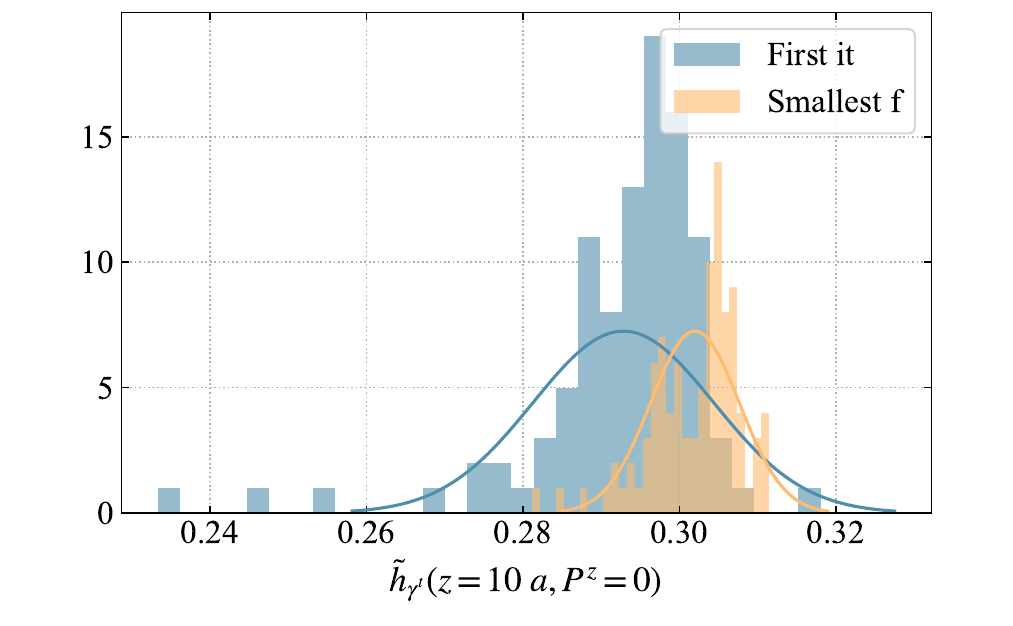}
    \includegraphics[width=.4\textwidth]{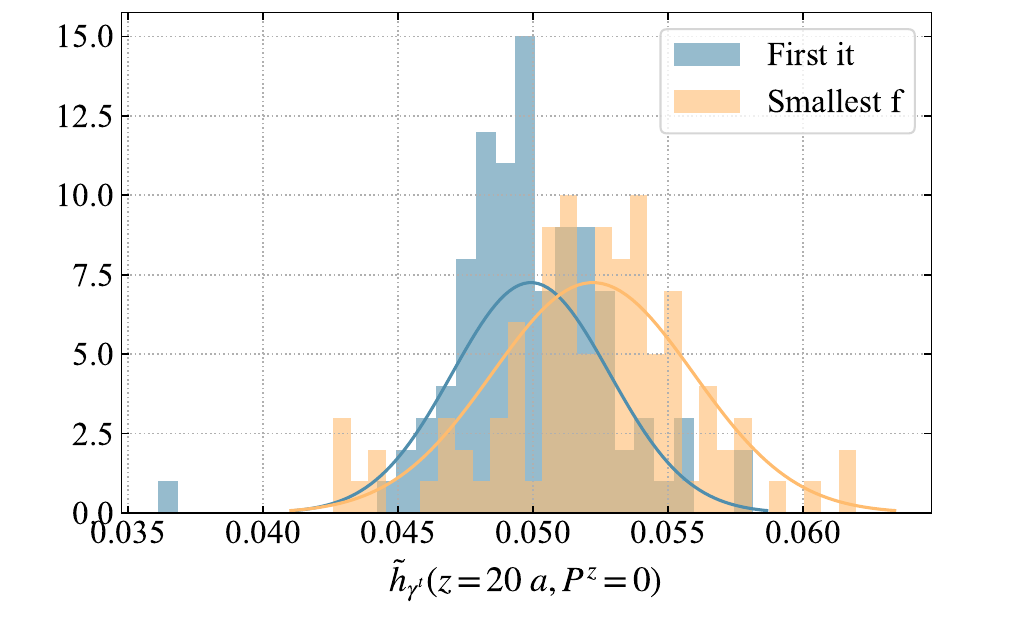}
    \caption[Measurement distortion from Gribov copies]{Histograms of the matrix element $\tilde h_{\gamma^t}(z, P^z=0)$ from the bootstrap samples of the 100 configurations, measured on the two different sets of Gribov copies.\label{fig:gribov_histogram}}  
\end{figure}

Our recent study~\cite{Gao:2024fbh} has investigated this issue by performing a systematic analysis of the Gribov copy effects on the lattice calculation of quark propagator and quasi-distribution matrix elements in Coulomb gauge. Considering the computational cost of the $3+1$ dimensional QCD simulation on the lattice, we study two strategies in this work: 
\begin{enumerate}
    \item {\bf First instance} (``First it"): choosing the first copy obtained in the first Gribov region;
    \item {\bf Smallest functional} (``Smallest f"): choosing the copy with the smallest functional value among all the obtained copies in the first Gribov region.
\end{enumerate}
Using these two strategies, we should obtain two different sets of configurations after fixing the CG. The measurement distortion introduced by Gribov copies is estimated via the comparison between these two sets. As shown in Fig.~\ref{fig:gribov_histogram}, we select four representative separations $z=\{-20,-10,10,20\}a$ and plot histograms of bootstrap samples for the quasi-distribution matrix elements. No statistically significant difference is observed between the two strategies, indicating that any measurement distortion from Gribov copies is smaller than the statistical errors. The study finds that, although Gribov copies can introduce measurable distortions, these are negligible relative to the current statistical precision of the quasi-distribution matrix elements. It is consistent with the argument in Ref.~\cite{Gao:2023lny} and supports the effectiveness of the CG method in the calculation of parton physics. If future calculations achieve substantially smaller statistical uncertainties, however, measurement distortions from Gribov copies will need to be examined more carefully.

\subsection{General Analysis Procedure}

This chapter presents several lattice calculations of partonic observables using the LaMET framework, including the pion DA, nucleon PDF, pion TMDs, and nucleon TMDs. The general analysis workflow common to these studies is summarized below.

\subsubsection{Operator Definition}

In lattice QCD calculations, hadron states are projected out by the interpolating operators. The choice of the interpolating operator is not unique, and different operators can have different overlaps with the ground state and excited states. Therefore, it is important to choose an operator that has a good overlap with the ground state and suppresses the excited state contamination. 

A standard choice for pion interpolating operator is the pseudoscalar operator
\begin{eqnarray}
    \pi_s(\vec{x},t)=\overline{d}_s(\vec{x},t)\gamma_5 u_s(\vec{x},t)~,
    \label{eq:pion_interpolator_standard}
\end{eqnarray}
and for nucleon, a standard interpolating operator is 
\begin{eqnarray}
    N^s_\alpha(\vec{x},t) = \epsilon_{abc} ~ u_{a \alpha}^s (\vec{x}, t)  \left( u^{s}_b (\vec{x}, t)^T ~ C \gamma_5 ~ d_c^s (\vec{x}, t) \right) ~,
    \label{eq:nucleon_interpolator_standard}
\end{eqnarray}
where $C=\gamma^t \gamma^y$ is the charge conjugation operator, the subscript $a, b, c$ are the color indices, $\alpha$ is the spin index, and the superscript $s$ indicates whether the quark field is smeared ($s=S$) or not ($s=P$).

Although the above interpolating operators have good overlap with static pion and nucleon states, they are not necessarily optimal for highly boosted hadrons, which are important in LaMET framework. In the large-momentum limit, hadrons are more naturally described in terms of light-cone dynamics, where the dominant contributions arise from the ``plus'' components of the quark fields. 

For a hadron boosted in the $z$-direction, one can decompose the Dirac field into light-cone components,
\begin{equation}
    \psi = \psi_+ + \psi_- ~, \qquad 
    \psi_\pm = \frac{1}{2} \gamma^\mp \gamma^\pm \psi ~,
\end{equation}
where $\gamma^\pm = \frac{1}{\sqrt{2}}(\gamma^t \pm i \gamma^z)$. In the large $P_z$ limit, the $\psi_+$ component dominates the leading Fock state of the hadron, while $\psi_-$ is suppressed by powers of $M/(E+P_z)$. 

As a result, interpolating operators constructed from the plus components have enhanced overlap with the ground state of a highly boosted hadron. In a recent work~\cite{Zhang:2025hyo}, new interpolating operators were proposed and referred to as \emph{kinematically-enhanced} interpolating operators. The central idea is to construct operators aligned with the boost direction so that their overlap with the hadron ground state acquires a kinematic enhancement at large momentum.

For the pion, instead of the standard pseudoscalar operator $\bar{d}\gamma_5 u$, one may consider the axial-vector operator $\bar{d} \gamma^\mu \gamma_5 u$ with $\mu$ chosen parallel to the boost direction. In the large-$P_z$ limit, this operator couples predominantly to the leading light-cone Fock component built from the plus spinor fields. Its overlap with the boosted ground state is enhanced by a factor proportional to $P_\mu$, while the variance of the correlator does not receive a corresponding kinematic enhancement. Consequently, the signal-to-noise ratio is parametrically improved at large momentum. 

A similar construction applies to the nucleon case, where interpolators involving $\gamma^\mu \gamma_5$ or $\gamma^\mu$ structures aligned with the boost direction lead to analogous kinematic enhancement for highly boosted states.

\subsubsection{Asymptotic Extrapolation}

One step in LaMET analysis is to perform a Fourier transform of the quasi-distribution matrix elements from coordinate space to momentum space, which formally requires an infinite range of spatial separations $z$. However, the lattice data for the quasi-distribution matrix elements are only available at a finite range of spatial separations $z$. Theoretically, lattice data should decay exponentially at large separations, implying that contributions from the long-distance tail to FT should be negligible. While in practice, while the values are statistically consistent with zero, the statistical uncertainties increase (or persist at a constant level for CG formalism) at large separations, which leads to non-physical fluctuations of the uncertainty band in a discrete Fourier transform. Therefore, in many cases, an asymptotic extrapolation is performed to extend the lattice data to large separations. 

Many efforts have been made to control the systematic uncertainties associated with the Fourier transform in the literature, such as Refs.~\cite{Ji:2026vir,Chen:2025cxr,Gao:2021dbh,Ji:2020brr}. In some earlier projects, a commonly used extrapolation form is
\begin{eqnarray}
	  \tilde{h}(\lambda = z P^z) &= \Big[\frac{c_1}{(-i\lambda)^{n_1}} + e^{i\lambda}\frac{c_2}{(i \lambda)^{n_2}}\Big]e^{-\lambda/\lambda_0}~,
	  \label{eq:early_asymptotic_extrapolation}
\end{eqnarray} 
where the algebraic terms in the square bracket account for a power law behavior $x^a (1-x)^b$ in the momentum space, and the exponential term comes from the expectation that at finite momentum the correlation function has a finite correlation length (denoted as $\lambda_0$), which becomes infinite when approaching light-cone. In the very recent work~\cite{Ji:2026vir}, more sophisticated derivations and results on the asymptotic behavior of the quasi-distribution matrix elements at large separations are presented. For example, for nucleon quark quasi-PDFs in GI formalism, the asymptotic behavior is found to be
\begin{eqnarray}
   \tilde{h} (\lambda = z P^z) =
   \left[
      A_2 \cdot e^{i \mathrm{sign}(\lambda)\phi_2}
      + \frac{A'_2}{|\lambda|}
      \cdot e^{i \mathrm{sign}(\lambda)\phi'_2}
   \right]e^{-m |\lambda|} ~.
   \label{eq:asymptotic_extrapolation}
\end{eqnarray}

\subsubsection{Matching}

According to LaMET~\cite{Ji:2020ect,Ji:2024oka}, when $P^z\gg \Lambda_{\rm QCD}$ the PDF can be obtained from the quasi-PDF through a perturbative matching~\cite{Izubuchi:2018srq,Gao:2023lny},
\begin{eqnarray}
	f_{\Gamma}^{\overline{\mathrm{MS}}}(x,\mu) &= \int_{-\infty}^{+\infty} {\mathrm{d}y\over |y|}\ C_{\tilde \Gamma}^R\left({x\over y}, {\mu\over |y|P^z}, z_sP^z\right) \tilde f_{\tilde\Gamma}^R(y,P^z,z_s)  + {\cal O}\left({\Lambda_{\rm QCD}^2\over x^2P_z^2},{\Lambda_{\rm QCD}^2\over(1-x)^2P_z^2}\right)\,,
    \label{eq:pdf_matching}
\end{eqnarray}
where $\Gamma$ is the gamma matrix depending on the polarization, $C_{\tilde \Gamma}^R$ is the matching coefficient corresponding to the quasi-PDF $\tilde f_{\tilde\Gamma}^R(y,P^z,z_s)$ renormalized in a renormalization scheme labelled $R$, and the power corrections are suppressed by the active and spectator parton momenta. After resumming the linear renormalon~\cite{Zhang:2023bxs}, the power corrections in Eq.~(\ref{eq:pdf_matching}) start from the quadratic order. 

In GI case, the matching coefficient in $\overline{{\rm MS}}$ scheme is derived at NNLO in the literature~\cite{Li:2020xml,Chen:2020ody}. 

In CG case, the $\overline{{\rm MS}}$ matching coefficient is obtained as a series in the strong coupling $\alpha_s$,
\begin{eqnarray}
	C_{\tilde \Gamma}^{\overline{\mathrm{MS}}}\big(\xi, {\mu\over p^z}\big) = \delta(\xi-1) + {\alpha_sC_F\over 2\pi}C_{\tilde \Gamma(1)}^{\overline{\mathrm{MS}}}\big(\xi,{\mu\over p^z}\big) + {\cal O}(\alpha_s^2)\,,
    \label{eq:cg_pdf_matching_kernel}
\end{eqnarray}
where $C_F=4/3$.
At NLO the matching kernels are given by:
\begin{eqnarray}
	C_{\gamma^t(1)}^{\overline{\mathrm{MS}}}\big(\xi,{\mu\over p^z}\big) &=& C_{\gamma^t\gamma_5(1)}^{\overline{\mathrm{MS}}}\big(\xi,{\mu\over p^z}\big) \nonumber \\
    &=& C^{(1)}_r\big(\xi,{\mu\over p^z}\big) + {1\over 2|1-\xi|} + {1\over2}\delta(1-\xi) \left[ 1 + \ln { 4p_z^2\over\mu^2} - \int_0^2 d\xi' {1\over |1-\xi'|}\right]\,,\\
	C_{\gamma^z(1)}^{\overline{\mathrm{MS}}}\big(\xi,{\mu\over p^z}\big) &=& C_{\gamma^z\gamma_5(1)}^{\overline{\mathrm{MS}}}\big(\xi,{\mu\over p^z}\big) = C_{\gamma^t(1)}^{\overline{\mathrm{MS}}}\big(\xi,{\mu\over p^z}\big) + 2(1-\xi)_{+(1)}^{[0,1]} + \delta(1-\xi)\,,
\end{eqnarray}
where
\begin{eqnarray}
    C^{(1)}_r\big(\xi,{\mu\over p^z}\big) &= \left[\frac{1+\xi^2}{1-\xi}\ln{4p_z^2\over\mu^2} + \xi - 1\right]_{+(1)}^{[0,1]}  + \Bigg\{\frac{1+\xi^2}{1-\xi} \Big[\mathrm{sgn}(\xi)\ln|\xi| + \mathrm{sgn}(1-\xi)\ln|1-\xi|\Big] \nonumber \\
    &\qquad + \mathrm{sgn}(\xi) + {3 \xi-1\over \xi-1} \frac{\tan^{-1}\left(\sqrt{1-2 \xi}/|\xi|\right)}{\sqrt{1-2 \xi}} - {3\over 2|1-\xi|} \Bigg\}_{+(1)}^{(-\infty, \infty)}
\end{eqnarray}
satisfies particle number conservation. Here the plus functions on a domain $D$ are defined as
\begin{eqnarray}
	\left[g(x)\right]^D_{+(x_0)} &= g(x) - \delta(x-x_0)\int_D dx'\ g(x')\,.
\end{eqnarray}
Note that $C^{(1)}_r$ is analytical at $\xi=1/2$ despite its form.

For the transversity PDF case,
\begin{eqnarray}
	C_{\gamma^t\gamma_\perp^\alpha\gamma_5(1)}^{\overline{\mathrm{MS}}}\big(\xi,{\mu\over p^z}\big) &= C_{\gamma^z\gamma_\perp^\alpha\gamma_5(1)}^{\overline{\mathrm{MS}}}\big(\xi,{\mu\over p^z}\big) = C^{\perp (1)}_r\big(\xi,{\mu\over p^z}\big) \,,
\end{eqnarray}
where
\begin{eqnarray}
    C^{\perp (1)}_r\big(\xi,{\mu\over p^z}\big)&= \left[{2\xi \over 1-\xi}\ln{4p_z^2\over\mu^2} \right]_{+(1)}^{[0,1]}  + \Bigg\{{2\xi \over 1-\xi} \Big[\mathrm{sgn}(\xi)\ln|\xi| + \mathrm{sgn}(1-\xi)\ln|1-\xi|\Big] \nonumber \\
    &\qquad + {3 \xi-1\over \xi-1} \frac{\tan^{-1}\left(\sqrt{1-2 \xi}/|\xi|\right)}{\sqrt{1-2 \xi}} - {1\over |1-\xi|} \Bigg\}_{+(1)}^{(-\infty, \infty)}\,.
\end{eqnarray}

In the TMD case, it has been shown that quasi TMDPDFs have the same collinear degrees of freedoms as TMDPDFs~\cite{Ji:2019ewn}. Their differences from soft modes can be attributed to the intrinsic soft function and different rapidity scales. Also, contributions from highly off-shell modes are local~\cite{Ebert:2019okf}. Thus, TMDPDFs $f\left(x, b_{\perp}; \mu, \zeta \right) $ are connected to quasi TMDPDFs $\tilde{f}_{\Gamma}\left(x, b_{\perp}, P^z; \mu\right)$ via a multiplicative  factorization~\cite{Ji:2019ewn,Ebert:2022fmh}: 
\begin{eqnarray}
    \sqrt{S_I\left(b_{\perp} ; \mu\right) } \cdot \tilde{f}_{\Gamma}\left(x, b_{\perp}, P^z; \mu\right) &= f\left(x, b_{\perp}; \mu, \zeta \right)  H_f \left(x, P^z; \mu\right)  \exp \left[\frac{1}{2}\ln \frac{\left(2 x P^z\right)^2}{\zeta} \gamma^{\overline{\mathrm{MS}}}\left(b_{\perp}; \mu\right)\right] \nonumber \\
    &+ \mathcal{O}\left(\frac{\Lambda_{\mathrm{QCD}}}{x P^z}, \frac{1}{b_{\perp}\left(x P^z\right)}\right) ~,
    \label{eq:tmdpdf_matching}
\end{eqnarray}
where $S_I\left(b_{\perp} ; \mu\right)$ denotes the intrinsic soft function that has been calculated on the lattice in Refs.~\cite{LatticeParton:2020uhz,Li:2021wvl,LatticeParton:2023xdl,LatticePartonLPC:2023pdv,Bollweg:2025iol} and $\gamma^{\overline{\mathrm{MS}}}\left(b_{\perp}; \mu\right)$ denotes the Collins-Soper kernel. 

In GI case, the hard kernel $H_f$ as a function of $\zeta_z/\mu^2=\left(2xP^z\right)^2/\mu^2$, has been perturbatively determined up to next-to-leading order (NLO)~\cite{Ji:2019ewn,Ji:2018hvs,Ebert:2019okf, Ji:2020ect}
\begin{eqnarray}
    H^{(1)}\left(\frac{\zeta_z}{\mu^2}\right)=\frac{\alpha_s C_F}{2 \pi}\left(-2+\frac{\pi^2}{12}+\ln \frac{\zeta_z}{\mu^2}-\frac{1}{2} \ln ^2 \frac{\zeta_z}{\mu^2}\right),
\end{eqnarray}
as well as next-to-next-to-leading order (NNLO), calculated recently~\cite{delRio:2023pse, Ji:2023pba} 
\begin{eqnarray}
	H^{(2)}\left(\frac{\zeta_z}{\mu^2}\right)=\alpha_s^2\left[c_2-\frac{1}{2}\left(\gamma_C^{(2)}-\beta_0 c_1\right) \ln \frac{\zeta_z}{\mu^2} -\frac{1}{4}\left(\Gamma_{\mathrm {cusp }}^{(2)}-\frac{\beta_0 C_F}{2 \pi}\right) \ln ^2 \frac{\zeta_z}{\mu^2}-\frac{\beta_0 C_F}{24 \pi} \ln ^3 \frac{\zeta_z}{\mu^2}\right],
\end{eqnarray}
where $\zeta_z=\left(2xP^z\right)^2$ and $c_2 = 0.0725 C_F^2 - 0.0840 C_F C_A + 0.1453 C_F n_f / 2$.

In CG case, the hard kernel $H_f\left(x, P^z; \mu\right) = |C_{\rm TMD} (xP^z; \mu)|^2$, where the coefficient is calculated up to NLO as ~\cite{Zhao:2023ptv}
\begin{eqnarray}
    C_{\rm TMD} \left(x P^z, \mu \right) = 1 + \frac{\alpha_s C_F}{4 \pi} \left[ - \frac{L_p^2}{2} - 3 L_p - 12 + \frac{7 \pi^2}{12} \right] ~,
    \label{eq:wilson_fixed}
\end{eqnarray}
in which $L_p = \ln\frac{\mu^2}{(2 x P^z)^2}$ and $C_F = 4/3$.

\section{Calculation of Pion DAs}

As the pseudo Nambu-Goldstone boson of the spontaneous chiral symmetry breaking in QCD, the pion plays a special role in hadron physics. The pion distribution amplitude (DA) is a fundamental nonperturbative function that describes the momentum fraction distribution of the quark and antiquark inside the pion. It is an essential input for understanding exclusive pion processes, including those that probe the pion electromagnetic form factor and the pion-photon transition form factor~\cite{Lepage:1980fj}. The calculation of the pion DA from first principles using lattice QCD has been a long-standing challenge due to the need to access light-cone correlations. However, with the development of LaMET, it has become possible to calculate the pion DA on the lattice by constructing appropriate quasi-distributions and performing the perturbative matching to extract the physical DA.

\begin{table}
    \centering
    \begin{tabular}{cclccccccc}
    \hline
    Ensemble ~~& $a$(fm) & \ \!$L^3\times\  T$   & $c_{\mathrm{SW}}$ & $m_{u/d}$  & $m_{s}$  \\
    \hline
    a06m130  ~~& 0.057  ~~& $96^3\times$192   ~~& 1.03493  ~~&-0.0439         ~~&-0.0191            \\ 
    a09m130  ~~& 0.088  ~~& $64^3\times$~ \!96  ~~& 1.04239  ~~&-0.0580         ~~&-0.0174           \\ 
    a12m130  ~~& 0.121  ~~& $48^3\times$~ \!64  ~~& 1.05088  ~~&-0.0785         ~~&-0.0191           \\
    \hline
    \end{tabular}
    \caption[Lattice setup for the lattice calculation of pion DAs]{Details of the simulation setup. The light and strange quark mass (both valence and sea quark) of the clover action are tuned such that $m_{\pi}$=140 MeV and $m_{\eta_s}$=670 MeV.}
    \label{tab:pida_setup}
\end{table}

The calculations in this project~\cite{LatticeParton:2022zqc} are done using the clover valence fermion on three ensembles with 2+1+1 flavors of HISQ sea quarks generated by the MILC Collaboration~\cite{Follana:2006rc,MILC:2012znn}. The valence quark masses are tuned to $m_\pi = 140$ MeV and $m_{\eta_s} = 670$ MeV, and the three lattice spacings are 0.057, 0.088 and 0.121 fm. The detailed lattice setup is summarized in Table.~\ref{tab:pida_setup}. We use three different pion momenta, $P_z = \{1.29, 1.72, 2.15\}$ GeV, the corresponding Lorentz boost factors $\gamma$ for the pion at the physical mass are $\gamma = \{9.21, 12.29, 15.36\}$, to study the convergence of the LaMET expansion and to perform the extrapolation to the infinite-momentum limit. 

The leading-twist light-cone DA of a pseudoscalar meson is defined as Eq.~(\ref{eq:da_definition}). To extract this quantity, we calculate the following quasi correlation on the lattice with momentum $\vec{P}$ along the $z$-direction~\cite{Zhang:2017bzy}: 
\begin{eqnarray}
    C_{2}^{m}(z,\vec P, t)  = \int d^3y e^{-i\vec P\cdot \vec y} \times \langle 0 |{\cal O}_{\Gamma_1}(z;\vec{y},t) \bar  \psi_2(0,0) \Gamma_2 \psi_1(0,0)|0\rangle ~, 
    \label{eq:pida_2pt}
\end{eqnarray} 
where ${\cal O}_{\Gamma_1}(z;\vec{y},t)\equiv\bar \psi_1(\vec y, t) \Gamma_1 U(\vec y, \vec y {\color{blue}-} z\hat z) \psi_2(\vec y {\color{blue}-} z\hat z, t)$ is the quasi operator with $\hat z$ being the unit vector in the $z$-direction, $U(\vec{x}, \vec{x} {\color{blue}-} \vec z)$ is the spatial Wilson line connecting lattice sites $\vec{x}$ and $\vec{x} - \vec z$, $\psi_2\Gamma_2 \psi_1$ is the interpolating field of the meson $m$, and $\Gamma_{1,2}$ are chosen as $\Gamma_1 = \gamma_z \gamma_5,\, \Gamma_2=\gamma_5$ for the pseudoscalar meson. The ground-state matrix elements can be extracted from the following two-state fit formula:
\begin{eqnarray}
    \frac{ C^{m}_{2}(z,\vec P, t)  }{ C^{m}_{2}(z=0,\vec P, t)  } = \frac{ H_{m}^{\rm B}(z)(1+ c_{m}(z) e^{-\Delta Et}) }{(1+ c_{m}(0) e^{-\Delta Et}) }~,
\end{eqnarray}
where $H_{m}^{\rm B}(z)$ is the normalized ground-state matrix element, $c_m$ and $\Delta E$ are free parameters accounting for (one or more) excited state contamination, which are exponentially suppressed in the large time limit. Based on the comparison of one- and two-state fits, we use the one-state fit results in the analysis below with $t_{min}=0.72,0.54,0.42$ fm (for $P_z=1.29,1.72,2.15$~GeV) which is large enough to eliminate the excited-state contamination.

The bare quasi correlation calculated above contains both linear and logarithmic ultraviolet (UV) divergences which have to be removed by renormalization. On the lattice, the numerical subtraction of linear divergences is extremely delicate. In particular, such divergences may not be fully removed if the RI/MOM renormalization scheme is used~\cite{Zhang:2020rsx}. Thus, we adopt the self renormalization proposed in Ref.~\cite{LatticePartonLPC:2021gpi}, which amounts to fitting the bare quasi correlation and subtracting the relevant UV divergences. To be more precise, one fits the bare quasi correlation at given hadron momentum and multiple lattice spacings with a perturbative-QCD-dictated parametrization that contains a linear divergence, a logarithmic divergence, and discretization effects. After removing all the UV divergences and discretization effects, one is left with the renormalized quasi correlation encoding the intrinsic non-perturbative physics.

As suggested in Ref.~\cite{LatticePartonLPC:2021gpi}, the UV divergences in the quasi correlator can be determined by using, e.g., the pion PDF matrix elements $\mathcal{M}(z)\equiv \langle \pi|{\cal O}_{\gamma_t}|\pi\rangle $ in the rest frame at multiple lattice spacings, and fitting the {bare} data \textbf{$\mathcal{M}^{\rm B}$} to the following form~\cite{LatticePartonLPC:2021gpi} 
\begin{eqnarray}
    \mathcal{M}^{\rm B}(z,a) = Z^{\rm self}(z,a) \mathcal{M}^{\rm R}(z),
\end{eqnarray}
with the renormalization factor parametrized as~\cite{LatticePartonLPC:2021gpi} 
\begin{eqnarray}
    Z^{\rm self}(z,a)\equiv&&\mathrm{exp}\Big\{ \frac{k z}{a \ln[a \Lambda_{\rm QCD}]} 
    + m_0 z + f(z) a\nonumber\\
    &&+ \frac{3 C_{F}}{b_0} \ln \Big[\frac{\ln [1 /(a \Lambda_{\rm QCD})]}{\ln [\mu / \Lambda_{\rm QCD}]}\Big]+\ln\Big[1+\frac{d}{\ln (a \Lambda_{\rm QCD})}\Big]\Big\}~,
\end{eqnarray}
where the first term in the curly bracket is the linear divergence, $m_{0}$ denotes a finite mass contribution arising from renormalon ambiguity, etc., and $f(z) a$ accounts for the discretization effects (The $\mathcal{O}(a)$ correction here arises from the mixed action effect in using clover valence fermions on HISQ sea ones). The last two terms come from the resummation of leading and sub-leading logarithmic divergences, which only affect the overall normalization at different lattice spacings. {To partially account for higher-order perturbative effects as well as remaining lattice artifacts, we also treat $d$ and $\Lambda_{\rm QCD}$ as fitting parameters}~\cite{LatticePartonLPC:2021gpi}. The renormalized matrix element is required to be equal to the continuum perturbative $\overline{\rm MS}$ result at short distances (chosen to be $z\in z_s=[0.06, 0.18]$\,fm as defined in \cite{Ji:2020brr}),
\begin{eqnarray}
    \mathcal{M}^{\rm R}(z)|_{z \in z_s}=\mathcal{M}^{\overline{\rm MS} ,\ {\rm 1-loop}  }(z) \equiv 1+\frac{\alpha^{\overline{\rm MS}}_s C_F}{4\pi}\left[3 \ln\frac{z^2\mu^2}{4e^{-2\gamma_E}} + 5\right]+{\mathcal O}(\alpha^{2,\overline{\rm MS}}_s)~, 
\end{eqnarray}
which helps the determination of $m_0$ and $d$. In the calculation we use the $\overline{\rm MS}$ renormalization scale ${\mu=2\ \mathrm{GeV}}$ and $\Lambda_{\overline{\rm MS}}=0.24\ \mathrm{GeV}$.

In the present case, we follow the same strategy as above, except that 
the renormalized matrix element in the $\overline{\rm MS}$ scheme,
\begin{equation}
    H^{\overline{\rm MS}}_m(z)=H^{\rm B}_m(z,a)/ \tilde{Z}^{\rm self}(z,a)
\end{equation} 
is now required to be matched to the continuum perturbative $\overline{\rm MS}$ result of the normalized quasi-DA matrix element at short distances in the rest frame, which reads at one-loop
\begin{eqnarray}
    H^{\overline{\rm MS},\ {\rm 1-loop} }_m(z)\equiv 1 + \frac{\alpha^{\overline{\rm MS}}_s C_F}{4\pi} \left[3 \ln\frac{z^2\mu^2}{4e^{-2\gamma_E}} + 7\right] ~.
\end{eqnarray}
$\tilde{Z}^{\rm self}$ turns out to be the same as $Z^{\rm self}$ except for the value of $d$.

\begin{figure}[!th]
    \begin{center}
    \includegraphics[width=0.6\textwidth]{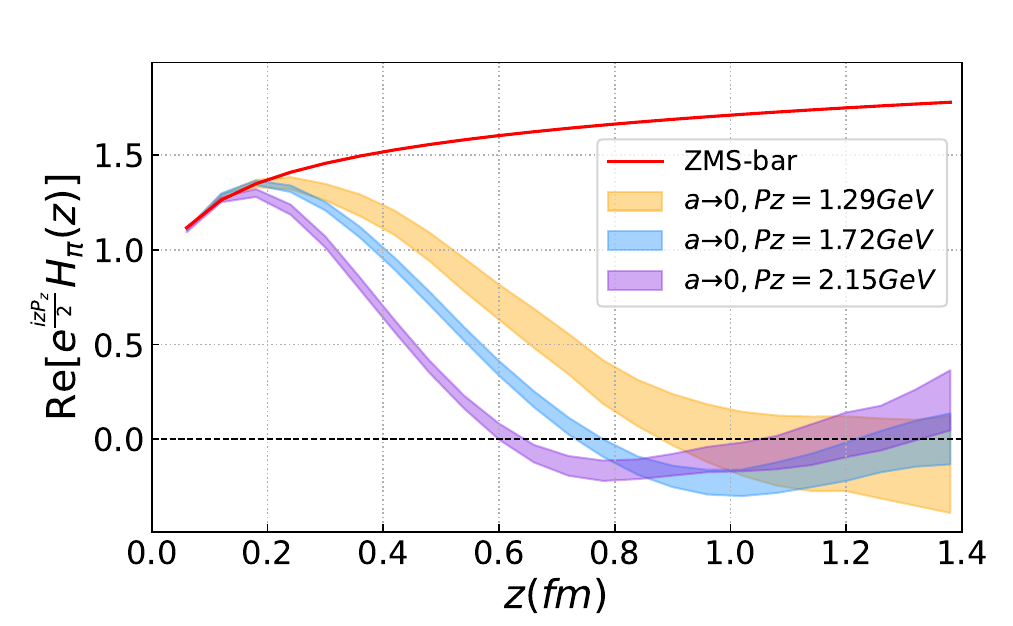}
    \caption[Self-renormalization of quasi-DA]{Comparison of self-renormalized quasi correlation Re[$e^{\frac{i z P_{z}}{2}}H^{\overline{\rm MS}}_m(z)$] of the pion with different momenta (bands), and the perturbative result in the $\overline{\rm MS}$ scheme $H^{\overline{\rm MS},\ {\rm 1-loop}}_m(z)$ (red curve).}
    \label{fig:pida_self_renorm}
    \end{center}    
\end{figure}

In Fig.~\ref{fig:pida_self_renorm}, we show a comparison between the self-renormalized quasi correlations Re[$e^{\frac{izP_{z}}{2}} H^{\overline{\rm MS}}_m(z)$] (after linear ${\cal O}(a)$ continuum extrapolation and phase rotation) with the perturbative one-loop result $H^{\overline{\rm MS},\ {\rm 1-loop}}_m(z)$. As can be seen from the figure, all quasi correlations agree well with the perturbative result for small $z$, indicating a mild $P_z$ dependence in that region. Besides, from the figure, we can also see that the uncertainty of the renormalized quasi correlation grows rapidly at large distance. As discussed before, an asymptotic extrapolation is needed to avoid non-physical fluctuations of the uncertainty band in a discrete Fourier transform. The extrapolation form in Eq.~(\ref{eq:early_asymptotic_extrapolation}) is adopted in this work. Note that the Lorentz boost factor $\gamma$ for the pion at the physical point is very large $\{9.21, 12.29, 15.36\}$, and thus the correlation length is very large. We therefore drop the $e^{-\lambda/\lambda_0}$ factor, and directly perform a polynomial extrapolation as suggested in~\cite{Ji:2020brr}.

\begin{figure}[!th]
    \begin{center}
    \includegraphics[width=0.6\textwidth]{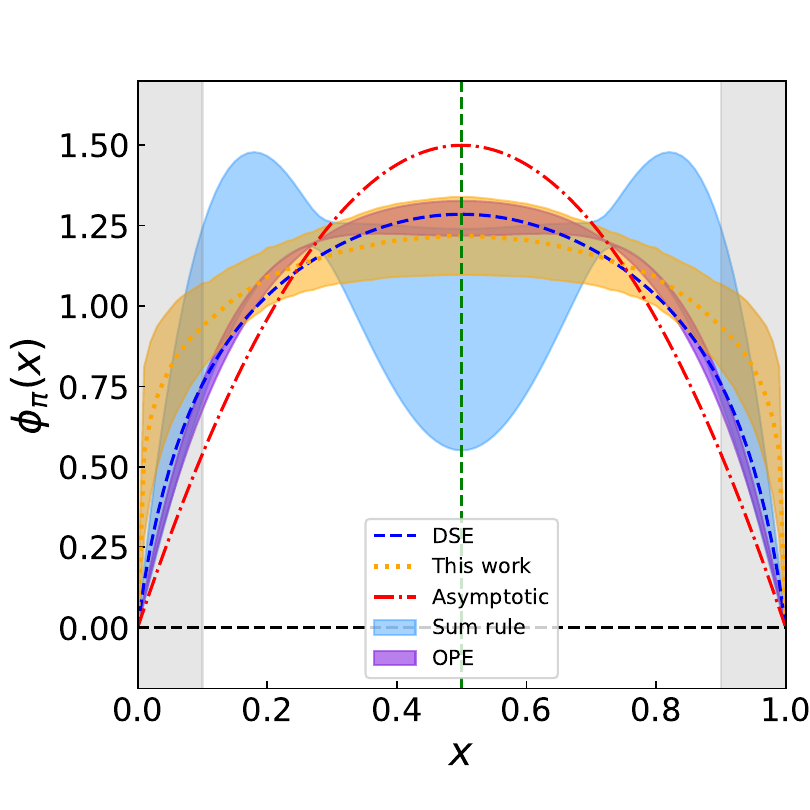}
    \caption[DAs of pion]{DAs of $\pi$ meson, extrapolated to the continuum $(a\to 0)$ and infinite momentum limit $(P_z \to \infty)$. For the kaon, $x$ is the momentum fraction carried by the light quark.}\label{fig:pida_lcdapik}
    \end{center}
\end{figure}

With the results for $P_z = 1.29, 1.72, 2.15$ GeV above, we can perform an extrapolation to $P_z \to \infty$ using the functional form:
\begin{eqnarray}
	\phi(x, P_z) = \phi(x, P_z \to \infty)  + \frac{c_2(x)}{P^2_z} +   \mathcal{O}\Big(\frac{1}{P^4_z}\Big).
\end{eqnarray}
The final results of the pion DAs are given in Fig.~\ref{fig:pida_lcdapik}, where systematic uncertainties from renormalization scale $\mu$ dependence, large $\lambda$ extrapolation, continuum and infinite momentum extrapolation have been taken into account. While the systematic uncertainty from continuum extrapolation dominates for the kaon, the statistical (larger than for the kaon due to the lighter quark mass) and continuum extrapolation uncertainties are comparable and dominate for the pion. Therefore, we expect that the kaon DA uncertainty can get significantly suppressed when data at smaller lattice spacings are available which enable us to do a more reliable continuum extrapolation, whereas for the pion, besides data at smaller lattice spacings, we also need increasing statistics to reduce the statistical uncertainty.
In the endpoint region which cannot be reliably predicted by LaMET, we adopt a phenomenological $x^a(1-x)^b$ extrapolation (taken as $0<x<0.1 \,\, \& \,\, 0.9<x<1$). The unreliable region is expected to shrink if a larger $P_z$ can be reached in future calculations.
For comparison, we also plot the asymptotic form $6x(1-x)$ and results from QCD sum rules~\cite{Ball:2007zt}, Dyson-Schwinger equations (DSE)~\cite{Roberts:2021nhw}, and reconstructed from moments calculations (OPE)~\cite{RQCD:2019osh}. As can be seen from the figure, the $\pi$ DAs deviate significantly from the asymptotic form, but are close to the results from DSE and OPE calculations. The shape of $\pi$ DA is much broader than the asymptotic form, manifesting the impact of DCSB at such a low scale. A similar behavior has also been observed in a recent analysis using QCD sum rules with nonlocal condensates~\cite{Stefanis:2014nla}. 

In summary, this is a state-of-the-art lattice calculation of the pion DA at the physical point. The renormalization is done in the hybrid scheme with self renormalization proposed recently. Based on the results at physical light and strange quark masses with three lattice spacings and momenta, we perform an extrapolation to the continuum and infinite momentum limit. The final results exhibit a significant deviation from the asymptotic form, while they are close to the DSE and OPE results, especially in the middle $x$ region where our method is reliable. However, there are still some significant differences in the endpoint regions. This could be due to missing higher-power/ high-order corrections in LaMET which can be improved in future calculations, or due to effects of higher moments ignored in the OPE and DSE calculations. A more accurate determination of the endpoint behavior of the DAs would be important towards a better understanding of quantities like the pion-photon transition form factor.

\section{Calculation of Nucleon PDFs}

Compared with the pion DA, which encodes the longitudinal momentum sharing between a valence quark and antiquark in an exclusive meson channel at small transverse separation, the parton distribution function (PDF) characterizes the probability density of finding a quark or gluon carrying a given longitudinal momentum fraction inside a fast-moving hadron in inclusive processes. In other words, the DA is a light-cone wave-function–like object relevant for exclusive reactions and involves a projection onto Fock states, whereas the PDF is an inclusive density that integrates over transverse dynamics and sums over all allowed final states.

Moving from the pion to the nucleon further increases both theoretical and computational complexity. While the pion is a comparatively simple two-body bound state with a relatively clean spectral structure, the nucleon is a three-valence-quark system embedded in a dense spectrum of excited states with significant sea-quark and gluon contributions. In lattice QCD calculations, this richer excitation spectrum leads to substantially more severe excited-state contamination in nucleon correlation functions. The energy gaps above the ground-state nucleon are typically smaller, and the state multiplicity is higher, making it considerably more challenging to isolate the ground state within feasible Euclidean time separations.

In addition, nucleon correlators generally suffer from a poorer signal-to-noise ratio, which decays exponentially with Euclidean time in a way governed by the difference between the nucleon and multi-pion energies. This statistical degradation limits the temporal window available for reliable ground-state extraction and exacerbates systematic uncertainties. The presence of more intricate spin and flavor structures in baryonic operators, as well as enhanced operator mixing under renormalization, further complicates the analysis.

Despite these theoretical and numerical challenges, nucleon PDFs remain among the most fundamental nonperturbative quantities in QCD. Owing to their universality in QCD factorization, PDFs extracted from a subset of processes can be used consistently as nonperturbative inputs for cross-section predictions across a broad class of hard-scattering reactions. Over recent decades, high-energy scattering experiments at HERA, COMPASS, SLAC, Fermilab, and the LHC have delivered state-of-the-art constraints on PDFs, enabling global analyses of the collinear structure of both unpolarized and polarized nucleons, with percent-level uncertainties in well-constrained kinematic regions~\cite{Alekhin:2017kpj,Hou:2019efy,Bailey:2020ooq,NNPDF:2021njg,Cruz-Martinez:2025ahf}. Looking ahead, the future Electron-Ion Collider~\cite{Accardi:2012qut,AbdulKhalek:2021gbh} will probe PDFs and other partonic observables with unprecedented precision, opening new opportunities to map the multidimensional structure of hadrons.

In our recent work~\cite{Gao:2026hix}, we present the first comprehensive benchmark of the CG method for nucleon PDFs through a lattice QCD calculation of the isovector unpolarized, helicity, and transversity distributions. We perform a numerical lattice QCD calculation on a 2+1 flavor gauge ensemble generated by the HotQCD Collaboration~\cite{HotQCD:2014kol} with Highly Improved Staggered Quarks (HISQ)~\cite{Follana:2006rc}. The ensemble has a lattice spacing of $a=0.0601(5)$ fm and a volume of $L_s^3 \times L_t = 48^3 \times 64$. We used tadpole-improved clover Wilson valence fermions on a hypercubic (HYP) smeared~\cite{Hasenfratz:2001hp} gauge background. We set the clover coefficient $c_{\rm sw} = u_0^{-3/4}$, in which $u_0$ is the average plaquette after HYP smearing; the parameter is set as $c_{\rm sw} = 1.0336$ for both time and spatial directions. For the masses of the valence quarks, we used hopping parameter $\kappa = 0.12623$ for the light quarks $u$ and $d$, corresponding to a valence pion mass $m_\pi = 300$ MeV. By boosting the nucleon along an off-axis direction $\vec{n}=(1,1,0)$, we are able to reach momenta of $P^{z}=2.43$ and $3.04$ GeV at a pion mass of $m_{\pi}=300$ MeV, which are among the largest boosts achieved in nucleon PDF calculations at a comparable pion mass. Nevertheless, the CG method faces similar systematic uncertainties as the GI approach, including excited-state contamination and power-suppressed corrections. A detailed analysis is carried out to address excited-state contamination in the extraction of matrix elements, where we find that imaginary parts of matrix elements are significantly more sensitive to excited-state contamination, consistent with previous observations for GI operators~\cite{Alexandrou:2020qtt}.
Besides excited-state contamination, we also perform a systematic investigation of other sources of uncertainty in the CG framework, including the choice of Dirac gamma structure in quasi-PDF operators, the renormalization schemes, uncertainties associated with the Fourier transform, and power-suppressed corrections in the LaMET factorization. 
For the real parts of the quasi-PDF matrix elements, we utilize a hybrid scheme~\cite{Ji:2020brr} to renormalize the matrix elements. To renormalize the imaginary part of the quasi-PDF matrix elements, we argue that the hybrid scheme is not optimal and instead propose a static version of the RI$^\prime$-MOM scheme~\cite{Martinelli:1994ty}, where we find that the systematic uncertainties are under control. 
While the CG formulation does not eliminate the intrinsic challenges of nucleon PDF calculations, such as excited-state contamination due to the dense excited-state spectra, our results demonstrate reasonable agreement with existing lattice and phenomenological studies. 
This work thus validates the CG method as a promising approach for precision studies of the nucleon partonic structure and provides a solid foundation for its future extensions, including the computation of TMDs.

\subsection{Theoretical Framework}

The quasi PDFs in the CG are defined as~\cite{Gao:2023lny}
\begin{eqnarray}
    \tilde{f}_{\tilde \Gamma}(x, P^z; \mu) &=& P^z \int \frac{d z}{2 \pi} e^{i z (x P^z)} \tilde{h}_{\tilde\Gamma}(z, P^z, \mu) ~,
    \label{eq:quasi_operator_momentum}\\
    \tilde{h}_{\tilde\Gamma}(z, P^z; \mu) &\equiv& \frac{1}{2 N_{\tilde\Gamma}} \langle PS | \left. \bar{\psi}(z) \tilde\Gamma \psi (0) \right|_{\vec{\nabla} \cdot \vec{A} = 0} | PS \rangle ~,
\end{eqnarray}
where the normalization factor $N_{\tilde{\Gamma}}$ is given by:
\begin{eqnarray}
    N_{\tilde\Gamma} &=\left\{ \begin{array}{cc}
    {P^t} : & \tilde\Gamma \in \{ \gamma^t, \gamma^z\gamma_5  , \gamma^t\gamma_\perp^\alpha\gamma_5 \}\,, \\
    {P^z} : & \tilde\Gamma\in \{ \gamma^z, \gamma^t\gamma_5, \gamma^z\gamma_\perp^\alpha\gamma_5 \} \,.
    \end{array} \right.
\end{eqnarray}
The Dirac matrices $\tilde\Gamma = \{\gamma^t/\gamma^z,\gamma^t\gamma_5/\gamma^z\gamma_5, \gamma^t\gamma_\perp^\alpha\gamma_5/\gamma^z\gamma_\perp^\alpha\gamma_5\}$ in the quasi-PDF corresponds to $\Gamma=\{\gamma^+, \gamma^+\gamma_5, \gamma^+\gamma_\perp^\alpha \gamma_5\}$ in the PDF.

In the LaMET framework, several nonperturbative renormalization schemes are used in practice to renormalize the gauge-invariant quasi-PDF matrix element, including ratio-type schemes~\cite{Constantinou:2017sej,Stewart:2017tvs,Alexandrou:2017huk,Chen:2017mzz,Orginos:2017kos,Braun:2018brg,Li:2020xml}, regularization independent momentum subtraction schemes~\cite{Constantinou:2017sej,Stewart:2017tvs,Alexandrou:2017huk,Chen:2017mzz}, and the hybrid scheme~\cite{Ji:2020brr,LatticePartonLPC:2021gpi}. The main advantage of the hybrid scheme is that discretization effects are cancelled by taking appropriate ratios at short separations, while also avoiding introducing any nonperturbative artifacts in the long-distance matrix elements. 

As shown in Ref.~\cite{Gao:2023lny,Zhang:2024omt}, due to the absence of Wilson line the CG correlator is free from the linear divergence and the associated renormalon~\cite{Liu:2023onm}. As a result, the renormalization of the nonlocal CG operator is purely multiplicative and can be defined as
\begin{equation}
    \bar{\psi}_0 (z) \tilde\Gamma \psi_0 (0) = Z^R_\psi (a) \left[ \bar{\psi}_R (z) \tilde\Gamma \psi_R(0) \right] ~,
    \label{eq:cg_renormalization}
\end{equation}
where $Z^R_\psi (a)$ is the renormalization factor of the quark wave-function into some renormalization scheme $R$ as a function of the lattice spacing $a$. $\psi_0$ and $\psi_R$ represent the bare and renormalized quark fields, respectively.

In the standard definition of the hybrid scheme~\cite{Gao:2021dbh,LatticeParton:2022xsd,Holligan:2024wpv}, the renormalized quasi-PDF matrix element is defined by the following ratios: 
\begin{eqnarray}
    \tilde{h}_{\tilde\Gamma}^{\mathrm{hyb.}} \left(z, P^z; z_s \right)&=N \frac{\tilde{h}^0_{\tilde\Gamma} \left(z, P^z; a\right)} {\tilde{h}^0_{\tilde\Gamma} (z, 0; a)} \theta\left(z_s-|z|\right) +N \frac{\tilde{h}^0_{\tilde\Gamma} \left(z, P^z; a \right)} {\tilde{h}^0_{\tilde\Gamma} \left(z_s, 0; a \right)} \theta\left(|z|-z_s\right) ~,
\end{eqnarray}
where $\tilde{h}^0_{\tilde{\Gamma}}/\tilde{h}^{\mathrm{hyb.}}_{\tilde{\Gamma}}$ denotes the bare/hybrid-renormalized matrix element respectively, the normalization factor is defined by $N = \tilde{h}^0_{\tilde\Gamma} \left(0, 0; a\right) / \tilde{h}^0_{\tilde\Gamma} (0, P^z; a)$, and $z_s$ ($a\ll z_s \ll \Lambda_{\rm QCD}^{-1}$) denotes the point that separates the short- and long-distance regimes. Since the matrix elements are usually strongly correlated at neighboring lattice sites, this normalization can reduce their statistical uncertainties at small $z$. 
However, a drawback of this treatment is that the discretization errors of $z=0$ matrix elements are unnecessarily propagated to all $z\neq0$, where such effects are expected to be suppressed as powers of $a/|z|$. An ideal solution would be a $z$-dependent normalization that accurately removes the discretization effects mostly at small $z$, which might be achieved with lattice perturbation theory. However, it is beyond the scope of this work.
After all, the discretization effects can be systematically eliminated through continuum extrapolation, so this normalization may still be favored for the purpose of reducing statistical uncertainties.

On a single lattice ensemble, there is no clear preference for applying or omitting the above normalization, since the discretization effects cannot be quantified. However, since the $z=0$ matrix elements are expected to be independent of $P^z$ due to Lorentz covariance, a significant deviation of $N$ from unity signals sizable lattice artifacts; in this case, a blind normalization would rescale the quasi-PDF by the same factor. Therefore, in this work we adopt a hybrid scheme without the normalization at $z=0$, namely,
\begin{equation}
    \tilde{h}^{\mathrm{hyb.}}_{\tilde \Gamma} \left(z, P^z; z_s \right)= \frac{\tilde{h}^0_{\tilde \Gamma} \left(z, P^z; a\right)} {\tilde{h}^0_{\tilde \Gamma} (z, 0; a)} \theta\left(z_s-|z|\right) + \frac{\tilde{h}^0_{\tilde \Gamma} \left(z, P^z; a \right)} {\tilde{h}^0_{\tilde \Gamma} \left(z_s, 0; a \right)} \theta\left(|z|-z_s\right) ~.
    \label{eq:hybrid_scheme_no_normalization}
\end{equation}
As a result, the discretization effects at $z\sim a$ remain uncorrected and propagate into momentum space as artifacts of order $(P^z a)^n$. Although this treatment spoils the normalization of the quasi-PDF and PDF, the effect vanishes in the continuum limit. The hybrid renormalization is applied to the real part of the quasi-PDF matrix elements with the separation point chosen as $z_s = 3 \sqrt{2}a$, where the factor of $\sqrt{2}$ originates from the off-axis momentum setup. 

For the imaginary part of the quasi-PDF matrix elements, however, it is likely that the ratios employed in the hybrid scheme do \textit{not} remove the leading lattice artifacts for $z \le z_s$, as the zero-momentum matrix elements in the denominator of the ratio are purely real. To address this issue, we instead employ a modification of the RI$^\prime$-MOM~\cite{Martinelli:1994ty} scheme for the quark wave function renormalization factor $Z_R$ in Eq.~\ref{eq:cg_renormalization}. Because the Coulomb gauge is not a complete gauge fixing due to the residual gauge freedom in the timelike direction, such a scheme based on the quark propagator either requires introducing \textit{additional} gauge-fixing conditions in the timelike direction~\cite{Burgio:2008jr,Burgio:2012ph}, or one must only use spatial correlation functions. 
Choosing the latter route leads to a Static RI$^\prime$-MOM scheme (SRI$^\prime$-MOM) which we define below. Given the CG static propagator:
\begin{equation}
    S(\vec{k})  = \int \mathrm{d}^3 \vec{x}\  e^{i \vec{x} \cdot \vec{k}} \left\langle \psi(\vec{0},t_E) \overline{\psi}(\vec{x},t_E) \right\rangle,
\end{equation}
computed from position-space propagators at a single Euclidean time $t_E$, we can define a renormalization constant such that the renormalized static propagator is equal to the tree-level value at $\vec{k}$:
\begin{equation}
    \psi_{0} = Z^{1/2}_{\mathrm{SRI'}}(|\vec{k}|) \psi_{\mathrm{SRI'}},
    \qquad Z_{\mathrm{SRI'}}(|\vec{k}|) =  \lim_{m \to 0} \frac{ \mathrm{Tr}[\slashed{\vec{k}} S_0(\vec{k})]}{ \mathrm{Tr}[\slashed{\vec{k}} S_\mathrm{tree}(\vec{k})]} ~,
    \label{eq:sri_mom_renormalization}
\end{equation}
where the chiral limit is taken in order to define a mass-independent renormalization scheme to match to $\overline{\mathrm{MS}}$.  
Using leading-logarithmic running and $O(\alpha_S)$ matching between SRI$^\prime$-MOM \ and $\mathrm{\overline{MS}}$, we find that the renormalization constant for the ensemble used in this work is:
\begin{equation}
    Z_{\overline{\mathrm{MS}}}(a,\mu = 2 \mathrm{GeV}) = 0.915(2)\,.
\end{equation}

As mentioned above, the PDF can be obtained from the CG quasi-PDF through a perturbative matching, as shown in Eq.~(\ref{eq:pdf_matching}). In the CG method, due to the absence of linear renormalon, the leading renormalon resummation (LRR) is not required. 

The $\overline{\mathrm{MS}}$ matching coefficient is given in Eq.~(\ref{eq:cg_pdf_matching_kernel}), which can be converted to the hybrid scheme as follows:
\begin{eqnarray}
	C_{(1)}^{\mathrm{hyb.}}\left(\xi,{\mu\over p^z},z_sp^z\right) = C^{(1)}_r\left(\xi,{\mu\over p^z}\right) + \delta C_{(1)}^{\mathrm{hyb.}}\left(\xi,z_sp^z\right)\,,
\end{eqnarray}
where
\begin{eqnarray}
	\delta C_{\gamma^t(1)}^{\mathrm{hyb.}}\big(\xi,z_sp^z\big) &=& \delta C_{\gamma^z(1)}^{\mathrm{hyb.}}\big(\xi,z_sp^z\big) =  \delta C_{\gamma^t\gamma_5(1)}^{\mathrm{hyb.}}\big(\xi,z_sp^z\big) =  \delta C_{\gamma^z\gamma_5(1)}^{\mathrm{hyb.}}\big(\xi,z_sp^z\big) \nonumber\\
    &=& {1\over2}\left[{1\over |1-\xi|} - {2{\rm Si}[(1-\xi)z_sp^z]\over \pi(1-\xi)}\right]_{+(1)}^{(-\infty, \infty)}\,,\\
    \delta C_{\gamma^t\gamma_\perp^\alpha\gamma_5(1)}^{\mathrm{hyb.}}\big(\xi,z_sp^z\big) &=& \delta C_{\gamma^z\gamma_\perp^\alpha\gamma_5(1)}^{\mathrm{hyb.}}\big(\xi,z_sp^z\big)= 0\,.
\end{eqnarray}

The matching coefficient includes the logarithm $\ln {\mu^2\over (2yP^z)^2}$ which follows the DGLAP evolution equation. However, since $y$ is an integration variable, the matching would be ill defined for all $x$ if we naively resum such logarithms. Instead, the physically meaningful logarithms should be given by the external scales after the convolution integral. For Eq.~(\ref{eq:pdf_matching}), they are identified as $\ln {\mu^2\over (2xP^z)^2}$~\cite{Su:2022fiu}, which indicates that the resummation becomes significant at small $x$. When $\alpha_s(2xP^z)\gtrsim 1$, the matching loses perturbative control, which provides an estimate of the smallest $x_{\rm min}$ that LaMET can reach.

The procedure for resumming small-$x$ logarithm is as follows: first, for each $x$ we match the quasi-PDF to the PDF at the initial scale $\mu_0(x)=2\kappa xP^z$ where $\kappa\sim 1$. Then, we evolve the PDF $f(x, \mu_0(x))$ to a fixed $\overline{\mathrm{MS}}$ scale $\mu=2$ GeV for $x\in [x_{\rm min}, 1]$. For matching accuracy at NLO, the two-loop DGLAP evolution kernel~\cite{Curci:1980uw} is needed, which makes the resummation at next-to-leading logarithmic (NLL) order. Since the DGLAP evolution is closed for $x\in [x_{\rm min}, 1]$, such an evolution kernel can be calculated uniquely following the strategies in Refs.~\cite{Su:2022fiu,Ding:2024saz}.

Apart from the small-$x$ logarithms, the matching coefficient also contains the singular terms $\ln(1-\xi)/(1-\xi)$ and $1/(1-\xi)$~\cite{Gao:2021hxl}, which become important as $\xi\to1$. These are the standard threshold logarithms that also appear in the matching kernel for the GI quasi-PDFs~\cite{Ji:2023pba,Ji:2024hit}. The factorization and resummation of such logarithms have been derived for the GI quasi-PDFs~\cite{Ji:2023pba,Ji:2024hit} and for the quasi generalized parton distributions~\cite{Holligan:2025baj}, and the same methodology can be applied to the CG quasi distributions, although this has not yet been carried out in the literature. According to Ref.~\cite{Ji:2024hit}, the relevant external scale is $2(1-x)P^z$, implying that threshold resummation becomes significant---and eventually loses perturbative control---at large $x$. For this reason, we postpone its implementation in the present work and instead estimate the maximum value $x_{\rm max}$ allowed by LaMET by imposing the condition $\alpha_s(2(1-x_{\rm max})P^z)\sim 1$.

\subsection{Numerical Results}

The two-point correlator with hadron momentum $\vec{P}$ is defined as  
\begin{equation}
    C^{s s'}_{\mathrm{2pt}} \left(t_{\rm sep} \right) = \sum_{\vec{x}} e^{-i \vec{P} \cdot \vec{x}} ~ \mathcal{P}^{\mathrm{2pt}}_{\alpha  \beta} \langle N^{s}_\alpha (\vec{x}, t_{\rm sep}) \bar{N}^{s'}_\beta (\vec{0}, 0) \rangle ~,
\end{equation}
where the nucleon interpolating operator $N$ is defined as Eq.~(\ref{eq:nucleon_interpolator_standard}), the source is positioned at the origin for simplicity, and the positive parity is projected using $\mathcal{P}^{\mathrm{2pt}} = \frac{1}{2}(1 + \gamma^t)$. In this work, we use a Gaussian-smeared source and sink ($ s = s' = S $), following the setup in Ref.~\cite{Izubuchi:2019lyk}. The superscript $SS$ will be removed in the following text for simplicity. By inserting a complete set of states, the two-point correlator can be expressed as a sum over energy eigenstates:
\begin{equation}
    C_{\mathrm{2pt}}\left(t_{\rm sep}\right) = \sum_{n=0}^{n_{\mathrm{s}}-1} \frac{\left|z_n\right|^2}{2 E_n} \left(e^{-E_n t_{\rm sep}}+e^{-E_n\left(L_t-t_{\rm sep}\right)}\right) ~.
    \label{eq:2pt}
\end{equation}  
The overlap amplitude $ z_n = \langle \Omega | N_\alpha \mathcal{P}^{\mathrm{2pt}}_{\alpha  \beta} | n \rangle $ quantifies the projection of the nucleon interpolator onto the $ n $th energy eigenstate, where $|\Omega \rangle$ denotes the QCD vacuum state and $ n_s $ denotes the number of excited states with the same quantum numbers as the nucleon, which are considered within the fit function.

\begin{table}[h]
\renewcommand{\arraystretch}{1.2} 
    \centering
    \begin{tabular}{|c|c|c|c|}
        \hline\hline
        Fit & $n_s$ & $t_{\rm sep}$ range & $\tau$ range \\
        \hline
        $C_{\rm 2pt}(t_{\rm sep})$ & 2 & $t_{\rm sep} \in [4, 12]$ & / \\
        \hline
        $R(t_{\rm sep}, \tau)$ & 2 & $t_{\rm sep} \in [8, 12]$ & $\tau \in [3, t_{\rm sep} - 3]$ \\
        \hline
        $\mathrm{FH}(t_{\rm sep}, \tau_{\rm cut} = 3)$ & 1 & $t_{\rm sep} \in [8, 12]$ & $\tau \in [3, t_{\rm sep} - 3]$ \\
        \hline\hline
    \end{tabular}
    \caption[Collection of ground state fit settings for nucleon PDFs]{Collection of ground state fit settings. \#state = 1 means that only one ground state is included in the fit functions.}
    \label{tab:nucleon_pdf_gsfit}
\end{table}

The bare quasi-PDFs can be extracted from the three-point function with hadron momentum $\vec{P}$
\begin{equation}
    C_{\mathrm{3pt}} (t_{\mathrm{sep}}, \tau) = \sum_{\vec{x}, \vec{z}_0} e^{-i \vec{P} \cdot \vec{x}} ~ \mathcal{P}^{\mathrm{3pt}}_{\alpha \beta} \left\langle N_\alpha  (\vec{x}, t_{\mathrm{sep}}) ~ \mathcal{O}_{\tilde \Gamma} (\vec{z}_0, z, \tau) ~ \bar{N}_\beta (\vec{0}, 0) \right\rangle ~,
\end{equation}
where nucleon polarization projection operator $\mathcal{P}^{\mathrm{3pt}}$ is chosen to be $\frac{1}{2}(1+\gamma_t)$, $\frac{1}{2}(1+\gamma_t)(-i\gamma_5\gamma_z)$, and $\frac{1}{2}(1+\gamma_t)(-i\gamma_5 \gamma_\perp)$
for the unpolarized, helicity and transversity quasi-PDF respectively. Throughout, we take the hadron momentum direction as $\hat{z}$ and the transverse spin direction as $\hat{x}$. The insertion operator is given as isovector bilinear quark operators without Wilson line
\begin{equation}
    \mathcal{O}_{\tilde \Gamma} (\vec{z}_0, z, \tau) = \bar{u}(\vec{z}_0 + z \hat{z}, \tau) \tilde \Gamma u (\vec{z}_0, \tau) - \bar{d}(\vec{z}_0 + z \hat{z}, \tau) \tilde \Gamma d (\vec{z}_0, \tau) ~,
\end{equation}
where the corresponding Dirac structures $\tilde \Gamma$ appropriate for the three channels are $\gamma^t$ (unpolarized), $i \gamma^z \gamma^5$ (helicity) and $-i \gamma^z \gamma^y$ (transversity).  
In terms of the spectral decomposition, it can be expanded as
\begin{eqnarray}
    C_{\mathrm{3pt}}\left(t_{\mathrm{sep}}, \tau \right) = \sum^{n_s -1}_{n, m = 0} \frac{z_n^\dagger O_{n m} z_m}{4 E_n E_m} e^{-E_n\left(t_{\mathrm{sep}}- \tau \right)} e^{-E_m \tau} ~,
    \label{eq:qtmdpdf_3pt}
\end{eqnarray}
where $ O_{n m}=\langle n| \mathcal{O}_{\tilde \Gamma} |m \rangle$ are the matrix elements of the insertion operator. The ground state matrix element $O_{00}$, determined by our selection of $\tilde \Gamma$, is related to the bare quasi-PDFs $\tilde{h}^0_{\tilde \Gamma}$ through the relationship $O_{00} = 2 E_0 \tilde{h}^0_{\tilde \Gamma}$.

We employ the Feynman–Hellmann (FH) method~\cite{Bouchard:2016heu,He:2021yvm} to suppress excited-state contamination in the extraction of the ground-state matrix elements. The analysis is carried out using a fully correlated Bayesian least-squares procedure. We first fit the two-point correlator $C_{\rm 2pt}$ from which the ground- and excited-state energies and amplitudes ($E_0$, $E_1$, $z_0$, $z_1$) are obtained. The posterior distributions of these parameters are then used as Bayesian priors in a subsequent joint fit of the ratio
\begin{equation}
    R\left(t_{\mathrm{sep}}, \tau \right) = \frac{C_{\mathrm{3pt}}\left(t_{\mathrm{sep}}, \tau \right)}{C_{\rm 2pt} \left(t_{\mathrm{sep}} \right)} ~,
\end{equation}
together with the FH correlators. The FH correlator is defined as
\begin{eqnarray}
    S\left(t_{\mathrm{sep}}, \tau_{\mathrm{cut}} \right) &\equiv \sum_{t=\tau_{\mathrm {cut }}}^{t=t_{\mathrm {sep }}-\tau_{\mathrm {cut }}} R\left(t_{\mathrm {sep}}, \tau\right) \\
    \mathrm{FH}\left(t_{\mathrm{sep}}, \tau_{\mathrm{cut}}, d\tau \right) &\equiv \frac{ S\left(t_{\mathrm{sep}} + d\tau, \tau_{\mathrm{cut}} \right) - S\left(t_{\mathrm{sep}}, \tau_{\mathrm{cut}} \right) }{d\tau} ~,
\end{eqnarray}
where $d\tau$ is set to unit. In the large-source–sink separation limit $t_{\rm sep} \to \infty$, both $R$ and $\mathrm{FH}$ converge asymptotically to the same bare quasi-PDF matrix element, $\tilde{h}^0_{\tilde \Gamma}$. The entire chained procedure—first fitting $C_{\rm 2pt}$ and then performing the joint fit of $R$ and FH with the two-point posteriors as priors—is repeated on each jackknife sample to ensure robust and reliable uncertainty estimates. The detailed setting can be found in Table~\ref{tab:nucleon_pdf_gsfit}.

\begin{figure}[!th]
    \begin{center}
    \includegraphics[width=0.3\textwidth]{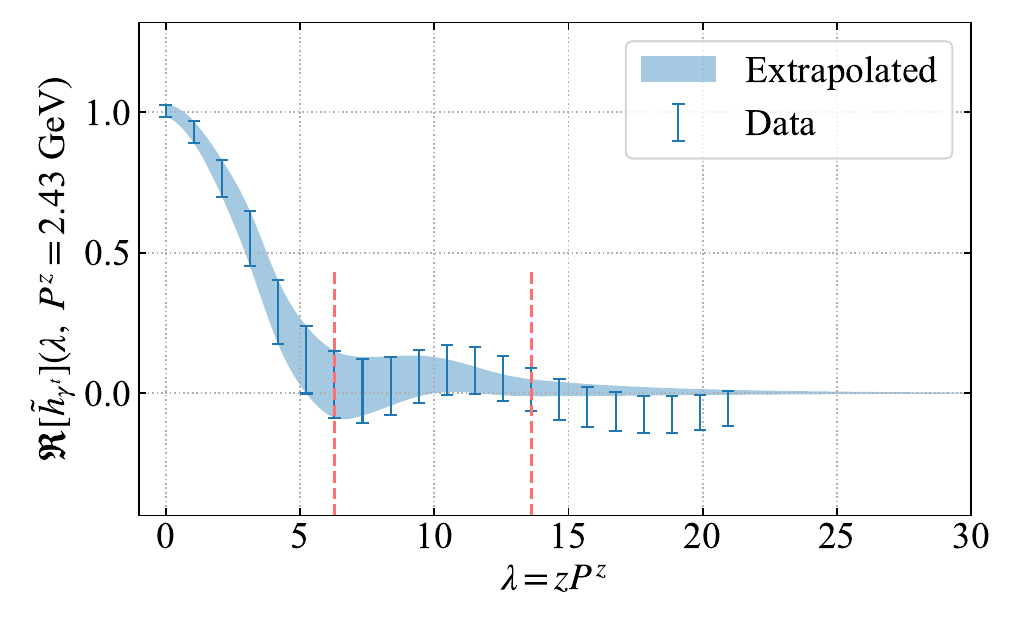}
    \includegraphics[width=0.3\textwidth]{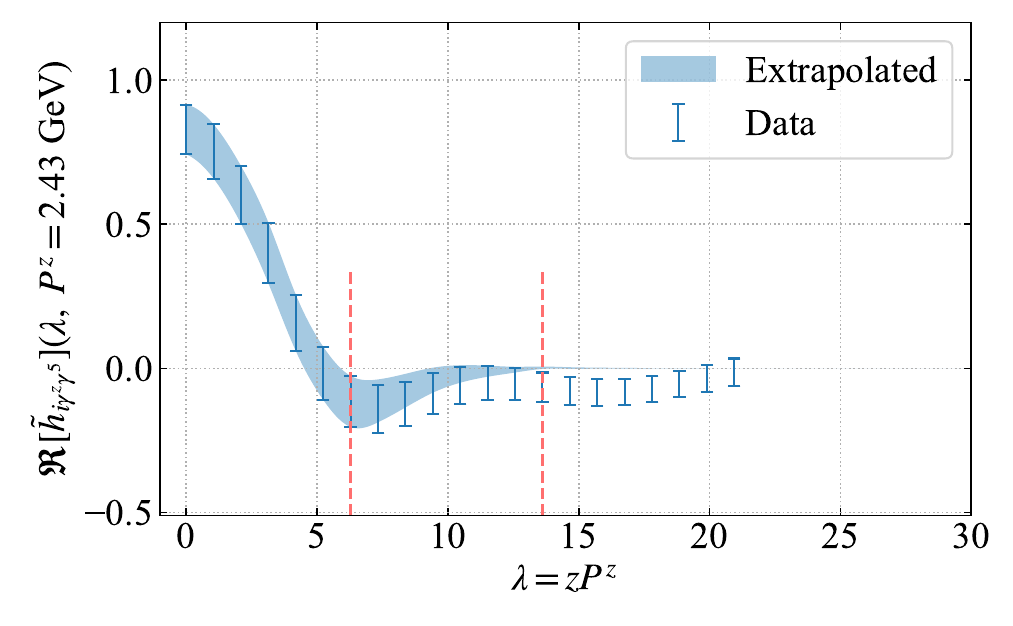}
    \includegraphics[width=0.3\textwidth]{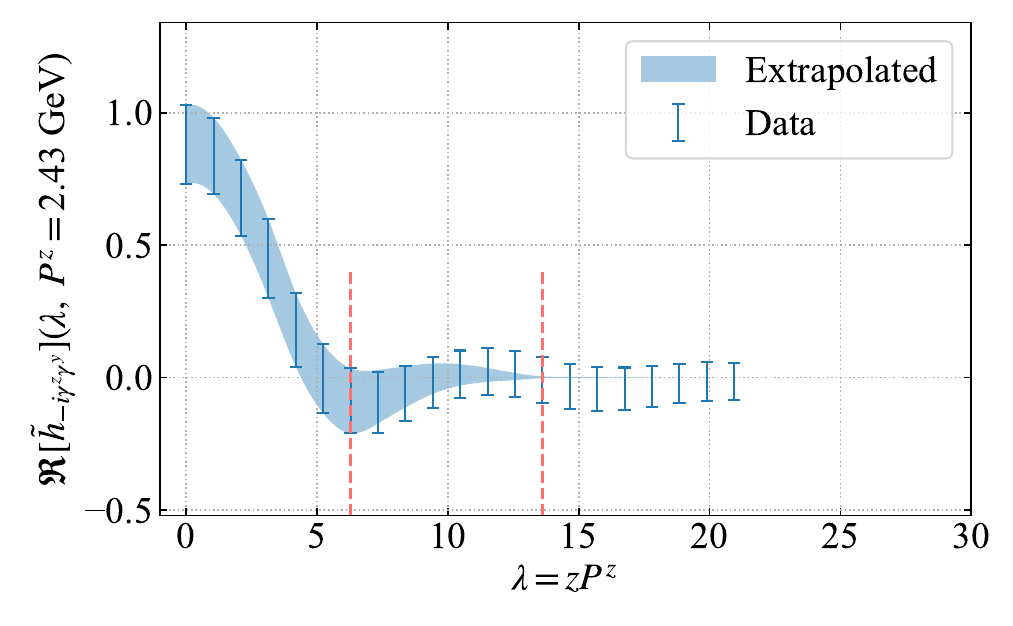}
    \includegraphics[width=0.3\textwidth]{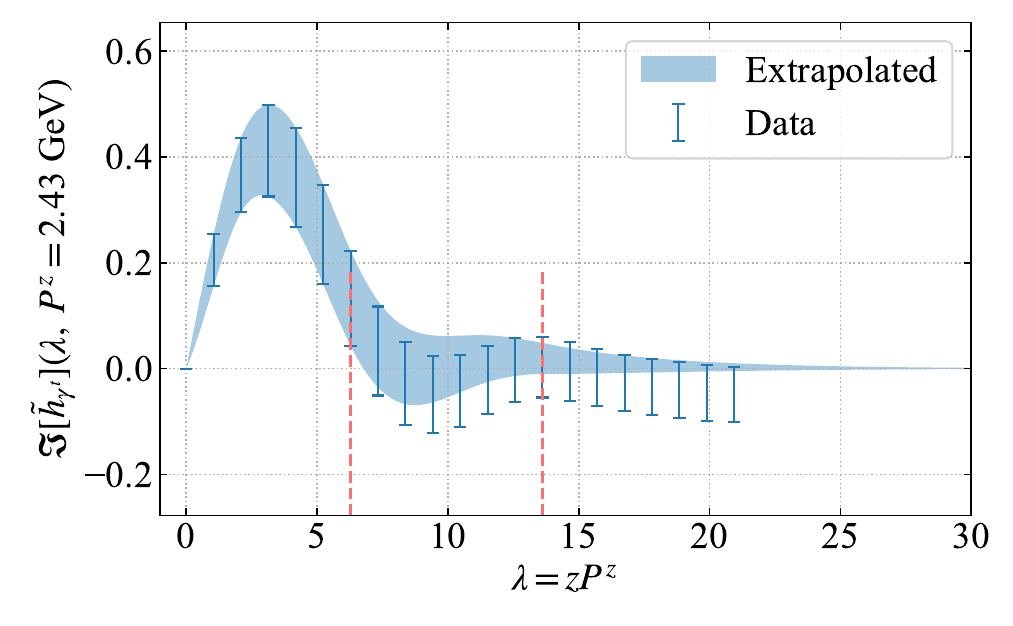}
    \includegraphics[width=0.3\textwidth]{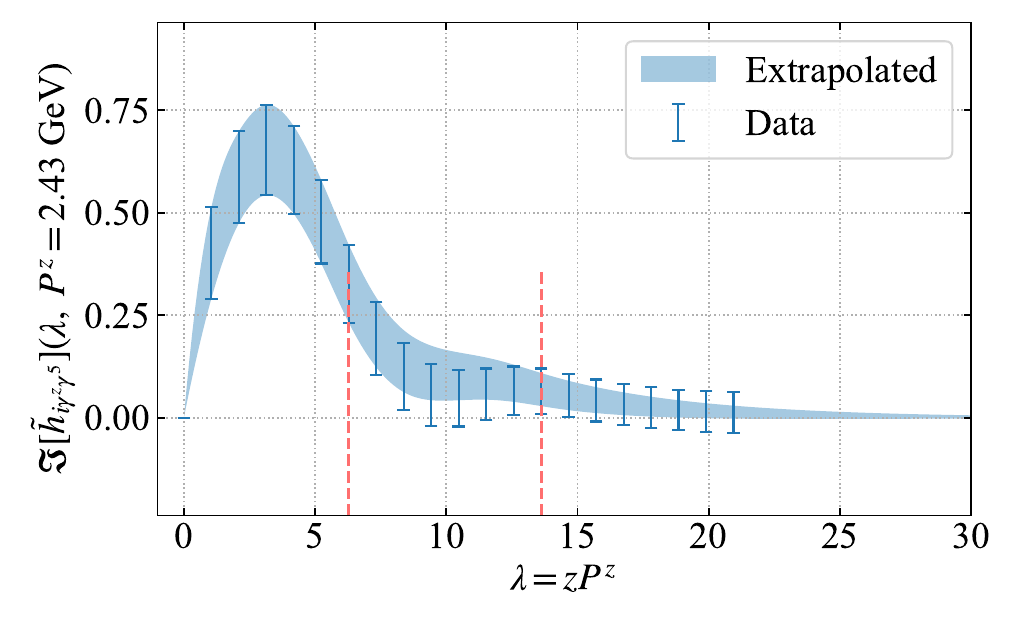}
    \includegraphics[width=0.3\textwidth]{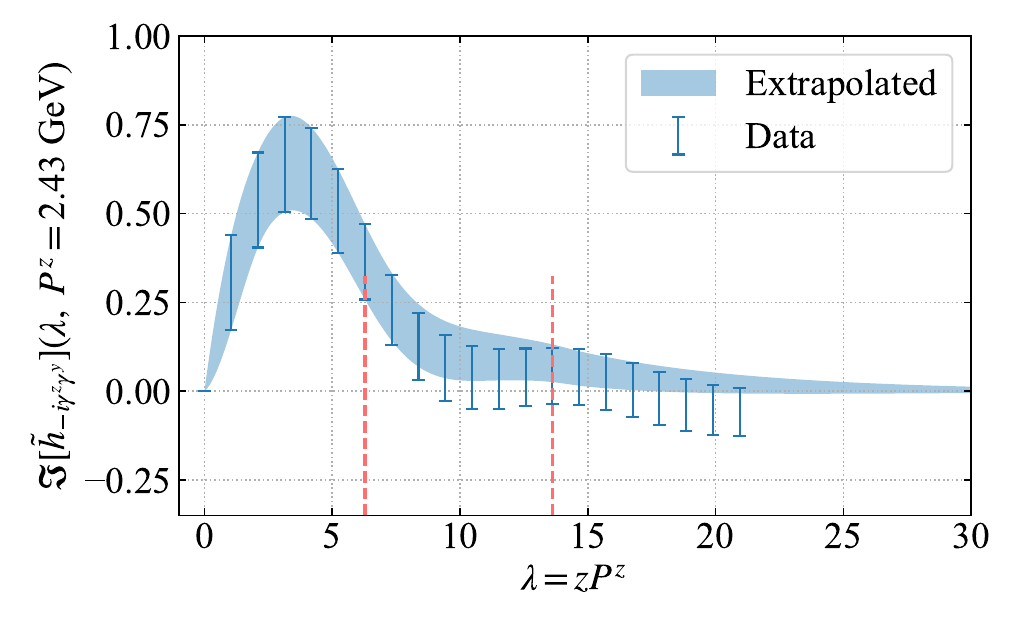}
    \caption[Renormalized quasi-PDF matrix elements]{Extrapolation of the renormalized quasi-PDF matrix elements in coordinate space as functions of $\lambda = z P^z$. The three columns from left to right correspond to the unpolarized, helicity, and transversity channels, while the upper (lower) row shows the results of real (imaginary) part at $P^z = 2.43$ GeV. The data points are shown with error bars, and the shaded bands represent the extrapolation using the asymptotic form. The two red dashed vertical lines indicate the sub-asymptotic region used in the fit.}\label{fig:pdf_extrapolation}
    \end{center}
\end{figure}

As discussed above, we employ the hybrid scheme as Eq.~(\ref{eq:hybrid_scheme_no_normalization}) to renormalize the real part of the quasi-PDF matrix elements, with the separation point is chosen as $z_s=3 \sqrt{2} a$; while the SRI$^\prime$-MOM scheme as Eq.~(\ref{eq:sri_mom_renormalization}) is used for the imaginary part. In addition, to preserve the antisymmetry of the imaginary part, its local contribution at $z=0$ is explicitly set to zero. The renormalized matrix elements in the coordinate space are shown in Fig.~\ref{fig:pdf_extrapolation} for three polarization channels.

The quasi-PDF matrix elements decrease rapidly as a function of $\lambda = zP^z$, approaching zero for $\lambda \gtrsim 5$. Nevertheless, as discussed above, an asymptotic extrapolation remains necessary to suppress unphysical fluctuations in the uncertainty band. A sub-asymptotic form for nucleon PDFs in GI formulations was derived in Ref.~\cite{Ji:2026vir}, as given in Eq.~(\ref{eq:asymptotic_extrapolation}). In this form, the large-$z$ behavior is characterized by an exponential suppression controlled by the lightest intermediate state, accompanied by a power-law prefactor and a phase factor that depends on the sign of $\lambda$.

While the exponential decay structure is robust, the precise power-law behavior depends on the details of the theoretical formulation. In particular, for the CG formalism, the corresponding algebraic dependence has not yet been fully understood from first principles. Motivated by these considerations, we adopt the following parameterization for the sub-asymptotic behavior in the CG,
\begin{eqnarray}
   \tilde{h} (\lambda) =
   \left[
      A_2 \cdot e^{i \mathrm{sign}(\lambda)\phi_2}
      + \frac{A'_2}{|\lambda|}
      \cdot e^{i \mathrm{sign}(\lambda)\phi'_2}
   \right]
   \frac{e^{-m |\lambda|}}{|\lambda|^n} ,
\end{eqnarray}
where the overall factor $1/|\lambda|^n$ is introduced to parametrize the unknown algebraic decay. This form is used to fit the data in the sub-asymptotic region
$0.5~\mathrm{fm} \leq z \leq 1.1~\mathrm{fm}$,
as indicated by the two red dashed lines in
Fig.~\ref{fig:pdf_extrapolation}. In the fit function above, there are six parameters: $A_2$, $A'_2$, $\phi_2$, $\phi'_2$, $m$ and $n$, where $m$ is set to be larger than $0.1$ GeV / $P^z$~\cite{Gao:2021dbh}. To get a smooth change of the error band, the extrapolated results are given in the form as
 \begin{eqnarray}
     \tilde{h}^{\rm ext} = w \cdot \tilde{h}^{\mathrm{data}} + (1 - w) \cdot \tilde{h}^{\rm fit} ~,
    \label{eq:extrapolation_weight}
 \end{eqnarray}
where $w$ is a weight function that transitions linearly from 1 to 0 within the fit range, $\tilde{h}^{\mathrm{data}}$ is the renormalized quasi-distribution, and $\tilde{h}^{\rm fit}$ refers to the fit results of the asymptotic fit. After applying this extrapolation, the uncertainty bands smoothly converge to zero, mitigating non-physical fluctuations. The model uncertainty of this asymptotic extrapolation is bounded from above and can be improved systematically~\cite{Gao:2021dbh,Chen:2025cxr}.

\subsubsection{Unpolarized PDF}

\begin{figure}[th!]
    \centering
    \includegraphics[width=0.4\linewidth]{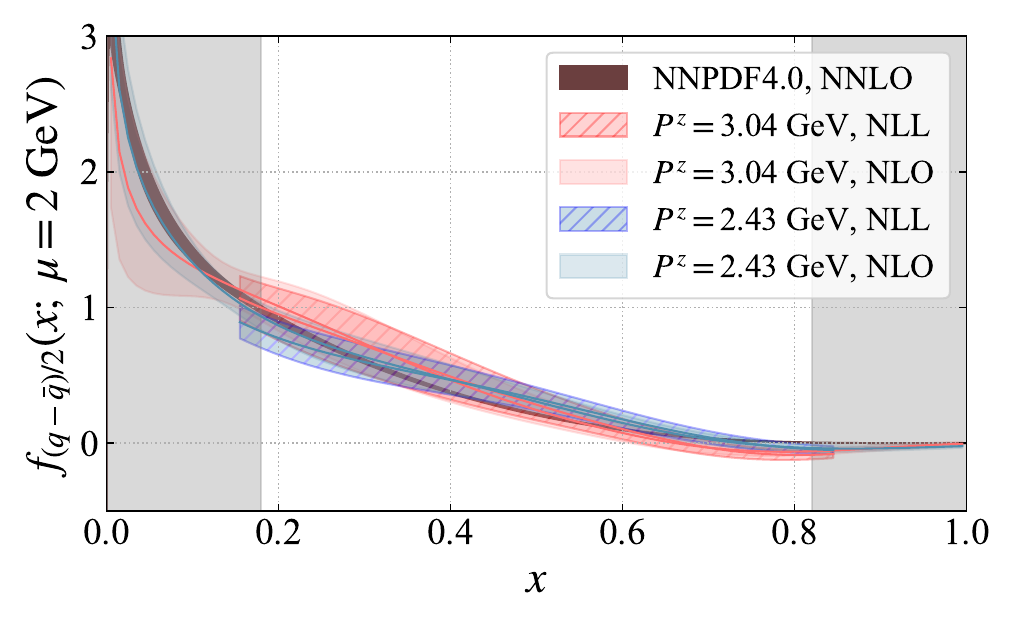}
    \includegraphics[width=0.4\linewidth]{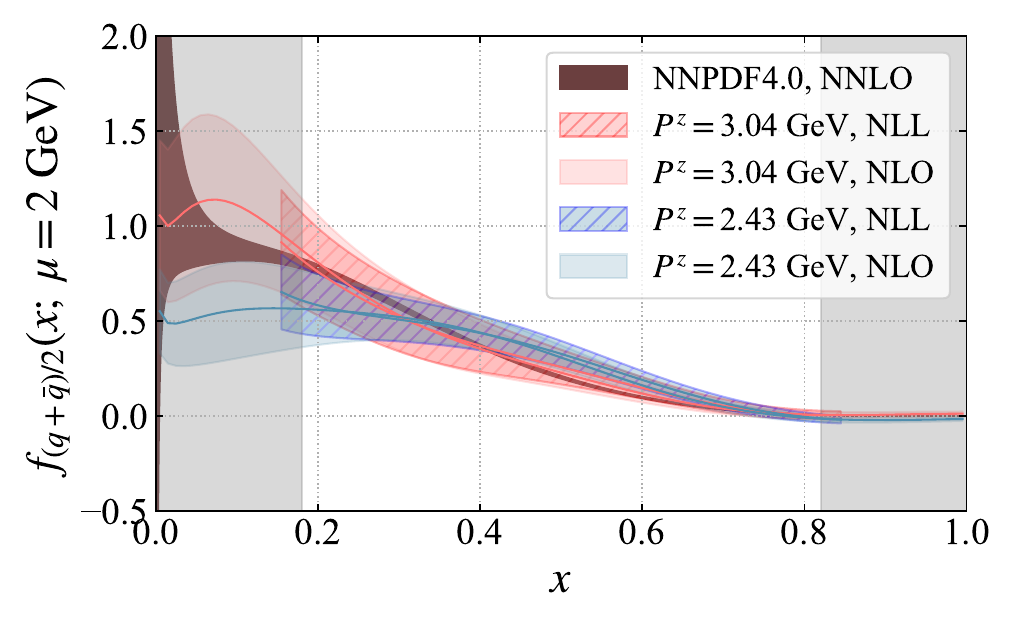}
    \caption[Valence and full contribution of unpolarized nucleon PDF]{Unpolarized quark PDFs of the nucleon at momenta $P^z = 2.43$ and $3.04$~GeV, extracted from lattice QCD and compared with the NNLO phenomenological results from NNPDF4.0~\cite{NNPDF:2021njg}. The NLO and NLL labels indicate the perturbative accuracy of matching coefficients. The left and right panels show the valence contribution $(q-\bar q)/2$ and the full distribution $(q+\bar q)/2$, respectively. Results obtained using matching coefficients at NLO and NLL are shown to illustrate the perturbative convergence. The shaded gray regions ($x < 0.18$ or $x > 0.82$) denote kinematic domains where the strong coupling $\alpha_s(\mu = 2 x P^z)$ or $\alpha_s(\mu = 2 (1-x) P^z)$ at NLO exceeds 0.5, indicating a poor convergence of perturbation theory.}
    \label{fig:unpol_pdf_xdep_sep}
\end{figure}

As discussed above, the real and imaginary parts of the matrix elements are renormalized in different schemes, and their contributions correspond to the valence and full quark channels of the nucleon PDFs, respectively. Therefore, these two contributions can be independently compared with existing results in the literature. After performing the large-momentum expansion, the matched results from the real and imaginary parts are shown in the left and right panels of Fig.~\ref{fig:unpol_pdf_xdep_sep}, respectively. The unpolarized nucleon PDFs at hadron momenta $P^z = 2.43$ and $3.04$~GeV are presented together with the NNLO phenomenological determination from NNPDF4.0~\cite{NNPDF:2021njg}. By employing matching coefficients at different perturbative accuracies, we further obtain results at next-to-leading order (NLO) as well as at next-to-leading logarithmic (NLL) accuracy with DGLAP evolution. The shaded gray regions in Fig.~\ref{fig:unpol_pdf_xdep_sep} indicate kinematic domains where the strong coupling $\alpha_s(\mu = 2 x P^z)$ evaluated at NLO exceeds 0.5. In these regions, the perturbative expansion underlying the matching procedure is expected to have poor convergence, and the corresponding lattice results should therefore be interpreted with caution.

\begin{figure}[th!]
    \centering
    \includegraphics[width=0.6\linewidth]{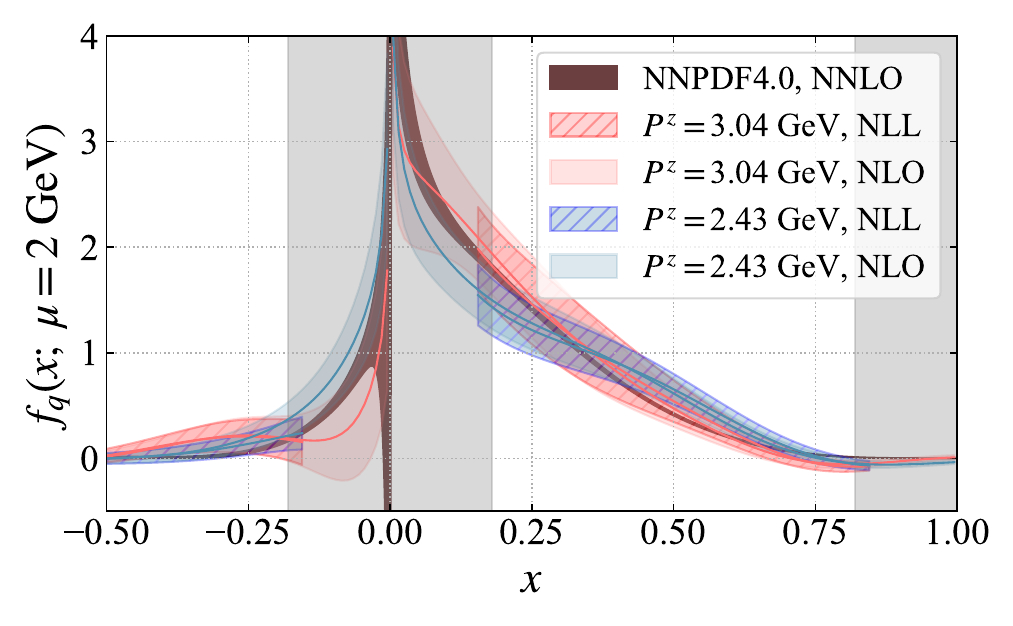}
    \caption[Unpolarized nucleon PDF]{Unpolarized quark PDFs of the nucleon at momenta $P^z = 2.43$ and $P^z = 3.04$~GeV. The lattice results are shown together with the NNLO phenomenological determination from NNPDF4.0 for comparison.}
    \label{fig:unpol_pdf_complete}
\end{figure}

The results in Fig.~\ref{fig:unpol_pdf_xdep_sep} show that the lattice calculations at hadron momentum $P^z = 3.04$~GeV are in good agreement with the determination of NNPDFs4.0 for both the valence and the full components of the unpolarized PDFs. The comparison between the NLO and NLL cases further indicates a stable perturbative behavior and supports the convergence of the matching expansion. In the right panel, corresponding to $(q + \bar q)/2$, the difference between the two lattice momenta is more pronounced than in the $(q - \bar q)/2$ channel, which is likely caused by residual excited-state contamination in the extraction of the bare quasi-matrix elements. By combining the valence and sea contributions shown in Fig.~\ref{fig:unpol_pdf_xdep_sep}, we present the resulting unpolarized nucleon PDFs in Fig.~\ref{fig:unpol_pdf_complete}, together with the results of NNPDF4.0~\cite{NNPDF:2021njg} for comparison.

\subsubsection{Helicity PDF}

\begin{figure}[th!]
    \centering
    \includegraphics[width=0.4\linewidth]{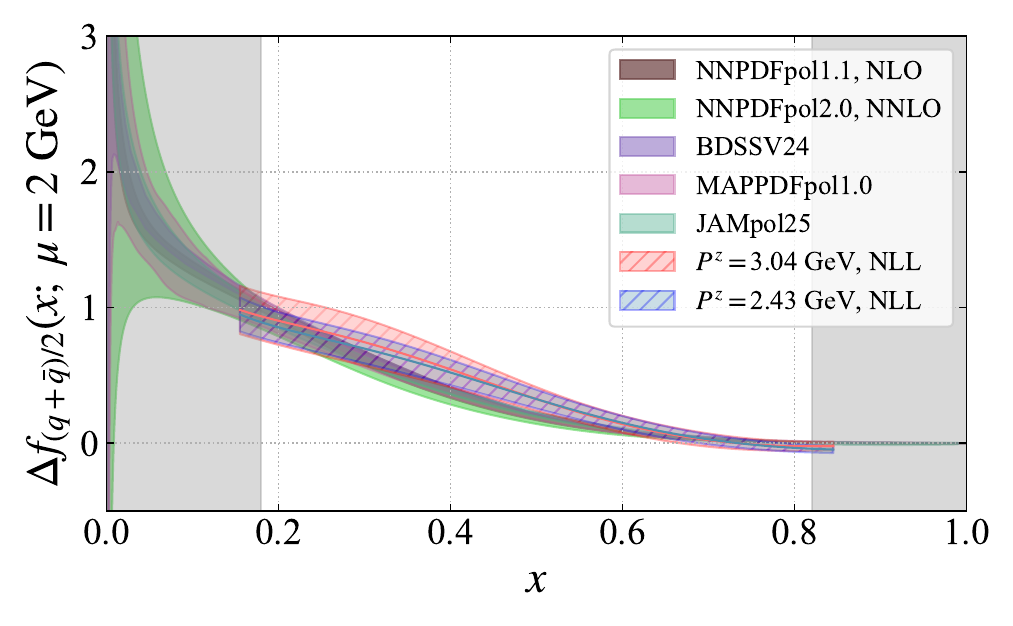}
    \includegraphics[width=0.4\linewidth]{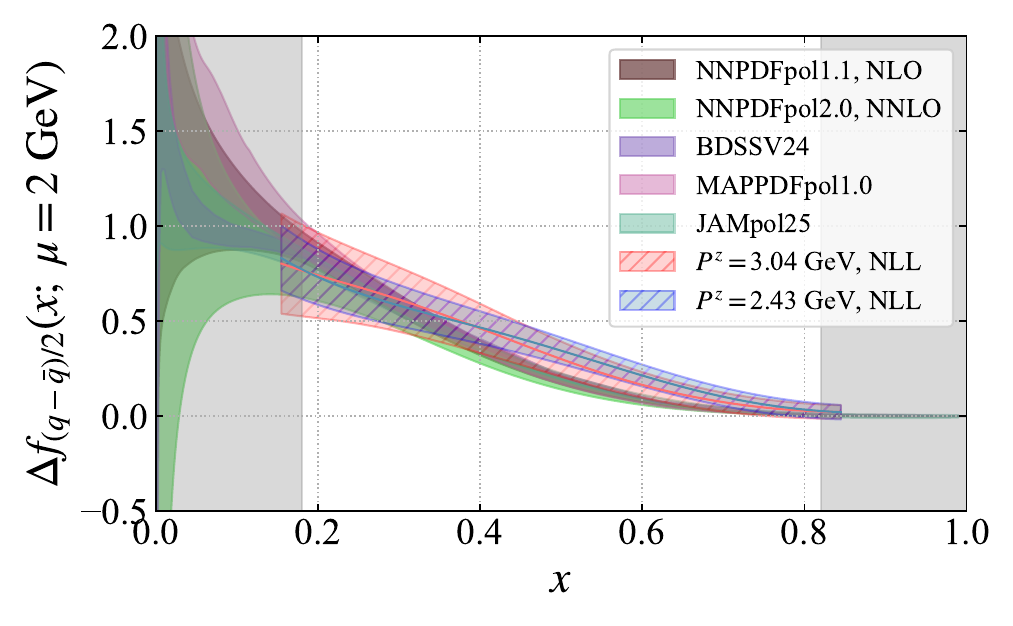}
    \caption[Valence and full contribution of helicity nucleon PDF]{Helicity quark PDFs of the nucleon at momenta $P^z = 2.43$ and $P^z = 3.04$~GeV, compared with phenomenological determinations. The left and right panels show the full $(q+\bar q)/2$ and valence $(q-\bar q)/2$ components, respectively. The lattice results are presented at NLL accuracy, while the phenomenological bands include the NLO NNPDFpol1.1~\cite{Nocera:2014gqa}, NNLO NNPDFpol2.0~\cite{Cruz-Martinez:2025ahf}, BDSSV24~\cite{Borsa:2024mss}, MAPPDFpol1.0~\cite{Bertone:2024taw}, and JAMpol25~\cite{Cocuzza:2025qvf} analyses.}
    \label{fig:heli_pdf_xdep_sep}
\end{figure}

Similar to the unpolarized case, the full $(q + \bar q)/2$ and valence $(q - \bar q)/2$ components of the helicity PDFs are shown in the left and right panels of Fig.~\ref{fig:heli_pdf_xdep_sep}, respectively. Note that the full helicity distribution is obtained from the real part of the matrix elements, since the helicity quasi-distribution is even under charge conjugation. The helicity nucleon PDFs at hadron momenta $P^z = 2.43$ and $3.04$~GeV are compared with phenomenological determinations, including the NLO results from NNPDFpol1.1~\cite{Nocera:2014gqa}, the NNLO results from NNPDFpol2.0~\cite{Cruz-Martinez:2025ahf}, as well as the BDSSV24~\cite{Borsa:2024mss}, MAPPDFpol1.0~\cite{Bertone:2024taw}, and JAMpol25~\cite{Cocuzza:2025qvf} analyses. All phenomenological results are normalized by their respective axial charges $g_A$, while the valence $(q - \bar q)/2$ components of the lattice results are normalized using the $g_A$ value from JAMpol25~\cite{Cocuzza:2025qvf}. 

\begin{figure}[th!]
    \centering
    \includegraphics[width=0.6\linewidth]{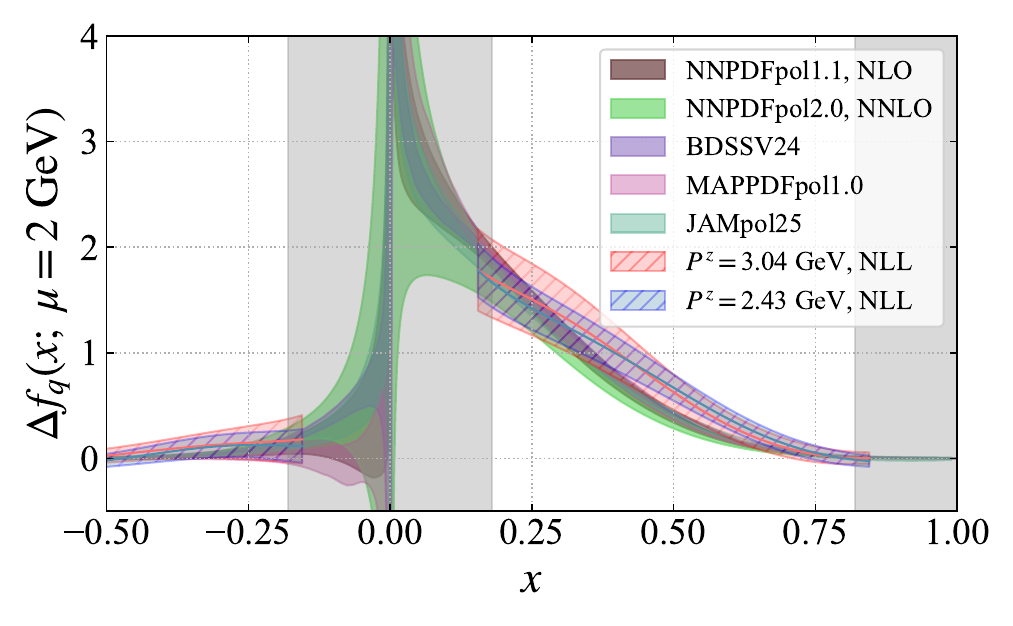}
    \caption[Helicity nucleon PDF]{Helicity quark PDFs of the nucleon at momenta $P^z = 2.43$ and $P^z = 3.04$~GeV. The lattice-extracted distributions are shown together with phenomenological determinations for comparison.}
    \label{fig:heli_pdf_complete}
\end{figure}

From the plots in Fig.~\ref{fig:heli_pdf_xdep_sep}, we find that our results at $P^z = 3.04$ GeV are consistent with the phenomenological ones within one standard deviation for both the full and valence parts. Moreover, as in the unpolarized case, the discrepancy between the two momenta is more pronounced in the contribution from the imaginary part of the matrix elements. Combining the full and valence parts, the helicity nucleon PDFs are presented in Fig.~\ref{fig:heli_pdf_complete}.

\subsubsection{Transversity PDF}

\begin{figure}[th!]
    \centering
    \includegraphics[width=0.4\linewidth]{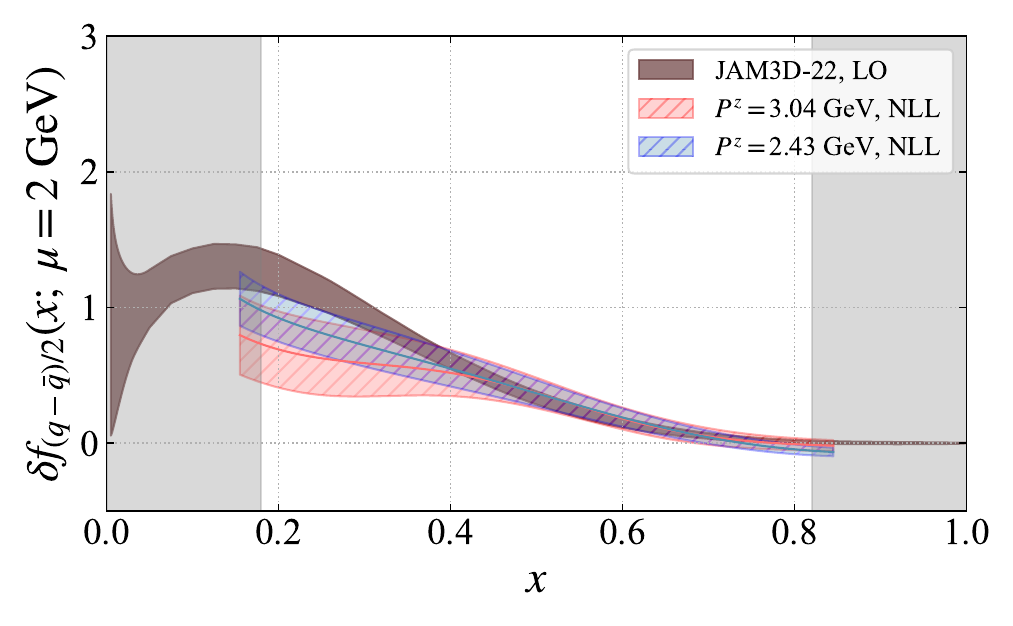}
    \includegraphics[width=0.4\linewidth]{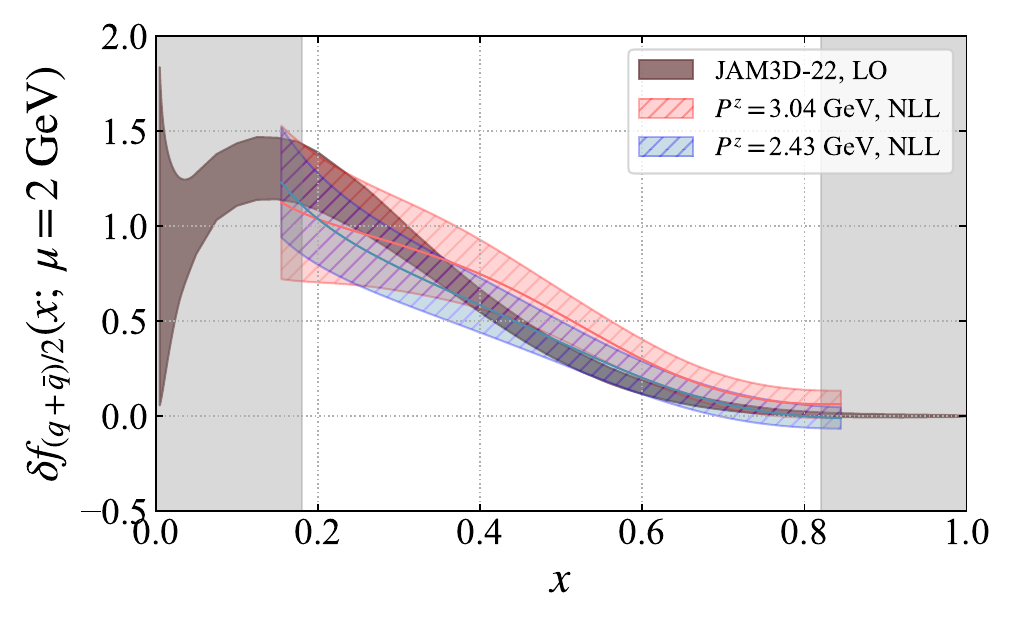}
    \caption[Valence and full contribution of transversity nucleon PDF]{Transversity quark PDFs of the nucleon at momenta $P^z = 2.43$ and $P^z = 3.04$~GeV, compared with phenomenological and lattice determinations. The left and right panels show the valence $(q-\bar q)/2$ and full $(q+\bar q)/2$ components, respectively. The lattice results at NLL accuracy are compared with the leading-order phenomenological extraction from JAM3D-22~\cite{Gamberg:2022kdb}.}
    \label{fig:trans_pdf_xdep_sep}
\end{figure}

In analogy with the unpolarized and helicity cases, we also present our results for the transversity PDFs. The valence $(q - \bar q)/2$ and full $(q + \bar q)/2$ contributions are shown in the left and right panels of Fig.~\ref{fig:trans_pdf_xdep_sep}, respectively. The transversity nucleon PDFs at hadron momenta $P^z = 2.43$ and $3.04$~GeV are compared with the phenomenological extraction from JAM3D-22~\cite{Gamberg:2022kdb} at leading order (LO). The JAM3D-22 distributions are normalized by their own tensor charge $g_T$, and the full $(q + \bar q)/2$ contribution of the lattice results are normalized using the same value to ensure a consistent comparison. 

\begin{figure}[th!]
    \centering
    \includegraphics[width=0.6\linewidth]{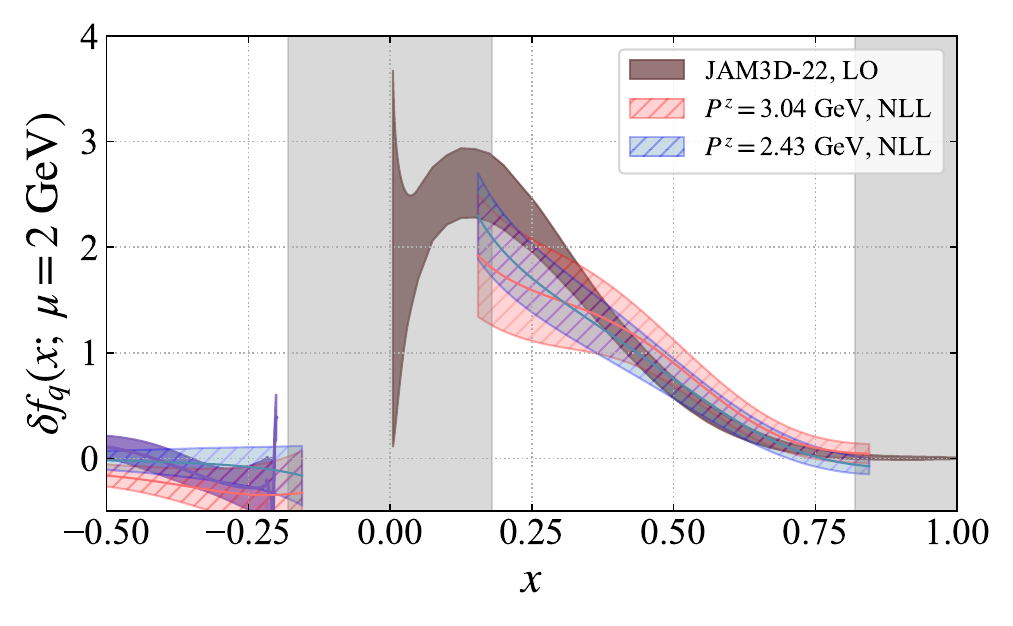}
    \caption[Transversity nucleon PDF]{Transversity quark PDFs of the nucleon at momenta $P^z = 2.43$ and $P^z = 3.04$~GeV, shown together with phenomenological results for comparison. The lattice-extracted PDFs are compared with the JAM3D-22 extraction at LO.}
    \label{fig:trans_pdf_complete}
\end{figure}

As illustrated in Fig.~\ref{fig:trans_pdf_xdep_sep}, we observe that our lattice results show good agreement with the JAM3D-22 phenomenological extractions in the moderate-$x$ region. The level of consistency is even more evident in the comparison of the full part of the transversity PDFs. Similar to the unpolarized and helicity cases, the discrepancy between the two momenta is more pronounced in the contribution from the imaginary part of the matrix elements, underlining once again the need to carefully control excited-state contamination in this sector. The combined valence and full parts of the transversity nucleon PDFs are presented in Fig.~\ref{fig:trans_pdf_complete}. After all, since the phenomenological extraction of the transversity PDF still carries substantial uncertainties, caution is required in our comparison.

\subsection{Conclusion}

In this work, we have carried out the first lattice QCD calculation of isovector nucleon PDFs using the CG method, including all three polarization channels. Employing an off-axis momentum setup, we achieved large hadron momenta up to $P^z > 3$ GeV, which allows for a controlled application of the LaMET framework. A systematic treatment of excited-state contamination was performed, and we demonstrated that the CG method produces nucleon PDFs in the moderate-$x$ region that are broadly consistent with phenomenological determinations. These findings validate the CG method as a promising tool for extracting precision-controlled nucleon parton distributions from lattice QCD.

More specifically, at the largest momentum in this work, $P^z=3.04$ GeV, we observed an encouraging agreement between our lattice results and the existing phenomenological extractions for all three polarizations of nucleon PDFs: unpolarized, helicity, and transversity. The consistency is more evident in the contributions from the real part of the matrix elements. 
In contrast, the contributions from the imaginary part of the matrix elements display larger discrepancies across different hadron momenta. This suggests that the imaginary sector is more sensitive to residual excited-state contamination and may require even more careful analyses in future studies.

The renormalization structure of the Coulomb-gauge quasi-PDF operators is significantly simpler than that of the GI operator due to the lack of the Wilson line (and hence, lack of linear divergence and associated renormalon). 
At short separations $z \sim a$, the operator $\bar{\psi}_0(z)\tilde\Gamma \psi_0(0)$ picks up large discretization effects. The hybrid scheme should remove the leading artifacts for the real part of the matrix elements, which, however, does not necessarily work for the imaginary part. Therefore, we propose to renormalize the latter using the SRI$^\prime$-MOM scheme, which is a simple method for renormalizing the Coulomb-gauge quark fields. 
To further understand the lattice artifacts, a full study of these schemes is planned by studying the quark mass and lattice spacing dependence of the renormalization constants. 

Together, these observations highlight both the promise and the challenges of extracting partonic information from the lattice at large momentum. The path forward is clear. With access to greater computational resources, it will become possible to carry out simulations at multiple lattice spacings, thereby enabling controlled continuum extrapolations that systematically reduce discretization effects. In parallel, accumulating higher statistics for correlators at large source–sink separations will improve the stability of the extracted observables and enhance the overall robustness of the final results. Further progress can be achieved by adopting more refined and effective strategies for treating excited-state contamination, such as the Lanczos algorithm~\cite{Wagman:2024rid,Hackett:2024xnx,Hackett:2024nbe} and the current-inspired Generalized Eigenvalue Problem (GEVP)~\cite{Barca:2024hrl,Barca:2025det}, which is expected to be particularly important for reliably resolving the imaginary components of matrix elements. Finally, the application of new lattice technologies aimed at improving the signal-to-noise ratio (SNR), such as kinematically enhanced interpolating operators~\cite{Zhang:2025hyo} and more optimized smearing schemes, will be essential to extend these studies to higher momenta and increased precision. Combining these improvements will place the CG method on a firmer foundation and allow precise lattice determinations of nucleon PDFs that can stand alongside phenomenological and experimental results. In this way, our present work provides not only a first demonstration of the CG method for nucleon PDFs, but also a roadmap for systematically improving the approach toward a future of quantitatively precise lattice QCD parton distributions.

\section{Calculation of Pion TMDs}

In high-energy scatterings, transverse-momentum-dependent parton distribution functions (TMDPDFs) provide a fundamental description of the transverse momentum and polarization degrees of freedom of quarks and gluons within hadrons~\cite{Boussarie:2023izj}. These distributions play a crucial role in understanding the intricate dynamics of quark-gluon interactions and the phenomenon of color confinement. Knowledge of TMDPDFs is essential for predicting observables in multi-scale, non-inclusive high-energy processes, such as semi-inclusive deep-inelastic scattering (SIDIS) and Drell-Yan (DY), based on QCD factorization. Experiments at facilities including the Jefferson Lab 12 GeV Program~\cite{Dudek:2012vr}, the EIC~\cite{Accardi:2012qut,AbdulKhalek:2021gbh}, and the LHC~\cite{Amoroso:2022eow} rely heavily on accurate knowledge of the TMDPDFs.

Significant progress has been made in the phenomenological parameterizations of TMDPDFs, particularly for the nucleon, through global analyses of experimental data~\cite{Davies:1984sp,Ladinsky:1993zn,Landry:2002ix,Konychev:2005iy,Sun:2014dqm,DAlesio:2014mrz,Bacchetta:2017gcc,Scimemi:2017etj,Bertone:2019nxa,Scimemi:2019cmh,Bacchetta:2019sam,Hautmann:2020cyp,Bury:2022czx,Bacchetta:2022awv,Isaacson:2023iui,Aslan:2024nqg,Bacchetta:2024yzl,Moos:2023yfa,Yang:2024drd,Bacchetta:2024qre,Moos:2025sal}. While these studies have significantly improved our understanding of the transverse-momentum structure of quarks and gluons within the nucleon, they remain in their early stages due to the limited availability of experimental data sensitive to the non-perturbative region. Compared to nucleons, much less is known about the transverse momentum structure of the pion~\cite{Vladimirov:2019bfa,Cerutti:2022lmb,Barry:2023qqh}. As the lightest QCD bound state, the pion plays a fundamental role in hadron structure and non-perturbative QCD. Its TMDPDFs provide crucial insights into the internal dynamics of quarks and gluons in a strongly interacting relativistic system, with direct implications for hadronization processes and the underlying mechanisms of non-perturbative QCD. However, the scarcity of experimental data on pion TMDPDFs, combined with the inherent nature of the parameterization dependence of global analyses, leaves significant gaps in our non-perturbative first-principles understanding. Lattice QCD calculations provide a natural approach to gain insight into non-perturbative TMD structures of the pion starting from first-principles QCD.

In our recent work~\cite{Bollweg:2025iol}, we present the first lattice QCD calculation of the unpolarized valence-quark pion TMDPDF using the CG method. We compute the quasi-TMDPDF and quasi-TMDWF matrix elements of the pion in CG. These matrix elements enable the extraction of the CS kernel, the intrinsic soft function, the TMDWF, and ultimately the unpolarized valence TMDPDF of the pion. Furthermore, we improve the perturbative accuracy of the matching procedure, particularly in the extraction of the intrinsic soft function, by resumming logarithmic terms in the Sudakov kernel, an aspect that has not been accounted for in previous lattice QCD studies. Finally, the comparison of our results with global fits of the experimental data shows encouraging consistency, demonstrating the bright potential of the CG approach for TMD physics from lattice QCD.

\subsection{Theoretical Framework}

As proposed in Ref.~\cite{Zhao:2023ptv}, the light-cone TMDPDF can be derived from the CG quasi-TMD beam function defined as
\begin{equation}
    \tilde{f}_{\Gamma}(x, b_\perp, P^z; \mu) = P^z \int \frac{d z}{2 \pi} e^{i z (x P^z)} \tilde{h}_{\Gamma}(z, b_\perp, P^z; \mu) ~,
\end{equation}  
where the matrix elements are given by
\begin{equation}
    \tilde{h}_{\Gamma}(z, b_\perp, P^z; \mu) = \frac{1}{2 P_\Gamma} \langle P | \left. \bar{q}(z, b_\perp) \Gamma q (0) \right|_{\vec{\nabla} \cdot \vec{A} = 0} | P \rangle ~,
\end{equation}  
where $P_\Gamma$ is the kinetic factor for normalization. In this work, we choose $ \Gamma = \gamma^t $ for quasi-TMD beam function and $P_\Gamma = P^t$. The parameter $\mu$ represents the $\overline{\mathrm{MS}}$ renormalization scale, and the space-like separation between the quark and the antiquark is denoted as $\vec{b} \equiv (z, \vec{b}_\perp)$. The hadron state carries momentum $P = (P^t, 0, 0, P^z)$ and satisfies the relativistic normalization condition $\langle P | P \rangle = 2 P^t (2\pi)^3 \delta^{(3)}(\vec{0})$. When the hadron moves with a large momentum $P^z$, the quasi-TMD beam function factorizes as
\begin{equation}
    \tilde{f}_{\Gamma}(x, b_\perp, P^z; \mu) = H_f\left(x, P^z; \mu\right) B (x, b_\perp, xP^+; \mu, \nu) \times S^0_C(b_\perp;\mu, \nu) ~,
    \label{eq:qtmdpdf_sep}
\end{equation}  
where $B (x, b_\perp, xP^+; \mu, \nu)$ is the light-cone beam function with zero-bin subtraction, following the procedure outlined in Refs.~\cite{Manohar:2006nz,Stewart:2009yx}. The term $S^0_C(b_\perp;\mu, \nu)$ represents the zero-bin contribution, and the operator definitions of both functions are detailed in Ref.~\cite{Zhao:2023ptv}. The parameters $\mu$ and $\nu$ correspond to the renormalization scales associated with the ultraviolet (UV) and rapidity divergences, respectively. The function $H_f\left(x, P^z; \mu\right) = |C_{\rm TMD} (xP^z; \mu)|^2$ represents the hard kernel that matches the QCD quark field operator to Soft-Collinear Effective Theory (SCET)~\cite{Vladimirov:2020ofp}.  

The light-cone TMDPDF can be defined from the light-cone beam function $B $ and the soft function $S$ as
\begin{equation}
    f (x, b_\perp; \mu, \zeta) = B (x, b_\perp, xP^+; \mu, \nu) S (b_\perp, y_n;\mu, \nu) ~,
    \label{eq:tmdpdf_sep}
\end{equation}  
where the dependence on the rapidity renormalization scale $\nu$ cancels out on the right-hand side, leaving only a dependence on the Collins-Soper scale $\zeta \equiv 2 (x P^+)^2 e^{-2 y_n}$. Combining Eqs.~(\ref{eq:qtmdpdf_sep}) and (\ref{eq:tmdpdf_sep}), the factorization formula connecting the quasi-TMDPDF and the light-cone TMDPDF is given by~\cite{Ebert:2018gzl,Ebert:2019okf,Ji:2019sxk,Ji:2019ewn,Zhao:2023ptv}  
\begin{eqnarray}
    \sqrt{ S_I\left(b_{\perp}, y_n ; \mu\right)} \cdot \tilde{f}_{\Gamma}(x, b_\perp, P^z; \mu) =f (x, b_\perp; \mu, \zeta) H_f\left(x, P^z; \mu\right) + \mathrm{p. c.} ~,
    \label{eq:factorization_tmdpdf_short}
\end{eqnarray}  
where p.c. indicates the power corrections, and the intrinsic soft function $S_I$ is defined as  
\begin{equation}
    S_I\left(b_{\perp}, y_n ; \mu\right) \equiv \left( \frac{S (b_\perp, y_n;\mu, \nu)}{S^0_C(b_\perp;\mu, \nu)} \right)^2 ~\,.
\end{equation}
Here $S^0_C$ is the CG quasi-soft function, as defined in Ref.~\cite{Zhao:2023ptv}, which exactly cancels the rapidity divergences in $S$. The above factorization formula is of the same form as that for the GI quasi-TMDPDFs~\cite{Ebert:2018gzl,Ebert:2019okf,Ji:2019sxk,Ji:2019ewn}, while the method to calculate $S_I$ was first proposed in Ref.~\cite{Ji:2019sxk}, which enables a complete determination of the TMDs from the lattice.

The intrinsic soft function satisfies the evolution equation  
\begin{eqnarray}
    \frac{d}{d y_n} \ln S_I\left(b_{\perp}, y_n ; \mu\right) = -2 \gamma^{\overline{\rm MS}}(b_\perp; \mu) ~,
\end{eqnarray}  
where $\gamma^{\overline{\rm MS}}(b_\perp; \mu)$ is the Collins-Soper kernel. Therefore, the intrinsic soft function at any $y_n$ satisfies 
\begin{eqnarray}
    S_I\left(b_{\perp}, y_n ; \mu\right) = S_I\left(b_{\perp}, y_n=0 ; \mu\right) \cdot \exp(-2 y_n \cdot  \gamma^{\overline{\rm MS}}(b_\perp; \mu)) ~.
\end{eqnarray}
One can redefine the intrinsic soft function as $S_I\left(b_{\perp}; \mu\right) \equiv S_I\left(b_{\perp}, y_n = 0; \mu\right)$, allowing us to rewrite Eq.~(\ref{eq:factorization_tmdpdf_short}) in a more explicit form, as shown in Eq.~(\ref{eq:tmdpdf_matching}).
This factorization framework provides a systematic approach to relating quasi-TMDPDFs extracted from lattice QCD to their light-cone counterparts, ensuring a well-defined separation between perturbative and non-perturbative contributions. However, power corrections of order 
$\mathcal{O}\left(\frac{\Lambda_{\mathrm{QCD}}}{x P^z}, \frac{1}{b_{\perp} x P^z}\right)$~\cite{Liu:2023onm} introduce systematic uncertainties, particularly in the small-$x$ region. Despite these limitations, the framework offers valuable first-principles constraints in the large-$x$ regime, where power corrections are expected to be relatively small.

Similar to the quasi-TMD beam function, the light-cone TMDWF can be derived from the CG quasi-TMDWF, which is defined as  
\begin{eqnarray}
    \tilde{\phi}_\Gamma (x, b_\perp, P^z; \mu) = P^z \int \frac{d z}{2 \pi} e^{i z (x P^z)} \tilde{\varphi}_{\Gamma}(z, b_\perp, P^z; \mu)~,
    \label{eq:qtmdwf_def_1}
\end{eqnarray}  
with the quasi-TMDWF matrix elements given by  
\begin{eqnarray}
    \tilde{\varphi}_{\Gamma}(z, b_\perp, P^z; \mu) = \frac{e^{-i z P^z / 2}}{f_\pi P_\Gamma} \langle \Omega | \left. \bar{q}(z/2, b_\perp/2) \Gamma q (-z/2, -b_\perp/2) \right|_{\vec{\nabla} \cdot \vec{A} = 0} | P \rangle ~,
    \label{eq:qtmdwf_def_2}
\end{eqnarray}  
which corresponds to a pion-to-vacuum matrix element, the $| \Omega \rangle$ represents the QCD vacuum state, the $f_\pi$ is the pion decay constant, and $P_\Gamma$ is the kinetic factor for normalization. In this work, we choose $ \Gamma = \gamma^z \gamma^5 $ for quasi-TMDWFs with nonzero momenta and $P_\Gamma = i P^z$, while for zero-momentum quasi-TMDWFs, we set $ \Gamma = \gamma^t \gamma^5 $ and $P_\Gamma = P^t$.   

In the large-momentum limit, the quasi-TMDWF can be matched to the light-cone TMDWF via the factorization formula
\begin{eqnarray}
    \sqrt{ S_I\left(b_{\perp}, y_n ; \mu\right)} \cdot \tilde{\phi}_{\Gamma}\left(x, b_{\perp}, P^z ; \mu\right)  =\phi (x, b_\perp, y_n; \mu, \zeta, \bar{\zeta}) H_\phi\left(x, \bar{x}, P^z; \mu\right) + \mathrm{p. c.} ~,
    \label{eq:factorization_tmdwf_short}
\end{eqnarray}  
with $\bar{x} = 1 - x$, $\bar{\zeta} \equiv 2 (\bar{x} P^+)^2 e^{-2 y_n}$, and the hard kernel 
\begin{eqnarray}
    H_\phi (x, \bar{x}, P^z; \mu) = C_{\rm TMD} (xP^z; \mu) \cdot C_{\rm TMD} (\bar{x} P^z; \mu)~,
\end{eqnarray}
calculated up to NLO in Ref.~\cite{Zhao:2023ptv}.

An essential component of quasi-TMD factorization is the CS kernel $ \gamma^{\overline{\rm MS}}(b_\perp; \mu) $, which governs the rapidity evolution of TMDs and is crucial to achieve consistent matching between quasi-TMDPDFs and light-cone TMDPDFs. The CS kernel can be extracted from the ratios of quasi-TMD matrix elements computed at different hadron momenta~\cite{Ji:2014hxa, Ebert:2018gzl,Ji:2019sxk}. In this work, we employ the ratio of CG quasi-TMDWFs~\cite{Ji:2019sxk,Ji:2021znw} as
\begin{eqnarray}
    \gamma^{\overline{\rm MS}}(b_\perp; \mu) &&=\gamma^{\overline{\rm MS}}(b_\perp, P_1, P_2; \mu) + \mathrm{p.c.} \nonumber \\
    &&=\frac{1}{\ln \left(P_2 / P_1\right)} \ln \frac{H_\phi \left(x, \bar{x}, P_1; \mu\right) \tilde{\phi}_{\gamma^z \gamma^5}\left(x, b_{\perp}, P_2;  \mu\right)}{H_\phi \left(x, \bar{x}, P_2; \mu\right) \tilde{\phi}_{\gamma^z \gamma^5}\left(x, b_{\perp}, P_1; \mu \right)} \nonumber \\
    &&\quad + \mathcal{O}\left(\frac{\Lambda_{\mathrm{QCD}}}{x P^z}, \frac{1}{b_{\perp}\left(x P^z\right)}, \frac{\Lambda_{\mathrm{QCD}}}{\bar{x} P^z}, \frac{1}{b_{\perp}\left(\bar{x} P^z\right)}\right) ~.
    \label{eq:cs_kernel_calc}
\end{eqnarray}
This method offers a key advantage over direct extractions from quasi-TMDPDFs, as quasi-TMDWFs naturally peak around $ x = 0.5 $, a region where power corrections are significantly suppressed. On the other hand, the quasi-TMDPDFs usually decrease quickly at moderate to large $x$, making the extraction of the CS kernel more susceptible to power corrections.

Another essential component of the factorization of quasi-TMDs is the intrinsic soft function, which accounts for soft gluon radiation in the process. Its calculation is particularly challenging for two reasons. First, as a non-perturbative quantity, it cannot be computed using standard perturbative techniques. Second, because the soft function involves correlations along two light-cone directions, it cannot be directly simulated on the lattice, even with a large-momentum boost.

A practical approach to determining the intrinsic soft function, as proposed in Refs.~\cite{Ji:2019sxk}, is to use the TMD factorization of a pseudoscalar light-meson form factor, which is defined as
\begin{eqnarray}
    F\left(b_{\perp}, P_1, P_2, \Gamma, \Gamma^\prime \right) \equiv -4N_c \frac{\left\langle P_2 \right| \bar{q}\left(b_{\perp}\right) \Gamma q\left(b_{\perp}\right) \bar{q}(0) \Gamma^{\prime} q(0)\left| P_1 \right\rangle}{f_\pi^2 (P_1\cdot P_2)} ~,
    \label{eq:ff_def}
\end{eqnarray}
where $| P_1 \rangle $ and $ | P_2 \rangle $ denoting the pion states with momentum $P_1$ and $P_2$. The coefficient $-4 N_c = -12$ is a normalization factor~\cite{Deng:2022gzi}. This form factor can be extracted using the following ratio
\begin{eqnarray}
    R_F\left(b_{\perp}, P_1, P_2, \Gamma, \Gamma^{\prime}\right) \equiv -4N_c \frac{\left\langle P_2\right| \bar{q}\left(b_{\perp}\right) \Gamma q\left(b_{\perp}\right) \bar{q}(0) \Gamma^{\prime} q(0)\left|P_1\right\rangle}{\langle 0| \bar{q}(0) \gamma^\mu \gamma^5 q(0)\left|P_1\right\rangle\left\langle P_2\right| \bar{q}(0) \gamma_\mu \gamma^5 q(0)|0\rangle} ~.
    \label{eq:ff_ratio}
\end{eqnarray}
To extract the leading-twist contribution, one can choose the Dirac matrices as 
\begin{equation}
    \Gamma = \Gamma^\prime \in \{ I, \gamma^5, \gamma^\perp, \gamma^\perp \gamma^5 \}
\end{equation}
as suggested in Refs.~\cite{Deng:2022gzi, LatticePartonLPC:2023pdv}. In this work, we choose $\Gamma = \Gamma^\prime \in \{ \gamma^\perp, \gamma^\perp \gamma^5 \}$. The denominator of the ratio $R_F$ consists of two-point functions used to normalize the form factor, and the index $\mu$ should be summed. The momenta in the external pion states are taken as $ P_1 = (P^t, 0, 0, P^z) $ and $ P_2 = (P^t, 0, 0, -P^z) $, moving back-to-back to achieve a large momentum transfer $ Q^2 = (2P^z)^2 $. In this kinematic regime, the form factor can be factorized in terms of the TMDWF $ \phi (x, b_\perp, y_n; \mu, \zeta) $~\cite{Ji:2019sxk, Ji:2021znw, Deng:2022gzi} as
\begin{eqnarray}
    F (b_\perp, P^z) = \int dx_1 dx_2 H_F (x_1, x_2, P^z; \mu) \phi^\dagger (x_1, b_\perp, y_n; \mu, \zeta_1, \bar{\zeta}_1) \phi (x_2, b_\perp, -y_n; \mu, \zeta_2, \bar{\zeta}_2) ~.
    \label{eq:factorization_ff}
\end{eqnarray}
where $ H_F $ is the hard kernel encapsulating the short-distance dynamics of the process. The $y_n$-dependence cancels between $\phi$ and $\phi^\dagger$, and the hard kernel can be expressed as the product of two Sudakov kernels~\cite{Deng:2022gzi}:  
\begin{eqnarray}
    H_F (x_1, x_2, P^z; \mu)= C_{\rm Sud} (x_1, x_2, P^z; \mu) \cdot C_{\rm Sud} (\bar{x}_1, \bar{x}_2, P^z; \mu) ~,
\end{eqnarray}
where $ C_{\rm Sud} $ represents the Sudakov kernel, which has been computed at one-loop order in the literature~\cite{Manohar:2003vb, Deng:2022gzi}. 

By combining the factorization of the quasi-TMDWF in Eq.~(\ref{eq:factorization_tmdwf_short}) with the form factor factorization in Eq.~(\ref{eq:factorization_ff}), the intrinsic soft function at $ y_n = 0 $ can be extracted as  
\begin{eqnarray}
    S_I (b_\perp; \mu) = \frac{F(b_\perp, P^z)}{\int dx_1 dx_2 H_F(x_1, x_2, P^z; \mu) \tilde{\Phi}^\dagger (x_1) \tilde{\Phi} (x_2) } ~,
    \label{eq:soft_func_calc}
\end{eqnarray}  
where the reduced quasi-TMDWF $ \tilde{\Phi} (x) $, is defined as  
\begin{eqnarray}
    \tilde{\Phi} (x) \equiv \frac{\tilde{\phi}_{\Gamma}\left(x, b_{\perp}, P^z ; \mu\right)}{H_\phi \left(x, \bar{x}, P^z; \mu\right)} ~.
    \label{eq:reduced_qtmdwf_def}
\end{eqnarray} 
Using different renormalization schemes for quasi-TMDWF, one can get different intrinsic soft functions, which can be perturbatively converted to each other at small $b_\perp$. However, the scheme dependence will eventually cancel between the renormalized quasi-beam and intrinsic soft functions. 

\subsection{Numerical Results}

We perform a numerical lattice QCD calculation using a $2+1$-flavor gauge ensemble generated by the HotQCD Collaboration~\cite{HotQCD:2014kol}. The ensemble employs the Highly Improved Staggered Quark (HISQ) action~\cite{Follana:2006rc} with a lattice spacing of $ a = 0.06 $ fm and a volume of $ L_s^3 \times L_t = 48^3 \times 64 $. The valence sector is treated using tadpole-improved clover Wilson fermions on a hypercubic (HYP) smeared~\cite{Hasenfratz:2001hp} gauge background. The clover coefficient is set to $ c_{\rm sw} = u_0^{-3/4} $, where $ u_0 $ is the average plaquette after HYP smearing. For this ensemble, we use $ c_{\rm sw} = 1.0336 $ in both the time and spatial directions. The valence quark masses are tuned to yield a valence pion mass of 300 MeV, with a corresponding hopping parameter of $ \kappa \approx 0.12623 $.

The key requirement for the factorization of quasi-TMDs is a sufficiently large hadron momentum. To achieve a higher boost factor, we take advantage of the three-dimensional rotational symmetry of the CG approach and adopt off-axis momenta for the quasi-TMD, choosing momentum directions along $\vec{n} = (n^x, n^y, 0)$. The hadron momenta on the lattice are given by $P^z = \frac{2\pi |\vec{n}|}{L_s a}$. In this study, we consider $ n^x = n^y \in \{3, 4, 5\} $, which allows us to reach a maximum hadron momentum of $ P^z = 3.04 $ GeV for the quasi-TMD calculations. To optimize the signal-to-noise ratio and enhance overlap with large-momentum hadron ground states, we employ boosted Gaussian smearing~\cite{Bali:2016lva}, using the same setup as in Ref.~\cite{Gao:2023lny}. To extract the ground-state contribution, we compute quasi-TMD three-point functions for multiple source-sink separations, choosing $ t_{\rm sep} / a = 6, 8, 10, 12 $. Calculations are performed on 553 gauge configurations and we apply the All-Mode Averaging (AMA) technique~\cite{Shintani:2014vja} to further improve the signal. The stopping criteria for the sloppy and exact inversions are set to $10^{-4}$ and $10^{-10}$, respectively, aligning with the settings in Ref.~\cite{Gao:2020ito}. For the smaller separation $ t_{\rm sep} / a = 6 $, we compute 1 exact source and 32 sloppy sources per configuration, while for the larger separations $ t_{\rm sep} / a = 8, 10,12 $, we compute 4 exact sources and 128 sloppy sources. 

In addition, we compute the large-momentum pion form factor following the strategy outlined in Ref.~\cite{LatticeParton:2020uhz}. Specifically, we evaluate the two-current three-point function  
\begin{eqnarray}
    C_{\mathrm{F}}(t_{\rm sep},\tau) = \left\langle \pi_{\rm w}(-\vec{P},t_{\rm sep})\, j_u (b_\perp, \tau)\, j_d(0, \tau)\, \pi_{\rm w}^\dagger(\vec{P},0)\right\rangle,
\end{eqnarray}
where the two currents $j_u (b_\perp, \tau) \equiv \bar{u}(b_\perp, \tau) \Gamma u(b_\perp, \tau)$ and $j_d(0, \tau) \equiv \bar{d}(0, \tau) \Gamma d(0, \tau)$ are separated by a transverse distance $ b_\perp $, inserted at the same time slice $ \tau $. The Dirac structures are chosen as $\Gamma \in \{ \gamma^\perp, \gamma^\perp \gamma^5 \}$ to extract the leading-twist contribution~\cite{Deng:2022gzi, LatticePartonLPC:2023pdv}. Unlike the two-point and quasi-TMD correlator calculations, where Gaussian-smeared sources are used, here we adopt wall sources $ \pi_{\rm w} $ for the pions. Given the significant computational demands of this calculation and the universality of the soft factor, insensitive to the choice of hadron species, we choose to employ a heavier valence pion mass of 670 MeV. Consequently, our analysis of the pion form factor is limited to a subset of 100 gauge configurations using wall sources in $L_t=64$ time slices. We compute the form factor at a single momentum $ \vec{n} = (0, 0, 6) $, corresponding to $ P^z = 2.58 $ GeV. To ensure consistency in the extraction of the intrinsic soft function, we also compute the quasi-TMDWF using the same valence pion mass. For quasi-TMDWF, we use momenta in the $ z $-direction with $ \vec{n} = (0, 0, n^z) $ and $ n^z \in \{8, 9, 10\} $, reaching a maximum momentum of $ P^z = 4.30 $ GeV. This part of the calculation is performed on the same 553 HotQCD gauge configurations used in the main quasi-TMD analysis, with 1 exact source and 32 sloppy sources per configuration.

To extract the ground-state quasi-TMD matrix elements, it is essential to first determine the energy spectrum associated with the pion interpolator $ \pi^\dagger $. The two-point correlator is defined as  
\begin{equation}
    C^{s s'}_{\mathrm{2pt}} \left(t_{\rm sep}, P^z \right) = \langle \pi_{s'} (P^z, t_{\rm sep}) \pi^\dagger_{s} (P^z, 0) \rangle ~,
\end{equation}
where $ \pi^\dagger_s $ and $ \pi_{s'} $ represent the pion source and sink defined from Eq.~(\ref{eq:pion_interpolator_standard}) as
\begin{equation}
    \pi_s(\vec{P},t_{\rm sep})= \sum_{\vec{x}} \pi_s(\vec{x},t_{\rm sep})  e^{-i \vec{P}\cdot\vec{x}} ~.
\end{equation}
In this work, we use a Gaussian-smeared source and sink ($ s = s' = S $), following the setup in Ref.~\cite{Izubuchi:2019lyk}. The superscript $SS$ will be removed in the following text for simplicity.

By inserting a complete set of states, the two-point correlator can be expressed as a sum over energy eigenstates:
\begin{equation}
    C_{\mathrm{2pt}}\left(t_{\rm sep}\right) = \sum_{n=0}^{N_{\mathrm{s}}-1} \frac{\left|z_n\right|^2}{2 E_n} \left(e^{-E_n t_{\rm sep}}+e^{-E_n\left(L_t-t_{\rm sep}\right)}\right) ~.
\end{equation}  
The overlap amplitude $ z_n = \langle n | \pi^\dagger | \Omega \rangle $ quantifies the projection of the pion interpolator onto the $ n $th energy eigenstate, and $ N_s $ denotes the number of excited states with the same quantum numbers as the pion, which are considered within the fit function.

The quasi-TMD beam function is extracted from the three-point correlator 
\begin{equation}
    C_{\tilde{h}}\left(t_{\mathrm{sep}}, \tau \right)=\left\langle \pi_s(\vec{P},t_{\rm sep}) \bar{q}(z, b_\perp, \tau) \gamma^t q (\vec{0}, \tau) \pi_s^\dagger(\vec{P},0)\right\rangle~,
\end{equation}  
which can be expressed in terms of a spectral decomposition: 
\begin{equation}
    C_{\tilde{h}}\left(t_{\mathrm{sep}}, \tau \right) = \sum^{N_s -1}_{n, m = 0} \frac{z_n^\dagger O_{n m} z_m}{4 E_n E_m} e^{-E_n\left(t_{\mathrm{sep}}- \tau \right)} e^{-E_m \tau} ~,
\end{equation}  
where $ O_{n m}=\langle n|\bar{q}(z, b_\perp) \gamma^t q (0)|m \rangle$ is proportional to the matrix elements of the quasi-TMD operator. To extract the ground-state matrix element $ O_{00} = 2 E_0 \tilde{h}_{\gamma^t}$, we employ a two-state chained fit using the two-point correlator $ C_{\rm 2pt} $ and the ratio  
\begin{equation}
    R_{\tilde{h}}\left(t_{\mathrm{sep}}, \tau \right) = \frac{C_{\tilde{h}}\left(t_{\mathrm{sep}}, \tau \right)}{C_{\rm 2pt} \left(t_{\mathrm{sep}} \right)} ~.
\end{equation}  
In the limit $ t_{\rm sep} \to \infty $, the ratio $ R_{\tilde{h}} $ asymptotically converges to the bare quasi-TMD matrix element $ \tilde{h}_{\gamma^t}$. The chained fit procedure first fits the two-point correlator $ C_{\rm 2pt} $, then uses the posterior distributions of $ E_0 $, $ E_1 $, $ z_0 $, and $ z_1 $ as priors in the subsequent fit of $ R_{\tilde{h}} $. 

As discussed in Ref.~\cite{Gao:2023lny}, the absence of Wilson lines in the CG correlator eliminates linear divergences, allowing the renormalized operator to be defined as  
\begin{equation}
    \bar{q}_0 (z,b_\perp) \Gamma q_0 (0) = Z_q (a) \left[ \bar{q} (z,b_\perp) \Gamma q(0) \right] ~,
\end{equation}  
where $Z_q$ is the CG quark wave function renormalization factor. This renormalization is independent of both the external hadron states and the spatial separation of the quark bilinear operator. Consequently, an appropriate ratio can be constructed to cancel out the renormalization factor.

Following the approach in Ref.~\cite{Bollweg:2024zet}, we renormalize the quasi-TMD matrix elements using the ratio  
\begin{eqnarray}
    \tilde{h}_{\gamma^t}(z, b_\perp, P^z; \mu) = \frac{\tilde{h}^0_{\gamma^t}(z, b_\perp, P^z; a)}{\tilde{\varphi}^0_{\gamma^t \gamma^5}(z=0, b_\perp, P^z=0; a)} ~,
\end{eqnarray}  
and similarly for the quasi-TMDWF, it is renormalized as
\begin{equation}
    \tilde{\varphi}_{\gamma^z \gamma^5}(z, b_\perp, P^z; \mu) = \frac{\tilde{\varphi}^0_{\gamma^z \gamma^5}(z, b_\perp, P^z; a)}{\tilde{\varphi}^0_{\gamma^t \gamma^5}(z=0, b_\perp, P^z=0; a)} ~.
    \label{eq:renorm_qtmdwf}
\end{equation}  
Here, $\tilde{\varphi}^0$ denotes the bare quasi-TMDWF matrix elements defined in Eq.~(\ref{eq:qtmdwf_def_2}). 

\begin{figure}[th!]
    \centering
    \includegraphics[width=.3\linewidth]{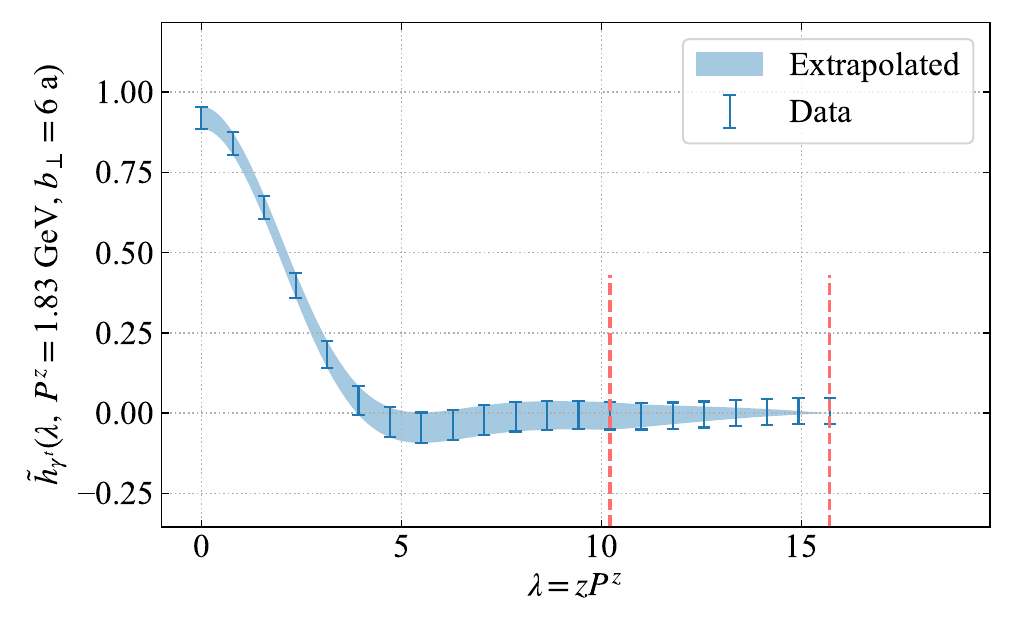}
    \includegraphics[width=.3\linewidth]{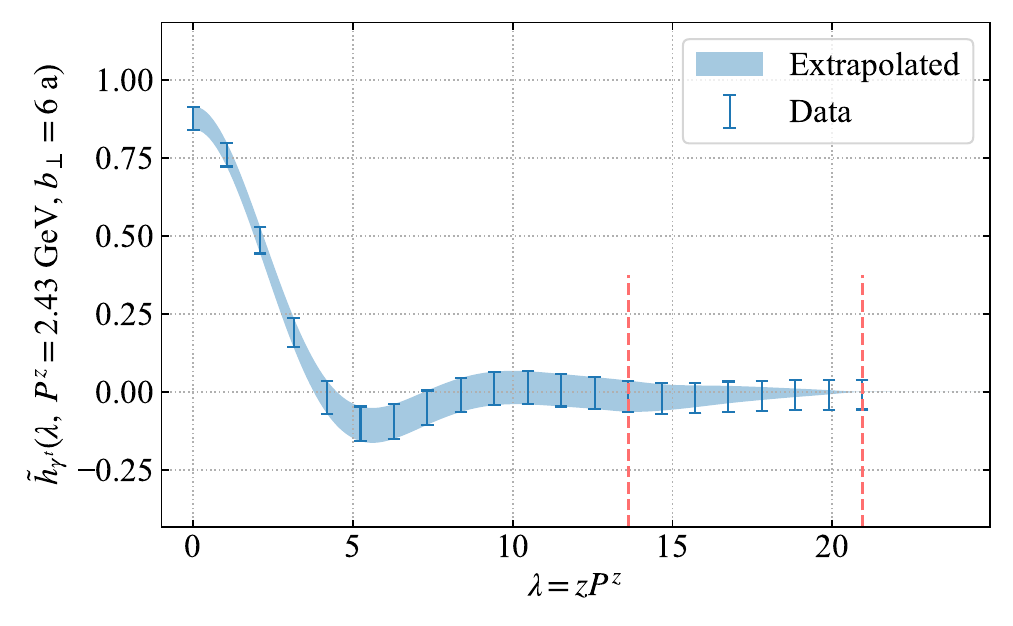}
    \includegraphics[width=.3\linewidth]{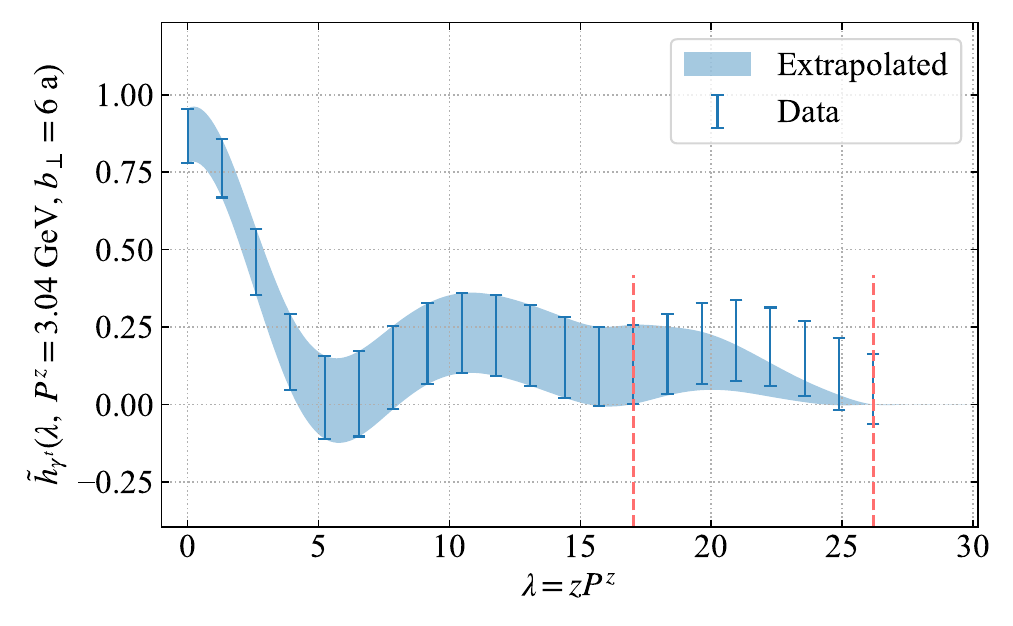}
    \caption[Renormalized quasi-TMD matrix elements]{The extrapolation of the renormalized quasi-TMD in coordinate space for various hadron momenta is shown in three panels. From left to right, they correspond to hadron momenta $P^z=1.83$, 2.43 and 3.04 GeV, respectively. The regions between the two red dashed lines indicate the extrapolation range. For different hadron momenta, the same starting point of $z = 0.78$ fm is chosen for the extrapolation.
    \label{fig:extra_qtmdpdf}}
\end{figure}

The renormalized matrix elements in the coordinate space are presented in Fig.~\ref{fig:extra_qtmdpdf}. As demonstrated in Ref.~\cite{Gao:2021dbh}, the extracted quasi-distributions in momentum space within the moderate $x$ region remain largely insensitive to the choice of extrapolation strategy. Therefore, we apply a non-fit extrapolation method to smooth the uncertainties of the renormalized quasi-TMD matrix elements at long distances. The extrapolation is performed using the following form:  
\begin{equation}
    \tilde{h}^{\rm ext} = w \cdot \tilde{h} + (1 - w) \cdot 0 ~,
\end{equation}  
where $w$ is a weight function that transitions linearly from 1 to 0 within the range $\lambda \in [z_{\rm ext}P^z, z_{\rm max} P^z]$, the $z_{\rm ext} = 0.78$ fm is the starting point of extrapolation and $z_{\rm max} = 1.2$ fm is the largest longitudinal separation of quasi-TMD. It is expected that $z > z_{\rm ext}$ is large enough to see the exponential decay behavior of quasi-TMD.

\subsubsection{The Collins-Soper Kernel}

The CS kernel can be extracted from the ratio of quasi-TMDWFs with different momenta, as described in Eq.~(\ref{eq:cs_kernel_calc}). For the matching procedure, we use the fixed-order one-loop results for the CG matching coefficient $ C_{\mathrm{TMD}} $ and the corresponding hard kernel $ H_\phi $, as provided in Ref.~\cite{Zhao:2023ptv}. Furthermore, we account for large logarithmic terms in the hard kernel by applying renormalization group evolution to improve the accuracy of the perturbative matching up to NLL.

\begin{figure}[th!]
    \centering
    \includegraphics[width=.6\linewidth]{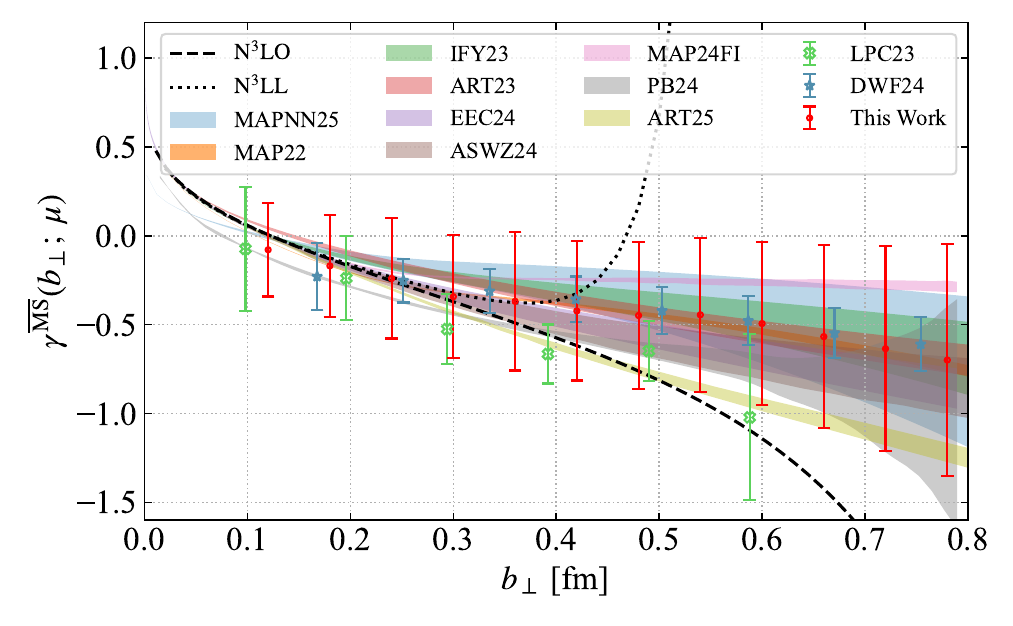}
    \caption[CS kernel]{The CS kernel at $\mu = 2$ GeV in this work is presented using red points with error bars. The 3-loop perturbative results~\cite{Li:2016ctv, Vladimirov:2016dll} for the CS kernel are denoted as $\rm N^3LO$ and $\rm N^3LL$ in the figure. The CS kernels from recent phenomenological parameterizations of experimental data are shown from MAP22~\cite{Bacchetta:2022awv}, IFY23~\cite{Isaacson:2023iui}, ART23~\cite{Moos:2023yfa},
    MAP24FI~\cite{Bacchetta:2024qre},
    EEC24~\cite{Kang:2024dja},
    PB24~\cite{Martinez:2024mou},
    MAPNN25~\cite{Bacchetta:2025ara} and
    ART25~\cite{Moos:2025sal}. 
    In addition, some recent lattice calculations are presented from LPC23~\cite{LatticePartonLPC:2023pdv}, ASWZ24~\cite{Avkhadiev:2024mgd} and DWF24~\cite{Bollweg:2024zet}.
    \label{fig:cs_kernel}}
\end{figure}

After incorporating both statistical and systematic uncertainties, Fig.~\ref{fig:cs_kernel} presents our final results for the CS kernel at $\mu = 2$ GeV. In the small $ b_\perp $ region, our results align well with the three-loop perturbative calculations from Refs.~\cite{Li:2016ctv, Vladimirov:2016dll}, labeled $ \rm N^3LO $ and $ \rm N^3LL $. Moreover, our calculation remains reliable in the large $ b_\perp $ region, where perturbative methods break down.

To further contextualize our findings, we compare them with CS kernels extracted from recent global fits of experimental data, including MAP22~\cite{Bacchetta:2022awv}, IFY23~\cite{Isaacson:2023iui}, ART23~\cite{Moos:2023yfa},
MAP24FI~\cite{Bacchetta:2024qre},
EEC24~\cite{Kang:2024dja},
PB24~\cite{Martinez:2024mou},
MAPNN25~\cite{Bacchetta:2025ara}
and ART25~\cite{Moos:2025sal}. Additionally, we include recent lattice QCD results from LPC23~\cite{LatticePartonLPC:2023pdv}, ASWZ24~\cite{Avkhadiev:2024mgd}, and DWF24~\cite{Bollweg:2024zet}. 

A particularly important observation is the agreement between our calculation and the DWF24 result from chirally symmetric domain-wall fermion configurations, both of which employ the CG framework despite using different lattice actions. Moreover, it is observable that the two most recent global analysis results (MAP24FI and ART25) exhibit a deviation from their prior results in different directions. Nevertheless, both remain consistent with this work due to the large uncertainty in our CS kernel, primarily stemming from power corrections at such a heavy pion mass. In our future study, the precision of our CS kernel could be refined by adopting a smaller valence pion mass.

\subsubsection{Intrinsic Soft Function}

\begin{figure}[th!]
    \centering
    \includegraphics[width=.3\linewidth]{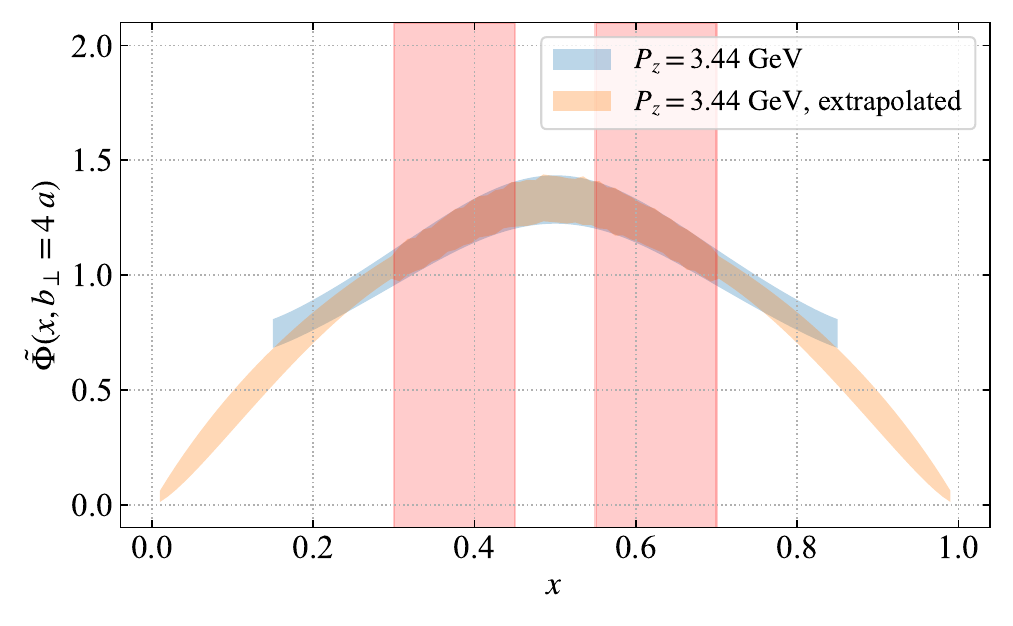}
    \includegraphics[width=.3\linewidth]{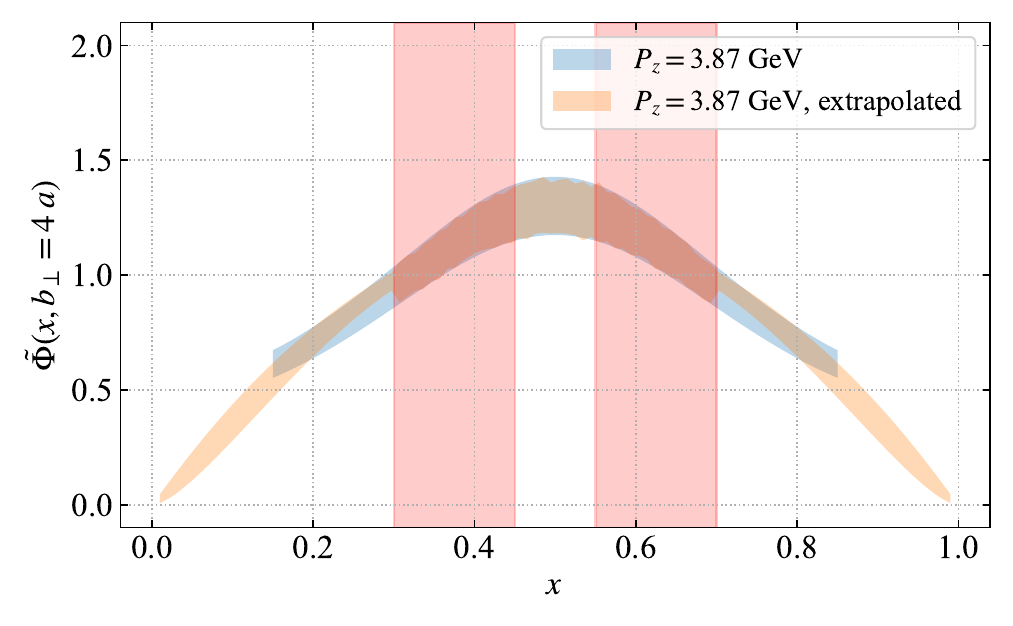}
    \includegraphics[width=.3\linewidth]{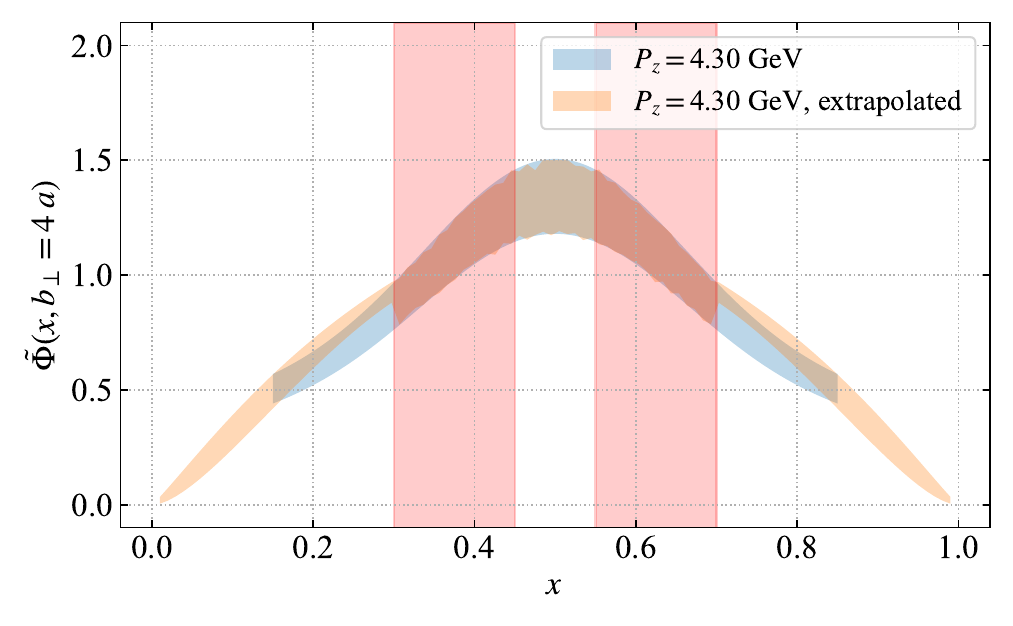}
    \caption[Extrapolations of reduced quasi-TMDWF]{
    Extrapolations of reduced quasi-TMDWF using Eq.~(\ref{eq:reduce_wf_extra}). From left to right, the three panels correspond to hadron momenta $P^z=3.44$, 3.87 and 4.30 GeV, respectively. All of three cases are evolved to $P_{\rm FF} = 2.58$ GeV using the rapidity evolution of quasi-TMDWF. The shaded red bands indicate the regions constitute the input for the extrapolation fit.
    \label{fig:reduced_qtmdwf_extra}}
\end{figure}

\begin{figure}[th!]
    \centering
    \includegraphics[width=.6\linewidth]{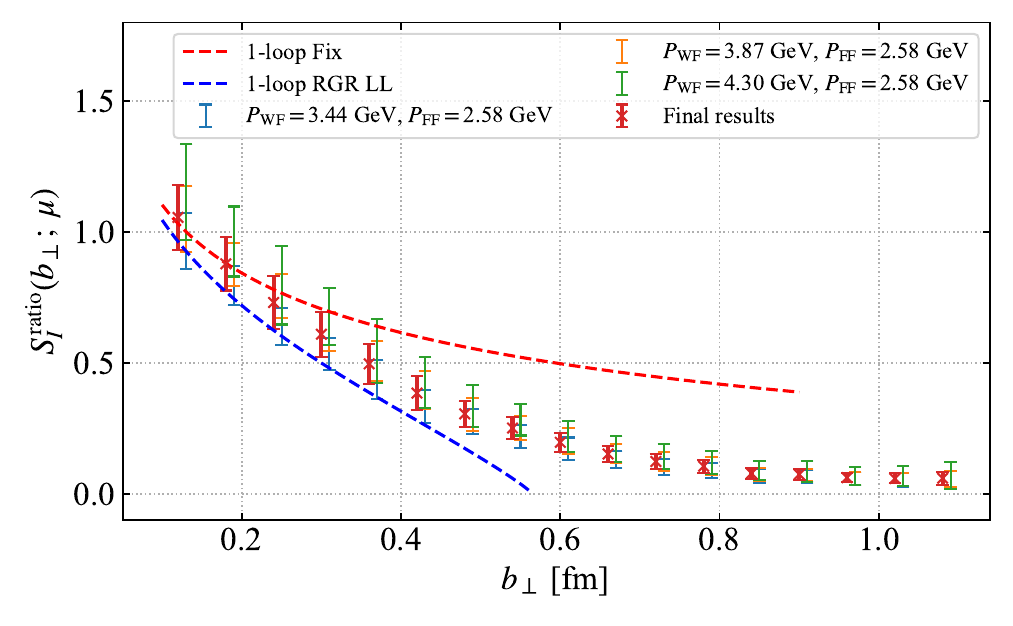}
    \caption[Intrinsic soft function]{Intrinsic soft function in CG at $\mu = 2$ GeV in the ratio scheme, for different hadron momentum pairs of the quasi-TMDWF, $P_\mathrm{WF}$,  and form factor, $P_\mathrm{FF}$. The final results are shown as red points with error bars (see text for details). The corresponding one-loop fixed-order perturbative results are the red dashed line, and the one-loop RG-resummed (RGR) perturbative results are the blue dashed line. 
    \label{fig:softf}}
\end{figure}

To extract the intrinsic soft function using Eq.~(\ref{eq:soft_func_calc}), we integrate the quasi-TMDWF over $ x_1 $ and $ x_2 $. However, near the endpoint regions ($ x \approx 0 $ and $ x \approx 1 $), quasi-distributions suffer from significant power corrections. To mitigate this issue, we extrapolate the reduced quasi-TMDWF $ \tilde{\Phi} $ using the functional form  
\begin{equation}
    \tilde{\Phi}(x) = c x^n (1-x)^n ~,
    \label{eq:reduce_wf_extra}
\end{equation}
where $ c $ and $ n $ are fitting parameters. A comparison of the reduced quasi-TMDWF before and after extrapolation is shown in Fig.~\ref{fig:reduced_qtmdwf_extra}, where the three panels correspond to hadron momenta $P^z=3.44$, 3.87 and 4.30 GeV, respectively. All three cases are evolved to $P_{\rm FF} = 2.58$ GeV using the rapidity evolution of quasi-TMDWF. The red shaded bands indicate the regions that constitute the input for the extrapolation fit, with the selected fit range of $ x \in [0.3, 0.45]$ and $x \in [0.55, 0.7]$, where the results of three different momenta show consistency. The extrapolation form derived from the fitting process is applied to the endpoint regions with $x<0.3$ or $x>0.7$. Fig.~\ref{fig:reduced_qtmdwf_extra} shows that the extrapolated results outside the fit range are in alignment with the observed trends. Furthermore, since the Sudakov kernel diverges near the endpoints ($x \approx 0$) after the RG resummation, the integral in Eq.~(\ref{eq:soft_func_calc}) is restricted to the range $ x_1, x_2 \in [0.05, 0.95] $. This cutoff has a small impact on the results of the intrinsic soft function, since the integrand in the denominator of Eq.~(\ref{eq:soft_func_calc}) converges near the endpoint regions.

To assess systematic effects of power corrections, we performed integration with three different momentum pairs, formed by the momentum of the form factor $P_{\rm FF} = 2.58$ GeV and three momenta of the quasi-TMDWF $P_{\rm WF} = 3.44$, 3.87 and 4.30 GeV. In order to bridge the gap between $P_{\rm FF}$ and $P_{\rm WF}$, the quasi-TMDWF is evolved to $P_{\rm FF}$ by solving the rapidity evolution equation. The extracted intrinsic soft function, according to Eq.~(\ref{eq:soft_func_calc}), is shown in Fig.~\ref{fig:softf}, where comparisons across different momentum pairs indicate that power corrections have only a minor impact. Note that the superscript ``ratio" of $S_I^{\rm ratio}$ indicates the ratio scheme in the renormalization of the quasi-TMDWF.

For the final results of $ S_I^{\rm ratio}$, the central values and statistical uncertainties are determined from a correlated average over the three momentum pairs formed by $P_{\rm FF}$ and $P_{\rm WF}$. The systematic uncertainties are determined by the spread of the mean values between these three momentum pairs, quantified as half the absolute difference between the maximum and minimum values. The extracted intrinsic soft function at $\mu = 2$ GeV, shown as red points with error bars in Fig.~\ref{fig:softf}, is compared to perturbative predictions of fixed order and RG resummed (RGR) at leading logarithmic (LL) precision. In particular, at small $ b_\perp $, our lattice results exhibit reasonable agreement with perturbative predictions, highlighting the robustness of our approach. In the large $ b_\perp $ region, where perturbation theory becomes unreliable, our lattice results are expected to provide a more accurate description of the intrinsic soft function, offering valuable non-perturbative insights.

\subsubsection{Pion TMDWF}

\begin{figure}[th!]
    \centering
    \includegraphics[width=.4\linewidth]{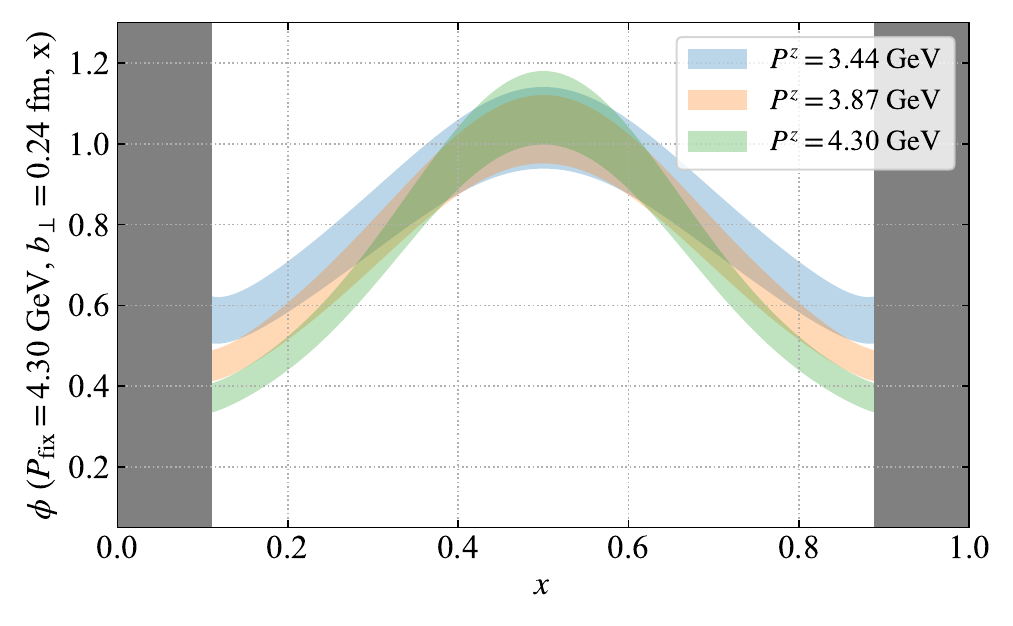}
    \includegraphics[width=.4\linewidth]{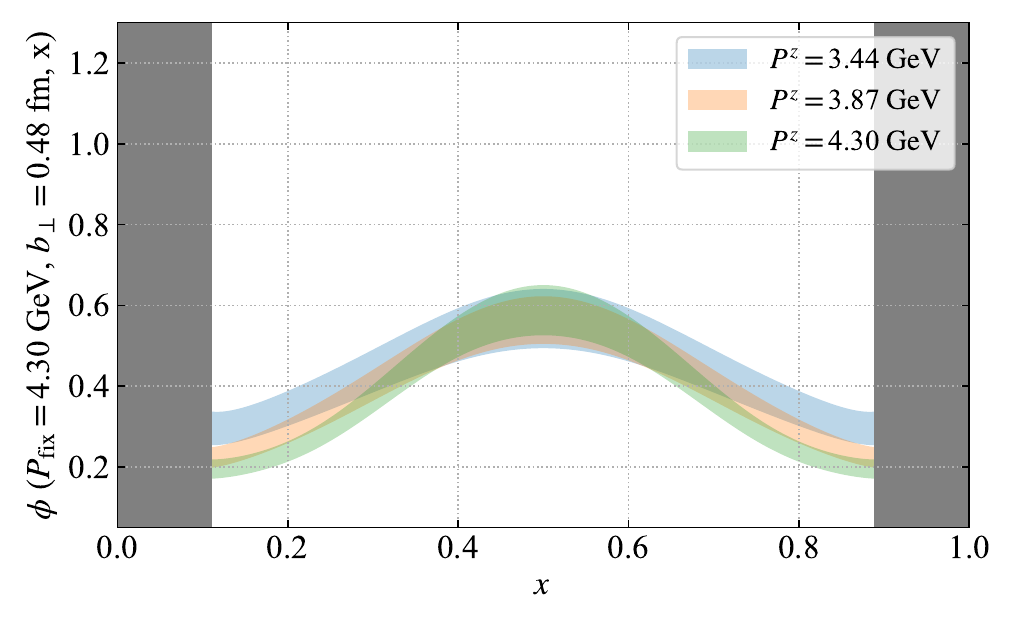}
    \caption[Pion TMDWF]{The light-cone pion TMDWF at $y_n = 0$, $P_\mathrm{fix} = 4.30$ GeV and $\mu = 2$ GeV of different hadron momenta, $P^z$, as the functions of momentum fraction, $x$, and for two transverse separations $b_\perp = 4a$ (upper panel) and $b_\perp = 8a$ (lower panel). The shaded gray bands ($x < 0.11$ and $x > 0.89$) indicate the endpoint regions where the estimated combined systematics are larger than $30\%$. 
    \label{fig:tmdwf_xdep}}
\end{figure}

By integrating the CG kernel, intrinsic soft function, and the renormalized quasi-TMDWF, the light-cone TMDWF can be extracted by employing the factorization formula found in Eq.~(\ref{eq:factorization_tmdwf_short}). Choosing scales $y_n=0$, $\zeta = (2x P_\mathrm{fix})^2$, and $\bar{\zeta} = (2\bar{x} P_\mathrm{fix})^2$, the factorization formula can be rewritten in a more explicit form:
\begin{eqnarray}
    &&\sqrt{S_I\left(b_{\perp} ; \mu\right) } \cdot \tilde{\phi}_{\Gamma}\left(x, b_{\perp}, P^z; \mu\right) = \phi\left(x, b_{\perp}; \mu, \zeta, \bar{\zeta} \right) H_\phi \left(x, \bar{x}, P^z; \mu\right)  \\ 
    && \qquad \qquad \qquad \times \exp \left[\ln \left( \frac{P^z}{P_\mathrm{fix}} \right) \gamma^{\overline{\mathrm{MS}}}\left(b_{\perp}; \mu\right)\right] +\mathcal{O}\left(\frac{\Lambda_{\mathrm{QCD}}}{x P^z}, \frac{1}{b_{\perp}\left(x P^z\right)}, \frac{\Lambda_{\mathrm{QCD}}}{\bar{x} P^z}, \frac{1}{b_{\perp}\left(\bar{x} P^z\right)}\right) ~,
    \label{eq:factorization_tmdwf}
\end{eqnarray}
where the CS scale is evolved from $\zeta_0 = (2x P^z)^2$ and $\bar{\zeta}_0$ to $\zeta = (2x P_\mathrm{fix})^2$ and $\bar{\zeta}$ using the CS kernel extracted from quasi-TMDWFs.

The final results for the TMDWF at $y_n = 0$, $P_\mathrm{fix} = 4.30$ GeV, and $\mu = 2$ GeV are shown in Fig.~\ref{fig:tmdwf_xdep} as a function of the momentum fraction $ x $ for three different hadron momenta. Two representative transverse separations, $ b_\perp = 4a $ (upper panel) and $ b_\perp = 8a $ (lower panel), are selected for illustration. As can be seen, the variation between different momenta remains mild in the moderate $ x $ region, demonstrating the validity of power expansion in large $ P^z $ within the quasi-TMD framework, where power corrections are small. The shaded gray bands ($x < 0.11$ and $x > 0.89$) indicate the endpoint regions where the estimated combined systematics are greater than $30\%$. The combined systematics include the variation of scales and power corrections, which are estimated by the variation of the initial scale $\mu_0$ of the RGR procedure by a factor of $\sqrt{2}$ and the spread in the central values of the TMDWFs with different momenta, respectively. 

\subsubsection{Pion TMDPDF}

\begin{figure}[th!]
    \centering
    \includegraphics[width=.4\linewidth]{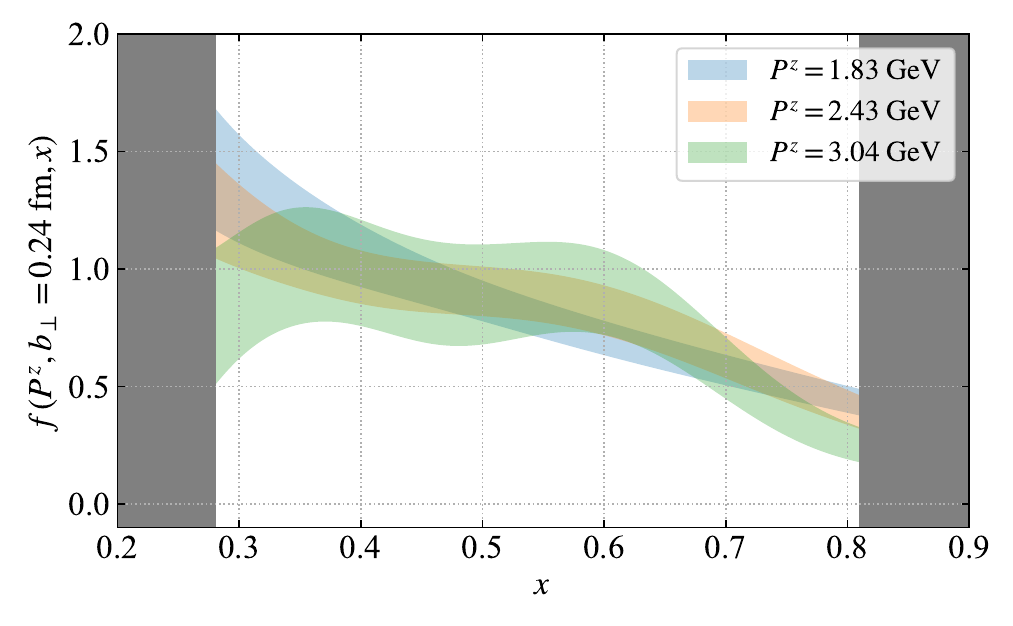}
    \includegraphics[width=.4\linewidth]{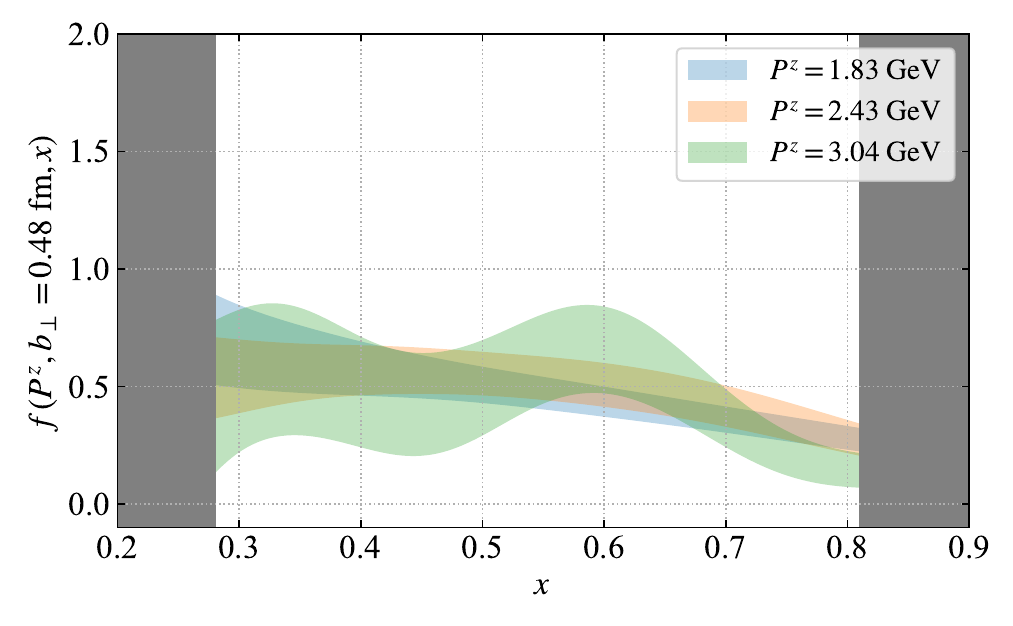}
    \caption[Pion TMDPDF in $x$ space]{The unpolarized light-cone pion TMDPDF at $\zeta = \mu^2 = 4~\mathrm{GeV}^2$ of different hadron momenta are shown as the function of momentum fraction, $x$, and for transverse separations, $b_\perp = 4a$ (upper panel) and $b_\perp = 8a$ (lower panel). The shaded gray bands ($x < 0.28$ and $x > 0.81$) indicate the endpoint regions where the estimated combined systematics are larger than $30\%$.
    \label{fig:tmdpdf_xdep}}
\end{figure}

\begin{figure}[th!]
    \centering
    \includegraphics[width=0.3\linewidth]{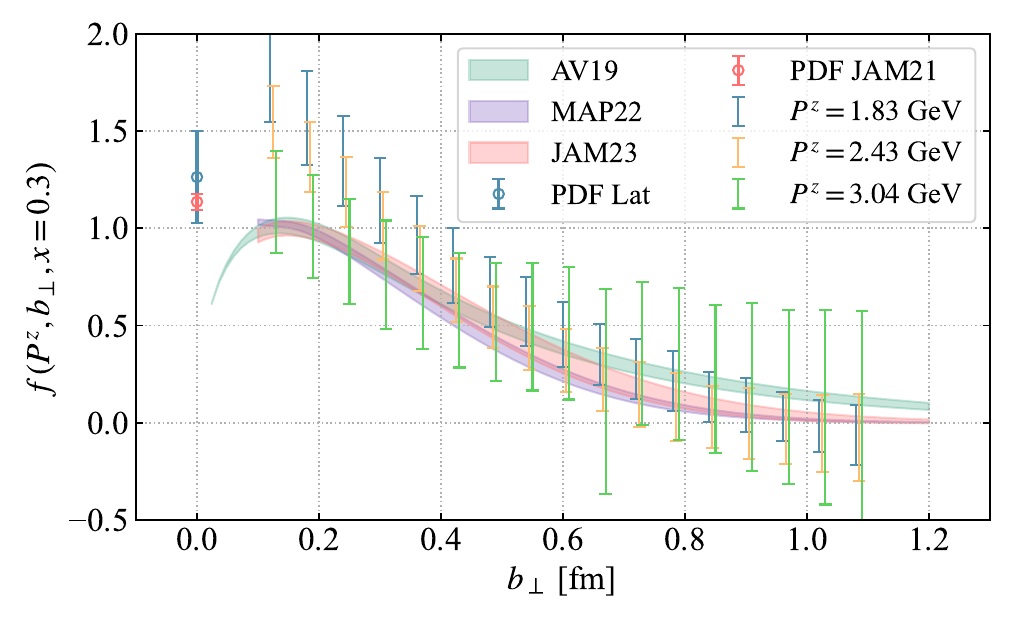}
    \includegraphics[width=0.3\linewidth]{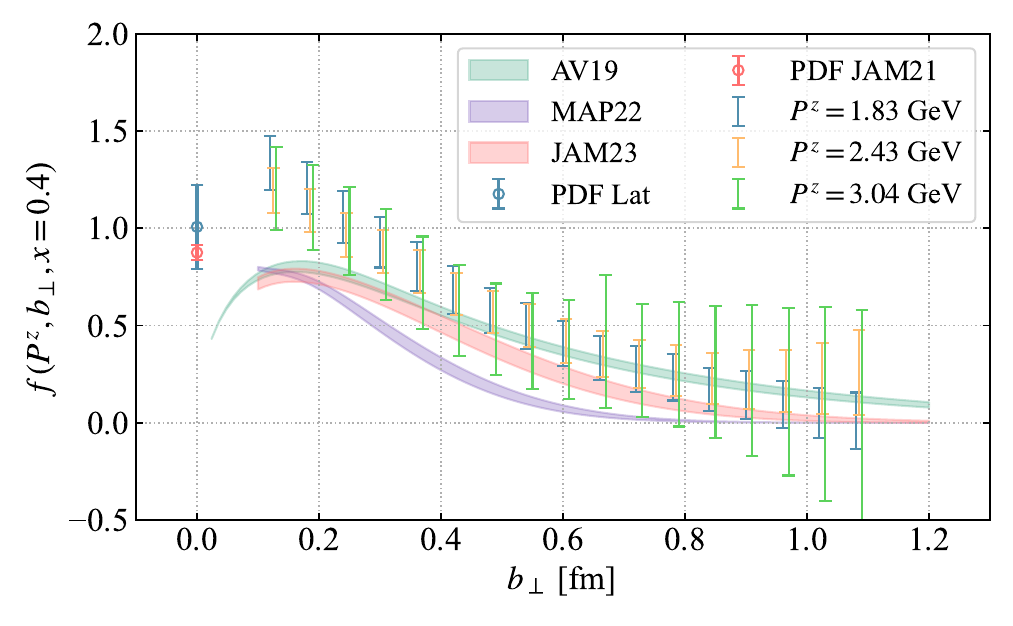}
    \includegraphics[width=0.3\linewidth]{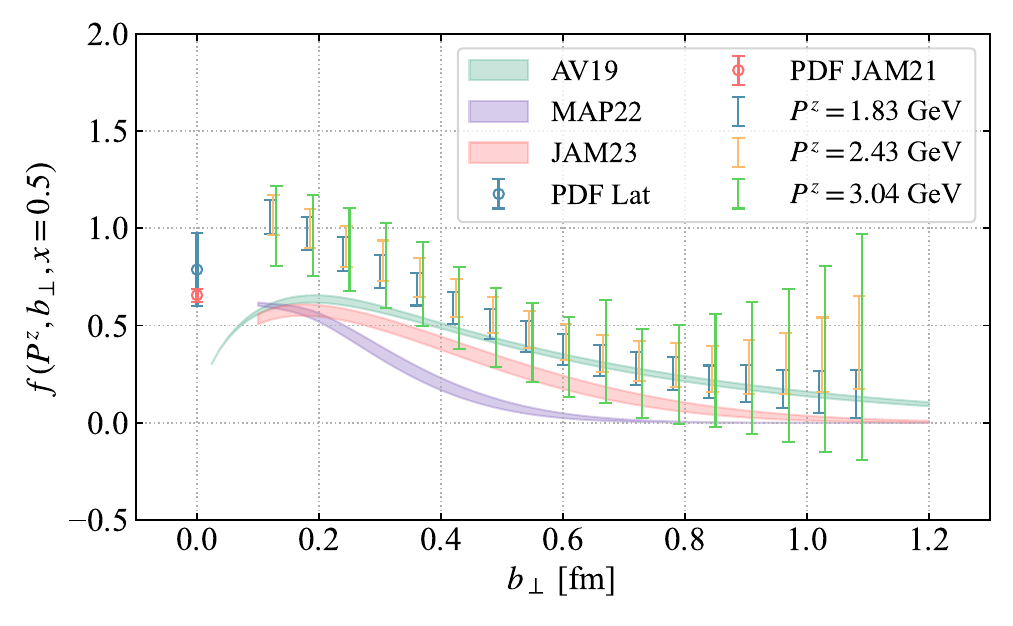}
    \caption[Pion TMDPDF in $b_\perp$ space]{Unpolarized light-cone TMDPDF of pion at $x=0.3$, $x=0.4$ and $x=0.5$ in $b_\perp$ space up to $b_\perp \gtrsim 1$ fm, and for different hadron momenta, $P^z$. The lattice results are illustrated alongside the phenomenological results from 
    AV19~\cite{Vladimirov:2019bfa},
    MAP22~\cite{Cerutti:2022lmb} and JAM23~\cite{Barry:2023qqh}. In addition, the collinear PDF from lattice calculation in CG (PDF Lat)~\cite{Gao:2023lny} and global fit (PDF JAM21)~\cite{Barry:2021osv} are plotted for comparison.}
    \label{fig:tmdpdf_bdep}
\end{figure}

The combination of the CG kernel, intrinsic soft function, and renormalized quasi-TMD beam functions enables the extraction of the light-cone TMDPDF using the factorization formula in Eq.~(\ref{eq:tmdpdf_matching}). The final results for the TMDPDF at $\zeta = \mu^2 = 4~\mathrm{GeV}^2$ are shown in Fig.~\ref{fig:tmdpdf_xdep} as a function of the momentum fraction $ x $ for three different hadron momenta. Two representative transverse separations, $ b_\perp = 4a $ (upper panel) and $ b_\perp = 8a $ (lower panel), are selected for illustration. As can be seen, the variation between different momenta remains mild in the moderate $ x $ region, demonstrating the validity of power expansion in large $ P^z $ within the quasi-TMD framework, where power corrections are small. The shaded gray bands ($x < 0.28$ and $x > 0.81$) indicate the endpoint regions where the estimated combined systematics are greater than $30\%$.

In Fig.~\ref{fig:tmdpdf_bdep}, we examine the $ b_\perp $ dependence of the TMDPDF at $ x = 0.3 $, $ x = 0.4 $, and $ x = 0.5 $. The results for different hadron momenta exhibit good consistency, which is improved as the $x$ value approaches $x=0.5$, thereby confirming our expectation of the behavior of power corrections. For comparison, we include recent global fits of the pion TMDPDF from AV19~\cite{Vladimirov:2019bfa}, MAP22~\cite{Cerutti:2022lmb} and JAM23~\cite{Barry:2023qqh}. Although slight deviations in magnitude are observed, the lattice results exhibit a qualitative agreement with global fits, especially at the smaller $x$ values, where the experimental data give a better constraint.

Since the collinear PDF and the $ b_\perp \to 0 $ limit of the unpolarized TMDPDF differ only by a perturbative expansion in $ b_\perp $~\cite{Moos:2023yfa}, we also compare our results to collinear PDFs. Specifically, we include the collinear PDF extracted from lattice calculations in CG (PDF Lat)~\cite{Gao:2023lny} and the global fit result from JAM21 (PDF JAM21)~\cite{Barry:2021osv}. As shown in the plots, the collinear PDFs closely match the TMDPDF in the small $ b_\perp $ region, supporting the expected theoretical behavior.

\begin{figure}[th!]
    \centering
    \includegraphics[width=0.3\linewidth]{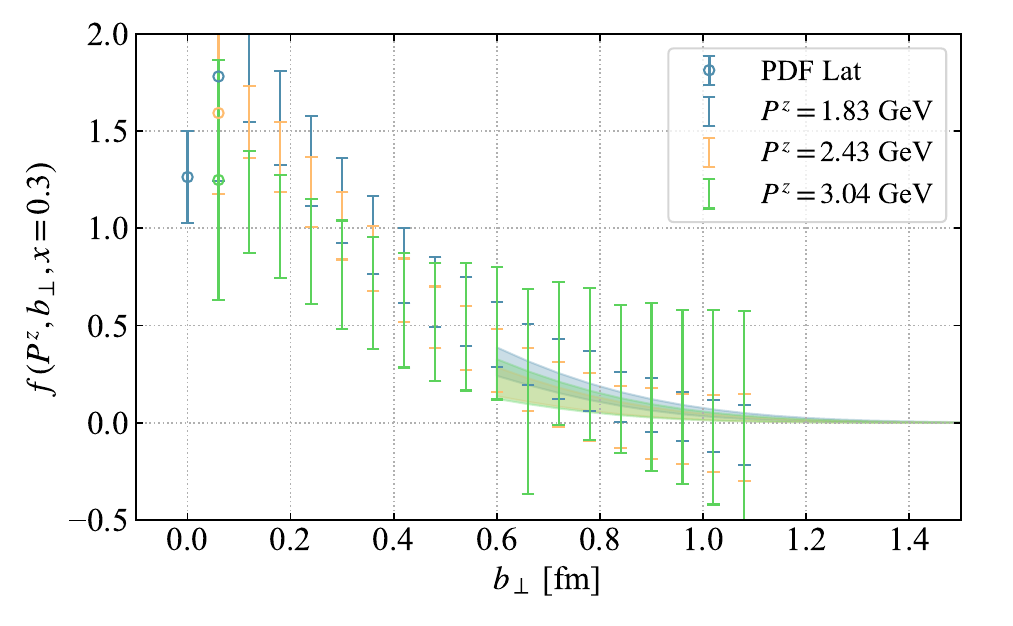}
    \includegraphics[width=0.3\linewidth]{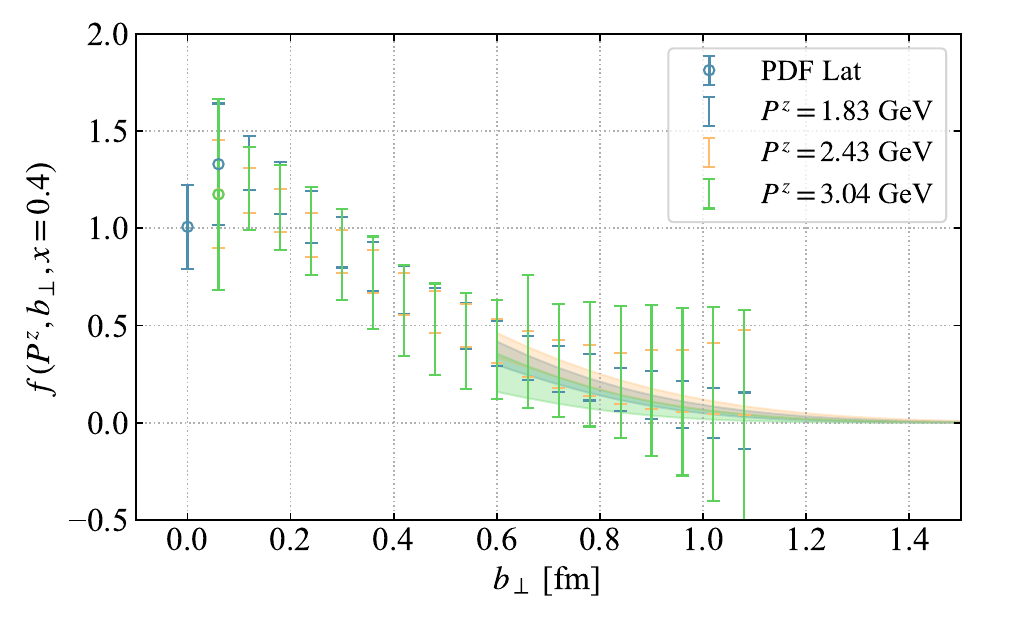}
    \includegraphics[width=0.3\linewidth]{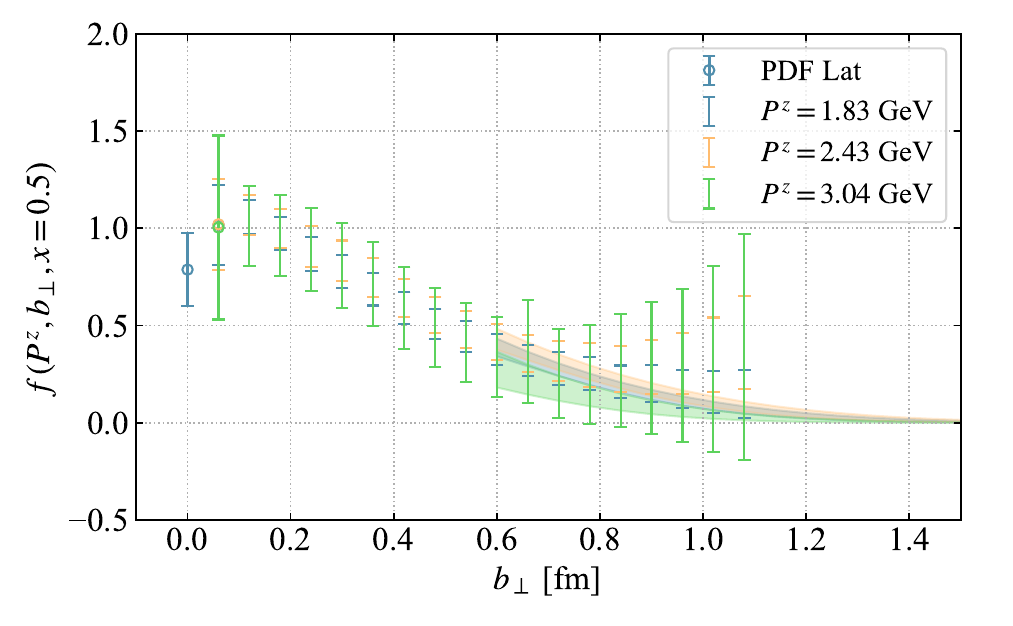}
    \caption[Extrapolated pion TMDPDF in $b_\perp$ space]{Unpolarized light-cone TMDPDF of pion at $x=0.3$, $x=0.4$ and $x=0.5$ in $b_\perp$ space after extrapolation to $b_\perp = 3$ fm. The collinear PDF from lattice calculation in CG (PDF Lat)~\cite{Gao:2023lny} is plotted at $b_\perp = $ 0 fm. The point at $b_\perp = 0.06$ fm is determined by cubic interpolation.}
    \label{fig:tmdpdf_bdep_extrapolated}
\end{figure}

An important advantage of the CG method is the absence of linear divergence, which allows us to probe large $ b_\perp $ regions with a reasonable SNR. At $ b_\perp \gtrsim 1 $ fm, the TMDPDF smoothly decays to values consistent with zero, facilitating a feasible Fourier transform into the $ k_\perp $ space. In this study, we combined our lattice result for the collinear pion PDF (PDF Lat) with the pion TMDPDF to interpolate the distribution at short distances around $ b_\perp \approx a $, and subsequently applied a Gaussian form to extrapolate the large $b_\perp$ regime up to $3$ fm, the extrapolation form is
\begin{equation}
    f (b_\perp) = A e^{- m \cdot b_\perp^2} ~,
    \label{eq:b_extrapolation_form}
\end{equation}
where $A$ and $m$ are fit parameters. This extrapolation form is inspired by the confinement in the transverse plane, while the model dependence of the results will be investigated in future works with improved data precision. The extrapolated unpolarized light-cone TMDPDF of three momenta are plotted in Fig.~\ref{fig:tmdpdf_bdep_extrapolated}.

\begin{figure}[th!]
    \centering
    \includegraphics[width=0.3\linewidth]{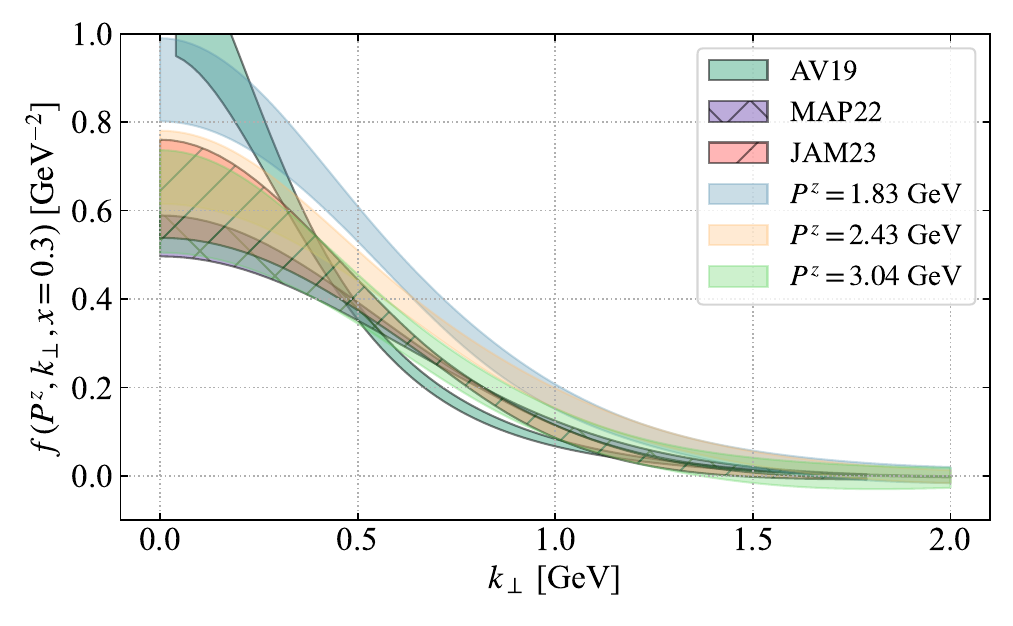}
    \includegraphics[width=0.3\linewidth]{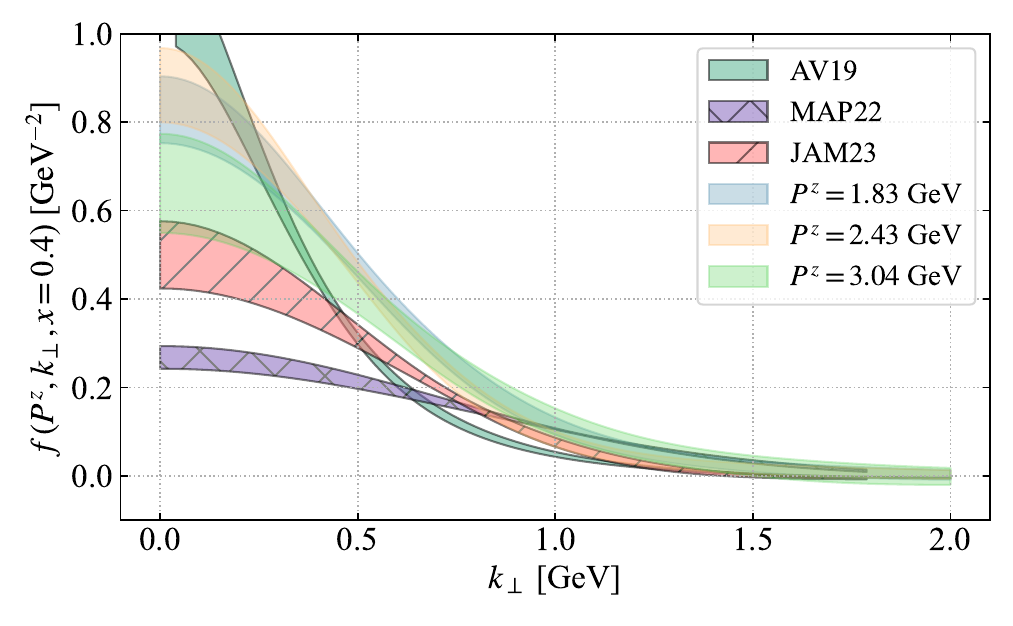}
    \includegraphics[width=0.3\linewidth]{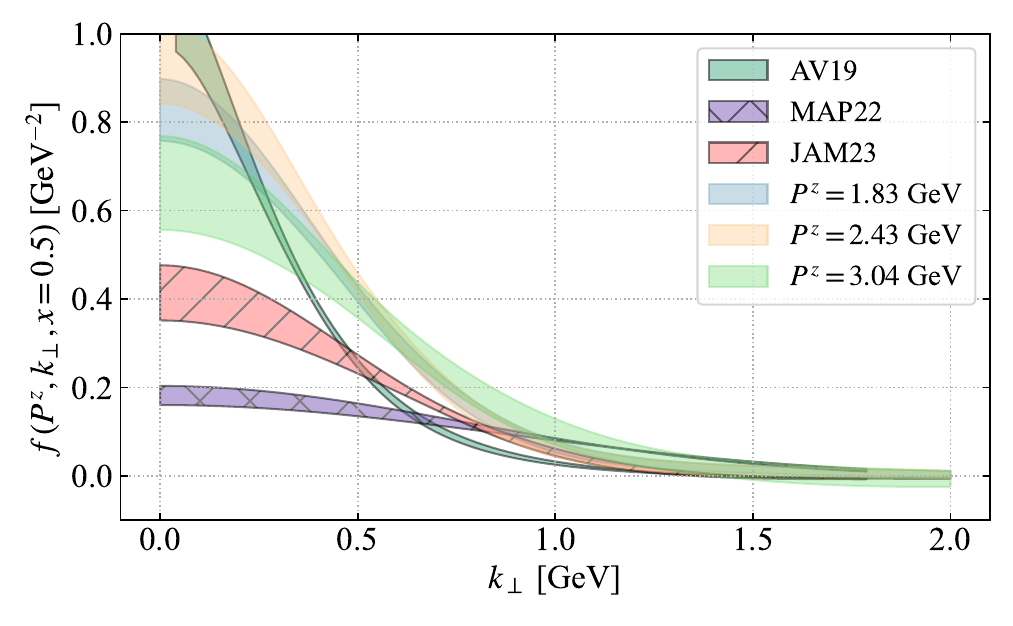}
    \caption[Pion TMDPDF in $k_\perp$ space]{Unpolarized light-cone TMDPDF of pion at $x=0.3$, $x=0.4$ and $x=0.5$ in $k_\perp$ space. The lattice results are illustrated alongside the phenomenological results from 
    AV19~\cite{Vladimirov:2019bfa},
    MAP22~\cite{Cerutti:2022lmb} and JAM23~\cite{Barry:2023qqh}.} 
    \label{fig:tmdpdf_kdep}
\end{figure}

After extrapolation, we performed a discretized Fourier transform, with the results presented in Fig.~\ref{fig:tmdpdf_kdep}. Although the extrapolation in $b_\perp$ introduces minor model dependencies, the resulting distributions exhibit qualitative consistency with global fits, especially at the smaller $x$ values, where the experimental data give a better constraint. This demonstrates the potential of lattice QCD to explore the transverse momentum structure of hadrons from first principles. In future studies with higher precision, the extrapolation modeling uncertainties, as well as other systematic effects, will be investigated more comprehensively to further improve the robustness of such calculations.

\subsection{Conclusion}

In this work, we have presented the first lattice QCD calculation of the pion unpolarized valence-quark TMDPDF within the framework of LaMET. The calculation was performed on a 2+1 flavor QCD ensemble with a lattice spacing of $a = 0.06$~fm and a pion mass of 300~MeV. By leveraging high statistics, off-axis momenta, a well-tuned boosted Gaussian smeared pion source, and the novel CG approach, we attained significant hadron momenta reaching 3~GeV. We computed the quasi-TMD beam function, extracted the CS kernel from quasi-TMDWF, and derived the intrinsic soft function. To enhance perturbative accuracy, we performed RGR of the Sudakov kernel at NLL order throughout our analysis. These advancements collectively enabled the determination of both the pion TMDWF and TMDPDF from the first principles.

Our results of the CS kernel and intrinsic soft function demonstrate consistency with perturbative calculations in the small-$b_\perp$ region, where perturbative methods remain valid. Additionally, the extracted CS kernel agrees with both previous lattice QCD studies and phenomenological fits of the experimental data.

By combining the renormalized quasi-TMDPDF, the CS kernel, and the intrinsic soft function, we obtained the pion valence TMDPDF across a range of $b_\perp$. Our results extend beyond $b_\perp\gtrsim1$ fm, allowing for a feasible Fourier transform into momentum space. The extracted TMDPDFs in the $b_\perp$ and $k_\perp$ space exhibit qualitative agreement with phenomenological fits. This agreement demonstrates the potential of lattice QCD to validate and complement global analyses.

The outcome of this study highlights the efficacy of the CG quasi-TMD approach in probing the transverse momentum structure of hadrons. Our results provide a valuable first-principles complement to global fits, especially in the non-perturbative region where experimental data remain scarce. Future improvements, such as the incorporation of finer lattice spacings, larger momenta, and improved control over systematic uncertainties, will further enhance the precision of lattice QCD determinations of TMDPDFs. Additionally, extending this methodology to other hadrons, including the nucleon, will provide deeper insights into the partonic structure of QCD bound states.

\section{Calculation of Nucleon TMDs}

In contrast to the pion, which serves as the lightest QCD bound state and a key probe of chiral dynamics, the nucleon represents a far more intricate composite system. As a spin-$1/2$ baryon with a rich flavor and spin structure, the nucleon encodes essential information about how quark and gluon degrees of freedom generate the mass, spin, and spatial structure of visible matter. While pion TMDs offer a comparatively clean environment to investigate transverse-momentum dynamics in a relativistic bound state, nucleon TMDs introduce additional layers of complexity associated with spin correlations, flavor asymmetries, and multi-parton interactions.

The TMDs of the nucleon therefore provide a multidimensional characterization of its internal structure, extending beyond the one-dimensional collinear picture to include correlations between transverse momentum and spin. In particular, spin-dependent TMDs encode nontrivial QCD dynamics such as spin-orbit correlations and time-reversal-odd effects, which have no direct analogue in the spinless pion. Consequently, nucleon TMDs play a central role in interpreting semi-inclusive deep-inelastic scattering and Drell–Yan measurements, where polarized observables probe subtle aspects of QCD dynamics~\cite{Collins:1981uk,Angeles-Martinez:2015sea,Collins:1984kg,Accardi:2012qut}.

Despite the extensive experimental program devoted to nucleon TMDs, their nonperturbative structure remains incompletely constrained. The Collins–Soper kernel governing rapidity evolution has been computed perturbatively to high orders~\cite{Moult:2022xzt,Duhr:2022yyp}, but the intrinsic transverse-momentum dependence at hadronic scales must be modeled or extracted from global fits. Although phenomenological analyses incorporate large datasets from low-$p_T$ SIDIS and Drell–Yan processes~\cite{Moos:2023yfa,Bury:2022czx,Bacchetta:2022awv,Scimemi:2019cmh,Bacchetta:2017gcc}, noticeable discrepancies among different extractions indicate substantial systematic uncertainties. Compared with the pion case, where experimental information is scarce, the challenge for the nucleon lies less in data availability and more in disentangling correlated spin, flavor, and kinematic effects within global analyses.

In our 2022 study~\cite{LatticePartonCollaborationLPC:2022myp}, we computed the quasi-TMDPDF matrix elements of a boosted nucleon within the GI framework. A systematic renormalization program was carried out by combining Wilson-loop subtractions with short-distance hadron matrix elements~\cite{Zhang:2022xuw}. The perturbative matching to light-cone TMDPDFs was performed including both one- and two-loop contributions, together with renormalization-group resummation of large logarithms. Our results constitute the first first-principles determination of the flavor isovector nucleon TMD momentum distributions in this framework. The extracted TMDPDFs show qualitative agreement with phenomenological extractions, demonstrating the robustness of the CG approach and its potential for a unified lattice description of pion and nucleon TMD structures.

\subsection{Theoretical Framework}

\begin{figure}[th!]
    \centering
    \includegraphics[width=0.6\linewidth]{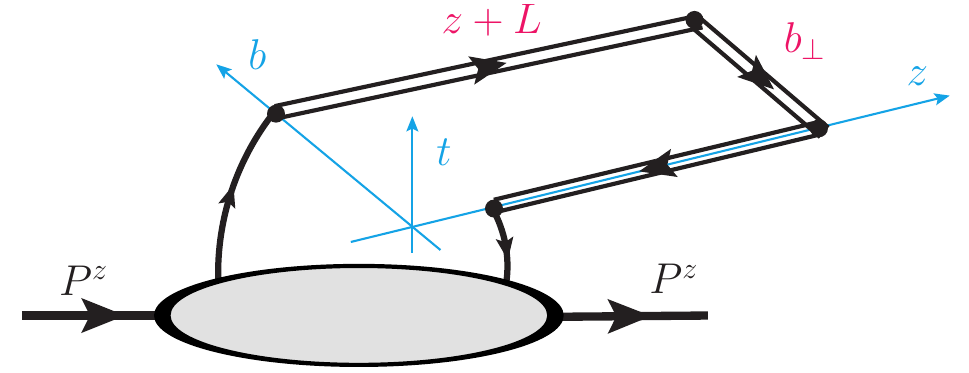}
    \caption[Quasi TMDPDFs in GI formalism]{Illustration of the quasi TMDPDFs in GI formalism.} 
    \label{fig:quasiTMDPDF}
\end{figure}

According to LaMET, the correlations with modes traveling along the light-cone can be extracted from distributions in a fast-moving nucleon through large-momentum expansion. 
On the lattice, the equal-time quasi TMDPDFs are constructed as
\begin{eqnarray}
    \tilde{f}_{\Gamma}\left(x, b_{\perp}, P^z, {\mu}\right) \equiv \lim_{a\to0 \atop L\to\infty} \int\frac{dz}{2\pi} e^{-iz\left(xP^z\right)} \frac{\tilde{h}^{0}_{\Gamma}\left(z, b_{\perp}, P^z, a, L\right)}{\sqrt{Z_E\left(2L+z, b_{\perp}, a\right)} Z_O\left(1/a, {\mu}, \Gamma\right)},
    \label{eq:momspacequasi}
\end{eqnarray}
where $a$ denotes the lattice spacing.  $\Gamma=\gamma^t$ or $\gamma^z$ is the Dirac matrix that can be projected onto $\gamma^+$ in the large momentum limit.  The $\tilde{h}^{0}_{\Gamma}\left(z, b_{\perp}, P^z, a, L\right)$ is built with a gauge-invariant nonlocal quark bilinear operator as
\begin{eqnarray}
    \tilde{h}^{0}_{\Gamma}\left(z, b_{\perp}, P^z, a, L\right)=\langle P^z |\tilde{O}^{0}_{\Gamma,\sqsubset}(z,b_{\perp},P^z;  L)|P^z\rangle,  \\
    \tilde{O}^{0}_{\Gamma,\sqsubset}(z,b_{\perp},  L)\equiv 
	\bar{\psi}(b_{\perp}\hat{n}_{\perp})\Gamma U_{\sqsubset, L}\left( b_{\perp}\hat{n}_{\perp},  z\hat{n}_z\right)\psi(z\hat{n}_z).  
\end{eqnarray}
 In the above, $|P^z\rangle$ denotes the unpolarized nucleon state and $L$ denotes the ``infinity"  that the gauge link can reach. The staple-shaped Wilson link   is chosen as
\begin{eqnarray}
	U_{\sqsubset, L}\left( b_{\perp}\hat{n}_{\perp},  z\hat{n}_z\right)&\equiv& U_{z}^{\dagger}\left(  (z+L)\hat{n}_z+b_{\perp}\hat{n}_{\perp},  b_{\perp}\hat{n}_{\perp}\right) \\
	&\quad&\times U_{{\perp}}\left( (z+L)\hat{n}_z+b_{\perp}\hat{n}_{\perp},  (z+L)\hat{n}_z \right) U_{z}\left( (z+L)\hat{n}_z, z\hat{n}_z\right), \label{eq:stapleshapedWilsonLink}
\end{eqnarray}
in which  $U_z$ is the path-ordered Euclidean gauge link along the $z$-direction, and $U_{\perp}$ is  along the transverse direction at the ``infinity" position $(z+L)$ on the finite lattice. Their explicit forms are:
\begin{eqnarray}
	&&U_{z}(\xi^z_1 \hat{n}_z+\xi_{\perp} \hat{n}_\perp, \xi^z_2\hat{n}_z+\xi_{\perp} \hat{n}_\perp) =\mathcal{P}\exp\left[ -ig\int_{\xi^z_1}^{\xi^z_2 } d\lambda \hat{n}_z\cdot A\left(\lambda \hat{n}_z+\xi_{\perp} \hat{n}_\perp\right) \right], \\
	&&U_{\perp}(\xi^z \hat{n}_z+\xi_{\perp 1} \hat{n}_\perp, \xi^z \hat{n}_z+\xi_{\perp 2} \hat{n}_\perp) = \mathcal{P}\exp\left[ -ig\int_{\xi_{\perp 1}}^{\xi_{\perp 2}} d\lambda \hat{n}_\perp \cdot A\left( \xi^z \hat{n}_z + \lambda \hat{n}_{\perp} \right) \right]. 
\end{eqnarray}

An illustration of the quasi TMDPDFs is given in Fig.~\ref{fig:quasiTMDPDF}, in which the staple-shaped Wilson link is depicted as double lines. 

The bare matrix elements of quasi TMDPDFs contain linear divergence, pinch-pole singularity and logarithmic divergence. Both the linear divergence and pinch-pole singularity can be renormalized by the square root of  Wilson loop $\sqrt{Z_E(2L+z,b_\perp,a)}$ \cite{Ji:2017oey, Ishikawa:2017faj, Green:2017xeu, Ji:2020brr,Shanahan:2019zcq}; and the logarithmic divergence can be renormalized by {a factor $Z_O(1/a,\mu)$, which is extracted from matching between the lattice and perturbative calculation of zero-momentum matrix elements in the perturbative region} \cite{Ji:1991pr, LatticePartonCollaborationLPC:2021xdx,Ji:2021uvr,Zhang:2022xuw}. 

The Wilson loop $Z_E(r=2L+z,b_{\perp},a)$ is defined as the vacuum expectation of a rectangular shaped space-like gauge links with size $r\times b_{\perp}$, which can be written as
\begin{eqnarray}
    Z_E(r,b_{\perp},a) &=& \left\langle U_{z}^{\dagger}\left(  (z+L)\hat{n}_z+b_{\perp}\hat{n}_{\perp},  (-z)\hat{n}_z+b_{\perp}\hat{n}_{\perp}\right) \right. \\
    &\quad& \times U_{{\perp}}\left( (z+L)\hat{n}_z+b\hat{n}_{\perp},  (z+L)\hat{n}_z \right)  U_{z}\left( (z+L)\hat{n}_z, (-z)\hat{n}_z\right) \\
    &\quad&  \times \left. U_{{\perp}}^\dagger\left( (-z)\hat{n}_z+b_{\perp}\hat{n}_{\perp},  (-z)\hat{n}_z \right) \right\rangle. \label{eq:Wilson_Loop}
\end{eqnarray}

The Wilson loop is introduced to eliminate the linear divergence of the form $e^{-\delta\bar{m}r}$, which comes from the self-energy corrections to the gauge link~\cite{Ji:2017oey,LatticePartonCollaborationLPC:2021xdx}, as well as the pinch-pole singularity, which is related to the heavy quark effective potential term $e^{-V(b_{\perp})L}$ describing the interactions between the two Wilson lines along the z direction in the staple link~\cite{Ji:2019ewn}.

The logarithmic divergence factor $Z_O$ can be extracted from the zero-momentum bare matrix elements $\tilde{h}_{\Gamma}^{0}\left(z, b_{\perp}, 0, a, L\right)$. In order to keep the renormalized matrix elements consistent with perturbation theory, $Z_O$ should be determined from the condition:
\begin{eqnarray}
    Z_O\left(\frac{1}{a}, \mu, \Gamma\right) =\lim _{L \rightarrow \infty} \frac{\tilde{h}_{\Gamma}^{0}\left(z, b_{\perp}, 0, a, L\right)}{\sqrt{Z_E\left(2L+z, b_{\perp}, a\right)} \tilde{h}_{\Gamma}^{\mathrm{\overline{MS}}}\left(z, b_{\perp}, \mu\right)}\label{eq:zo_definition}
\end{eqnarray}
in a specific window where $z \ll \Lambda_{\mathrm{QCD}}^{-1}$ so that perturbation theory is valid.

\subsection{Numerical Results}

We use the valence tadpole improved clover fermion on the hypercubic (HYP) smeared ~\cite{Hasenfratz:2001hp} $2+1+1$ flavors MILC configurations with highly improved staggered quark (HISQ) sea and 1-loop Symanzik improved gauge action~\cite{MILC:2012znn}. We analyze a single ensemble with lattice spacing $a=0.12$~fm and volume $n_s^3\times n_t=48^3\times64$ using physical sea quark masses, and two choices of light valence quark mass corresponding to $m_{\pi}^{\mathrm{val}}=\{220,~310\}$ MeV. The HYP smearing is also used for nonlocal correlation functions to improve the statistical signal. In order to explore  the momentum dependence, we employ three different nucleon momenta as $P^z=2\pi/(n_s a)\times\{8, 10, 12\}=\{1.72, 2.15, 2.58\}$~GeV.

We adopt momentum-smearing point source~\cite{Bali:2016lva}  at several time slices, and average  correlation functions for both the forward and backward directions in $z$  and transverse  space of  the gauge link. In total, there are  $1000$ (configurations) $\times 16$  (source time slices) $\times 4$ (forward/backward directions of the $z$ and transverse axes) measurements for the $m_{\pi}^{\mathrm{val}}=220$ MeV case and $1000\times 4\times 4$ measurements for the  $310$ MeV case.

The two-point correlation functions of the nucleon are defined as
\begin{equation}
    C_2(t, \vec{P}) = \Big\langle \sum_{\vec{y}} e^{i\vec{P}\cdot\vec{y}} T_{\alpha  \beta} N_\alpha(\vec{y},t) \bar{N}_\beta (\vec{0},0) \Big\rangle, 
\end{equation}
where $T=\left(1+\gamma^t\right)/2$ denotes the unpolarized projector, and the nucleon interpolating operator $N$ is defined as Eq.~(\ref{eq:nucleon_interpolator_standard}). For the three momenta $P^z=1.72, 2.15, 2.52$ GeV, the statistics of the two-point correlation functions are 1000 (configurations) $\times16$ (source time slices) for $m_{\pi}=220$~MeV and $1000\times4$ for $m_{\pi}=310$~MeV; while for the cases with a momentum smaller than $1.72$~GeV, there are $1000\times4$ measurements for $m_{\pi}=220$~MeV and $1000\times1$ measurements  for $m_{\pi}=310$~MeV.

To extract the quasi TMDPDFs, one constructs the three-point functions
\begin{equation}
    C_{3}^{\Gamma}\left(t, t_s\right) = \Big\langle \sum_{\vec{y}} e^{i\vec{P}\cdot\vec{y}} T N(\vec{y},t_s) \sum_{\vec{x}}  \tilde{O}_{\mathrm{TMD}}^{\Gamma}\left(\vec{x}, t\right) \bar{N} (\vec{0},0)     \Big\rangle.
\end{equation}
We adopt the sequential source method~\cite{Kilcup:1985fq} to reduce the number of propagators in three-point functions, $t_s$ denotes the time position of the sequential source. The operator $\tilde{O}_{\mathrm{TMD}}^{\Gamma}\left(\vec{x}, t\right)$ is short for the TMD nonlocal quark bilinear operator $\tilde{O}^{0}_{\sqsubset}(\vec{x}+z\hat{n}_z, \vec{x}+b_{\perp}\hat{n}_{\perp},\Gamma,  L)$ at discrete time slice $t\in[0,t_s]$. The three-momentum is chosen as $\vec{P}=(0,0, P^z)$. 

After inserting the single particle intermediate states, one can parametrize the ratio of three- and two-point functions as 
\begin{equation}
    \frac{C_{3}^{\Gamma}\left(t, t_s\right)}{C_2(t_s)} = \frac{\tilde{h}^0_{\Gamma}+c_2\big( e^{-\Delta Et} + e^{-\Delta E( t_s - t)} \big) +c_3e^{-\Delta Et_s}}{1+c_1e^{-\Delta Et_s}},
    \label{eq:ratio_of_2pt_and_3pt}
\end{equation}
in which $\tilde{h}^0_{\Gamma}\equiv\tilde{h}^0\left(z, b_{\perp}, P^z, \Gamma\right)$,  $\Delta E$ is the mass gap between the ground-state and excited state, and $c_{1, 2, 3}$ are parameters for the excited-state contaminations. Combining the parametrization form of 2pt functions: $C_2(t)=c_0e^{- E_0t}\left(1+c_1e^{-\Delta Et} \right)$, one can extract the values of $\tilde{h}^0_{\Gamma}$ at fixed $(z, b_{\perp}, P^z)$ through a correlated joint fit. 

In our calculation, we use bootstrap resampling to establish correlations among all datasets, and the correlations are maintained consistently throughout the entire analysis. For the three-point functions, we have excluded the contact points ($t=0$ and $t=t_{\mathrm{seq}}$) for the five values of the source-sink separation time range $t_{\mathrm{seq}}\in (0.48 \sim 0.84)$~fm. For small $b_\perp$ ($b_\perp\leq3a$), we have further removed two points near the source and sink($t=1$ and $t=t_{\mathrm{seq}}-1$).

\begin{figure}[!th]
    \centering
    \includegraphics[width=0.4\linewidth]{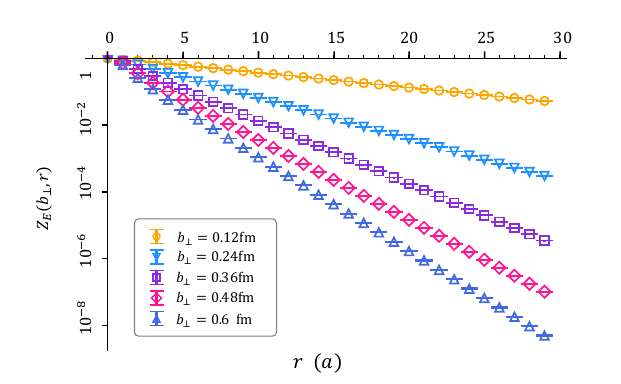}
    \includegraphics[width=0.4\linewidth]{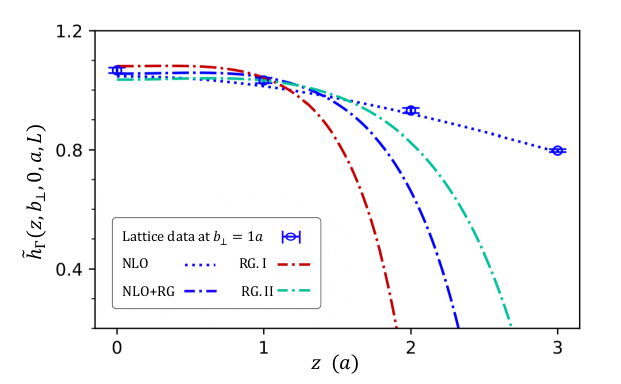}
    \caption[Wilson loop and renormalized zero-momentum matrix element]{Upper panel: Wilson loop with $r=2L+z$. 
    Lower panel: renormalized zero-momentum matrix element $\tilde{h}_{\Gamma}\left(z, b_{\perp}, 0, a, L\right)$ at $b_{\perp}=1a$, compare with 1-loop results before (NLO) and after (NLO+RG) RG evolution to $\mu=2$ GeV in the $\overline{\mathrm{MS}}$ scheme. Besides, we vary the starting point of evolution from  $0.8\mu_0$ (``RG. I'') to $1.2\mu_0$ (``RG. II'') to estimate the higher-order correction effects. }
    \label{fig:Fig_WilsonLoop_and_ZO}
\end{figure}

As mentioned above, we use the square root of the Wilson loop $\sqrt{Z_E}$ and logarithmic divergence factor $Z_O$ to renormalize the bare quasi-TMD matrix elements. In practice, the signal-to-noise ratio of $Z_E(r,b_{\perp},a)$ defined in Eq.~(\ref{eq:Wilson_Loop}) decreases fast with $r$ and $b_{\perp}$ such that it is hardly available for large $r$ or $b_{\perp}$. To address this, we fit the effective energies of the Wilson loop, which gives the QCD static potentials, and then extrapolate them to large $r$ and/or $b_{\perp}$, as in Ref.~\cite{Zhang:2022xuw}. Numerical results for the Wilson loop are shown in the upper panel of Fig.~\ref{fig:Fig_WilsonLoop_and_ZO}.

The logarithmic divergence factor $Z_O$ can be extracted from the matching between lattice and perturbative results in the region where both two theories work well. To preserve a good convergence of the perturbation theory before and after RG evolution, we choose the region where $b_\perp = 1a$, and $z = 0\ \mathrm{or}\ 1a$, where both perturbation theory and lattice calculations work. The extracted $Z_O$ values at these two points are $1.0734 (93)$ and $1.0509 (92)$, respectively. Averaging these data points yields an aggregated result of $Z_O = 1.0622(87)$.

With Eq.~(\ref{eq:zo_definition}), it can be concluded that after dividing the bare matrix elements $\tilde{h}_{\Gamma}^{0}$ by $\sqrt{Z_E}$ and $Z_O$, the renormalized matrix elements should approximately be equal to the RG evolved perturbation results $\tilde{h}_{\Gamma}^{\mathrm{\overline{MS}}}$. The lower panel of Fig.~\ref{fig:Fig_WilsonLoop_and_ZO} shows the consistency at points where we extract the $Z_O$ factor, which serves as a check of the numerical result for $Z_O$. { The points at $z=0$ and $z=1$ depicted in Fig.~\ref{fig:Fig_WilsonLoop_and_ZO} demonstrate that, following renormalization with the extracted factor $Z_O = 1.0622(87)$, the renormalized matrix elements on the lattice show a consistent trend with the perturbative results obtained in $\overline{\mathrm{MS}}$ scheme at short distance. }
To verify the accuracy of scale evolution from the intrinsic physical scale in lattice calculation to $\overline{\mathrm{MS}}$ scale $\mu=2$GeV, we vary the origin of the evolution from $0.8\mu_0$ (shown as ``RG. I'') to $1.2\mu_0$ (shown as ``RG. II''), which is sensitive to  higher-order corrections. 
One can see that the perturbative results from different $\mu_0$ are consistent with the lattice data in the matching region ($b_\perp = 1a$, $z = 0\ \mathrm{or}\ 1a$).

For a well-defined quasi TMDPDF, the length of the Wilson link $L$ should be large enough to ensure that the final results are independent of $L$.  
In the definition of the staple-shaped Wilson link in Eq~.(\ref{eq:stapleshapedWilsonLink}), $L$ should be extended to infinity. The $L$-dependence is included in the linear divergence of the Wilson link self-energy and pinch-pole singularity, which can be subtracted by the square root of the Wilson loop. Therefore, the subtracted quasi TMD matrix elements
\begin{eqnarray}
	\tilde{h}_{\Gamma}(z,b_{\perp},a,L)\equiv \frac{\tilde{h}^{0}_{\Gamma}\left(z, b_{\perp}, P^z, a, L\right)}{\sqrt{Z_E\left(2L+z, b_{\perp}, a\right)} Z_O\left(1/a, {\mu}, \Gamma\right)}
\end{eqnarray}
will saturate to a plateau at large enough $L$ to ensure that the link can extend outside the region of the parton, and exhibit independence of $L$.

\begin{figure}[!t]
    \centering
    \includegraphics[width=0.9\linewidth]{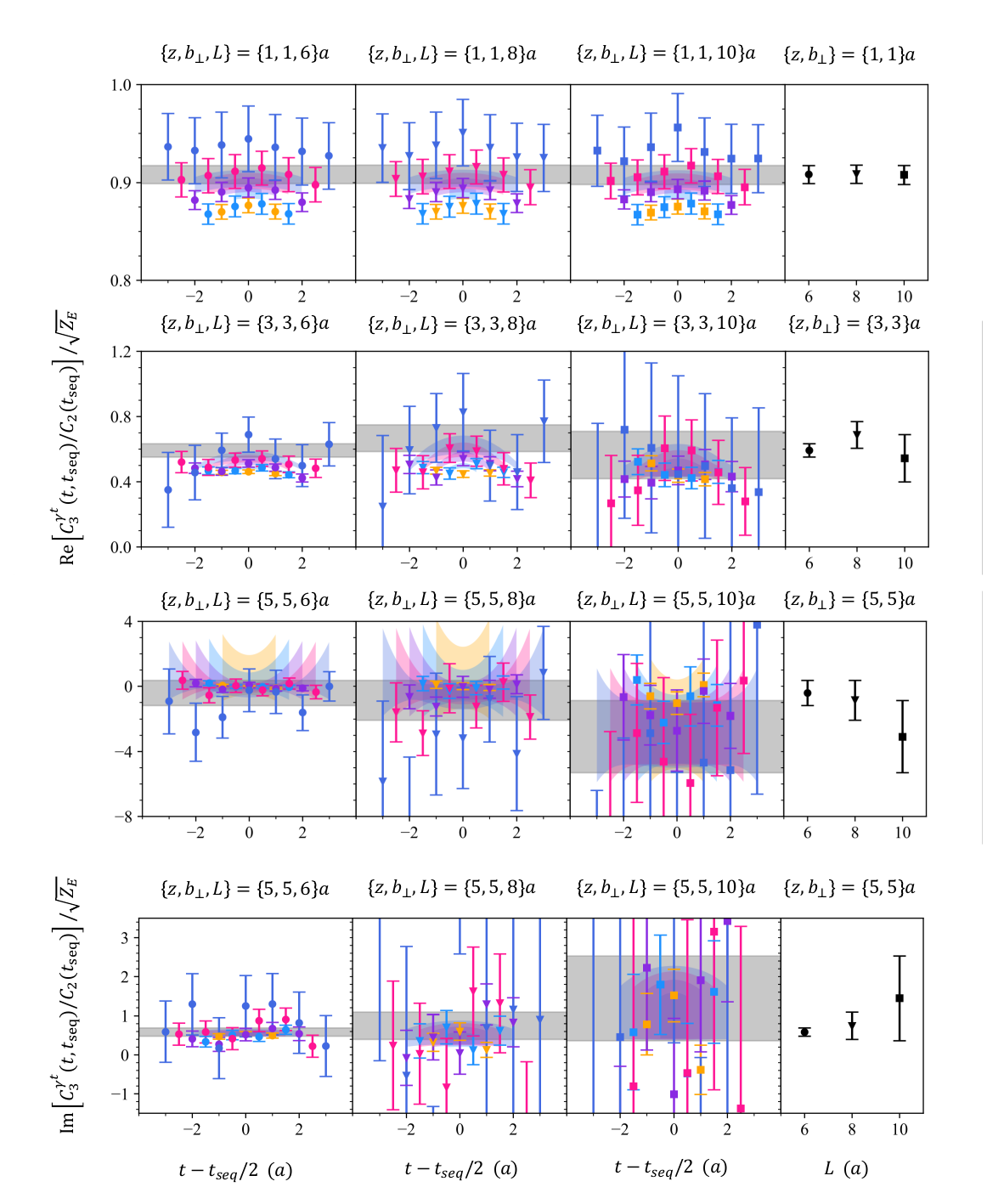}
    \caption[$L$-dependence of subtracted quasi TMDPDF matrix elements]{The ratios defined in Eq.~(\ref{eq:ratio_of_2pt_and_3pt}) divided by the Wilson loop (three left panels) and the $L$ dependence of the ground state results after fitting (right panels) at $\{z,b_{\perp}\}=\{1,1\}, \{3,3\}$ and $\{5,5\}a$, with $P^z=1.72$ GeV, $m_{\pi}=220$~MeV and $\Gamma=\gamma^t$. Since the real parts for $\{z,b_{\perp}\}=\{5,5\}a$ are close to zero, we also give the imaginary parts in this example. }
    \label{fig:Fig_Ldependence}
\end{figure}

Ref.~\cite{Zhang:2022xuw} explored the $L$-dependence of renormalized TMD matrix elements and reached the conclusion that setting $L=6a$ for MILC12 is sufficient, aligning well with the ensemble we have employed. To verify this, we further examine the $L$-dependence of the subtracted quasi TMD matrix elements, illustrated in Fig.~\ref{fig:Fig_Ldependence}.

In Fig.~\ref{fig:Fig_Ldependence}, the three panels on the left show the real parts of the ratios defined in Eq.~(\ref{eq:ratio_of_2pt_and_3pt}) divided by the square root of the Wilson loop with various combinations of $z$ and $b_{\perp}$ and $L=\{6, 8, 10\}a$. The right panel gives the fit results of subtracted quasi TMD matrix elements with different $L$. From the comparison, one can see that the fitted results for $L=\{6, 8, 10\}a$ are consistent with each other. 
 
There is a phenomenological explanation for the earlier saturation of the Wilson line; however, it is essential to note that a definitive proof is currently unavailable. When calculating the TMDPDF of a proton, which reflects the quark correlation within the proton, a reasonable estimate for the correlation length could be the size of a proton, approximately 1 fm. Beyond this distance, quarks and gluons may escape the proton, potentially diminishing their impact. This suggests a typical saturation length of around 1 fm, with $L=6a$ (or more precisely, $L=8a$) being in proximity to this length scale.

\begin{figure}[!th]
    \centering
    \includegraphics[width=0.9\linewidth]{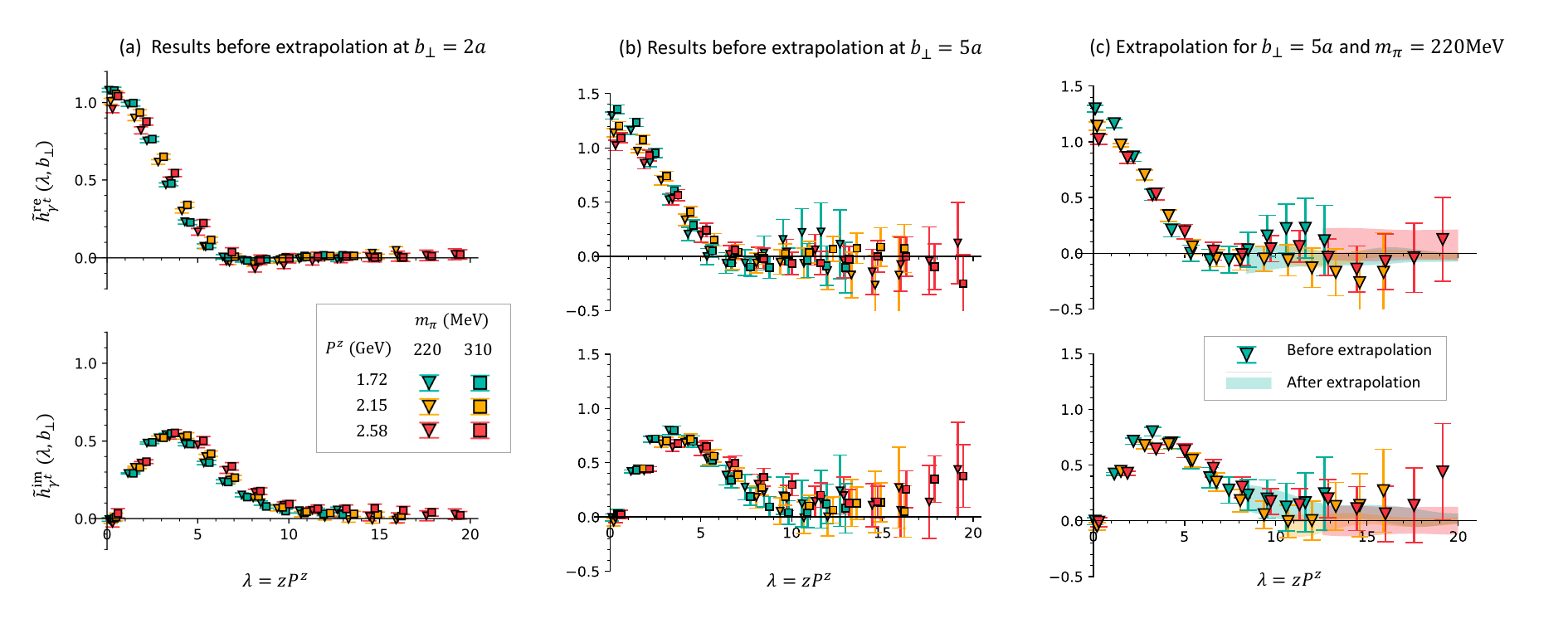}
    \caption[Renormalized quasi TMDPDFs of nucleon]{(a, b): Renormalized quasi TMDPDFs in coordinate space as function of $\lambda=zP^z$ with various $m_{\pi}$ and $P^z$ at $b_{\perp}=2a$ and $5a$; (c): Comparison of original (data points) and extrapolated results (colored bands) at $b_{\perp}=5a$. }
    \label{fig:Fig_quasiMEbeforeExtra}
\end{figure}

Combining the bare quasi TMDPDFs matrix elements with the corresponding Wilson loop and renormalization factor, we obtain  numerical results for renormalized matrix elements at different $\lambda=zP^z$. The two panels on the left in Fig.~\ref{fig:Fig_quasiMEbeforeExtra} exhibit the $\lambda$ dependence of the renormalized matrix elements with $b_{\perp}=2a$ and $5a$ with various $P^z$ and $m_{\pi}$. Similar to the previous calculations, we apply an asymptotic extrapolation in $\lambda$ to address the large $\lambda$ behavior of the quasi TMDPDFs, using the form given in Eq.~(\ref{eq:early_asymptotic_extrapolation}). It should be noticed that this asymptotic form was initially proposed for the one-dimensional parton distribution functions~\cite{Ji:2020brr}. For TMDPDFs,  whether this form holds and whether the involved parameters depend on the transverse separation remain to be found out.

\begin{figure}[!th]
    \centering
    \includegraphics[width=0.6\linewidth]{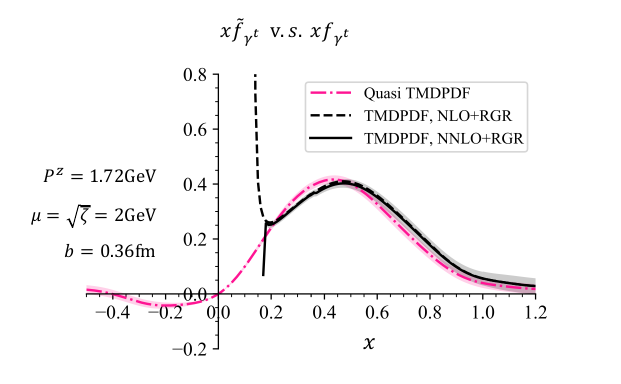}
    \caption[Quasi TMDPDF v.s. TMDPDF of nucleon]{Quasi TMDPDF (dashed-red line) with nucleon boosted momentum $P^z=1.72$~GeV and matched TMDPDF from NLO kernel (dashed-black line, no error) and NNLO kernel (solid-black line) with RG running to scale $\mu=\sqrt{\zeta}=2$~GeV at $b_{\perp}=3a$. Only statistical errors are included in the bands. Deviations between NLO and NNLO results at $x<0.2$ indicate that the perturbative matching fails in the small-$x$ region.}
    \label{fig:Fig_compare_quasi_lc}
\end{figure}

After Fourier transforming the renormalized quasi TMDPDFs in coordinate space to momentum space, one can obtain the quasi TMDPDF $\tilde{f}_{\Gamma}(x,b_{\perp}, \zeta_z, \mu)$, shown as the  dashdotted (magenta) line in Fig.~\ref{fig:Fig_compare_quasi_lc}. Combing the updated results of Collins-Soper kernel ~\cite{LPC:2022ibr} and intrinsic soft function \cite{LatticePartonLPC:2023pdv} calculated on the same ensembles, we obtain the TMDPDFs through the matching formula in Eq.~(\ref{eq:tmdpdf_matching}). Fig.~\ref{fig:Fig_compare_quasi_lc} shows an example of matched TMDPDF from the NLO~\cite{Ji:2019ewn, Ji:2020ect} and NNLO~\cite{delRio:2023pse, Ji:2023pba} kernel with RG running to scale $\mu=\sqrt{\zeta}=2$~GeV. One can see that the results agree within uncertainties, except for the endpoint region. 
 
It should be noted that the TMDPDFs can be obtained after employing the factorization formula, while residual dependence on the lattice inputs (such as $P^z$ and $m_\pi$) may still reside in the obtained results. To  diminish  the dependence, we extrapolate the above results to the physical $m_{\pi}$ value ($135$~MeV) and  infinite momentum using the following ansatz:  
\begin{eqnarray}
    f(m_{\pi}, P^z) = f_{\mathrm{phy}}\bigg[1+d_0 \left( m_{\pi}^2-m_{\pi, \mathrm{phy}}^2 \right)   +d_0' \ln(m_{\pi}^2/m^2_{\pi, \mathrm{phy}})   + \frac{d_1}{\left(P^z\right)^2}+ \frac{d_1'}{P^z} \bigg] ~, \label{eq:mpiPzextrapolation-general}
\end{eqnarray}
where the $d_0^{(')}$ term characterizes the pion mass dependence, and $d_1^{(')}$ accounts for the momentum-dependent discretization error.
An interesting analysis has recently derived the chiral logarithms for quasi-PDFs~\cite{Liu:2020krc}. In this approach, one first performs an operator product expansion and constructs the moments of quasi PDFs with derivative operators. Using chiral perturbation theory, one can find hadron-level operators which have the same symmetry. Calculating the one-loop perturbative contributions and summing these contributions for moment operators will lead to the chiral logarithms in quasi-PDFs. However, this analysis has not been generalized to TMDPDFs at present. In addition, since TMDPDFs are three-dimensional distributions, expanding the nonlocal operators will require the derivatives not only in the $z$ direction, but also in the transverse direction. This will likely introduce additional complexities. 

\begin{figure}[!th]
    \centering
    \includegraphics[width=0.6\linewidth]{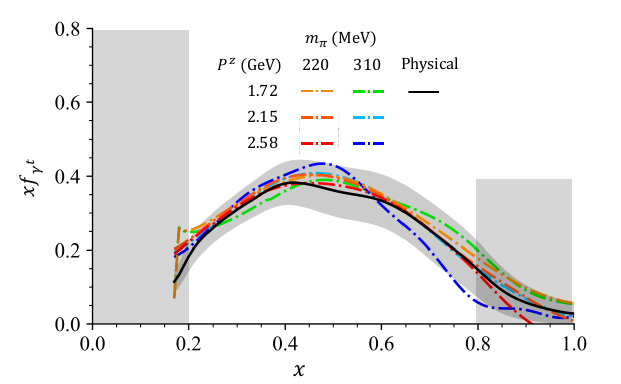}
    \caption[Physical limit of nucleon TMDPDF]{TMDPDFs obtained from different $m_{\pi}$ and $P^z$ (dashed-color lines) at $b_{\perp}=3a$, and their physical results after extrapolation. The shaded grey band indicates the endpoint regions where LaMET predictions are not
    reliable. Only statistical errors of physical results are exhibited. }
    \label{fig:Fig_TMDPDF_mPz_extra_both}
\end{figure}

\begin{figure}[!th]
    \centering
    \includegraphics[width=0.4\linewidth]{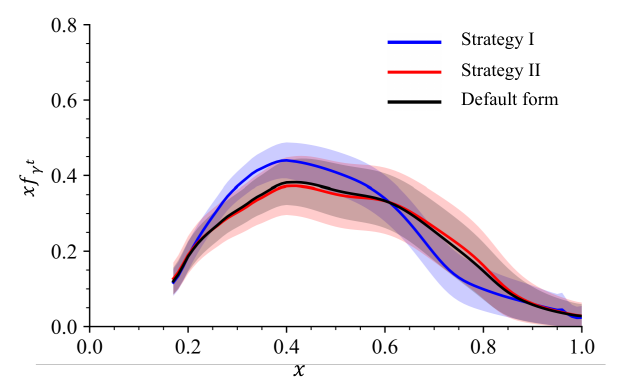}
    \includegraphics[width=0.4\linewidth]{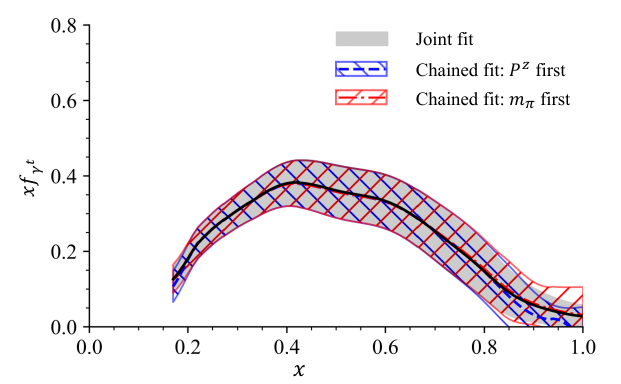}
    \caption[Nucleon TMDPDF with different physical limit]{Left panel: Default extrapolation form in Eq.(\ref{eq:mpiPzextrapolation}), considering the systematic uncertainties from chiral extrapolation in Eq.(\ref{eq:varyingmpi}) (strategy I) and from the large-momentum extrapolation form in Eq.(\ref{eq:varyingPz}) (strategy II).
    Right panel: Comparison of the results from joint fit and chained fits with different order.}
    \label{fig:Fig_chi2_chiral_Pz}
\end{figure}

Taking $d_0'=d_1'=0$ as our default case, we obtain the physical TMDPDF from a joint fit of $f(m_{\pi}, P^z)$ shown in Fig.~\ref{fig:Fig_TMDPDF_mPz_extra_both}. In order to explore the systematic bias in the extrapolation ansatz, we also estimate the $m_\pi$ dependence by using the $d_0'$ term (with $d_0=0$) and examine the $1/P^z$ contributions by adding the $d_1'$ term, shown as the following two different strategies:
\begin{itemize}
\item Default form:
\begin{equation}
    	f(m_{\pi}, P^z) = f_{\mathrm{phy}}\left[1+d_0 \left( m_{\pi}^2-m_{\pi, \mathrm{phy}}^2 \right) + \frac{d_1}{\left(P^z\right)^2} \right], \label{eq:mpiPzextrapolation}
\end{equation}
\item Strategy I: 
\begin{equation}
    	f(m_{\pi}, P^z) = f_{\mathrm{phy}}\left[1+d_0' \log\frac{m_{\pi}^2}{m_{\pi, \mathrm{phy}}^2}+ \frac{d_1}{\left(P^z\right)^2} \right], \label{eq:varyingmpi}
\end{equation}
\item  Strategy II: 
\begin{equation}
    	f(m_{\pi}, P^z) = f_{\mathrm{phy}}\left[1+d_0 \left( m_{\pi}^2-m_{\pi, \mathrm{phy}}^2 \right) + \frac{d_1}{\left(P^z\right)^2} +\frac{d_1'}{P^z}  \right]. \label{eq:varyingPz}
\end{equation}
\end{itemize}
Specifically, we introduce the chiral log term as well as the linear $1/P^z$ term into the fit formula, and consider the deviation between them and the default form as the systematic uncertainty.  A comparison of them is shown in the left panel in Fig.~\ref{fig:Fig_chi2_chiral_Pz} from which one can see that the results agree with each other. 

Furthermore, we have examined two different sequences of extrapolation: First, extrapolation to the physical $m_{\pi}$ followed by the $P^z\to \infty$  extrapolation, and second, the $P^z\to \infty$  extrapolation followed by the extrapolation to the physical $m_{\pi}$. A comparison of the results is presented in the right panel of Fig.~\ref{fig:Fig_chi2_chiral_Pz}. It is evident that the results are in good agreement with each other.

\begin{figure}[!th]
    \centering
    \includegraphics[width=0.6\linewidth]{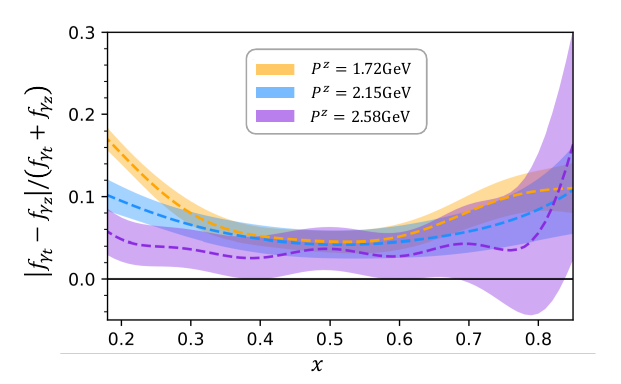}
    \caption[Difference of nucleon TMDPDF from different Lorentz structures]{Ratio of $|{f}_{\gamma^t}-{f}_{\gamma^z}|/({f}_{\gamma^t}+{f}_{\gamma^z})$ with $m_{\pi}=220$~MeV and different $P^z$ at $b_{\perp}=3a$
    }
    \label{fig:Fig_TwistCorr}
\end{figure}

In quasi TMDPDFs, both Lorentz structures $\Gamma=\gamma^t$ and $\gamma^z$ can  project onto $\gamma^+$ in the large momentum limit. 
An interesting observation  in Ref.~\cite{Constantinou:2019vyb} is that the $\gamma^t$ case has fewer operator mixing effects. 
Therefore, we use the $\Gamma=\gamma^t$ in Eq.~(\ref{eq:momspacequasi}) to extract the TMDPDFs.
For the $\gamma^z$ case, in addition to the sizable operator mixing effects, it is anticipated that its deviations from the $\gamma^t$ results may come from power corrections from the operator product expansion of quasi correlators. These corrections are of order $\mathcal{O}(1/(P^z)^2)$ with opposite signs at large $P^z$.   In order to analyze the impact caused by different structures, we plot the ratio $|{f}_{\gamma^t}-{f}_{\gamma^z}|/({f}_{\gamma^t}+{f}_{\gamma^z})$ with different $P^z$ in Fig.~\ref{fig:Fig_TwistCorr}. One can see that with increasing nucleon momentum, the ratio becomes smaller, except for the endpoint region with large uncertainties.  The differences will be incorporated as a systematic uncertainty.  

\begin{figure}[!th]
    \centering
    \includegraphics[width=0.6\linewidth]{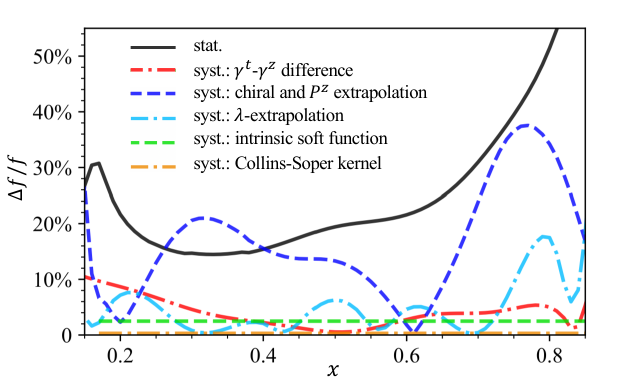}
    \caption[Systematic uncertainties of nucleon TMDPDF]{Ratios of various uncertainties and central value of final TMDPDF at  $b_{\perp}=3a$, include the statistical one and systematic one from: (1) $\lambda$ extrapolation, (2) chiral and $P^z$ extrapolation, (3) $\gamma^t-\gamma^z$ differences, (4) intrinsic soft function~\cite{LatticePartonLPC:2023pdv} and (5) Collins-Soper kernel~\cite{LPC:2022ibr}.  
    }
    \label{fig:Fig_AllUncertainties}
\end{figure}

To control the precision of the final results, in this work, we consider the following sources of systematic uncertainties:
\begin{itemize}
	\item different fitting ranges in the $\lambda$-extrapolation;
	\item different strategies used for the chiral and $P^z$ extrapolation;
	\item the difference between $\gamma^t$ and $\gamma^z$ results.
\end{itemize}
In addition, we also consider the error propagating from the intrinsic soft function~\cite{LatticePartonLPC:2023pdv} and Collins-Soper kernel~\cite{LPC:2022ibr}, which are calculated on the same configurations. The comparison of the statistic as well as all systematic uncertainties are shown in Fig.~\ref{fig:Fig_AllUncertainties}.

\begin{figure}[!th]
    \centering
    \includegraphics[width=0.9\textwidth]{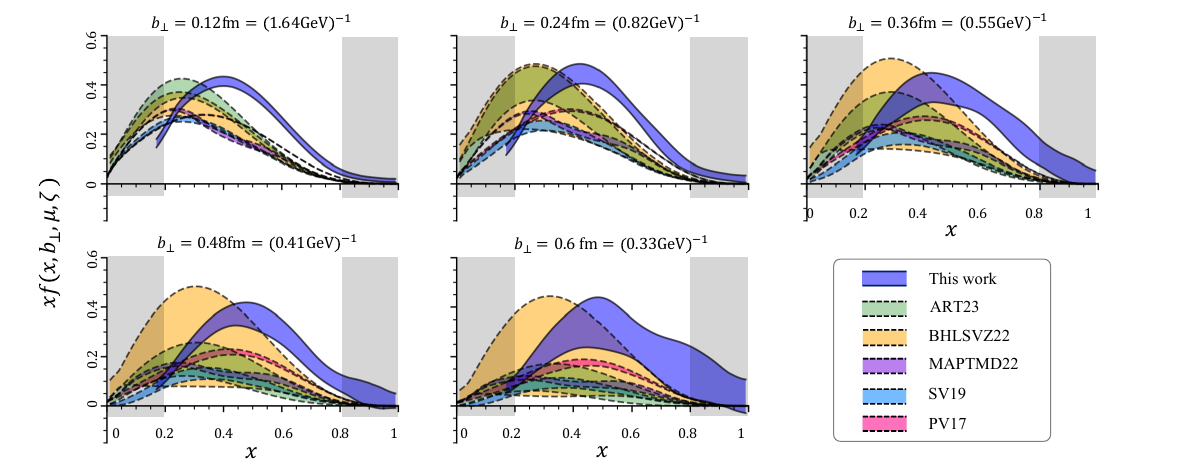}
    \caption[Nucleon TMDPDF in $x$ space]{Our final results for unpolarized nucleon's isovector TMDPDFs $xf(x, b_{\perp}, \mu, \zeta)$ at renormalization and rapidity scales at $\mu=\sqrt{\zeta}=2~$GeV, extrapolated to physical pion mass 135~MeV and infinite momentum limit $P^z\to\infty$, compared with ART23 \cite{Moos:2023yfa}, BHLSVZ22 \cite{Bury:2022czx}, MAPTMD22 \cite{Bacchetta:2022awv}, SV19 \cite{Scimemi:2019cmh}and PV17 \cite{Bacchetta:2017gcc} global fits. 
    The colored bands denote our results with both statistical and systematic uncertainties, the shaded grey regions imply the endpoint regions where LaMET predictions are not reliable. 
    }
    \label{fig:Fig_finalTMDPDFs}
\end{figure}

\begin{figure}[!th]
    \centering
    \includegraphics[width=0.9\textwidth]{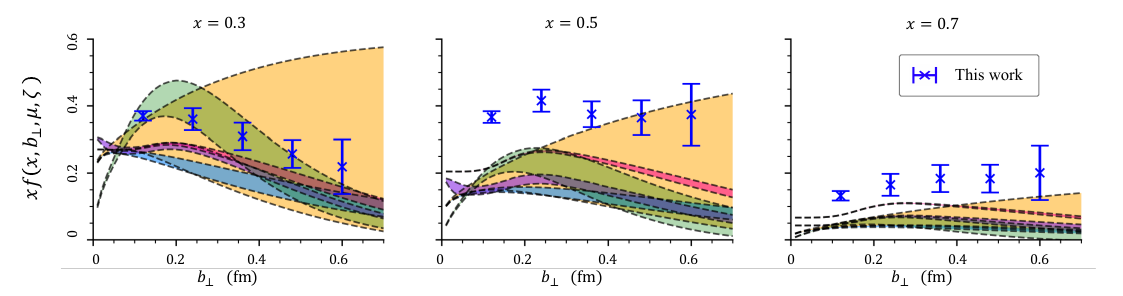}
    \caption[Nucleon TMDPDF in $b_{\perp}$ space]{The TMDPDFs $xf(x, b_{\perp}, \mu=2~\mathrm{GeV}, \zeta=4~\mathrm{GeV}^2)$ with longitudinal momentum fraction $x=\{0.3, 0.5, 0.7\}$, together with the phenomenological results. The labels of the latter are the same as Fig.~\ref{fig:Fig_finalTMDPDFs}. }
    \label{fig:Fig_finalTMDPDFs_with_b}
\end{figure}

Combing all the known uncertainties indicated in Fig.~\ref{fig:Fig_AllUncertainties}, we obtain numerical results of unpolarized nucleon's isovector TMDPDFs from our lattice simulation. Fig.~\ref{fig:Fig_finalTMDPDFs} shows $xf(x, b_{\perp}, \mu, \zeta)$ at renormalization and rapidity scales $\mu=\sqrt{\zeta}=2$~GeV as a function of $x$, together with the phenomenological results from global analyses~\cite{ Moos:2023yfa, Bury:2022czx, Bacchetta:2022awv, Scimemi:2019cmh, Bacchetta:2017gcc}.
From the comparison one can see that, our results are in qualitative agreement with  phenomenological results and share a similar behavior in $b_{\perp}$ space: the central values slowly decrease and uncertainties are gradually increasing  with the increase of $b_{\perp}$. 

In Fig.~\ref{fig:Fig_finalTMDPDFs}, a bump in the $x$ distribution can be observed, representing the highest probability for the longitudinal momentum distributions. It is worth noting that the peak positions do not align precisely between the lattice results and the respective phenomenological results. Further investigations are required to elucidate this phenomenon.

The two shaded bands at the endpoint regions ($x<0.2$ and $x>0.8$) in each subplot of Fig.~\ref{fig:Fig_finalTMDPDFs} indicate that LaMET predictions are not reliable there, {which is estimated from the power correction terms $\Lambda_{\mathrm{QCD}}^2/(xP^z)^2$ and threshold logarithms $\ln((1-x)P^z)$~\cite{Braun:2018brg,Gao:2021hxl} .}  
Besides, since only one lattice spacing is used,  discretization uncertainties are not  properly handled at this stage. Especially the $b_{\perp}\sim a$ case might suffer sizable discretization effects, which can be improved by a more detailed analysis in the future. 

Fig.~\ref{fig:Fig_finalTMDPDFs_with_b} shows the results for $xf(x, b_{\perp}, \mu, \zeta)$ with $x=0.3, 0.5, 0.7$ as a function of spatial separation $b_{\perp}$. The spatial distributions reflect correlations between the partons at transverse interval $b_{\perp}$ inside a highly boosted nucleon, and will reveal aspects of nucleon structure. It could be conjectured that the distributions vanish when $b_{\perp}$  is larger than the  nucleon radius, however,  neither the present day lattice results nor the phenomenological results are precise enough to draw such a conclusion.

\subsection{Conclusion}

In summary, we have performed the first calculation of TMDPDFs within a nucleon using the LaMET framework. State-of-the-art techniques in renormalization and extrapolation on the lattice have been employed, including the consideration of the perturbative kernel up to NNLO with RG resummation. We investigate the dependence on pion mass and hadron momentum, incorporating both statistical and systematic errors to provide a robust characterization of the inner structure of nucleons through the parton distributions.

While the current findings are promising, further improvements are required. Firstly, a comprehensive analysis of multiple ensembles with varying lattice spacings, pion masses, and volumes is essential. This approach will not only reassess all uncertainties addressed in this study but also systematically explore additional significant factors. Secondly, our results exhibit notable theoretical uncertainties in the endpoint regions, which could stem from power corrections or threshold logarithms. Thirdly, we expect a decay of the TMDPDF with increasing $b_{\perp}$ which is not (yet) visible in our data. Going to larger $b_{\perp}$ is thus imperative. Moreover, the presence of chiral logarithms in the analysis of quasi PDFs (as demonstrated in Ref.~\cite{Liu:2020krc}) highlights the potential for similar logarithms in our case, warranting dedicated theoretical investigation.

%% file: Chapter5.tex
%Chapter 1

\renewcommand{\thechapter}{5}

\vspace*{-1.35in}

\chapter{Applications of Machine Learning in Lattice Gauge Theory}
\label{ch:ml}

\vspace*{-.5in}

In recent years, machine learning has become a powerful paradigm in scientific research, enabling data driven discovery and significantly accelerating traditional computational and experimental workflows. In biology and chemistry, a prominent example is AlphaFold~\cite{jumper2021highly,abramson2024accurate}, which demonstrated that deep learning models trained on large structural databases can predict protein three-dimensional structures with near experimental accuracy, transforming structural biology and molecular design. In materials science, the GNoME (Graph Networks for Materials Exploration) models represent a significant leap; by scaling deep learning to predict the stability of crystalline compounds, it has discovered over 2.2 million new structures below the current convex hull, many of which escaped previous human chemical intuition~\cite{merchant2023scaling}. In astronomy, the shift toward foundation models is exemplified by AION-1 (Astronomical Omni-modal Network)~\cite{parker2025aion}, a multi-billion parameter model developed by PolymathicAI. Unlike previous CNN-based classifiers, AION-1 integrates heterogeneous data—including imaging, spectroscopy, and metadata—across multiple sky surveys, enabling a unified approach to tasks such as spectral super-resolution and galaxy property estimation without the need for task-specific retraining. Collectively, these advancements illustrate how AI is evolving from a specialized tool into a foundational methodological layer across different scientific domains.

Building on these representative successes, it is important to clarify the types of scientific problems for which machine learning is particularly well suited. In practice, such problems are typically outcome oriented rather than mechanism centered, and their solutions can be independently verified without introducing additional, uncontrolled systematic uncertainties from the learning model itself. In these settings, neural networks function primarily as high dimensional function approximators that map inputs to outputs with high accuracy, while the ultimate scientific validity of the result is established through external benchmarks such as experiments, higher fidelity simulations, or well-defined evaluation metrics. This separation between prediction and validation mitigates concerns about limited interpretability, since correctness does not rely on understanding the internal representations of the model.

AlphaFold~\cite{jumper2021highly,abramson2024accurate} provides a paradigmatic example. Protein structure prediction is intrinsically predictive: given an amino acid sequence, the goal is to determine its three-dimensional structure. The predicted structures can then be directly compared against independent experimental measurements, including X-ray crystallography and cryo electron microscopy, using clear quantitative metrics. Because validation is external and objective, the use of a deep neural network does not introduce uncontrolled systematic bias into the final scientific claim. The transformative impact of AlphaFold on structural biology and molecular design illustrates why such verifiable, result driven problems are especially well-matched to modern AI methods, and why this achievement was recognized with the 2024 Nobel Prize in Chemistry.

Building on this perspective, meaningful applications of machine learning in lattice gauge theory should satisfy the same criteria outlined above: they should be outcome oriented and admit independent validation through established consistency checks. Such caution is essential because lattice gauge theory provides a nonperturbative, first-principles formulation of quantum field theory on a discrete spacetime. Its predictive power rests on systematically controllable uncertainties, including finite lattice spacing effects, finite volume effects, and statistical errors, which can be quantified and removed through well-defined extrapolation procedures. Thus, a good application of machine learning should preserve this rigorous framework and avoid introducing additional, uncontrolled sources of systematic uncertainty.

Broadly speaking, current applications of machine learning in lattice gauge theory can be grouped into three categories. The first concerns gauge configuration generation, where learned proposals or generative models are used to accelerate sampling and mitigate long autocorrelation times. The second involves data analysis, where machine learning supports tasks such as operator optimization and parameter inference from correlators or ensembles. The third emerging direction encompasses AI agents and foundation models, which go beyond simple automation by learning to accelerate end to end workflows and propose improved strategies. In this chapter, we will discuss these three classes of applications in turn.

\section{Applications in Gauge Configuration Generation}

The generation of gauge configurations constitutes the computational foundation of lattice gauge theory simulations. In this framework, physical observables are defined through Euclidean path integrals over gauge fields, which are discretized on a spacetime lattice to provide a nonperturbative regularization. In practice, these high dimensional integrals are evaluated by importance sampling of gauge field configurations according to the Boltzmann weight of the lattice action. The central task of lattice gauge theory simulations is therefore to construct a Markov chain whose stationary distribution is proportional to $e^{-S(U)}$, so that expectation values of observables can be approximated by ensemble averages over sampled configurations.

Formally, the partition function is given by
\begin{equation}
    Z = \int \mathcal{D}U \; e^{-S(U)} ~,
\end{equation}
where U denotes the set of link variables and S(U) the lattice action. Expectation values are estimated as Monte Carlo averages over configurations generated via Markov Chain Monte Carlo (MCMC) methods. A variety of MCMC algorithms are employed in practice, all designed to ensure exactness through detailed balance and ergodicity. As the continuum limit is approached, however, these algorithms exhibit critical slowing down: autocorrelation times grow rapidly with decreasing lattice spacing and increasing physical volume. This effect is particularly severe for topological observables, whose dynamics may effectively freeze over long Monte Carlo histories. The resulting computational cost represents one of the central algorithmic bottlenecks in high-precision lattice QCD.

\subsection{Overview}

Motivated by the algorithmic limitations of conventional MCMC methods, a variety of machine learning approaches have recently been proposed to accelerate gauge configuration generation. These developments can be broadly categorized into two directions. The first aims at direct generative modeling of the target distribution, while the second focuses on integrating learned components into existing MCMC methods.

Among generative approaches, normalizing flow~\cite{Albergo:2019eim,Albergo:2021vyo} have received particular attention. A normalizing flow constructs an invertible and differentiable map between a simple base distribution and the target gauge-field distribution, with a tractable Jacobian determinant. In principle, such flows can serve as exact bridges between distributions, enabling either reweighting or independence Metropolis sampling. If a sufficiently accurate flow were available, one could perform approximate direct sampling of gauge configurations, thereby eliminating long autocorrelation times and potentially resolving topological freezing. 

In practice, however, applying normalizing flow to four-dimensional non-abelian gauge theories presents substantial challenges. The dimensionality of configuration space scales with the lattice volume, gauge equivariance must be preserved, and locality constraints must be respected. Recent progress has improved model expressivity through gauge-equivariant architectures and learned loop constructions, and has demonstrated advantages in correlated ensemble generation and replica exchange frameworks. Nevertheless, fully replacing HMC at state-of-the-art lattice sizes remains an open challenge.

These considerations motivate hybrid approaches in which learned transformations are integrated into established MCMC frameworks. By retaining the exactness guarantees of traditional algorithms while reshaping the geometry of configuration space through machine-learned maps, one may achieve algorithmic acceleration without abandoning theoretical control.

\subsection{Neural Field Transformations for Hybrid Monte Carlo}

A particularly promising direction along this line is the construction of invertible neural field transformations to modify the geometry of configuration space prior to HMC. Conceptually, this idea is closely related to Lüscher's notion of trivializing maps~\cite{Luscher:2009eq}, where a change of variables in the path integral is introduced to simplify the effective action and smooth the energy landscape explored by HMC. Although exact trivializing maps are not known for interacting non-abelian gauge theories, approximate transformations can still reshape the distribution in a way that reduces force fluctuations and mitigates critical slowing down.

Based on this insight, our recent work~\cite{He:2025inl} presents a systematic of neural network transformed HMC (NTHMC)~\cite{Jin:2022bgq,Jin:2023shj}, focusing on how neural architectural designs such as residual connections, attention mechanisms, channel-dependent activation, and batch size affect performance and scaling.
Testing on a two-dimensional U(1) gauge theory, we evaluate both small-lattice performance and the transferability to larger volumes and smaller lattice spacing.
Our results suggest that wider receptive fields and channel-dependent activation improve efficiency while maintaining favorable scaling, thereby informing extensions to four-dimensional SU(3) gauge fields and large-scale lattice QCD.

The NTHMC framework uses an invertible, gauge-covariant field transformation $\mathcal{F}: \tilde{U}\mapsto U$ parameterized by a neural network to perform a change of variables in the integration over the Boltzmann weight $\exp(-S(U))$,
leading to an effective action,
\begin{equation}
    S_{\mathrm{FT}}(\tilde{U}) = S\big(\mathcal{F}(\tilde{U})\big) - \log\big| \det \mathcal{F}{*}(\tilde{U}) \big| ~,
\end{equation}
where $\mathcal{F}{*}(\tilde{U})$ is the Jacobian matrix of the transformation.
HMC is then performed in the space of auxiliary variables $\tilde{U}$
with the Boltzmann weight $\exp(-S_{\mathrm{FT}}(\tilde{U}))$.

Intuitively, the field transformation acts like a learned preconditioner. We construct the field transformation from gauge-covariant, locally equivariant convolutional neural networks (CNNs). For each link variable $\tilde{U}_{x,\mu}$ (a gauge field element going from site $x$ to $x+\hat{\mu}$), the update takes the form
\begin{equation}
    \mathcal{F}: \tilde{U}_{x,\mu} \mapsto U_{x,\mu} = e^{\Pi_{x,\mu}} \tilde{U}_{x,\mu}, \quad
    \Pi_{x,\mu} = \sum_{l} \epsilon_{x,\mu,l}\, W_{x,\mu,l} ~.
\end{equation}
The coefficients are given by
$\epsilon_{x,\mu,l} = \phi \!\left[ N_l(X,Y,\dots) \right]$,
with $N_l$ a CNN mapping local gauge-invariant inputs $(X,Y,\dots)$ to the coefficients for the loop functions $W_{x,\mu,l}$, and $\phi$ a bounded activation function (e.g.\ $\tanh$ or $\arctan$) that enforces the invertibility of the transformation. 
$W_{x,\mu,l}$ is the Lie algebra projection of a specific (labeled by $l$) Wilson loop (path ordered products of gauge fields along a closed loop)
that starts and ends at site $x$ and first moves in the $\mu$ direction
(the first on the right of the product).
For the $U(1)$ gauge theory, the projection just takes the imaginary part.

\begin{figure}[ht!]
    \centering
    \includegraphics[width=0.6\linewidth]{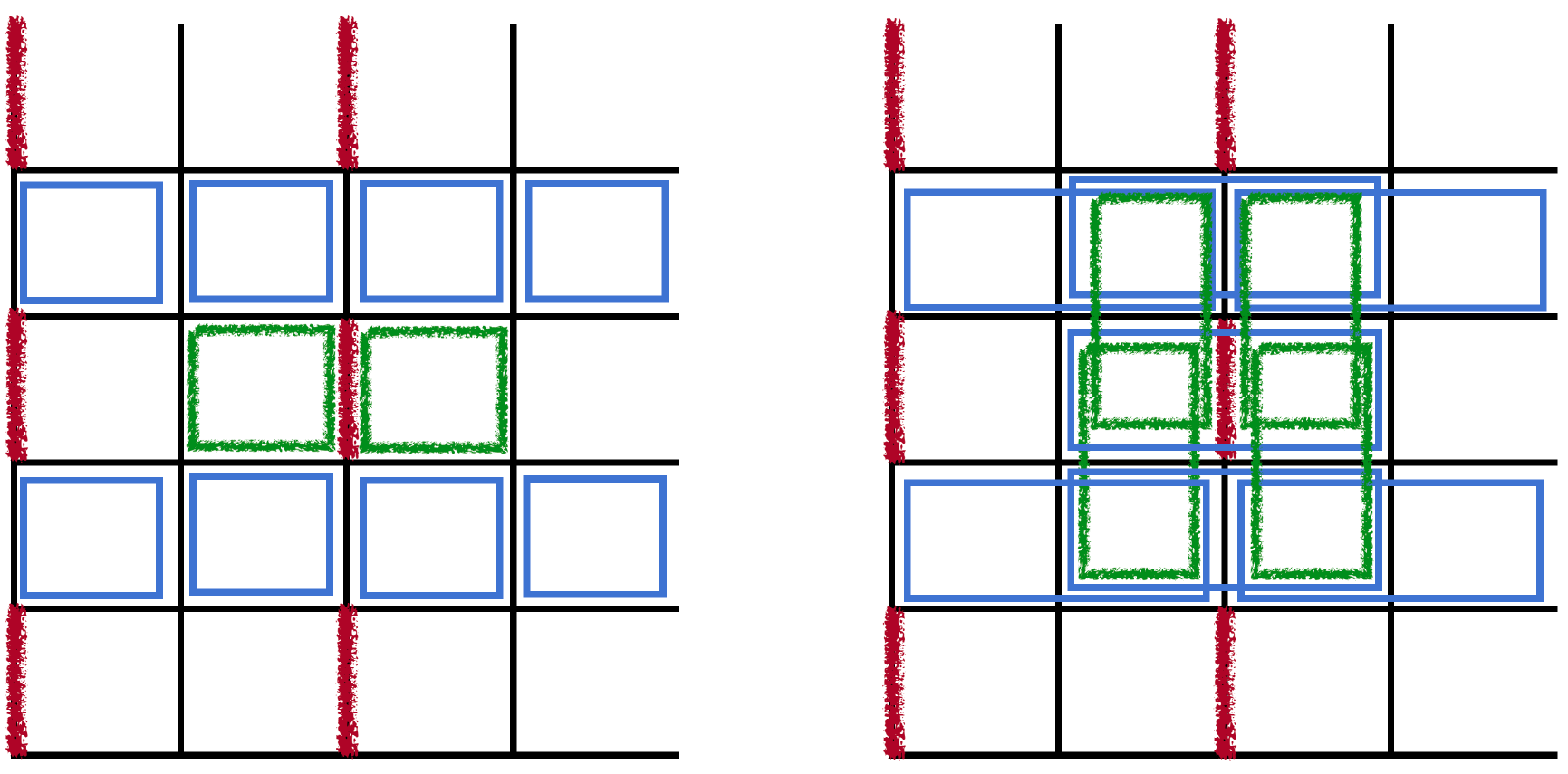}
    \caption[Cropped section of a lattice configuration illustrating the field transformation]{Cropped section of a lattice configuration illustrating the field transformation. Red lines mark one of the eight link subsets updated sequentially, green lines indicate Wilson loops $W_{x,\mu,l}$,
    and blue lines show the gauge-invariant features $(X,Y,\dots)$ used as CNN inputs.}
    \label{fig:subsets}
\end{figure}

As suggested in Ref.~\cite{Luscher:2009eq}, the lattice links are partitioned into eight disjoint subsets and updated sequentially. This ensures that the Jacobian determinant of each update step is reduced to the product of diagonal elements. Moreover, by choosing CNN inputs $(X,Y,\dots)$ to be independent of the links in the active subset, these diagonal elements can be written in a closed form, making the Jacobian determinant explicitly tractable. In this work, we take the $l$ index to run over $1\times1$ plaquettes and $1\times2$ rectangles, as illustrated in Fig.~\ref{fig:subsets}, where one link subset (red) is updated, while the corresponding loops (green) and local gauge-invariant CNN inputs (blue) are highlighted.

For U(1) gauge theory,
we use $U=e^{i\theta}$, and the Wilson plaquette action~\cite{Wilson:1974sk} as $S(U)$.
The transformation reduces to exponentials of sine terms of plaquette and rectangle loops,
\begin{equation}
    \mathcal{F}:~\tilde{U}_{x,\mu} \;\mapsto\; U_{x,\mu} = e^{\sum_{l} \epsilon_{x,\mu,l} \, W_{x,\mu,l}} \, \tilde{U}_{x,\mu} = e^{ \sum_{p} i \epsilon_{x,\mu,p} \sin(\theta_p) + \sum_{r} i \epsilon_{x,\mu,r} \sin(\theta_r)} \tilde{U}_{x,\mu} ~,
\end{equation}
with $p$ and $r$ denoting individual plaquette and rectangle loops, respectively, and $\theta_p$, $\theta_r$ the phase angle
of the corresponding Wilson loop.
The Jacobian has the matrix elements
\begin{equation}
    \mathcal{F}_* (\tilde{U})_{x,\mu,x',\mu'} = 
    \delta_{x,x'} \delta_{\mu,\mu'} \left[
    1 + \sum_p \epsilon_{x,\mu,p} \cos(\theta_p) + \sum_r \epsilon_{x,\mu,r} \cos(\theta_r) \right]~.
\end{equation}
In order to have an invertible transformation, the Jacobian determinant may not change sign.
The invertibility condition is then
\begin{equation}
    \sum_p |\epsilon_{x,\mu,p}| + \sum_r |\epsilon_{x,\mu,r}| < 1 ~.
    \label{eq:invertibility}
\end{equation}

The field transformation is trained on configurations generated from regular HMC by minimizing the transformed action force:
\begin{equation}
    \mathcal{L}(\tilde{U}) = 
    \sum_{p \in \{2,4,6,8\}} 
    \frac{c_{p}}{V^{1/p}} \;
    \left\| \frac{\partial S_{\mathrm{FT}}(\tilde{U})}{\partial \tilde{U}} 
    \right\|_{p} ,
    \label{eq:loss}
\end{equation}
where $V$ is the lattice volume and $c_{p}$ are hyperparameters that weight different force norms, thereby balancing the emphasis between average and peak forces. In our experiments, we set $ c_{2} = c_{4} = c_{6} = c_{8} = 1$. Different from Ref.~\cite{Jin:2022bgq}, which matched the transformed force to that at a larger lattice spacing, we directly minimize the transformed force itself, effectively training the transformation to smooth the action landscape, reduce the stiff directions and improve the sampling efficiency.

Inside the HMC update we employ the 2nd order minimum norm symplectic integrator of Omelyan et al.~\cite{omelyan2003symplectic} with $10$ MD steps per trajectory, tuning the step size to achieve $\sim 80\%$ acceptance. All HMC simulations were verified by checking the plaquette values of the generated configurations, which were found to agree with the theoretical expectations within statistical uncertainties. We benchmark five architectures: a baseline (\texttt{Base}) model similar to the design in Ref.~\cite{Jin:2022bgq}, and four variants (\texttt{Tanh}, \texttt{Resn}, \texttt{Attn}, \texttt{Comb}) introducing different modifications.

\textbf{Base}
The baseline model is a two-layer CNN that processes six input channels (two plaquette, four rectangle). A first $3\times3$ convolution maps them to 12 hidden channels with GELU activation, followed by a second $3\times3$ convolution producing coefficients, which are scaled by $\arctan$ to $(-1/6,1/6)$, satisfying the invertibility condition in Eq.~\ref{eq:invertibility}.
This design yields an effective $5\times5$ receptive field with $\sim$2,000 parameters.

\textbf{Tanh}
Same as \texttt{Base}, but replaces the $\arctan$ scaling with a channel-dependent $\tanh$ activation, ensuring $|\epsilon_{x,\mu,p}| < 0.4$ and $|\epsilon_{x,\mu,r}| < 0.05$, with $\sim 2,000$ parameters.

\textbf{Resn}
Adds two residual blocks (each has two $3\times3$ convolutions with skip connections) before the output, and replaces the second $3\times3$ convolution of \texttt{Base} with a $1\times1$ projection. This enlarges the receptive field to $11\times11$ and increases the parameters to $\sim 6,000$.

\textbf{Attn}
Augments \texttt{Base} with a channel-attention module between the two convolutions, reweighting channels without changing the $5\times5$ receptive field. The parameter count is $\sim 2,100$.

\textbf{Comb}
Combines residual blocks, channel attention, and channel-dependent $\tanh$ activation ($|\epsilon_{x,\mu,p}|<0.4$, $|\epsilon_{x,\mu,r}|<0.05$). The resulting architecture has an $11\times11$ receptive field and $\sim 6,100$ parameters.

All models are trained with the $p$-norm force loss (Eq.~(\ref{eq:loss})) on $2^{12}$ independent configurations at $\beta=3$, $V=32^2$ using an 80/20 train–test split.
The input gauge coupling, $\beta$, enters in the action $S(U)$, and in this theory, the lattice spacing is proportional to $\beta^{-1/2}$.
The default batch size is $64$, except that \texttt{Base} and \texttt{Comb} are also trained with $32$ (denoted \texttt{Base32}, \texttt{Comb32}) to study batch-size effects. Each model is trained for 32 epochs in PyTorch~\cite{paszke2017automatic}, with one single run taking about 12–20 minutes on a single NVIDIA A100 GPU for batch size 64, and roughly twice as long for batch size 32.

To assess the sampling efficiency of regular HMC and NTHMC, we consider two complementary metrics. The first is $\gamma(\delta)$, which measures the relative independence of the configuration~\cite{Jin:2022bgq}:
\begin{equation}
    \gamma(\delta) = \frac{1}{1 - \Gamma_t(\delta)}~, \quad \Gamma_t(\delta) \equiv \frac{\left\langle Q_\tau Q_{\tau+\delta}\right\rangle}{\left\langle Q^2\right\rangle}  =1-\frac{\langle\left(Q_\tau-Q_{\tau+\delta}\right)^2\rangle}{2 V \chi_t^{\infty}(\beta)} ~.
    \label{eq:gamma}
\end{equation}
Here $\Gamma_t(\delta)$ is the autocorrelation of the topological charge $Q$ at separation $\delta$ in HMC steps, and $\chi_t^\infty(\beta)$ is the topological susceptibility~\cite{Jin:2022bgq}. The larger $\gamma(\delta)$ indicates stronger autocorrelations and therefore less efficient sampling. 

The second metric is the average topological change $\Delta Q$, the mean absolute shift in topological charge per HMC step, which directly probes the tunneling between sectors; larger values of $\Delta Q$ indicate more frequent topological transitions and thus more efficient sampling. All training and evaluation experiments are repeated with 8 random seeds. Statistical uncertainties are estimated and consistently propagated through derived quantities using Jackknife resampling~\cite{miller1974jackknife} across the seed variability.

\begin{figure}[ht!]
    \centering
    \includegraphics[width=0.4\linewidth]{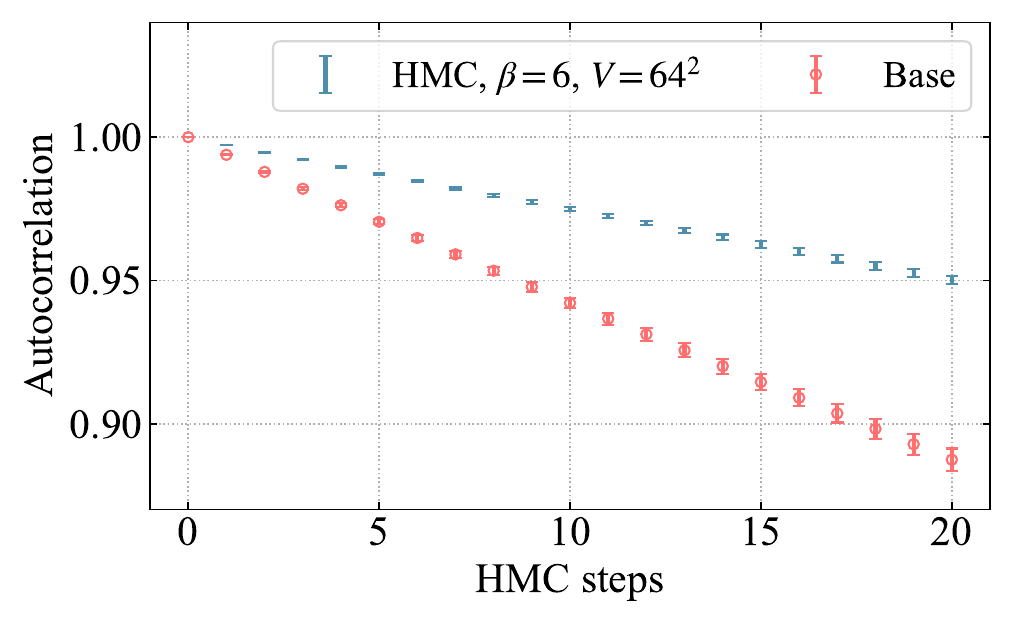}
    \includegraphics[width=0.4\linewidth]{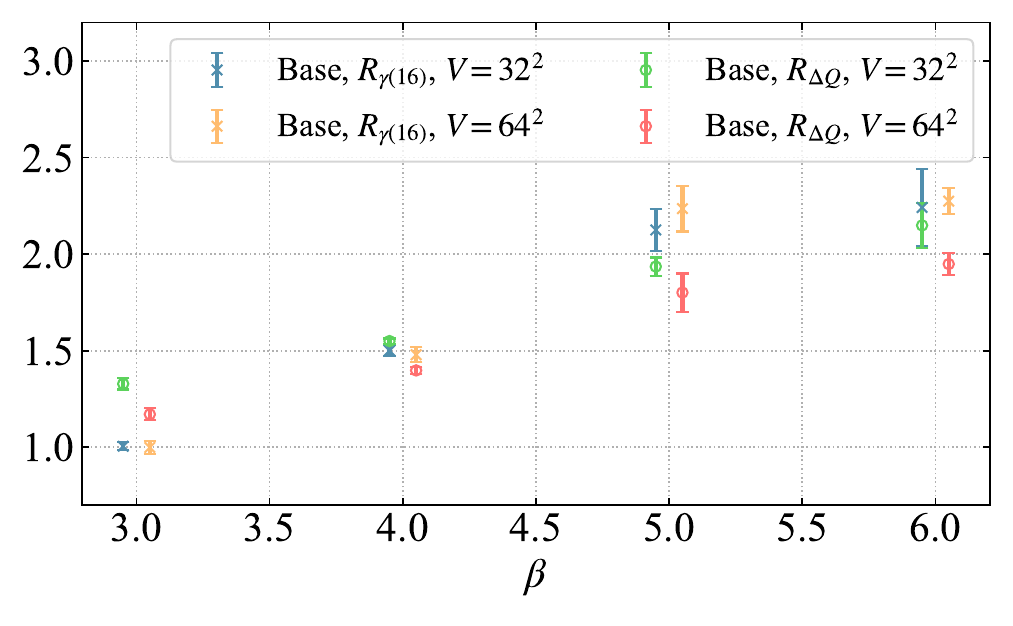}
    \caption[Comparison of regular HMC and NTHMC with the \texttt{Base} model]{Comparison of regular HMC and NTHMC with the \texttt{Base} model. 
    Left: autocorrelation decay at $\beta=6.0$, $V=64^2$. 
    Right: scaling of two efficiency ratios across $\beta$ and volume, with 
    $R_{\gamma(16)} \equiv \gamma(16)_{\mathrm{HMC}} / \gamma(16)_{\mathrm{Base}}$ 
    and $R_{\Delta Q} \equiv \Delta Q_{\mathrm{Base}} / \Delta Q_{\mathrm{HMC}}$.}
    \label{fig:scaling}
\end{figure}

As shown in the left panel of Fig.~\ref{fig:scaling}, NTHMC with the \texttt{Base} model achieves a markedly faster decay of autocorrelations than regular HMC. The right panel reports the efficiency ratios $R_{\gamma(16)} = \gamma(16)_{\mathrm{HMC}} / \gamma(16)_{\mathrm{Base}}$ and $R_{\Delta Q} = \Delta Q_{\mathrm{Base}} / \Delta Q_{\mathrm{HMC}}$. Despite being trained only at $\beta=3$, $V=32^2$, NTHMC yields even larger relative gains at higher $\beta$ (smaller lattice spacing) and larger volumes, demonstrating favorable scaling and generalization. At $\beta=3$, where regular HMC is already efficient, both ratios remain close to unity.

\begin{figure}[ht!]
    \centering
    \includegraphics[width=0.4\linewidth]{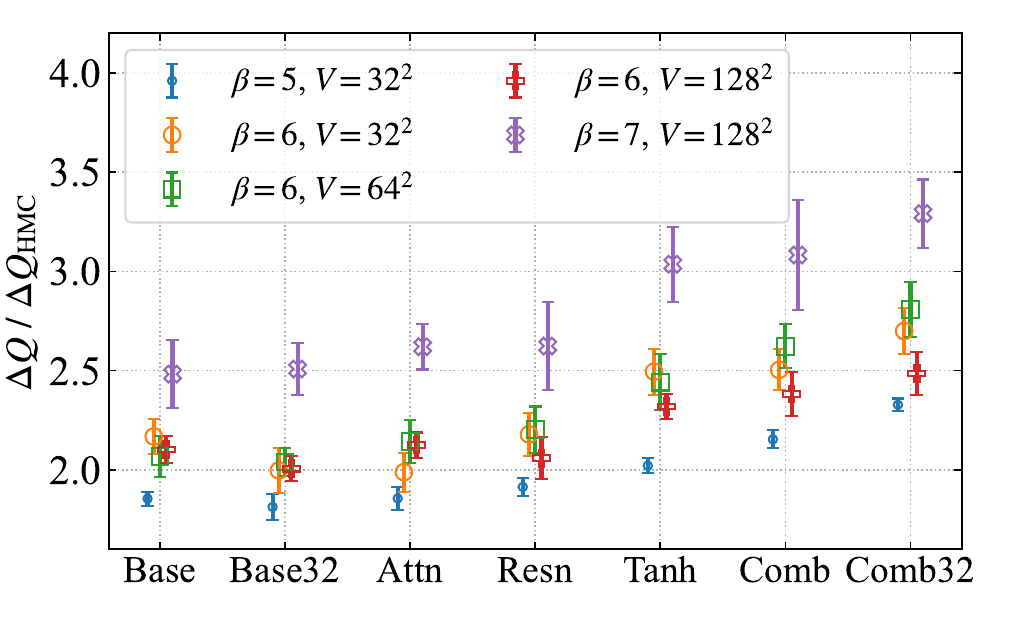}
    \includegraphics[width=0.4\linewidth]{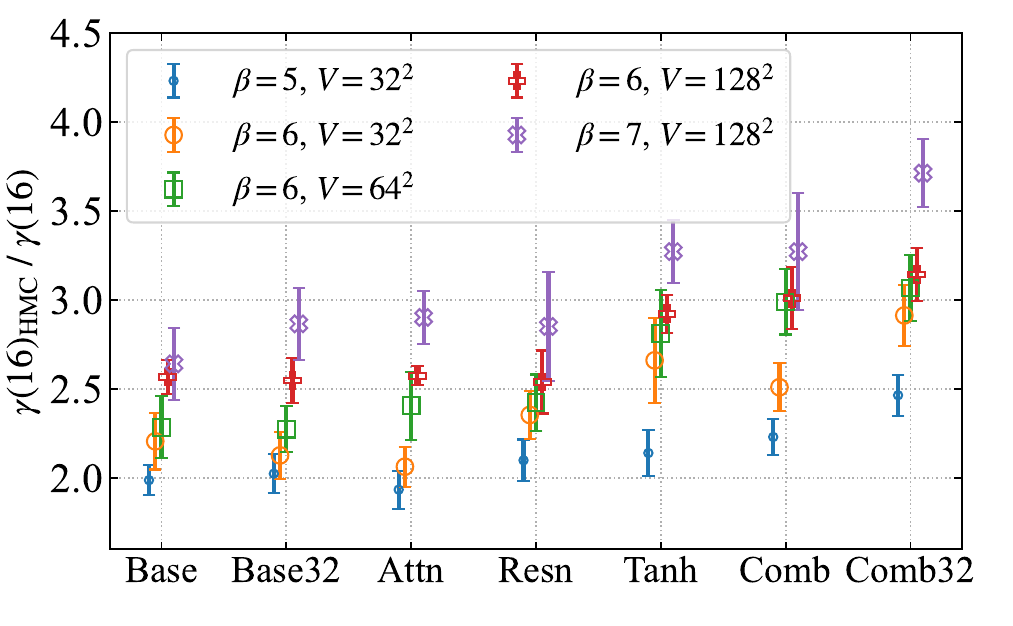}
    \caption[Comparison of NTHMC with various network architectures]{Comparison of NTHMC with various network architectures across different lattice couplings and volumes. The plots show $\Delta Q$ (left panel) and $\gamma(16)$ (right panel), both normalized to standard HMC results.}
    \label{fig:models}
\end{figure}

Fig.~\ref{fig:models} summarizes the performance of different CNN architectures within NTHMC across several lattice couplings and volumes. The comparison includes results at $\beta = 5$, 6, and 7 with lattice sizes $V = 32^2$, $64^2$, and $128^2$, allowing us to examine both coupling and volume dependence relative to standard HMC. Residual blocks and channel-dependent activation lead to improvements over the \texttt{Base} model, and the combined design (\texttt{Comb}) yields the strongest effect, achieving nearly a $40\%$ reduction in $\gamma(16)$ with a corresponding increase in the average topological change $\Delta Q$. Moreover, as the coupling $\beta$ increases, all NTHMC models exhibit a higher sampling efficiency compared to standard HMC, and the efficiency does not show statistically significant changes with increasing lattice volume, indicating that our method can be effectively extended to larger volumes and finer lattices.

Interestingly, the \texttt{Tanh} model achieves competitive gains despite having the same parameter count as \texttt{Base}, which we attribute to the faster saturation of $\tanh$ compared to $\arctan$ and, more importantly, to the larger magnitude of plaquette coefficients $\epsilon_{x,\mu,p}$ enabled by channel-dependent activation. Likewise, residual connections expand the receptive field, which might be expected to hinder transferability, yet we find that an appropriate enlargement does not compromise scaling or generalization and may even be beneficial. A possible explanation is that, as the continuum limit is approached in LGT, infrared (long distance) structures are preserved while ultraviolet (short distance) details become increasingly refined, so the transferable features are precisely the long-range correlations that benefit from a relatively wider receptive field.

Finally, batch size interacts mildly with architecture. Although the baseline shows little sensitivity, the richer \texttt{Comb} model benefits from smaller batches, consistent with gradient noise acting as an implicit regularizer that helps larger capacity models generalize more effectively.

Our preliminary results demonstrate that neural field transformations can substantially reduce autocorrelations, yet several open directions remain. Future work includes exploring integrator choices such as varying the number of MD steps at fixed trajectory length, refining training objectives beyond minimizing the force norms, and systematically analyzing training regimes (e.g. number of epoches, learning rate, training set size and data from different $\beta$ and volumes). Finally, extending from U(1) to SU(3) will be essential to connect this approach with large-scale lattice QCD applications.

\section{Applications in Data Analysis}

Beyond configuration generation, machine learning has found increasingly important applications in the data analysis stage of lattice gauge theory. In contrast to sampling algorithms, which operate at the level of the path integral measure, data-driven methods act on correlation functions or ensembles that have already been generated. These approaches can be broadly divided into two categories: operator optimization and parameter inference from correlators or gauge ensembles.

\subsection{Operator Optimization}

The extraction of physical observables from lattice simulations relies critically on the choice of interpolating operators. An ideal operator has strong overlap with the ground state and suppressed overlap with excited states, thereby reducing statistical noise and systematic contamination. Traditionally, operator construction is guided by symmetry considerations and physical intuition, supplemented by variational techniques such as the GEVP.

Machine learning provides a complementary strategy: treat operator construction as an optimization problem. Neural networks can be trained to maximize ground-state overlap or minimize excited-state contamination, effectively learning optimal linear combinations of operator bases. Related ideas have been explored in the context of neural-network-assisted spectroscopy and variational optimization, where data-driven methods improve the efficiency of spectral extraction~\cite{Shi:2022yqw}. More recently, automated operator discovery and structure learning techniques have been proposed to systematically search large operator spaces~\cite{Bellscheidt:2026rjh}. These developments suggest that operator optimization is a natural interface between traditional lattice methodologies and modern representation learning.

\subsection{Parameter Inference from Correlators and Ensembles}

A second major application concerns parameter inference. In lattice QCD, physical observables such as hadron masses, form factors, and matrix elements are obtained from Euclidean correlation functions through multi-exponential fits, Bayesian model averaging, or spectral reconstruction techniques. This inverse problem is inherently ill-conditioned: correlators decay exponentially in Euclidean time, and statistical noise grows at large separations.

Machine learning approaches have been proposed to assist in this inference task. Neural networks can be trained to extract spectral parameters directly from correlator data, effectively learning non-linear regression mappings between correlation functions and physical parameters. In addition, Bayesian neural networks and Gaussian-process-based methods provide flexible frameworks for incorporating prior information and quantifying uncertainties. Ensemble-level inference tasks, such as identifying phase transitions or estimating renormalization constants, have similarly benefited from ML-based regression and classification techniques.

Importantly, these data-analysis applications do not alter the underlying field theory or sampling algorithm. Instead, they enhance the extraction of physics from already-generated data, often reducing human intervention and improving robustness against model selection ambiguities. As lattice calculations continue to increase in precision and complexity, systematic and automated data analysis tools will become increasingly valuable.

\section{AI Agent and Foundation Model}

Recent advances in artificial intelligence suggest a broader paradigm beyond task-specific machine learning models: the development of AI agents and foundation models capable of operating across complex scientific workflows. An AI agent may be understood as a system that interacts with an environment, makes sequential decisions under specified objectives, and adapts its strategy based on feedback. In computational science, such an agent would not merely perform a single regression or classification task, but rather orchestrate multi-step pipelines, select among competing algorithms, tune hyperparameters, and iteratively refine intermediate results. A foundation model, by contrast, refers to a large-scale model trained on broad and diverse data within a domain, capable of learning general-purpose representations that transfer across tasks. Rather than solving one narrowly defined problem, a foundation model encodes structural regularities of the domain and serves as a reusable backbone for downstream applications.

Lattice gauge theory provides a particularly well-structured environment in which to explore these ideas. Euclidean correlation functions possess strict symmetry constraints, dimensional consistency, and well-defined spectral decompositions. The transformation from correlation functions to physical observables such as spectra, matrix elements, and form factors follows a systematic and theoretically controlled pipeline. Unlike many empirical domains, the correctness of intermediate and final results can be verified through known identities, scaling relations, and continuum-limit consistency. These characteristics make lattice QCD an ideal testbed for evaluating the reasoning capabilities, robustness, and limitations of AI-based scientific systems.

Initial exploratory work has begun to examine the role of large models and AI systems in structured physics problems~\cite{miao2025physmaster}. However, a comprehensive framework tailored to lattice gauge theory remains largely undeveloped. A natural first step toward such a framework is the construction of quantitative benchmarks. The lattice community possesses substantial collections of correlation function data, including ensembles that were generated but not fully analyzed or not publicly disseminated. For many of these datasets, the extracted ground-state spectra and matrix elements are already known from conventional analyses. This setting enables the creation of controlled benchmark problems in which an AI system is provided with raw correlators and evaluated on its ability to reproduce established results, quantify uncertainties, and choose appropriate fitting strategies. Because the spectral decomposition of two- and three-point functions is well understood, such benchmarks can be scaled in difficulty by varying noise levels, operator bases, or excited-state contamination.

Beyond benchmarking, one may envision the construction of an AI agent capable of automating the full analysis workflow from ensemble to physical observables. Such a workflow would encompass the design and optimization of interpolating operators, the choice of smearing schemes and lattice actions, the implementation of Wick contractions, the extraction of spectra using multi-state fits or generalized eigenvalue methods, the application of Feynman-Hellmann techniques for matrix elements, and the extrapolation to physical limits through continuum and chiral fits. Many of these steps involve model selection, numerical stability considerations, and trade-offs between statistical precision and systematic control. An effective agent would not only reproduce established procedures but also evaluate alternatives, compare competing methods such as GEVP-based approaches, model averaging, Lanczos-type analyses, or current-inspired constructions, and identify strategies that improve stability or efficiency. In this sense, the goal is not merely automation, but the possibility of creative methodological exploration guided by physical constraints.

Foundation models offer a complementary long-term perspective. By training on large collections of correlators and associated metadata across different lattice spacings, volumes, hadron channels, and actions, a foundation model could learn universal representations of Euclidean observables. Such representations might enable transfer across related tasks, facilitate anomaly detection in new datasets, or accelerate parameter inference by providing strong domain-aware priors. Crucially, the theoretical rigidity of lattice gauge theory provides natural validation criteria: symmetries, Ward identities, scaling behavior, and renormalization properties serve as internal consistency checks. This allows the assessment of foundation models not only in terms of predictive accuracy, but also in terms of physical consistency and generalization across regimes.

In summary, lattice gauge theory occupies a unique position at the intersection of high-performance computing, structured statistical inference, and first-principles theoretical control. The development of AI agents and foundation models in this domain is both technically feasible and scientifically meaningful. By establishing rigorous benchmarks and progressively integrating learned systems into established workflows, one may systematically evaluate the capabilities and limitations of artificial intelligence in a controlled, theory-driven environment.

%% file: Chapter6.tex
%Chapter 6

\renewcommand{\thechapter}{6}

\vspace*{-1.35in}

\chapter{Summary and Outlook}
\label{ch:summary}

\vspace*{-.5in}

This dissertation has been motivated by a question: how to determine partonic structures from QCD in a framework that remains first-principles while being quantitatively relevant for precision phenomenology. Across the previous chapters, the work has been organized around a consistent strategy that combines lattice gauge theory, LaMET framework, control of systematic uncertainties, and machine-learning-based methods. The objective is to present a coherent computational pathway from Euclidean matrix elements to light-cone partonic information in a model-independent manner with controlled precision. 

This thesis presents a set of successful lattice calculations of collinear and transverse partonic observables, demonstrating that the LaMET framework has evolved into a practically implementable and systematically improvable approach. For meson structure, the lattice extraction of pion DAs demonstrates that the LaMET framework can now be carried through with mature renormalization and extrapolation strategies, yielding distributions that exhibit physically interpretable deviations from asymptotic forms and broad consistency with complementary nonperturbative approaches in the reliable kinematic regions. For baryon structure, the studies of nucleon PDFs and TMDs have extended this logic to substantially more demanding systems, where dense excitation spectra, noisier correlators, and richer operator structures intensify the analysis burden. The results show that state-of-the-art treatments of renormalization, finite-separation behavior, and momentum dependence can produce stable nucleon-structure signals that align qualitatively with existing phenomenological results while preserving visibility of unresolved tensions, especially in TMD cases and in sectors where higher-order or power-suppressed contributions are still numerically important. Taken together, these outcomes indicate that lattice partonic physics has moved beyond proof-of-principle demonstrations toward a regime of structured precision building, where the key questions are less about feasibility and more about how systematically one can suppress the remaining uncertainty budget across observables.

In parallel with the results of lattice calculations, this dissertation emphasizes that computational methodology has become an increasingly important component of research in lattice gauge theory. The machine-learning chapter presents a pragmatic framework for identifying where learning-based tools can contribute in first-principles calculations: acceleration of gauge-configuration generation, enhancement of data-analysis workflows, and the longer-term development of AI systems that can assist in multi-stage scientific inference. In the configuration-generation domain, neural field transformations integrated with hybrid Monte Carlo illustrate a promising route to reducing autocorrelation and improving sampling efficiency while retaining physical distributions through the underlying Markov-chain framework. In data analysis, learning-based methods can support operator optimization and parameter inference when they are embedded in externally verifiable pipelines and evaluated against established physical consistency checks. At a broader horizon, the discussion of AI agents and foundation-model concepts has framed lattice gauge theory as a suitable environment for benchmark-driven development, because symmetry constraints, controlled limits, and cross-checkable intermediate quantities provide strict criteria for model assessment. The central conclusion is that machine learning is most useful in this domain when it supports controlled calculations rather than replacing theoretical structure or systematic uncertainty analysis.

At the same time, the thesis makes clear that several limitations remain decisive for the next stage of progress, and these limitations should be viewed as concrete research targets in future investigations. Excited-state contamination continues to be a dominant challenge in nucleon observables and can produce nontrivial biases if fit-model choices, source-sink separations, and analysis windows are not jointly constrained. Continuum, infinite-volume, and physical-mass control remains uneven across different quantities, as finite computational resources make precision studies near these limiting regimes technically demanding. In the LaMET context, achievable hadron boost factors, and residual higher-power corrections continue to limit the quantitative precision and systematic control of the extracted distributions, and these effects can couple to practical choices in non-perturbative renormalization, and perturbative matching. None of these issues invalidates the present results; instead, they define the concrete boundary between current precision and future precision, and they emphasize that progress depends on coordinated improvements in simulation design, analysis methodology, and theoretical control.

Looking ahead, further progress should proceed in clearly differentiated stages. In the near term, the priority is the systematic deployment of several recently developed innovations, including kinematically enhanced interpolating operators, the current-inspired GEVP method~\cite{Barca:2024hrl,Barca:2025det}, the Lanczos algorithm~\cite{Wagman:2024rid,Hackett:2024xnx,Hackett:2024nbe}, and asymptotic analyses at large distances~\cite{Chen:2025cxr,Ji:2026vir}. These developments aim to better control excited-state contamination and Fourier reconstruction systematics, while improving the SNR, with the overarching objective of enhancing the stability of the analysis and enabling more reliable estimates of systematic uncertainties. In the mid term, emphasis should shift toward structural optimization at both the lattice and perturbative interfaces, including optimized source profiles and improved momentum injection strategies to enhance overlap with highly boosted states, higher-order precision in perturbative matching to reduce truncation uncertainties, and more robust nonperturbative renormalization procedures—particularly within CG method—to achieve a reduction of systematic uncertainties, while multi-observable programs are developed within unified uncertainty frameworks to guarantee internal consistency across collinear and transverse sectors. In parallel, algorithmic acceleration and AI-based tools should continue to mature beyond controlled prototypes into full SU(3) production environments, with strict validation criteria tied to physical consistency, reproducibility, and systematic uncertainty propagation, so that computational efficiency gains translate into scientifically reliable outputs. In the long term, once lattice extractions and phenomenological analyses achieve mutually compatible precision, the objective is a genuinely joint program in which first-principles calculations and global data constraints are incorporated into coherent combined analyses, enabling iterative refinement on both sides and ultimately delivering quantitatively sharper and more internally consistent determinations of partonic structure.